\documentclass[12pt,oneside,english]{book}
\usepackage[T1]{fontenc}
\usepackage[utf8]{inputenc}
\usepackage[a4paper]{geometry}
\usepackage{color}
\usepackage{babel}
\usepackage{float}
\usepackage{amsmath}
\usepackage{amssymb}
\usepackage{mathdots}
\usepackage{graphicx}
\usepackage{setspace}
\usepackage{esint}

\usepackage[bookmarks=false]{hyperref}



\begin{document}

\title{{\large UNIVERSITY OF SCIENCE AND TECHNOLOGY OF CHINA }\\
{\large Heifei, CHINA}\textbf{\Large }\\
\textbf{\Large }\\
\textbf{\Large }\\
\textbf{\Large }\\
\textbf{\Large }\\
\textbf{\huge Relativistic fluid dynamics in heavy ion collisions}\textbf{\date{ }}}

\author{A dissertation submitted \\
for the degree of \\
Doctor of Philosophy in Physics\\
\\
by\\
\\
Shi Pu \\
\\
\\
\\
\\
\\
\\
\\
Supervisor: Qun Wang\\
\\
2011}

\maketitle
\renewcommand{\thepage}{\roman{page}}
\renewcommand{\bibname}{Reference}

\newpage

\begin{center}
\begin{minipage}[c][3.5\totalheight]{1\columnwidth}%
\begin{doublespace}
\begin{center}
\textcopyright   Copyright by
\par\end{center}

\begin{center}
Shi Pu
\par\end{center}

\begin{center}
2011
\par\end{center}

\begin{center}
{\large All Rights Reserved}
\par\end{center}\end{doublespace}
\end{minipage}
\par\end{center}

\newpage

\begin{center}
\begin{minipage}[c][15\totalheight]{1\columnwidth}%
\begin{doublespace}
\begin{center}
\textbf{\Large Dedicated to my dear family}
\par\end{center}\end{doublespace}
\end{minipage}
\par\end{center}

\newpage


\chapter*{Abstract}

Quantum Chromodynamics (QCD) is the fundamental theory for strong
interaction, one of the four forces in nature. Different from electromagnetic
interaction, due to its non-Abelian color symmetry, the QCD has the
property of asymptotic freedom at large momentum transfer, while remains
strongly coupled at low energies. This property leads to color confinement,
i.e. there are no free quarks and gluons carrying color degree of
freedom, and quarks and gluons are confined inside hadrons. In 1974-75,
Lee and Collins-Perry suggested that the deconfinement can be reached
through the ultra-relativistic heavy ion collisions, where the vacuum
can be excited to a new state of matter or a quark gluon plasma. In
the Relativistic Heavy Ion Collider (RHIC) at Brookhaven National
Lab (BNL), the gold nuclei are accelerated and collide head-to-head
at the center-of-mass energy of 200 GeV per nucleon. After the collisions,
the huge amount of energy is deposited in the central rapidity region
to excite vacuum and produce many quarks and gluons. In very short
time, these quarks and gluons collide to each other and the fireball
reaches the local thermal equilibrium, and a quark-gluon-plasma (QGP)
is then formed. After expansion and cooling, the temperature drops
substantially and the quarks recombine into hadrons which are finally
observed in detectors. The collective flows such as radial and elliptic
flows are observed at RHIC and can be well described by ideal fluid
dynamics. Further study of the elliptic flow using dissipative fluid
dynamics gives very low values of the ratio of shear viscosity to
entropy density, close to the lower bound $1/4\pi$ by AdS/CFT correspondence.
This is one of the most surprising observation at RHIC: the QGP is
strongly coupled or a nearly perfect liquid instead of a gas-like
weakly coupled system by convention. The relativistic fluid dynamics
provides a useful tool to bridge the gap between the initial state
of quarks and gluons and the experimental data. Pushed by heavy ion
collision experiments, the theory of relativistic fluid dynamics also
has a lot of new development in recent years. This thesis is about
the study of three important issues in the theory of relativistic
fluid dynamics: the stability of dissipative fluid dynamics, the AdS/CFT
application to shear viscosity in a Bjorken expanding fluid, and a
consistent description of kinetic equation with triangle anomaly.

Hydrodynamics is a long-wavelength effective theory for many particle
systems. The basic hydrodynamic equation consists of conservation
equations of energy-momentum and charge numbers. One can carry out
gradient expansion for energy-momentum tensor and charge currents,
or equivalently expansion in powers of the Knudsen number. The Knudsen
number is defined as the ratio of the macroscopic length scale (hydrodynamic
wave-length) to the microscopic one (mean free path). When the macroscopic
length scale is much larger than the microscopic one, the fluid dynamics
is a good effective theory. In expansion, the zeroth order corresponds
to the ideal fluid. The first order gives the Navier-Stokes equations
with dissipative terms like shear and bulk viscous terms. There are
a lot of candidates for the second order theory. We focus on the widely
used Israel-Stewart (IS) theory including the simplified and complete
version. The simplified version can be obtained by correspondence
to the macroscopic phenomena. To derive the complete version the kinetic
theory or the Boltzmann equation is necessary. In this thesis, we
will give the details as how to obtain the complete version. The connection
between the transport coefficients in the first and the second order
theories will be demonstrated.

In the first order theory, causality is violated since it takes no
time for a system in a non-equilibrium state to reach equilibrium,
i.e. the propagating speed of the signal is arbitrarily large. In
the second order theory, due to finite relaxation time introduced,
it takes finite time for the system to reach equilibrium. The propagating
speed of the signal is limited. Therefore the second order theory
is necessary for causality. However the causality cannot be guaranteed
for all parameters. The constraints for parameters are then given.
We also point out that the causality and the stability are inter-correlated.
Relativity requires that the signal propagate in light-cones. An acausal
propagating modes must be forbidden by instability or singularity.
The connection between causality and stability is also discussed.
For convenience we work in a general boost frame. It is found that
a causal system must be stable, but an acausal system in the boost
frame at high speed must be unstable.

The transport coefficients can be determined in kinetic theory. There
are two main techniques to compute the transport coefficients. One
is to employ the Boltzmann equation, the other is to use the Kubo
formula in field theory. We will firstly discuss about derivation
of the shear viscosity via variational method in the Boltzmann equation.
Secondly, after a short review of Kubo formula, we will compute the
shear viscosity via AdS/CFT duality.

Different from the work of Policastro, Son and Starinets (PSS), we
focus on strongly coupled QGP (sQGP) with the radial expansion and
Bjorken boost invariance. The information of the sQGP is encoded on
the boundary of AdS space via the holographic renormalization. Solving
the Einstein equation with the boundary condition given by the stress
tensor yields the metric of the AdS space. In this case, the metric
is Bjorken boost invariant and has the radial flow. The evolution
of the shear viscosity as a function of proper time can be obtained
via the Kubo formula for the retarded Green function given by the
AdS/CFT duality. It is found that the ratio of the shear viscosity
to entropy density is consistent with PSS.

As another application of AdS/CFT duality, we investigate the property
of baryons in sQGP in a Wilson-loops-like model. The quarks located
at the boundary of AdS space are connected to a probe D5 brane by
superstrings. By studying the configurations of baryons with different
spins, the screening length of baryons can be obtained as a function
of spin and temperature. We also study the relationship between the
angular momentum and energy for different kinds of baryons, which
shows the Regge-like behavior, i.e. the total angular momentum is
proportional to the energy squared.

As the last topic, we investigate the fluid dynamics with quantum
triangle anomalies. Generally the relativistic fluid dynamics does
not allow the vorticity due to parity conservation. Recently it is
pointed out that the vorticity has to be introduced to relativistic
fluid dynamics with anomalies to satisfy the second law of thermodynamics.
These new terms are also relevant to the Chiral Magnetic Effect (CME)
or Chiral Vorticity Effect (CVE). Such terms can be derived from the
kinetic approach. The coefficients of the vorticity in the case of
right-handed quarks (or left-handed anti-quarks) and quarks-antiquarks
of mixed chirality are evaluated.

\newpage

\chapter*{Acknowledgment}

I would like to express my gratitude to all those who helped me during
the writing of this thesis.

My deepest gratitude goes first and foremost to my supervisor, Prof.
Qun Wang, who provided me the opportunity to study at Frankfurt University
for one year. He has walked me through all the stages of the writing
of this thesis. Without his illuminating guidance, constructive suggestions
,as well as unwavering support in the past six years ,I cannot finish
this thesis.

I would like to express my heartfelt gratitude to Prof. Dirk. H. Rischlke,
who provided me a wonderful environment at ITP, Frankfurt University
and his enlightening suggestions, infinite patience, as well as consistent
encouragement, helped me a lot in the past three years.

My next thanks also go to Prof. Hongfan Chen, Prof. Junwei Chen, Prof.
Enke Wang, Prof. Peifei Zhuang , Prof. Mei Huang and other professors
and instructors from their inspiring ideas I have benefited immensely
and then finish my thesis smoothly.

I also owe my sincere gratitude to my cooperators, Dr. Tomoi Koide,
Dr. Yang Zhou, Dr. Jianhua Gao for their fruitful cooperation and
plentiful discussion. And thanks also go to Dr. Luan Chen for the
helpful advice on jet quenching and parton energy loss, Dr. He Song
for the useful suggestions on AdS/QCD.

In addition, I would acknowledge my friends at ITP, Dr. Xuguang Huang,
Dr. Zhe Xu , Dr. Xiaofang Chen, Tian Zhang, Dr. Nan Su, Martin Grahl,
Dr. Tomas Brauner, Dr. Harmen Warringa, Dr. Armen Sedrakian for lots
of interesting talk and the joyful time we had together.

And many thanks to all members of high energy group (HEPG) in USTC
and hydrodynamic group at ITP for helping me work out my problems
during the difficult course of the thesis. Such as Dr. Jian Deng,
Longgang Pang for their help on numerical simulations, Dr. Mingjie
Luo for the discussion on AdS/CFT, Jinyi Pang for his discussion on
the chiral perturbative theory, Haojie Xu for his help on the hard
thermal loops.

Last but not least, I would like to extend my special thanks to my
family who has been assisting, supporting and caring for me all of
my life. Their encouragement has sustained me through frustration
and depression. Without their support, the completion of this thesis
would be impossible.

\tableofcontents{}

\renewcommand{\thepage}{\arabic{page}}

\setcounter{page}{0}


\chapter{Introduction\label{chap:Introduction}}

\section{Quantum Chromodynamics and deconfinement phase transition}

\subsection{Asymptotic freedom}

\begin{figure}
\begin{centering}
\includegraphics[scale=0.5]{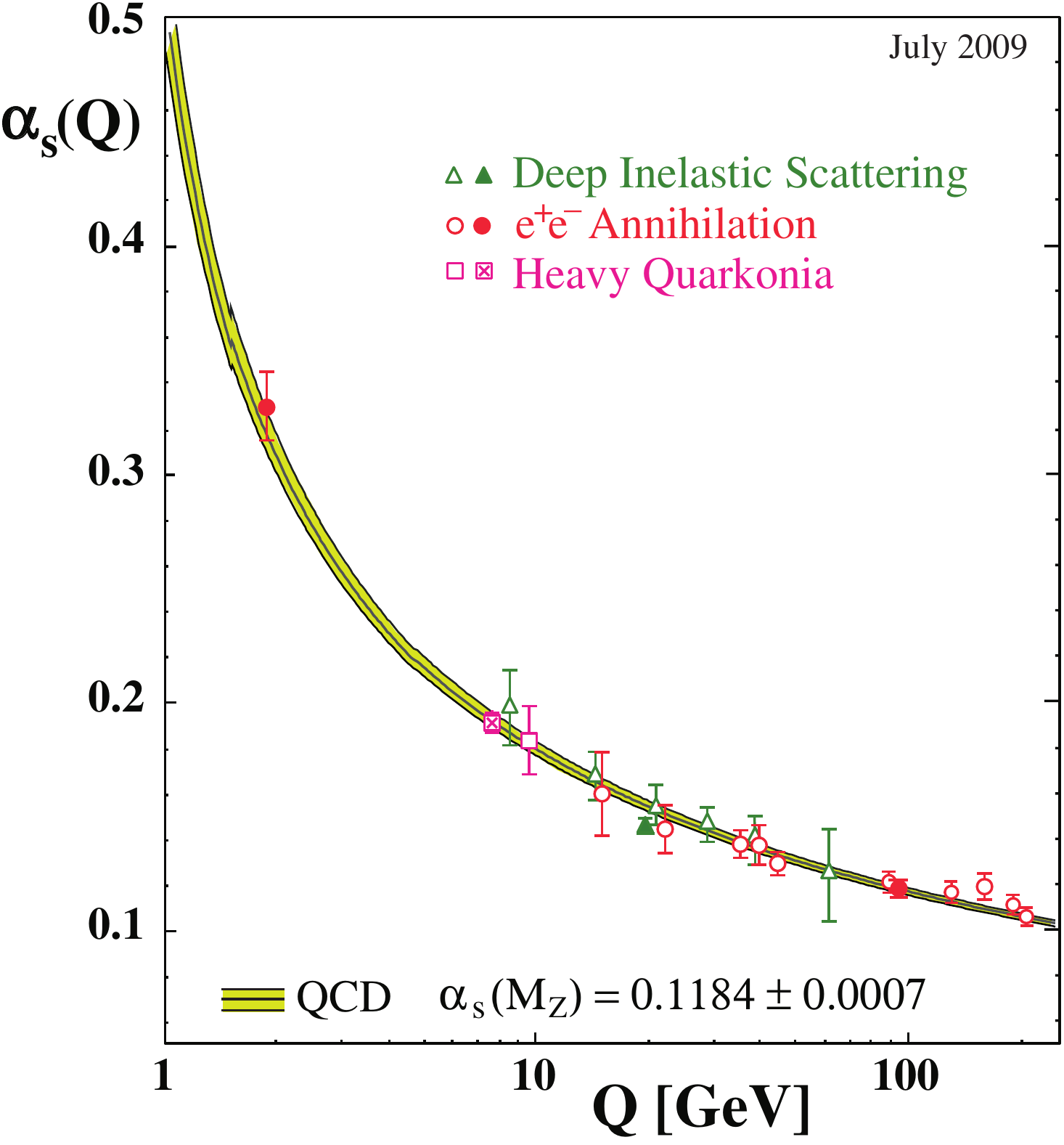}
\par\end{centering}

\caption{\label{fig:coupling_constant_QCD}Measurements of $\alpha_{s}$ as
a function of the momentum transfer $Q$ \cite{Nakamura:2010zzi}.}
\end{figure}

Quantum Chromodynamics (QCD) is a gauge theory for the strong interaction
which is one of the four fundamental interactions in the nature. In
contrast to photons in Quantum Electrodynamics (QED), the interaction
for gluons are complicated because of non-Abelian $SU(3)$ color symmetry.
The coupling constant $\alpha_{s}$ of renormalized QCD in one-loop
approximation is given by 
\begin{equation}
\alpha_{s}(Q^{2})=\frac{\alpha_{s}(M^{2})}{1+b_{0}\frac{\alpha_{s}(M^{2})}{2\pi}\ln(Q^{2}/M^{2})}\;,\label{eq:aymptot-freedom-01}
\end{equation}
where $Q$ is the momentum transfer scale, $M$ is the energy scale
and $b_{0}=\frac{33-2N_{f}}{3}$ is the first coefficient of $\beta$-function
given by the renormalization group equation with the quark flavor
$N_{f}$. Here $\alpha_{s}(M^{2})$ can be chosen as $\alpha_{s}(M_{Z}^{2})$
at Z-boson mass $M_{z}$ \cite{Bethke:2009jm,Nakamura:2010zzi} 
\begin{equation}
\alpha_{s}(M_{Z}^{2})=0.1184\pm0.0007\;.
\end{equation}
 The data for $\alpha_{s}(Q^{2})$ \cite{Nakamura:2010zzi} are shown
in Fig.\ref{fig:coupling_constant_QCD}. Equation (\ref{eq:aymptot-freedom-01})
indicates that the coupling constant decreases with the energy scale
$Q$. This property is the called asymptotic freedom \cite{Gross:1973id,Politzer:1973fx}.
The interaction for quarks and gluons will be strong in the low energy
scale which leads to confinement, i.e. quarks are limited inside hadrons
in vacuum or the ground state. 

\begin{figure}
\begin{centering}
\includegraphics[scale=0.6]{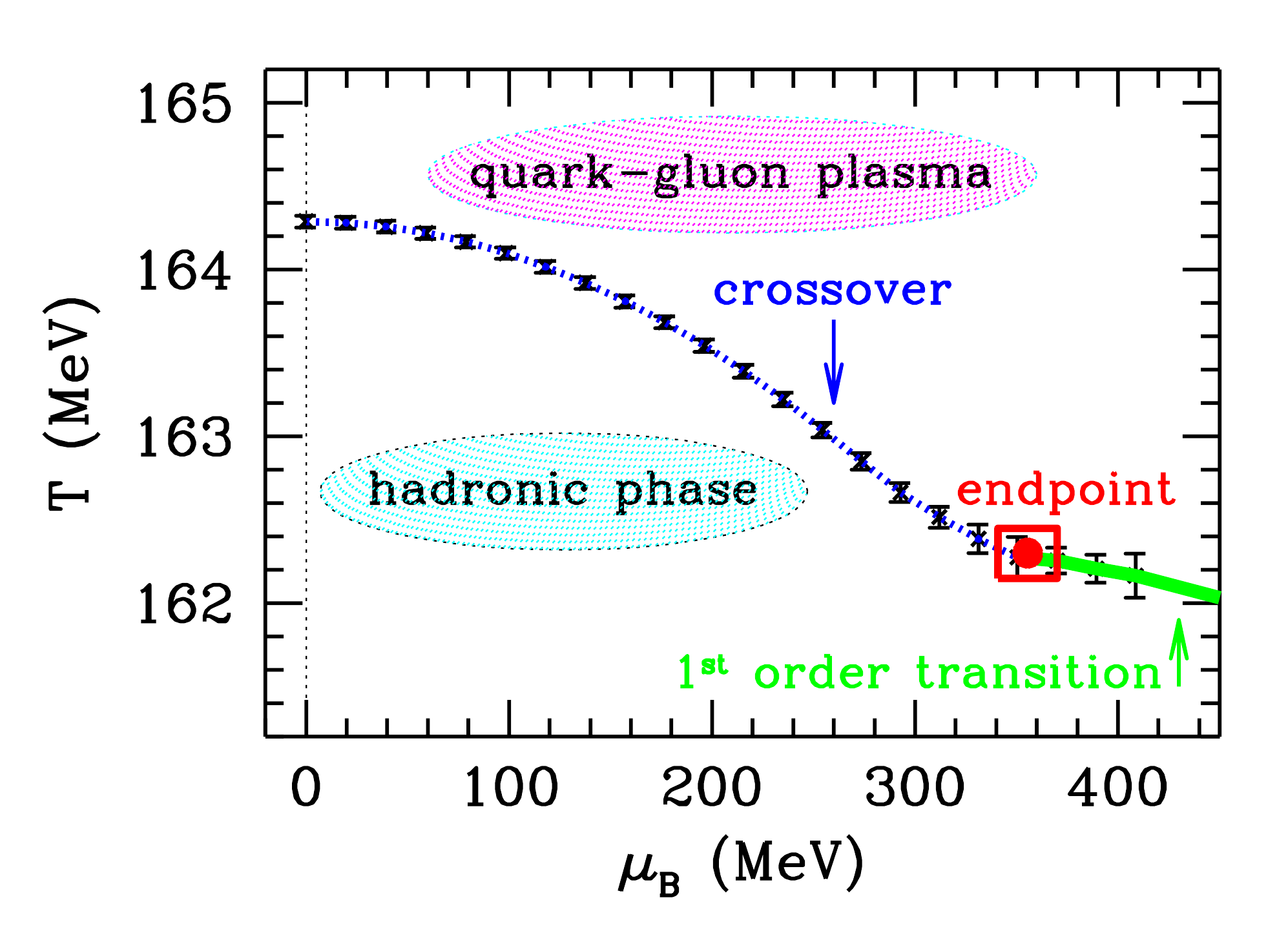}
\par\end{centering}

\caption{\label{fig:phase-diagram-01}Phase diagram of QCD from the lattice
calculations with non-zero chemical potential \cite{Fodor:2004nz}.}
\end{figure}

In the early 1970s Lee and Collins et al. \cite{Lee1974,Collins:1974ky}
proposed that the deconfinement can be reached through the ultra-relativistic
heavy ion collisions. According to the calculations from lattice QCD,
the confinement/deconfinement phase transition will take place at
temperature of about 170 MeV in three flavor case \cite{Karsch2000a}
as shown in Fig. \ref{fig:phase-diagram-01}.

\subsection{Experiments for high energy heavy ion collisions}

\subsubsection{RHIC experiments}

Relativistic Heavy Ion Collider (RHIC) at Brookhaven National Laboratory
(BNL) has been running since 2000. Two beams of nuclei (Au or Cu)
are accelerated to collide with the center-of-mass energy of 200 GeV/nucleon
or 62.4 GeV/nucleon. After the most part of nuclei pass through each
other, the huge amount of energy is deposited in the central rapidity
region, and thus excites the quarks and gluons from the vacuum. These
quarks and gluons form an expanding fireball, and reach the local
thermal equilibrium within a extremely short time of $1\sim2fm/c$.
It is believed that new state of matter, the quark-gluon-plasma (QGP)
has been formed \cite{Gyulassy:2004vg,Gyulassy:2004zy}. The QGP expands
and cools down with it freezes out at some critical temperature, below
which the quarks recombine into hadrons observed by the detectors. 

\begin{figure}
\begin{centering}
\includegraphics[scale=0.7]{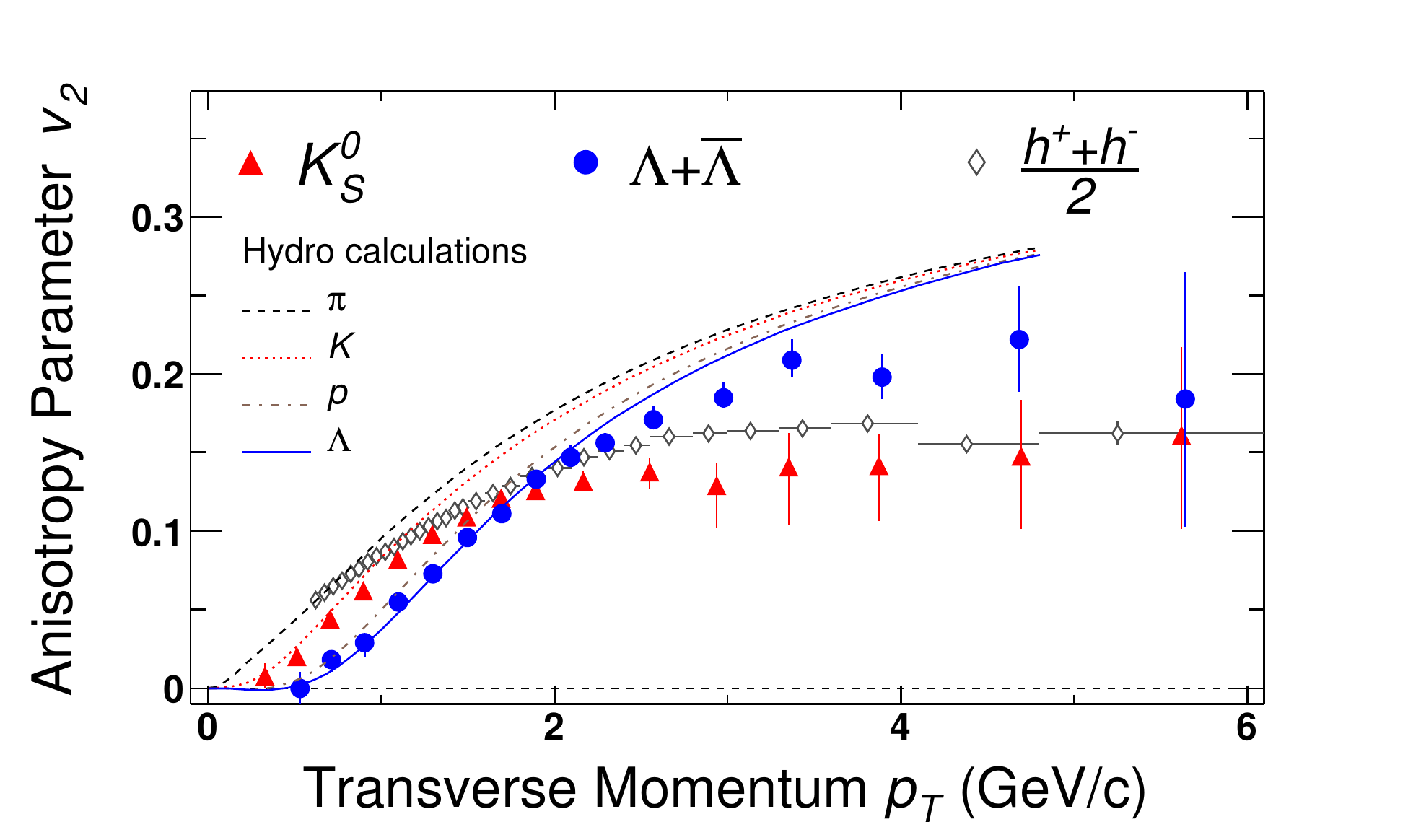}
\par\end{centering}

\caption{\label{fig:v2-01}The elliptic flow measured at RHIC with prediction
given by hydrodynamics \cite{Adams2004a}. }

\end{figure}

The collective flows such as radial and elliptic flows are observed
at RHIC and can be well described by ideal fluid dynamics. In non-central
collisions, the anisotropic momentum leads to the gradient of pressure.
This effect can be detected by analysis of the final particle spectrum
in momentum space. The Fourier transformation for the particle spectrum
in terms of particle azimuthal angle $\phi$ with respect to the reaction
plane $\psi_{r}$ gives 
\begin{equation}
E\frac{d^{3}N}{dp^{3}}=\frac{d^{2}N}{2\pi p_{T}dp_{T}dy}\left\{ 1+\sum_{n=1}^{\infty}2v_{n}\cos[n(\phi-\psi_{r})]\right\} \;.
\end{equation}
where the second coefficient $v_{2}$ is the anisotropy parameter,
which is also called elliptic flow. The data from RHIC have delivered
a surprising result that elliptic flow $v_{2}$ is very large \cite{Adams2004a,Adams2004h,Adler2002f,Sorensen2003}
and compatible with the numerical simulations of ideal fluid dynamics
\cite{Kolb:2003dz,Hama2005,Huovinen2006,Ollitrault2008}, see Fig.
\ref{fig:v2-01}. This indicates that the QGP is strongly coupled
in contrast to the assumption that QGP is a weakly coupled system. 

\begin{figure}
\begin{centering}
\includegraphics[scale=0.65]{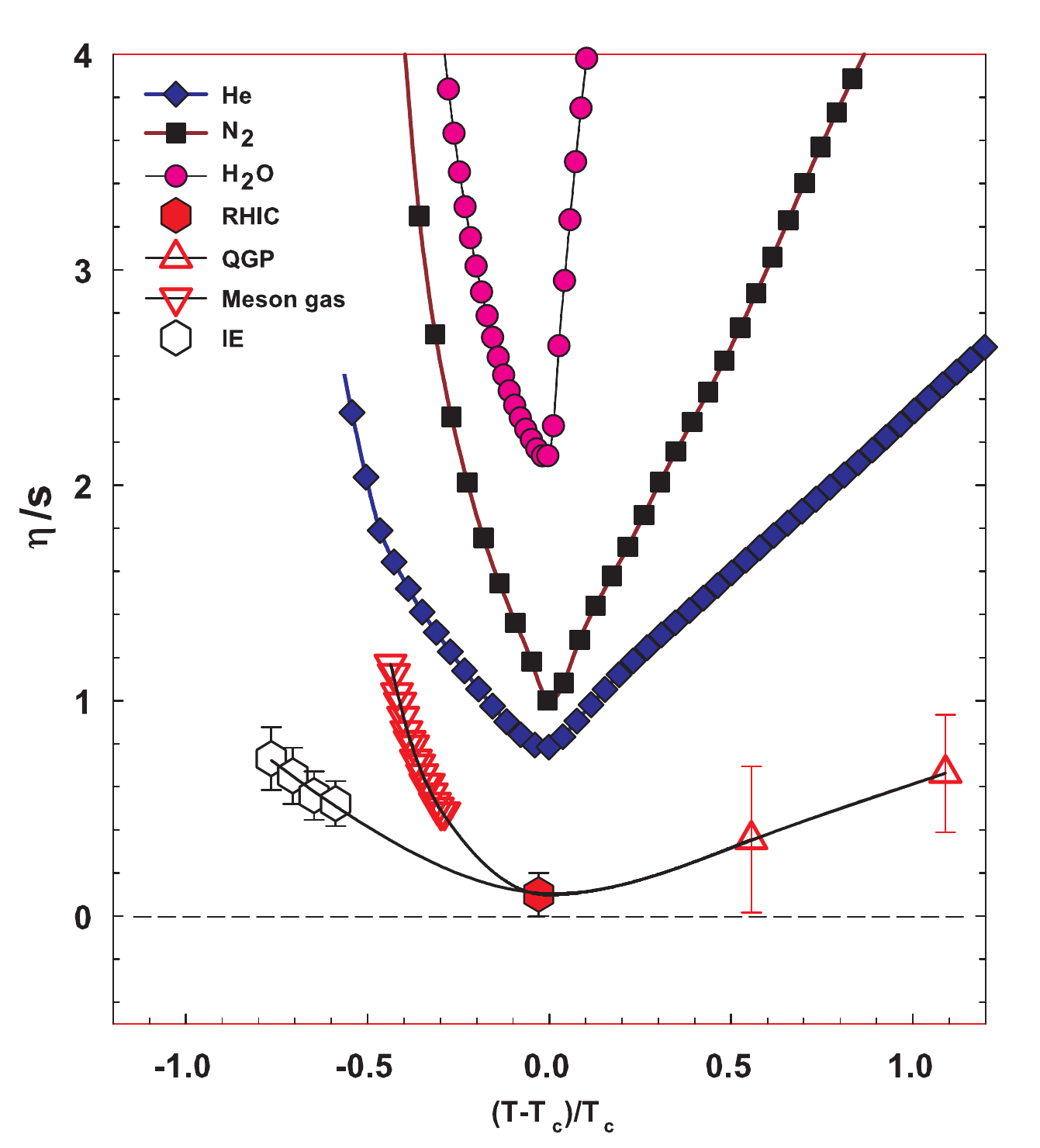}
\par\end{centering}

\caption{\label{fig:eta-s-01}Ratio of $\eta/s$ as a function of temperature
scaled by transition temperature $T_{c}$ in different system \cite{Lacey:2006bc}. }
\end{figure}

Further study for QGP gives very low values of the ratio of shear
viscosity to entropy density $\eta/s\sim0.1$\cite{Teaney:2003kp,Lacey:2006bc}
close to the theoretical lower bound given by AdS/CFT duality \cite{Policastro2001},
see Fig. \ref{fig:eta-s-01}. For a weakly coupled system with well-defined
quasi-particles, the ratio $\eta/s$ is very large \cite{Arnold2003a,Arnold2000}.
For the system with strong couplings, the lower bound from the AdS/CFT
duality is $1/(4\pi)$ \cite{Kovtun2005,Son2007b} in the comparison
with the $1/12$ evaluated from the uncertainty principle \cite{Danielewicz:1984ww}.
Therefore, the fact that $\eta/s$ of QGP is close to the lower bound
is one piece of most convincing evidence that QGP is strongly coupled.
The calculations for the ratio are shown in Chap. \ref{chap:Transport}
for a weakly coupled system and Chap. \ref{chap:Applications-of-AdS/CFT}
for a strongly coupled system.

There are a lot of other interesting and important phenomena at RHIC
experiment, e.g. the jet quenching, parton energy loss, heavy quark
production, etc. (see e.g. Ref. \cite{Kolb:2003dz,Baier:2000mf,Tomasik:2002rx,Gyulassy:2003mc,Rischke2004,Jacobs2005,Adams:2005dq}
for reviews). Those phenomena will also be mentioned in the relevant
sections.

\subsubsection{LHC experiments}

The main goal of the Large Hadron Collider (LHC) experiments at European
Organization for Nuclear Research (CERN) is to discover the Higgs
bosons, supersymmetic particles and other new physics. There are three
major experiments: A Toroidal LHC Apparatus (ATLAS), Compact Muon
Solenoid (CMS), and A Large Ion Collider Experiment (ALICE). The purpose
of ATLAS and CMS experiments is to hunt for Higgs boson and new physics,
while the ALICE experiment is to pin down and study the QGP. 

In the CMS experiment, the two-particle angular correlations for charged
particles in the proton-proton (pp) collisions at center-of-mass energies
of $0.9$, $2.36$, and $7$ GeV are measured. A long-range, near-side
feature in two-particle correlation functions have been observed in
pp collisions for the first time \cite{Khachatryan2010a}. A ridge-like
structure is observed in the two-dimensional correlation function
for particle pairs with intermediate transverse momentum of $1-3$GeV$/c$,
$2.0<\left|\Delta\eta\right|<4.8$ and $\Delta\phi\approx0$ with
$\eta$ the pseudorapidity and $\phi$ the azimuthal angle. Some authors
\cite{Shuryak2010,Dumitru2010} thought that this discovery might
imply the QGP has also been formed in the pp collisions. The collective
flow has also been analyzed \cite{Bozek:2010pb,Dremin2010}. The relativistic
fluid dynamics may become a powerful tool to investigate the new phenomena
in the ultra high energy pp collisions at LHC. 

The first Pb-Pb collision at center-of-mass energy of 2.7 TeV was
realized at LHC in November 2010 \cite{Aamodt2010,Aamodt2010a}. The
ALICE experiment is aimed to search for QGP at 2 to 3 times higher
temperatures than RHIC. Since the collisional energy at LHC is one
magnitude larger than at RHIC, the perturbative QCD (pQCD) is expected
to work much better than at RHIC. The semi-classical Boltzmann equation
with the collision terms given by the pQCD is a good tool to describe
non-equilibrium dynamics of the QGO formed in heavy ion collisions
at LHC.

\section{Relativistic fluid dynamics and kinetic theory\label{sec:Relativistic-Fluid-dynamics}}

In long wavelength or small-frequency limit, almost all theories can
be described by the fluid dynamics as effective theories. L.D. Landau
first suggested to apply fluid dynamics to the hadronic fireballs
\cite{Landau:1953gs}. Then Siemens and Rasmussen \cite{Siemens:1978pb}
attempted to use the collective transverse flow to describe date of
the low energy heavy ion collision experiment BEVALAC. Zhirov and
Shuryak \cite{Shuryak:1979ds} tried to explain the data of the high
energy proton-proton (pp) collisions at CERN-ISR using fluid dynamics.
Now fluid dynamics becomes a necessary tool to describe data of high
energy heavy ion collisions \cite{Kolb:2003dz,Rischke:1998fq,Shuryak:2003xe,Stock:2004cf}.

\subsection{Second order theory}

The basic equations for fluid dynamics consist of conservation equations
of energy-momentum and charges (\ref{eq:em_conserve},\ref{eq:current_conserve}).
The energy-momentum tensor and the conserved currents can be expanded
in the terms of the so-called Knudsen number \cite{Israel:1979wp,Betz:2008me,Betz:2009zz,Betz:2010cx},
which is defined by the ratio of mean free path to the macroscopic
characteristic length. The zeroth order of this expansion corresponds
to the ideal fluid. In the first order, the Navier-Stokes (NS) equations
(\ref{eq:NS-equation-01}) are obtained and the shear stress tensor
and bulk viscous pressure are introduced. The details for the expansion
in the power series of Knudsen number will be discussed in Sec. \ref{sub:Complete-IS-equations}. 

There are different representations for the dissipative second order
theories of the fluid dynamics, e.g. the the theory of the conformal
fluid \cite{Baier2008}, Israel-Stewart (IS) theory \cite{Israel:1979wp,Betz:2008me,Betz:2009zz,Betz:2010cx},
the memory function theory \cite{Denicol2009,Koide2007}, the extended
thermodynamics \cite{Denicol2009,Jou1999,Jou1988}, and others \cite{Carter1991,Grmela1997}.
They differ only in non-linear second order terms. In this dissortation,
we will focus on the Israel-Stewart theory only \cite{Israel:1979wp,Betz:2008me,Betz:2009zz,Baier2008}. 

The simplest IS theory is given by a combination of all irreducible
quantities in the first order theory, see in Sec. \ref{sub:Israel-Stewart-equations}.
However, this description does not demonstrate the fact that quantities
in the first and second order theories are related to each other.
It is not clear whether the simple IS equation (\ref{eq:2nd-equation-phe})
contains all possible quantities in second order theory. The conservation
equations can also be studied by the kinetic theory (relativistic
Boltzmann equation), or the Grad's 14 moment approximation \cite{Israel:1979wp}
(also see \cite{Muronga:2006zw,Muronga:2006zx} in a different metric).
A complete IS equations are given in a power counting scheme \cite{Betz:2008me,Betz:2009zz}.
The details will be shown in Chap. \ref{chap:complete}. The relationship
between the transport coefficients in the first and second order theory
is also discussed in Sec. \ref{sub:Complete-IS-equations}.

\subsection{Causality and stability}

It has been pointed out that the first order theory does not obey
the causality \cite{Hiscock1983,Hiscock1985,Hiscock1987,Olson:1989dp,Olson:1989eu}.
For instance, as will be shown in Eq.(\ref{eq:heat-conductivity-01}),
the heat conduction equation in the first order theory is
\begin{equation}
\frac{\partial T}{\partial t}=D\frac{\partial^{2}T}{\partial x^{2}}\;,\label{eq:heat-conduct-02}
\end{equation}
where $D$ is the heat conductivity. The dispersion relation in the
linear approximation is given by 
\begin{equation}
\omega=iDk^{2}\;,
\end{equation}
which implies that the group velocity of the signal $v_{g}=\partial\omega/\partial k$
is proportional to the wave-number $k$. For $k\rightarrow\infty$,
the group speed goes to infinite and violates causality \cite{Denicol:2008ha}.
Therefore, the second order theory is necessary. The second order
term of $\partial_{t}^{2}T$ has to be introduced in the Eq.(\ref{eq:heat-conduct-02}),
\begin{equation}
\tau_{q}\frac{\partial^{2}T}{\partial t^{2}}+\frac{\partial T}{\partial t}=D\frac{\partial^{2}T}{\partial x^{2}}\;,
\end{equation}
where $\tau_{q}$ is the relaxation time. Now the dispersion relation
becomes $\omega\sim\sqrt{D/\tau_{q}}k$. For $D<\tau_{q}$, the signal
propagating speed is smaller than the speed of light. For $\tau_{q}\rightarrow0$,
the system is acausal again. Therefore, there has to be constraint
condition for the $\tau_{q}$ and $D$, which is called asymptotic
causality condition \cite{Denicol:2008ha,Pu:2009fj,Pu:2010zz}. 

Stability is intimately related to causality \cite{Denicol:2008ha,Pu:2009fj,Pu:2010zz}.
A signal propagating faster than the light will move out of the light-cone.
The acausal propagating modes will lead to some non-physical results,
e.g. instability and singularities. In this case, for all parameters
considered the theory will be unstable if it becomes acausal. The
details will be given in Chap. \ref{cha:Causality-ans-stability}.

In the linear approximation, the discussion in Chap. \ref{cha:Causality-ans-stability}
will be universal since all candidates for second order theories \cite{Israel:1979wp,Betz:2008me,Baier2008,Denicol2009,Koide2007,Jou1999,Jou1988,Carter1991,Denicol:2008ha}
differ only by non-linear second order terms. 

\subsection{Boltzmann equation\label{sub:Boltzmann-equation}}

The transport coefficients can be determined by the microscopic transport
theories. There are two main techniques to compute these coefficients.
The first one is to employ the Kubo formula (\ref{eq:kubo01}) where
the transport coefficients are expressed by the commutator of operators,
e.g. the energy-momentum tensor or conserved currents \cite{kubo1984,Lifshitz}.
The commutator can be worked out through standard perturbation techniques
in field theory.

The second one is to employ the relativistic Boltzmann equation. If
the mean free path of the particles is much larger than the interaction
length, the quasi-particle is well-defined. In this case, a semi-classical
description for the equation of motion of the particle distributions,
the Boltzmann equation, works well. As will be shown in Sec. \ref{sub:Complete-IS-equations}
and \ref{sec:Jnk-Ink}, the energy-momentum tensor and charge currents
can be determined by the integrals of the distribution functions.
By a near equilibrium expansion, the transport coefficients can also
be related to the integrals of the distribution function via Boltzmann
equation.

At high temperature, the shear viscosity in a gauge theory has been
found in a leading-log form \cite{Arnold2003a,Arnold2000} 
\begin{equation}
\eta\propto\frac{T^{3}}{g^{4}\ln g^{-1}}\;,
\end{equation}
where $g$ is the coupling constant. For $g\rightarrow0$, $\eta\rightarrow\infty$
which indicates that small $\eta/s$ means a strongly coupled system.
Recently, the calculation of shear viscosity in a gluon gas with $2\rightarrow3$processes
has attracted attention from several authors \cite{Xu:2007jv,Chen:2009sm,Chen:2010xk}.
The same technique can also be applied to investigate the transport
coefficients of dense matter near phase transition \cite{Chen2010,Chen2010a}.
The detail will be presented in Chap. \ref{chap:Transport}. 

Along this lines, there are more many other on this topics, e.g. calculation
of the bulk viscosity \cite{Arnold:2006fz,Moore:2008ws,Lu:2011df}
and the second order transport coefficients \cite{York:2008rr}. There
are more applications of kinetic theory to the heavy ion collisions,
e.g. the collective flow \cite{Xu:2008dv,Xu:2008av,Uphoff:2010sy,Uphoff:2010bv},
the jet quenching and partons energy loss \cite{Gallmeister:2002nz,Gallmeister:2002us,Qin2009b,Qin:2009ff,Fochler:2010fq,Fochler:2010fe,Fochler:2010wn,Bouras:2010yw},
the shock wave and Mach cone \cite{Bouras:2010nt,Bouras:2010jd,Bouras:2010zz,Bouras:2011kt,Bouras:2008ip,Bouras:2009pa,Bouras:2009vs}
and the thermalization \cite{Xu:2005tm,Xu:2005wv,Greiner:2005ca,El:2006xj,El:2007vg,Arnold:2007pg,El:2009zz}.

\subsection{Numerical simulations of hydrodynamics}

\begin{figure}
\begin{centering}
\includegraphics[scale=0.7]{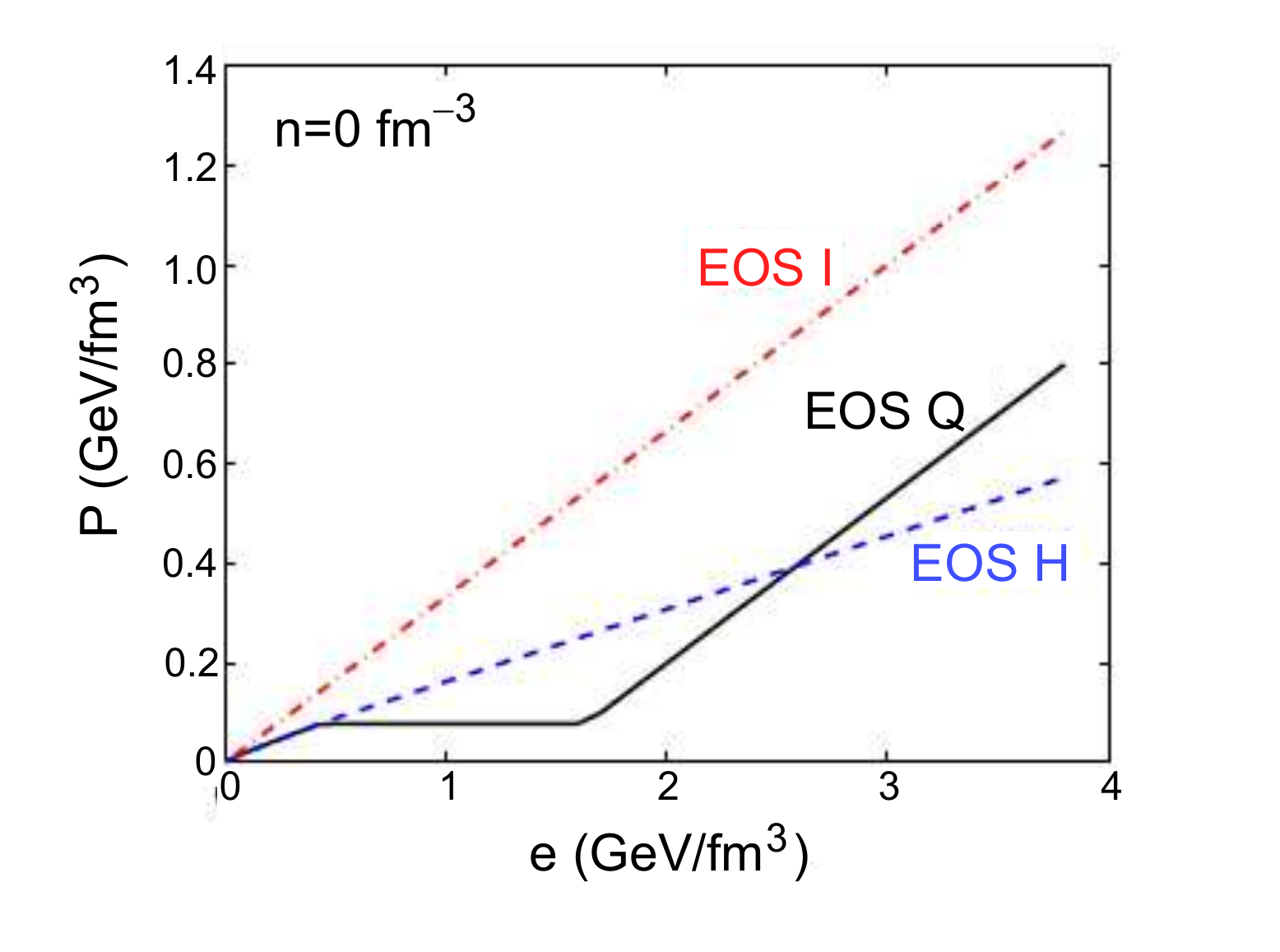}
\par\end{centering}

\begin{centering}
\includegraphics[scale=0.32]{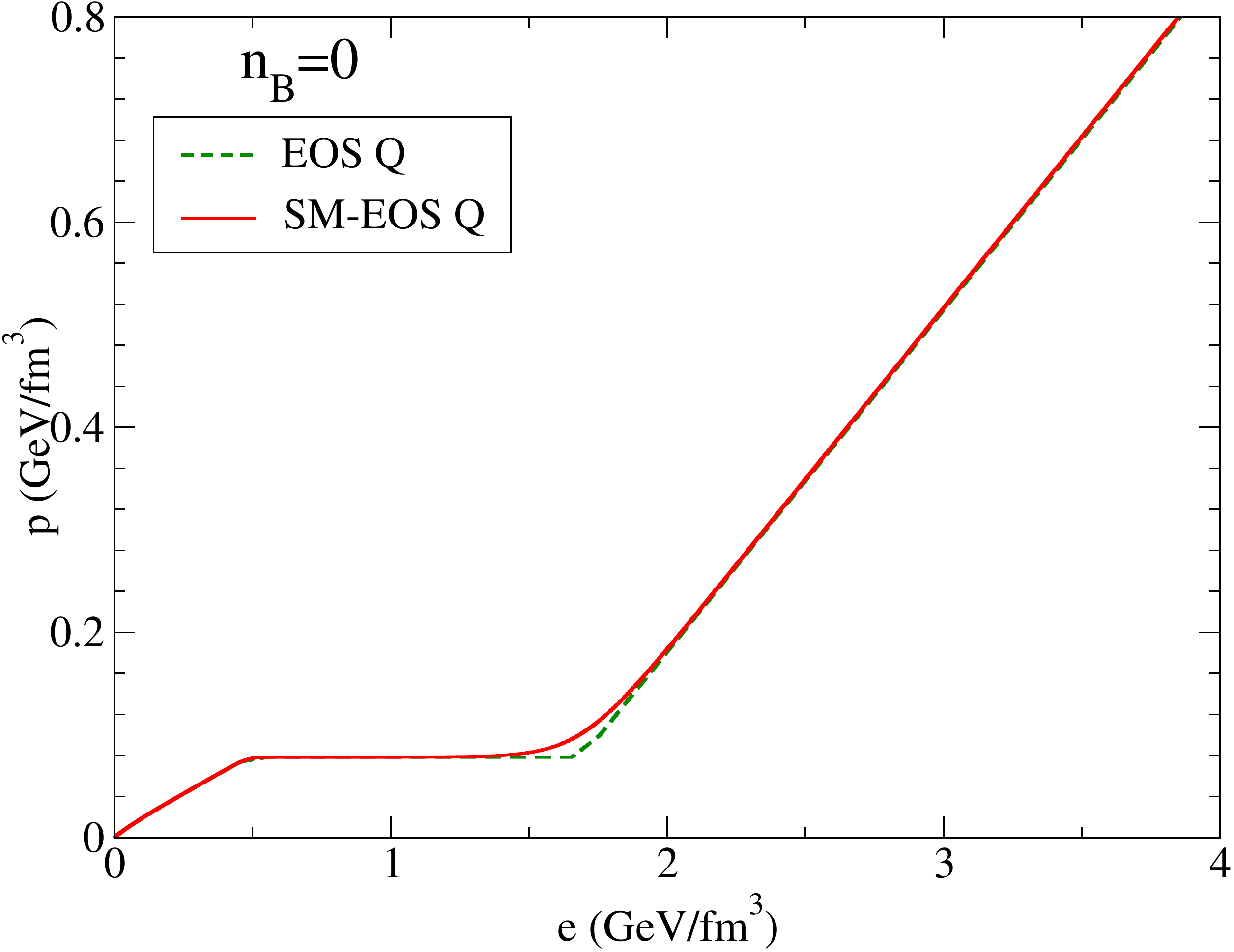}
\par\end{centering}

\caption{\label{fig:Equation-of-state-01}Equation of state for numerical simulations.}
\end{figure}

At RHIC experiment, the local thermal equilibrium is established in
a very short time after collisions. Therefore, the fireball will expand
over a sufficient evolution time, when the relativistic fluid dynamics
works well. In the earlier time of this field, the numerical simulations
was only for the ideal fluid, see e.g. \cite{Kolb:2003dz,Rischke:1998fq}
for reviews. However, the ratio of the shear viscosity $\eta$ to
the entropy density $s$ is found to be small but not zero. The simulation
for the dissipative fluid is necessary. Recently, the simulation for
the dissipative fluid dynamics has been developed \cite{Song2008,Song2008a}.
In comparison with the $v_{2}$ data at RHIC, the ratio $\eta/s$
is found to be \cite{Song:2010mg} 
\begin{equation}
1<4\pi\frac{\eta}{s}<2.5\;.
\end{equation}
The initial condition is usually given by the Glauber model \cite{Kolb:2003dz}
or Kharzeev-Levin-Nardi (KLN) approach \cite{Kharzeev:2000ph,Kharzeev:2002ei,Kharzeev:2004if}.
More extended models are also used, e.g. MC-Glauber model \cite{Miller:2007ri}
and fKLN \cite{Drescher:2006ca}. Generally, there are mainly three
kinds of EOS, see Fig. \ref{fig:Equation-of-state-01}, EOS I for
the ideal gas of massless partons, EOS H for hagedorn resonance gases,
EOS Q for a combination of the above two. Recently, the SM-EOS Q \cite{Kolb:1999it,Kolb:2000sd},
a smooth version of EOS Q, is also used \cite{Song2008}. 

\section{AdS/CFT\label{sec:AdS/CFT}}

In quantum field theory, the partition function is written as 
\begin{eqnarray}
Z_{QFT} & = & \int D\phi e^{i\int d^{4}x(\mathcal{L}_{0}+g\mathcal{L}_{1})}\nonumber \\
 & = & \int D\phi e^{i\int d^{4}x\mathcal{L}_{0}}[1+ig\int d^{4}\mathcal{L}_{1}+O(g^{2})]\;,\label{eq:parition function}
\end{eqnarray}
where $D$ denotes functional integrals of the fields $\phi$, $g$
is the coupling constant, $\mathcal{L}_{0}$ and $\mathcal{L}_{1}$
are Lagrangian for the free and the first order interacting parts,
respectively. For $g\ll1$, the theory is weakly coupled, and the
perturbation works well. If $g\gg1$, higher order contributions is
not negligible and all orders in the expansion should be summed. It
is difficult to describe a strongly coupled system due to its non-linear
and non-perturbative feature.

In recent years a new technique to deal with strongly coupled systems
in gauge theory has been developed, made use of the string/gauge duality
or the AdS/CFT duality proposed by Maldacena and many others \cite{Maldacena1998c,Gubser1998,Witten1998e}.
Here AdS and CFT are abbreviations for anti-de Sitter space and conformal
field theory, respectively. In the 't Hooft limit or large $N$ limit
(with $N$ the dimension of fundamental representation of $SU(N)$
group), the action of the open strings is equivalent to that of a
conformal $SU(N_{c})$ theory. In the classical limit (i.e. the gravitons
are almost free), the weakly coupled closed strings (free gravitons)
in a curved space correspond to the strongly coupled open strings
in a flat space. Finally, the strongly coupled field conformal theory
in a flat space is equivalent to the theory of closed strings in curved
space. 

Policastro, Son and Starinets first used AdS/CFT duality to compute
the transport coefficients of sQGP \cite{Policastro2001} and derived
the ratio of $\eta/s\gtrsim1/4\pi$, which is consistent to the RHIC
data. After that, there are more developments in this field. Up to
date three main applications of AdS/CFT duality have been explored.
The first one is to use pure AdS/CFT to compute quantities of QCD-like
CFT at very high temperatures (see e.g. Ref.\cite{Liu2006,Herzog:2006gh,Gubser:2006bz}
for the energy loss, Ref. \cite{Hou:2009qj,Hou:2007uk} for the potential
of heavy quarks, Ref. \cite{Hou:2007uk,Liu:2006nn,Peeters:2006iu,Li:2008py}
for the screening length ). The second one is to use the so-called
AdS/QCD to calculate the properties of hadrons (see e.g. Ref.\cite{Karch2006}
and Ref.\cite{Sakai2005,Sakai2005a} for the Sakai-Sugimoto model).
The third one is the correspondence between gravity and condense matter
theory (CMT), which is the so-called AdS/CMT, see e.g. Ref.\cite{Hartnoll2008b,Herzog2009,Chen:2009pt,Pu:2009wn}
for strongly coupled superconductivity and superfluidity, Ref.\cite{Liu2009,Faulkner2010a}
for the non-Fermi liquids. More reviews on applications of AdS/CFT
duality can be found in Ref. \cite{Aharony:1999ti,D'Hoker:2002aw,Maldacena:2003nj,Nastase:2007kj}
for the pure AdS/CFT, Ref.\cite{Hartnoll2009,Herzog:2009xv,Lee:2010fy}
for superconductivity and superfluidity, Ref. \cite{Iancu:2008sp}
for jets and partons, Ref. \cite{Gursoy:2009zza} for the deconfinement
phase transition and Ref. \cite{Baier2008,Rangamani:2009xk,Romatschke:2009im}
for the fluid dynamics.

In the pure AdS/CFT, two main quantities, the (retarded) Green functions
and Wilson loops, can be computed via the duality.

\subsection{Green functions}

The AdS/CFT duality means that the partition function $Z_{QFT}[J]$
in the conformal theory is equivalent to the partition function $Z_{string}[\phi]$
in the classical string theory. In this case, one finds 
\begin{eqnarray}
Z_{QFT}[J] & = & e^{iS_{string}[\phi_{cl}]}\;,
\end{eqnarray}
where $J$ the source coupled to the operator $O$ and $S_{string}$
is the action of the classical gravity and 
\[
Z_{QFT}[J]=\int D\phi exp\left(iS+i\int d^{4}xJO\right)\;.
\]

The Green functions in the CFT is associated to the derivatives of
the action in the AdS space, e.g. the two-point Green function of
$O$ is given by the functional derivatives of $S[\phi_{cl}]$ with
respect to the boundary value of $\phi$, 
\begin{eqnarray}
G(x-y) & = & \left.\frac{\delta^{2}Z_{QFT}[J\phi]}{i^{2}\delta J(x)\delta J(y)}\right|_{J=0}\nonumber \\
 & = & -\left.\frac{\delta^{2}S_{string}[\phi_{cl}]}{\delta J(x)\delta J(y)}\right|_{z\rightarrow0,\;\phi=J}\;.\label{eq:green_function}
\end{eqnarray}
The problem to compute Green functions in a strongly coupled quantum
field theory is made to compute the classical action of gravity. 

Based on the work of Ref.\cite{Policastro2001}, we study sQGP with
the radial expansion and Bjorken boost invariance. The information
of the sQGP is encoded on the boundary of AdS space via the holographic
renormalization. The evolution of the shear viscosity as a function
of proper time can be obtained via the Kubo formula for the retarded
Green function given by the AdS/CFT duality, see Chap. \ref{chap:Applications-of-AdS/CFT}
for detail.

\subsection{Wilson loops\label{sub:Wilson-loops}}

It is well-known that the gauge invariant Wilson loops for quark and
anti-quarks in QCD can be written in the form 
\begin{equation}
W[C]=e^{-V(R)T}=e^{-\sigma A(C)}\;,
\end{equation}
where $C$ is a contour which is usually chosen as a rectangle in
Euclidean space-time with the area $A(C)=RT$, $R$ is the distance
between quarks and anti-quarks, $T$ is the imaginary time, $V$ is
the potential and $\sigma=V(R)/R$ is the string tension. 

\begin{figure}
\begin{centering}
\includegraphics[scale=0.7]{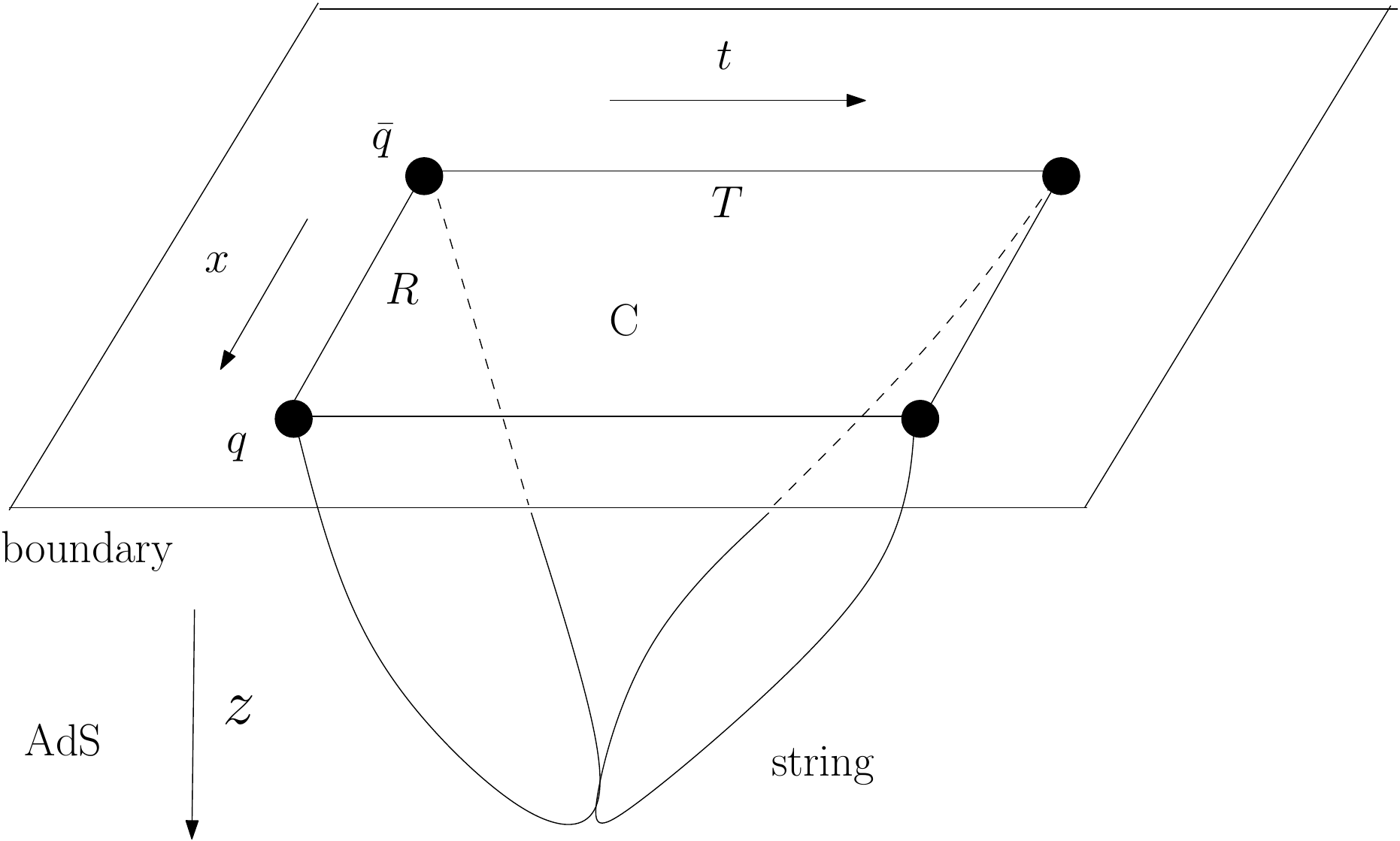}
\par\end{centering}

\caption{\label{fig:Wilson-loops-01}Wilson loops in AdS/CFT duality.}

\end{figure}

As shown in Fig.\ref{fig:Wilson-loops-01}, the area of the contour
$C$ can be obtained by integrating over of the world-volume of the
strings. These integrals are given by the Nambu-Goto action. In this
case, the Wilson loops are found to be 
\begin{equation}
W[C]=Z_{string}[C]=e^{-(S_{string}-l\phi)}\;,
\end{equation}
where $l$ is the length of the loop $C$ and $\phi$ is the mass
field. In order to renormalize the potential of heavy quarks, the
contributions from the mass of strings $l\phi$ must be removed. 

In the standard AdS metric, the potential for quarks and anti-quarks
is given by \cite{Nastase:2007kj},
\begin{equation}
V_{q\bar{q}}(R)=-\frac{4\pi^{2}}{\Gamma(1/4)^{4}}\frac{\sqrt{2g_{YM}^{2}N_{c}}}{R}\;,
\end{equation}
where $g_{YM}$ is the coupling constant of the super Yang-Mills theory
and also see Ref. \cite{Hou:2009qj,Hou:2007uk} for the higher order
contributions. This potential is strongly related to the phenomena
in high energy physic, e.g. the jet quenching \cite{Liu2006,Herzog:2006gh,Gubser:2006bz}. 

In this thesis, a Wilson-loop-like model will be used to investigate
the property of high spin baryons in the QGP in Chap. \ref{chap:Applications-of-AdS/CFT}.
The quarks located at the boundary of AdS space are connected to a
probe D5 brane by superstrings. The configurations of strings give
the screening length of baryons as a function of spin and temperature.
A Regge-like behavior, i.e. the total angular momentum is proportional
to the energy squared, is also found.

\section{Fluid dynamics with triangle anomalies}

The vorticity vanishes in the first order theory of fluid dynamics
due to the parity conservation. However, the analysis of the power
counting in Eq.(\ref{eq:omega-mu-nu}) indicates that the vorticity
has the same order as other dissipative terms (e.g. shear stress tensor
or bulk viscous pressure) in the first order theory. It implies that
some terms related to vorticity are absent in the previous treatment
(\ref{eq:complete}).

Recently, the relativistic fluid dynamics corresponding to charged
black-branes through the AdS/CFT duality was found to have two new
terms associated with the axial anomalies in the first order theory
\cite{Banerjee:2008th,Erdmenger2009} (see, e.g, Ref. \cite{Torabian2009a}
about the holographic model with multiple/non-Abelian symmetries,
or Ref. \cite{Rebhan2010} for the Sakai-Sugimoto model). The authors
of Ref. \cite{Son2009} have derived these new terms in relativistic
fluid dynamics with triangle anomalies. Also see Ref. \cite{Lublinsky2010}
about the similar result obtained in microscopic theory of the superfluid. 

Actually, the anomalous fluid dynamics is closely related to the Chiral
Magnetic Effect (CME) in heavy ion collisions \cite{Kharzeev2008,Fukushima2008,Fukushima2010a,Kharzeev2010}.
The separation of the left- and right-hand particles or anti-particles
leads to macroscopic currents in the case of strong electromagnetic
field. This effect is called CME. In comparison with CME, the separation
of left- and right-hand particles given by vorticity is also called
Chiral Vorticity Effect (CVE) \cite{Kharzeev2010}. The separation
of chirality can be realized by the change of the topological charge
\cite{Kharzeev2008} or the quantum triangle anomalies \cite{Son2009,Kharzeev2010}.
Therefore, it is necessary to introduce an additional term proportional
to the magnetic field in the conserved currents $j^{\mu}\propto B^{\mu}$.
And the vorticity related to the magnetic field will also be introduced
to the conserved currents $j^{\mu}\propto\omega^{\mu}$, where 
\begin{equation}
\omega^{\mu}=\epsilon^{\mu\nu\alpha\beta}u_{\nu}\partial_{\alpha}u_{\beta}\;,
\end{equation}
is defined in Ref. \cite{Son2009}. Here $\epsilon^{\mu\nu\alpha\beta}=1,-1$
for that the order of Lorentz indices $\left(\mu\nu\alpha\beta\right)$is
an even/odd permutation of $(0123)$. 

The anomalous fluid will be studied in a kinetic approach in Chap.
\ref{chap:vorticity}. The transport coefficients for the vorticity
will also be obtained.



\chapter{Basics of hydrodynamics and kinetic theory\label{cha:Preliminaries}}

Fluid dynamics is an effective theory for any interacting theories
in long wavelength limit. Recently, it is widely used to describe
the data in many aspects of the high heavy ion collisions \cite{Kolb:2003dz,Rischke:1998fq,Shuryak:2003xe,Stock:2004cf}. 

In this chapter, we will give a brief introduction to the basics of
the hydrodynamics and kinetic theory. The basic conservation equations
of fluid dynamics are given in Sec. \ref{sec:Conservation-equations}.
The first order theory of the fluid, called the relativistic Navier-Stokes
equations, is shown in Sec. \ref{sub:Navier-Stokes-equations}. Then
we introduce the classic and quantum Boltzmann equation in Sec. \ref{sec:BE}
and \ref{sec:Covariant-kinetic-equations}, respectively. In Sec.
\ref{chap:complete}, we give a short review to the complete second
order theory of the fluid dynamics via kinetic theory.

\section{Conservation equations\label{sec:Conservation-equations}}

The basic fluid dynamic equations are the conservation equations of
energy-momentum and charge
\begin{eqnarray}
\partial_{\mu}T^{\mu\nu} & = & 0\;,\label{eq:em_conserve}\\
\partial_{\mu}j^{\mu} & = & 0\;,\label{eq:current_conserve}
\end{eqnarray}
where $T^{\mu\nu}$ is the energy-momentum tensor and $j^{\mu}$ is
the charge current. Here we consider the homogenous fluid with only
one particle species. 

The tensor decomposition of $T^{\mu\nu}$ with respective to the fluid
velocity $u^{\mu}$ reads
\begin{equation}
T^{\mu\nu}=\epsilon u^{\mu}u^{\nu}-(P+\Pi)\Delta^{\mu\nu}+(h^{\mu}u^{\nu}+h^{\nu}u^{\mu})+\pi^{\mu\nu}\;,\label{eq:TEM_01_01_01}
\end{equation}
where $\epsilon$, $h^{\mu}$, $P$ , $\Pi$ and $\pi^{\mu\nu}$ are
the energy density, the heat flux current, the pressure, the bulk
pressure and the shear stress tensor, respectively, and $\Delta^{\mu\nu}=g^{\mu\nu}-u^{\mu}u^{\nu}$
is the projector onto the $3$-space orthogonal to $u^{\mu}$. Here
the velocity is time-like and normalized to $1$, i.e. $u^{\mu}u_{\mu}=1$.
In a Lorentz boosted frame, the velocity is given by 
\begin{equation}
u^{\mu}=\gamma(1,\,\overrightarrow{v})\;,
\end{equation}
where $\gamma$ is the Lorentz factor $\gamma=\sqrt{1-v^{2}}$. On
the other hand, these thermal quantities can also be expressed by
$T^{\mu\nu}$,
\begin{alignat}{2}
 & \epsilon=u_{\mu}u_{\nu}T^{\mu\nu}\;,\quad & P+\Pi=-\frac{1}{3}\Delta^{\mu\nu}T_{\mu\nu}\;,\nonumber \\
 & \pi^{\mu\nu}=P^{\mu\nu\alpha\beta}T_{\alpha\beta}\;,\quad & h^{\mu}=\Delta^{\mu\nu}T_{\nu\alpha}u^{\alpha}\;,\label{eq:T-decomp}
\end{alignat}
 where 
\begin{equation}
P^{\mu\nu\alpha\beta}=\frac{1}{2}\left(\Delta^{\mu\alpha}\Delta^{\nu\beta}+\Delta^{\mu\beta}\Delta^{\nu\alpha}\right)-\frac{1}{3}\Delta^{\mu\nu}\Delta^{\alpha\beta}\;,
\end{equation}
is the symmetric rank-four projection operator. By construction, $\pi^{\mu\nu}$
is traceless $\pi_{\mu}^{\mu}=0$ and $\pi^{\mu\nu}u_{\nu}=h^{\mu}u_{\mu}=0$
since $u^{\mu}\Delta_{\mu\nu}=0$.

The tensor decomposition of the conserved current $j^{\mu}$ reads
\begin{equation}
j^{\mu}=nu^{\mu}+\nu^{\mu}\;,
\end{equation}
where $n$ is the number density and $\nu^{\mu}$ is the diffusion
current. By construction, 
\begin{equation}
n=u_{\mu}j^{\mu}\;,\qquad\nu^{\mu}=\Delta^{\mu\nu}j_{\nu}\;.\label{eq:j-decomp}
\end{equation}

Using these tensor decompositions in Eq.(\ref{eq:em_conserve}, \ref{eq:current_conserve}),
we obtain
\begin{eqnarray}
\dot{n}+n\theta+\partial_{\mu}\nu^{\mu} & = & 0\;,\nonumber \\
\dot{\epsilon}+(\epsilon+P+\Pi)\theta-\partial_{\mu}h^{\mu}-h^{\mu}\dot{u}_{\mu}-\pi^{\mu\nu}\partial_{\mu}u_{\nu} & = & 0\;,\nonumber \\
\Delta^{\mu\alpha}\partial_{\alpha}(P+\Pi)-(\epsilon+P)\dot{u}^{\mu}-\Pi\dot{u}^{\mu}-\Delta^{\mu\alpha}\dot{h}_{\alpha}\nonumber \\
-h^{\mu}\theta-h^{\alpha}\partial_{\alpha}u^{\mu}-\Delta^{\mu\alpha}\partial^{\beta}\pi_{\alpha\beta} & = & 0\;,\label{eq:eom-01}
\end{eqnarray}
which represent the conservation of charges and energy, the acceleration
of the fluid, respectivley. Here $\dot{A}$ denotes $\frac{d}{d\tau}A=u^{\mu}\partial_{\mu}A$.
The $\theta=\partial^{\mu}u_{\mu}$ is the expansion scalar. 

For the sake of simplicity, people usually consider two special frames
for the fluid dynamics. The first one is the Landau frame or energy
frame \cite{Landau}. In this frame, the velocity of the fluid $u_{E}^{\mu}$
is defined to describe the energy flow 
\begin{equation}
u_{E}^{\mu}=\frac{T_{\nu}^{\mu}u_{E}^{\nu}}{\sqrt{u_{E}^{\alpha}T_{\alpha}^{\beta}T_{\beta\gamma}u_{E}^{\gamma}}}=\frac{1}{\epsilon}T_{\nu}^{\mu}u_{E}^{\nu}\;,
\end{equation}
which is related to $u^{\mu}$ and $h^{\mu}$ via
\begin{equation}
u_{E}^{\mu}=u^{\mu}+\frac{1}{\epsilon+P}h^{\mu}\;,
\end{equation}
Then the heat flux current $h^{\mu}$ vanishes. The second frame is
the Eckart frame \cite{Eckart1940}. In this frame, the velocity $u_{N}^{\mu}$
is used to describe the charge flow
\begin{equation}
u_{N}^{\mu}=\frac{j^{\mu}}{\sqrt{j^{\mu}j_{\mu}}}\;,
\end{equation}
with
\[
u_{N}^{\mu}=u^{\mu}+\frac{1}{n}\nu^{\mu}\;.
\]
Then the diffusion current vanishes. Throughout the thesis, only the
Landau frame will be used. For convenience, the following quantities
is also used 
\begin{equation}
q^{\mu}=h^{\mu}-\frac{n}{\epsilon+P}\nu^{\mu}\;,\label{eq:qmu-01}
\end{equation}
which is $h^{\mu}$ in Eckart frame and $-\frac{n}{\epsilon+P}\nu^{\mu}$
in Landau frame.

For a given fluid (the velocity $u^{\mu}$ is fixed), there are 15
unknown parameters in the equations of fluid dynamics. However, the
choice of the frame does not reduce the number of the parameters since
in Eckart frame the $\nu^{\mu}=0$ and in Landau frame $h^{\mu}=0$,
then the velocity $u^{\mu}$ is not fixed.

\section{Navier-Stokes approximation\label{sub:Navier-Stokes-equations}}

\subsection{Equilibrium state}

The first law of thermodynamics read
\begin{eqnarray}
d\epsilon & = & Tds+\mu dn\;,\nonumber \\
dP & = & sdT+nd\mu\;,\label{eq:Gibbs-01}
\end{eqnarray}
where $s$ is the entropy density, $T$ is the temperature and $\mu$
is the chemical potential. The entropy density is given by the Durham-Gibbs
relation 
\begin{equation}
s=\frac{\epsilon+P}{T}-\alpha n\;,
\end{equation}
where $\alpha=\mu/T$. Rewriting Eq.(\ref{eq:Gibbs-01}) with the
Durham-Gibbs relation, the following equation is obtained 
\begin{eqnarray}
d(P\beta) & = & nd\alpha-\epsilon d\beta\;.\label{eq:Gibbs-2}
\end{eqnarray}
By introducing the new variable $\beta^{\mu}$ 
\[
\beta_{\mu}=\beta u_{\mu}\;,
\]
with $\beta=T^{-1}$, the Eq. (\ref{eq:Gibbs-2}) becomes 
\begin{equation}
d(P\beta^{\mu})=j_{0}^{\mu}d\alpha-T_{0}^{\lambda\mu}d\beta_{\lambda}\;,
\end{equation}
where $j_{0}^{\mu}=nu^{\mu}$ and $T_{0}^{\mu\nu}=\epsilon u^{\mu}u^{\nu}-P\Delta^{\mu\nu}$
are quantities in the ideal fluid. The entropy flow is then 
\begin{eqnarray}
s_{0}^{\mu} & = & P\beta^{\mu}+T_{0}^{\lambda\mu}\beta_{\lambda}-\alpha j_{0}^{\mu}\;.\label{eq:entropy-flow-equil-01}
\end{eqnarray}
The differential of the entropy flow reads 
\begin{eqnarray}
ds_{0}^{\mu} & = & -\alpha dj_{0}^{\mu}+\beta_{\lambda}dT_{0}^{\lambda\mu}\;.\label{eq:entropy-equil-01}
\end{eqnarray}

\subsection{Off equilibrium state}

It is straightforward to assume that in an off equilibrium state Eq.(\ref{eq:entropy-equil-01})
becomes
\begin{equation}
ds^{\mu}=-\alpha dj^{\mu}+\beta_{\lambda}dT^{\lambda\mu}\;.\label{eq:entropy-d-01}
\end{equation}
if the system is in a state near the equilibrium one. The entropy
flow (\ref{eq:entropy-flow-equil-01}) becomes 
\begin{equation}
s^{\mu}=P\beta^{\mu}+T^{\lambda\mu}\beta_{\lambda}-\alpha j^{\mu}-Q^{\mu}\;,\label{eq:entropy-flow-covar}
\end{equation}
where $Q^{\mu}$ is the high order deviations $j^{\mu}-j_{0}^{\mu}$,
and $T^{\mu\nu}-T_{0}^{\mu\nu}$. Under the infinitesimal changes
in $\alpha$ and $\beta$, 
\[
\alpha\rightarrow\alpha^{\prime}=\alpha+\delta\alpha\;,\qquad\beta_{\lambda}\rightarrow\beta_{\lambda}^{\prime}=\beta_{\lambda}+\delta\beta_{\lambda}\;,
\]
the change of $Q^{\mu}$ is 
\begin{equation}
Q^{\mu}-Q^{\mu\prime}=(j^{\mu}-j_{0}^{\mu})\delta\alpha-(T^{\lambda\mu}-T_{0}^{\lambda\mu})\delta\beta_{\lambda}\;.\label{eq:delta_Q}
\end{equation}
Recalling Eqs. (\ref{eq:T-decomp}, \ref{eq:j-decomp}), we assume
that the charge and the energy densities do not change in the off-equilibrium
state, 
\begin{equation}
u_{\mu}(j^{\mu}-j_{0}^{\mu})=u_{\mu}u_{\nu}(T^{\mu\nu}-T_{0}^{\mu\nu})=0\;,\label{eq:dj-dT-01}
\end{equation}
but the entropy density flow changes in the way

\begin{equation}
u_{\mu}(s^{\mu}-s_{0}^{\mu})=-u_{\mu}Q^{\mu}\;,
\end{equation}
In principle the temperature and chemical potential will be changed
if taking the contributions from the second order theory into account,
i.e. they are not invariant under the transformation of $u^{\mu}$.
In an off-equilibrium state, the temperature and chemical potential
are not global, i.e. they are the functions of the location $x$. 

Taking the additional term $Q^{\mu}$ into account, Eq. (\ref{eq:entropy-d-01})
and (\ref{eq:entropy-flow-covar}) become 
\begin{eqnarray}
s^{\mu} & = & \frac{s}{n}j^{\mu}+\beta q^{\mu}-Q^{\mu}\;,\label{eq:entropy_q_1}
\end{eqnarray}
and 
\begin{eqnarray}
\partial_{\mu}s^{\mu} & = & -(j^{\mu}-j_{0}^{\mu})\partial_{\mu}\alpha+(T^{\mu\lambda}-T_{0}^{\lambda\mu})\partial_{\mu}\beta_{\lambda}-\partial_{\mu}Q^{\mu}\;,\label{eq:entorpy-d-02}
\end{eqnarray}
where $q^{\mu}$ is given in Eq.(\ref{eq:qmu-01}).

\subsection{Entropy principle}

Taking Eqs. (\ref{eq:T-decomp}, \ref{eq:j-decomp}) in Eq.(\ref{eq:entorpy-d-02})
and neglecting $Q^{\mu}$, the entropy production rate reads 
\begin{eqnarray}
\partial_{\mu}s^{\mu} & = & -\nu^{\mu}\partial_{\mu}\alpha-\frac{\Pi}{T}\partial_{\mu}u^{\mu}+\pi^{\mu\nu}\frac{1}{T}\partial_{\mu}u_{\nu}\;.
\end{eqnarray}
The second law of thermodynamics $\partial_{\mu}s^{\mu}\geq0$ requires
$\nu^{\mu},\Pi,\pi^{\mu\nu}$ be in the form 
\begin{eqnarray}
\pi_{NS}^{\mu\nu} & = & 2\eta\partial^{<\mu}u^{\nu>}\;,\nonumber \\
\Pi_{NS} & = & -\zeta\theta\;,\nonumber \\
\nu_{NS}^{\mu} & = & \kappa T\Delta^{\mu\nu}\partial_{\nu}\alpha\;,\label{eq:NS-equation-01}
\end{eqnarray}
which is called the Navier-Stokes (NS) approximation. Here $\eta,\zeta,$
and $\kappa$ are shear viscosity, bulk viscosity and heat conductivity,
respectively. Note that in the metric convention $g^{\mu\nu}=\textrm{diag}\{+,-,-,-\}$,
$\nu^{\mu}\partial_{\mu}\alpha$ must be negative, 
\begin{equation}
\nu^{\mu}\partial_{\mu}\alpha=-\kappa T(\nabla\alpha)^{2}\;.
\end{equation}

\subsection{Simple Israel-Stewart theory\label{sub:Israel-Stewart-equations}}

In Ref. \cite{Israel:1979wp}, the authors suggest that in phenomenology
the additional term $Q^{\mu}$ should include all the irreducible
quantities in the first order theory 
\begin{equation}
TQ^{\mu}=\frac{1}{2}u^{\mu}(\beta_{0}\Pi^{2}+\beta_{1}q^{2}+\beta_{2}\pi^{2})-\alpha_{0}\Pi q^{\mu}-\alpha_{1}\pi^{\mu\lambda}q_{\lambda}\;,
\end{equation}
where $\beta_{i}$ and $\alpha_{j}$ are constants with $j=0,1,2$
and $i=0,1$. 

The entropy production rate becomes \cite{Israel:1979wp}, 
\begin{eqnarray}
\partial_{\mu}s^{\mu} & = & \beta\Pi[-\theta-\dot{\beta}_{0}\Pi-2\beta_{0}\dot{\Pi}-\beta_{0}\Pi T\partial_{\alpha}(\beta u^{\alpha})+a_{1}T\partial_{\mu}(\alpha_{0}\beta)\nu^{\mu}+\alpha_{0}\partial_{\mu}\nu^{\mu}]\nonumber \\
 &  & +\beta\pi^{\mu\nu}[\partial_{\mu}u_{\nu}-\dot{\beta}_{2}\pi_{\mu\nu}-2\beta_{2}\dot{\pi}_{\mu\nu}-\beta_{2}T\partial_{\alpha}(\beta u^{\alpha})\pi_{\mu\nu}+a_{3}T\partial_{\mu}(\alpha_{1}\beta)\nu_{\nu}\nonumber \\
 &  & +\alpha_{1}\partial_{\mu}\nu_{\nu}]+\beta\nu^{\mu}[-T\partial_{\mu}\alpha-T\dot{\beta_{1}}\nu_{\mu}-2\beta_{1}\dot{\nu}_{\mu}-\beta_{1}T\partial_{\alpha}(\beta u^{\alpha})\nu_{\mu}\nonumber \\
 &  & +a_{2}\Pi T\partial_{\mu}(\alpha_{0}\beta)+\alpha_{0}\partial_{\mu}\Pi+a_{4}T\pi_{\mu}^{\nu}\partial_{\nu}(\alpha_{1}\beta)+\alpha_{1}\partial_{\nu}\pi_{\mu}^{\nu}]\label{eq:div-en-dens}
\end{eqnarray}
where the new coefficients $a_{i}$ ($i=1,2,3,4$) satisfy $a_{1}+a_{2}=1$
and $a_{3}+a_{4}=1$. Thus the general form for the irreducible quantities
are 
\begin{eqnarray}
\Pi & = & \zeta[-\theta-2\beta_{0}\dot{\Pi}+\alpha_{0}\partial_{\mu}\nu^{\mu}+a_{1}\alpha_{0}\nu^{\mu}\dot{u}_{\mu}]\;,\nonumber \\
\pi_{\mu\nu} & = & 2\eta[\partial_{<\mu}u_{\nu>}-2\beta_{2}\dot{\pi}_{<\mu\nu>}+a_{3}\alpha_{_{1}}\dot{u}_{<\mu}\nu_{\nu>}+\alpha_{1}\partial_{<\mu}\nu_{\nu>}]\;,\nonumber \\
\nu_{\mu} & = & -\kappa\Delta_{\mu}^{\nu}[-T\partial_{\nu}\alpha-2\beta_{1}\dot{\nu}_{\nu}+a_{2}\alpha_{0}\Pi\dot{u}_{\nu}+\alpha_{0}\partial_{\nu}\Pi\nonumber \\
 &  & +a_{4}\alpha_{1}\pi_{\nu}^{\alpha}\dot{u}_{\alpha}+\alpha_{1}\partial_{\alpha}\pi_{\nu}^{\alpha}]\;,\label{eq:2nd-equation-phe}
\end{eqnarray}
which are the called the simplest Israel-Stewart (IS) equations. Here
the approximation $\partial_{\mu}\beta\approx\beta\dot{u}_{\mu}$
has been used, and the coefficients $\beta_{i}$ and $\alpha_{i}$
are assumed independent of space-time. The complete IS equations have
been discussed in Ref. \cite{Betz:2008me,Betz:2009zz,Betz:2010cx}.

\section{Relativistic Boltzmann equation \label{sec:BE}}

The fluid dynamics is closely related to kinetic theory which is based
on the Boltzmann equation. The relativistic Boltzmann equation describes
the time evolution of the single particle distribution function $f(x,p)$,
which is based on the following assumptions \cite{Csernai}:
\begin{itemize}
\item Only two-particle collisions, or the so-called binary collisions are
considered.
\item {}``Sto$\beta$zahlansatz'' collision number ansatz, i.e., number
density of binary collisions at $x$ is proportional to $f(x,p_{1})\times f(x,p_{2})$.
\item $f(x,p)$ is a smoothly varying function compared to the mean free
path $l_{mfp}$.
\end{itemize}
For example, in the $\phi^{4}$ interaction the size of the collision
region $l_{sc}\sim\sigma^{1/2}$, where $\sigma$ is the cross section,
needs to be much smaller than the mean free path $l_{mfp}\sim1/(n\sigma)$.
In that case, the particles interact in a very small region and travel
freely in a long distance. For $l_{sc}/l_{mfp}\ll1$, the semi-classical
treatment of quasi-particle collisions in Boltzmann equations will
be allowed.

The relativistic Boltzmann equation can be written as 
\begin{equation}
\frac{df}{dt}=\frac{p^{\mu}}{E_{p}}\partial_{\mu}f(x,p)=C[f]\;,\label{eq:Boltzmann-eq-01}
\end{equation}
where $p^{\mu}$ and \textbf{$E_{p}$} are the four-momentum and the
energy of the particles respectively, and $C[f]$ is the collision
term. With the particle velocity as $\mathbf{v}_{p}=\mathbf{p}/E_{p}$,
the relativistic Boltzmann equation is in the same form as the non-relativistic
one
\begin{equation}
\frac{d}{dt}f=\partial_{t}f+\mathbf{v}_{p}\cdot\nabla f=C[f]\;.
\end{equation}
The collision term $C[f]$ describes the change of the distribution
function $f$ in a given time $dt$. For example, considering a two
particles scattering $12\rightarrow3p$, the collision term is given
by 
\begin{eqnarray}
C[f] & = & \frac{1}{2}\int_{123}d\Gamma_{12\rightarrow3p}[f_{1}f_{2}(1\pm f_{3})(1\pm f_{p})-(1\pm f_{1})(1\pm f_{2})f_{3}f_{p}]\;,
\end{eqnarray}
where 
\begin{equation}
d\Gamma_{12\rightarrow3p}=\frac{1}{2E_{p}}\left|T(p,\mathbf{k})\right|^{2}\underset{i=1}{\overset{3}{\prod}}\frac{d^{3}k_{i}}{(2\pi)^{3}(2E_{i})}(2\pi)^{4}\delta(k_{1}+k_{2}-k_{3}-p)\;,
\end{equation}
with $T(p,\mathbf{k})$ the amplitude of the scattering. More details
on the collision term $C[f]$ will be given in Sec. \ref{sec:Order-expansion}. 

In the local rest frame, the charge density and the number current
are given by 
\begin{eqnarray}
n & = & \int d^{3}pf(x,p)\;,\\
\overrightarrow{\mathbf{j}\,} & = & \int d^{3}p\overrightarrow{\mathbf{p}}f(x,p)\;.
\end{eqnarray}
 In a Lorentz boost frame, the above can be written in a compact form
as a Lorentz vector 
\begin{equation}
j^{\mu}=nu^{\mu}=\int_{p}p^{\mu}f(x,p)\;,
\end{equation}
where $\int_{p}=\int\frac{d^{3}p}{(2\pi)^{3}E_{p}}\;,$ and the four-velocity
vector is 
\begin{equation}
u^{\mu}=\gamma(1,\overrightarrow{\mathbf{v}})\;,
\end{equation}
with $\gamma=\sqrt{1-v^{2}}$ and $\overrightarrow{\mathbf{v}}$ is
fluid 3-velocity.

The energy-momentum tensor can be expressed in terms of $f$, 
\begin{equation}
T^{\mu\nu}=\int_{p}p^{\mu}p^{\nu}f(x,p)\;.
\end{equation}
The tensor decomposition of $T^{\mu\nu}$ and $j^{\mu}$ gives 
\begin{eqnarray}
n & = & u_{\mu}j^{\mu}=\int d^{3}pf(x,p)\;,\nonumber \\
\epsilon & = & T^{\mu\nu}u_{\mu}u_{\nu}=\int d^{3}pE_{p}f(x,p)\;,\nonumber \\
P & = & -\frac{1}{3}\Delta^{\mu\nu}T_{\mu\nu}=-\frac{1}{3}\int_{p}\left|\overrightarrow{\mathbf{p}}\right|^{2}f(x,p)\;,
\end{eqnarray}
If the particles are massless, we have the equation of state for the
prefect or conformal fluid is obtained 
\begin{equation}
\epsilon=3P\;.
\end{equation}

In relativistic fluid dynamics, the distribution function $f_{0}(x,p)$
in an equilibrium state is 
\begin{equation}
f_{0}(x,p)=\frac{A_{0}}{e^{\beta(u\cdot p)-\alpha}-a}\;,\label{eq:f0-01}
\end{equation}
where $a=0,\pm1$ are for Boltzmann, Bose and Fermi distributions,
respectively and 
\[
A_{0}=\frac{d_{g}}{(2\pi)^{3}}\;,
\]
with $d_{g}$ the particle's degree of freedom.

\subsection{Juttner distribution}

The relativistic Boltzmann distribution is also called the Juttner
distribution. For the sake of simplicity, all thermal quantities are
evaluated in the local rest frame. The number density is given by
\begin{equation}
n=A_{0}e^{\mu/T}4\pi m^{2}TK_{2}(z)\;,\label{eq:number_density_f_Jutter_01}
\end{equation}
where $K_{2}$ is the Modified Bessel function of the second kind
and $z=\frac{m}{T}$. The pressure is 
\begin{eqnarray}
P & = & A_{0}e^{\mu/T}4\pi m^{2}T^{2}K_{2}(z)\;.\label{eq:pressure_f_Jutter_01}
\end{eqnarray}
Then the Clapeyron equation $PV=nRT$ is obtained from Eqs.(\ref{eq:number_density_f_Jutter_01},
\ref{eq:pressure_f_Jutter_01}) 
\begin{equation}
P=nT\;,
\end{equation}
where $k_{B}=1$. The energy density is 
\begin{eqnarray}
\epsilon & = & A_{0}e^{\mu/T}4\pi[3m^{2}T^{2}K_{2}(z)+m^{3}TK_{1}(z)]\nonumber \\
 & = & 3nT+nm\frac{K_{1}(z)}{K_{2}(z)}\;,
\end{eqnarray}
which is identical to the relativistic Boltzmann-Gibbs statistics.

\subsection{Conservation laws}

The Boltzmann equation is used to describe a system in or close to
an equilibrium state. There are conserved quantities because of the
time reverse symmetry in the microscopic state. Assuming that the
quantity $\Psi$ is conserved during the binary collision $12\rightarrow3p$
:
\begin{equation}
\Psi_{1}+\Psi_{2}=\Psi_{3}+\Psi_{p}\;,\label{eq:conserved_0}
\end{equation}
then the macroscopic conservation law will be 
\begin{eqnarray}
 &  & \partial_{\mu}\left[\int_{p}\Psi_{1}p^{\mu}f(x,p)\right]\nonumber \\
 & = & \int d^{3}p\Psi_{1}C[f]\nonumber \\
 & = & \frac{1}{2}\int_{123}d\Gamma d^{3}p\Psi_{1}[f_{1}f_{2}(1\pm f_{3})(1\pm f_{p})-(1\pm f_{1})(1\pm f_{2})f_{3}f_{p}]\nonumber \\
 & = & \frac{1}{4}\int_{123}d\Gamma d^{3}p[\Psi_{1}+\Psi_{2}-\Psi_{3}-\Psi_{p}][f_{1}f_{2}(1\pm f_{3})(1\pm f_{p})]\nonumber \\
 & = & 0\;.
\end{eqnarray}
Choosing $\Psi=1,$ and $p^{\mu}$, the conservation equations for
the charge density and energy-momentum are obtained 
\begin{eqnarray}
\partial_{\mu}j^{\mu} & = & \int_{p}p^{\mu}\partial_{\mu}f(x,p)=0\;,\nonumber \\
\partial_{\mu}T^{\mu\nu} & = & \int_{p}p^{\mu}p^{\nu}\partial_{\mu}f(x,p)=0\;.
\end{eqnarray}

The conservation laws follow Eq.(\ref{eq:conserved_0}). If Eq.(\ref{eq:conserved_0})
is not fulfilled or the time reverse symmetry is broken (i.e. the
amplitude of the scattering $12\rightarrow3p$ is not equal to that
of the inverse scattering) , the charge and energy-momentum will not
be conserved. For instance, considering the triangle anomalies (i.e.
the right-hand quarks will become to left-handed quarks if the topological
charge is $1$), the time reverse symmetry is broken; therefore the
quark number is not conserved due to 
\begin{equation}
\partial_{\mu}j^{\mu}=C\mathbf{E}\cdot\mathbf{B}\;,
\end{equation}
where $\mathbf{E}$ and $\mathbf{B}$ are the electric and magnetic
field, respectively. Here $C$ is the constant determined by the quantum
field theory. More discussions on this topic will be presented in
Chap. \ref{chap:vorticity}.

\section{Kadanoff-Baym equation\label{sec:Covariant-kinetic-equations}}

The Boltzmann equation can be derived from the Kadanoff-Baym (KB)
equation in quantum field theory via gradient expansion. In this section
we will derive the Kadanoff-Baym equation from Dyson-Schwinger (DS)
equation in the closed-time-path formalism. Then we will show how
the Boltzmann equation can be derived from the KB equation. There
are many references about the closed-time-path formalism and derivation
of the Boltzmann equation from the KB equation, see e.g. Ref. \cite{Chou:1984es,Heinz:1984yq,Heinz:1985qe,Heinz:1985vf,Elze:1986qd,Elze:1986hq,Vasak:1987um,Elze:1987ii,Elze:1989un,Schonhofen:1994zf,Wang:2001dm,Wang:2002qe}.

\begin{figure}[H]
\begin{centering}
\includegraphics[scale=0.4]{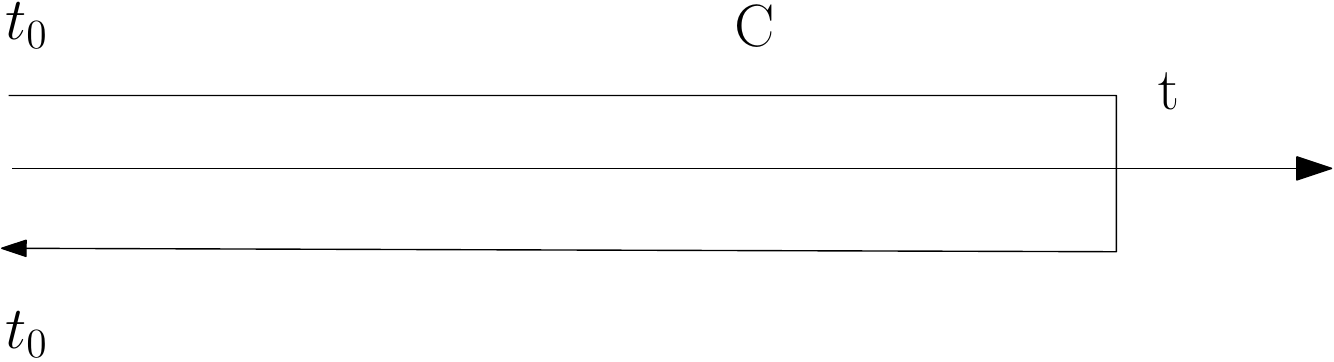}
\par\end{centering}

\caption{Closed time path.\label{fig:Closed-time-path.}}
\end{figure}

We consider a fermionic system in quantum field theory, the two-point
Green function for a fermion is defined as 
\begin{equation}
iG(x_{1},x_{2})\equiv iG(1,2)=\left\langle T_{t}[\psi_{H}(1)\overline{\psi}_{H}(2)]\right\rangle \;,
\end{equation}
where the operator $T_{t}$ is the causal time-ordering operator and
$H$ denotes the Heisenberg picture. In order to describe a non-equilibrium
system, the closed time path formalism is proposed by Keldysh and
Schwinger, see Fig. \ref{fig:Closed-time-path.}. Then the two-point
Green function on the contour $C$ is written as 
\begin{eqnarray}
iG(1,2) & = & \left\langle T_{C}[\psi_{H}(1)\overline{\psi}_{H}(2)]\right\rangle \nonumber \\
 & = & \left\langle T_{C}\left\{ \exp\left[i\int_{C}d^{4}x\mathcal{L}_{I}\right]\psi_{I}(1)\overline{\psi}_{I}(2)\right\} \right\rangle \;,
\end{eqnarray}
where $I$ denotes the interaction picture and $\mathcal{L}_{I}$
is the interacting part of the Lagrangian. The retarded and advanced
Green functions are given by 
\begin{equation}
G^{R}(1,2)\equiv G(1,2)-G^{<}(1,2)\;,\qquad G^{A}(1,2)\equiv G(1,2)-G^{>}(1,2)\;,
\end{equation}
where Feynman propagator $G(1,2)$ is defined by 
\[
G(1,2)=\theta_{C}(t_{1}-t_{2})G^{>}(1,2)+\theta_{C}(t_{2}-t_{1})G^{<}(1,2)
\]
 with $\theta_{c}(x)$ the Heaviside step function in the closed time
path $C$. 

The Dyson-Schwinger equation reads 
\begin{equation}
G^{-1}(1,2)=G_{0}^{-1}(1,2)+\Sigma(1,2)\;,
\end{equation}
where $G_{0}$ is the Green function for the free particles and $\Sigma$
is the self-energy (normally with an $i=\sqrt{-1}$ as $i\Sigma$)
which can be also written as 
\begin{equation}
\Sigma(1,2)=\Sigma^{\delta}(1,2)\delta(t_{1}-t_{2})+\theta_{C}(t_{1}-t_{2})\Sigma^{>}(1,2)+\theta_{C}(t_{2}-t_{1})\Sigma^{<}(1,2)\;.
\end{equation}
The solution of the Dyson-Schwinger equation can be formally written
as 
\begin{eqnarray}
G(1,2) & = & G_{0}(1,2)-\int_{C}d4\int_{C}d3G_{0}(1,4)\Sigma(4,3)G(3,2)\;,\nonumber \\
G(1,2) & = & G_{0}(1,2)-\int_{C}d4\int_{C}d3G(1,4)\Sigma(4,3)G_{0}(3,2)\;,\label{eq:Green-DS-1}
\end{eqnarray}
Using the Dirac equation and Eq.(\ref{eq:Green-DS-1}), in the case
of $t_{1}<t_{2}$, $G^{>,<}$ fulfill the following equations 
\begin{eqnarray}
(i\gamma_{\mu}\partial_{1}^{\mu}-m)G^{<}(1,2) & = & -\int_{C}d3\delta(1,3)\Sigma^{\delta}(1,3)G^{<}(1,2)\nonumber \\
 &  & +\int_{t_{0}}^{t_{1}}d3[\Sigma^{<}(1,3)-\Sigma^{>}(1,3)]G^{<}(3,2)\nonumber \\
 &  & +\int_{t_{0}}^{t_{2}}d3\Sigma^{<}(1,3)[G^{>}(3,2)-G^{<}(3,2)]\;,\nonumber \\
(-i\gamma_{\mu}\partial_{2}^{\mu}-m)G^{<}(1,2) & = & -\int_{C}d3\delta(1,3)\Sigma^{\delta}(1,3)G^{<}(1,2)\nonumber \\
 &  & +\int_{t_{0}}^{t_{1}}d3[G^{<}(1,3)-G^{>}(1,3)]\Sigma^{<}(3,2)\nonumber \\
 &  & +\int_{t_{0}}^{t_{2}}d3G^{<}(1,3)[\Sigma^{>}(3,2)-\Sigma^{<}(3,2)]\;,\label{eq:Green-<-1}
\end{eqnarray}
Note here $\Sigma^{\delta}$ is the self-energy depending on a single
time corresponding to the contribution from the mean field approximation
\cite{Schonhofen:1994zf}. Without loss of the generality, the $\Sigma^{\delta}$
can be neglected in normal situations. 

Taking traces and the Fourier transformation for Eqs.(\ref{eq:Green-<-1}),
we obtain 
\begin{eqnarray*}
 &  & \textrm{Tr}[(i\gamma_{\mu}\partial_{1}^{\mu}-m)G^{<}(X,q)]\\
 & = & \int_{0}^{\infty}dy^{\prime}dq_{1}e^{i(q-q_{1})y^{\prime}}\textrm{Tr}\{[\Sigma^{<}(X,q_{1})-\Sigma^{>}(X,q_{1})]G^{<}(X,q)\}\\
 &  & +\int_{-\infty}^{0}dy^{\prime}dq_{1}e^{i(q-q_{1})y^{\prime}}\textrm{Tr}\{\Sigma^{<}(X,q)[G^{>}(X,q_{2})-G^{<}(X,q_{2})]\}\;,\\
 &  & \textrm{Tr}[(-i\gamma_{\mu}\partial_{2}^{\mu}-m)G^{<}(X,q)]\\
 & = & \int_{0}^{\infty}dy^{\prime}dq_{1}e^{i(q-q_{1})y^{\prime}}\textrm{Tr}\{[G^{<}(X,q_{1})-G^{>}(X,q_{1})]\Sigma^{<}(X,q)\}\\
 &  & +\int_{-\infty}^{0}dy^{\prime}dq_{1}e^{i(q-q_{1})y^{\prime}}\textrm{Tr}\{G^{<}(X,q)[\Sigma^{>}(X,q_{2})-\Sigma^{<}(X,q_{2})]\}\;,
\end{eqnarray*}
where the gradient expansion variables 
\begin{alignat*}{2}
 & X\equiv X_{12}=\frac{x_{1}+x_{2}}{2}\;, & \qquad & y\equiv y_{12}=x_{1}-x_{2}\;,
\end{alignat*}
have been used. Combining the above two equations, we derive the covariant
Kadanoff and Baym equation \cite{Kadanoff1962}
\begin{equation}
i\partial_{X}^{\mu}\textrm{Tr}[\gamma_{\mu}G^{<}(X,q)]=\textrm{Tr}[G^{>}\Sigma^{<}-G^{<}\Sigma^{>}](X,q)\;.\label{eq:KB-equation}
\end{equation}

It is known that the two-point Green function of free particles must
be proportional to the distribution function $f_{0}$ of an equilibrium
state at the finite temperature. Therefore, the generalization to
the Green function of the interacting particles is straightforward.
One needs to use the complete distribution function $f$ to replace
$f_{0}$ in Eq.(\ref{eq:KB-equation}). On the left-hand side of KB
equation, it will give such a term as $p^{\mu}\partial_{\mu}f$ while
on the right-hand side, the self-energy $\Sigma^{<,>}$ are associated
with the amplitude of the interaction. Finally we end up with the
Boltzmann equation (\ref{eq:Boltzmann-eq-01}) 
\[
\frac{d}{dt}f=p^{\mu}\partial_{\mu}f=\mathcal{C}[f]\;.
\]

\section{Complete second order theory\label{chap:complete}}

In hydrodynamics, in a small departure from equilibrium, it is assumed
that $j^{\mu}$ and $T^{\mu\nu}$ can still provide a complete description
of the non-eqilibrium states. Then there are 14 independent parameters.
Correspondingly, in microscopic distribution function, a departure
from equilibrium state can also be characterized by 14 parameters
in $y$, the exponent in the distribution function $1/(\mathrm{Exp}(y)-a)$.
The comparison between the macroscopic and microscopic approach will
give part of constraints on these parameters. The method is called
Grad 14 moments approximation \cite{Israel:1979wp}.

\subsection{First order theory}

In the kinetic theory, the entropy flow is defined as 
\begin{eqnarray}
s^{\mu} & = & -\int_{p}p^{\mu}\psi(f)\;,\label{eq:entropy_kin_1}
\end{eqnarray}
where
\begin{eqnarray}
\psi(f) & = & f\ln(A_{0}^{-1}f)-a^{-1}A_{0}\Delta\ln\Delta\;,
\end{eqnarray}
with
\begin{eqnarray}
\Delta & = & 1+af\;,\nonumber \\
y(f) & \equiv & \psi^{\prime}(f)=\ln[A_{0}^{-1}f(x,p)/\Delta(x,p)]\;.
\end{eqnarray}

In an equilibrium state, we have 
\begin{equation}
y_{0}\equiv y(f_{0})=\alpha-\beta_{\mu}p^{\mu}\;.
\end{equation}
Since there are 14 independent parameters in $j^{\mu}$ and $T^{\mu\nu}$,
the derivation from equilibrium is given by $y-y_{0}$which can be
decomposed into
\[
y-y_{0}=\epsilon(x)-\epsilon_{\mu}(x)p^{\mu}+\epsilon_{\mu\nu}(x)p^{\mu}p^{\nu}\;,
\]
o
\begin{equation}
y=(\epsilon+\alpha)-(\beta_{\mu}+\epsilon_{\mu})p^{\mu}+\epsilon_{\mu\nu}p^{\mu}p^{\nu}\;,\label{eq:y_expansion}
\end{equation}
where $\epsilon,\epsilon_{\mu},\epsilon_{\mu\nu}$ are 14 small parameters
in the first order (with $\epsilon_{\mu}^{\mu}=0$). 

By using the $n$-th and auxiliary moments defined in Eq.(\ref{eq:IJ-full}),
the infinitesimal changes of $j^{\mu}$, $T^{\mu\nu}$ and the $3$-rd
moment $F^{\mu\nu\lambda}$ with the full distribution $f(x,p)$ under
an arbitrary variation of $f(x,p)$ are
\begin{eqnarray}
\delta j^{\mu} & = & j^{\mu}-j_{0}^{\mu}=\epsilon J^{\mu}-\epsilon_{\nu}J^{\mu\nu}+\epsilon_{\lambda\nu}J^{\mu\nu\lambda}\;,\nonumber \\
\delta T^{\mu\nu} & = & T^{\mu\nu}-T_{0}^{\mu\nu}=\epsilon J^{\mu\nu}-\epsilon_{\lambda}J^{\mu\nu\lambda}+\epsilon_{\lambda\rho}J^{\mu\nu\lambda\rho}\;,\nonumber \\
\delta F^{\mu\nu\lambda} & = & F^{\mu\nu\lambda}[f]-F^{\mu\nu\lambda}[f_{0}]=\epsilon J^{\mu\nu\lambda}-\epsilon_{\rho}J^{\mu\nu\lambda\rho}+\epsilon_{\rho\sigma}J^{\mu\nu\lambda\rho\sigma}\;.\label{eq:delta_j_T_F}
\end{eqnarray}

It is reasonable to assume that the charge and energy density in an
off equilibrium state should be the same as those in an equilibrium
state,
\begin{eqnarray*}
n & \equiv & \int_{p}(u\cdot p)f=n_{0}=\int_{p}(u\cdot p)f_{0}\;,\\
\epsilon & \equiv & \int_{p}(u\cdot p)^{2}f=\epsilon_{0}=\int_{p}(u\cdot p)^{2}f_{0}\;.
\end{eqnarray*}
Thus, the equation of state will be the same as before 
\[
P(\epsilon,n)=P_{0}(\epsilon_{0},n_{0})\;.
\]
By construction, Eq.(\ref{eq:dj-dT-01}) becomes 
\begin{equation}
u_{\mu}\delta j^{\mu}=u_{\mu}u_{\nu}\delta T^{\mu\nu}=0\;,\label{eq:constraint_j_T_1}
\end{equation}
or
\begin{eqnarray}
\epsilon & = & -\left(m^{2}+4\frac{J_{31}J_{30}-J_{41}J_{20}}{D_{20}}\right)\epsilon_{**}\;,\nonumber \\
\epsilon_{*} & = & 4\frac{J_{31}J_{20}-J_{41}J_{10}}{D_{20}}\epsilon_{**}\;,
\end{eqnarray}
where Eqs.(\ref{eq:delta_j_T_F}, \ref{eq:ide-I-J-1}) have been used
and $A_{*}$ and $A_{**}$ denote $A^{\mu}u_{\mu}$ and $A^{\mu\nu}u_{\mu}u_{\nu}$.

Using Eq.(\ref{eq:delta_j_T_F}), all quantities in the first order
theory can be written in the terms of $\epsilon_{\mu}$ and $\epsilon_{\mu\nu}$.
The bulk pressure, shear stress tensor and heat flow are given by
\begin{eqnarray}
\Pi & = & -\frac{1}{3}\Delta_{\mu\nu}\delta T^{\mu\nu}=\frac{4}{3}J_{42}\Omega\epsilon_{**}\;,\nonumber \\
\pi^{\mu\nu} & = & \delta T^{<\mu\nu>}=2J_{42}\epsilon^{<\lambda\rho>}\;,\nonumber \\
q^{\mu} & = & \Delta_{\nu}^{\rho}u_{\mu}\delta T^{\mu\nu}-\frac{\epsilon+P}{n}\Delta^{\rho\mu}\delta j_{\mu}=-2J_{21}\Lambda\epsilon_{\alpha*}\Delta^{\mu\alpha}\;,\label{eq:PI-pi-q}
\end{eqnarray}
where $q^{\mu}$ is defined in Eq.(\ref{eq:qmu-01}) and 
\begin{eqnarray}
\Lambda & = & \frac{D_{31}}{J_{21}^{2}}\;,\nonumber \\
\Omega & = & -3\left(\frac{\partial\ln I_{31}}{\partial\ln I_{30}}\right)+5\nonumber \\
 & = & -3\frac{J_{31}(J_{21}J_{30}-J_{20}J_{31})-J_{41}(J_{21}J_{20}-J_{10}J_{31})}{J_{42}D_{20}}+5\;.
\end{eqnarray}
Equivalently, $\epsilon$, $\epsilon_{\mu}$ and $\epsilon_{\mu\nu}$
in the Eq.(\ref{eq:PI-pi-q}) can be expressed in terms of $\Pi,\;\pi^{\mu\nu}$
and $q^{\mu}$,
\begin{eqnarray}
\epsilon_{\mu\nu} & = & \mathcal{A}_{2}\left(3u^{\mu}u^{\nu}-\Delta^{\mu\nu}\right)\Pi-\mathcal{B}_{1}u_{(\mu}q_{\nu)}+\mathcal{C}_{0}\pi_{\mu\nu}\;,\nonumber \\
\beta_{\mu}+\epsilon_{\mu} & = & \beta(u_{E})_{\mu}+\mathcal{A}_{1}\Pi u_{\mu}-\mathcal{B}_{0}q_{\mu}\;,\nonumber \\
\epsilon & = & \mathcal{A}_{0}\Pi\;,\label{eq:exp_epsilon_1}
\end{eqnarray}
where 
\begin{eqnarray}
\mathcal{A}_{0} & = & -3\mathcal{A}_{2}\left(m^{2}+4\frac{J_{31}J_{30}-J_{41}J_{20}}{D_{20}}\right)\nonumber \\
\mathcal{A}_{1} & = & -12\frac{J_{31}J_{20}-J_{41}J_{10}}{D_{20}}\mathcal{A}_{2}\;,\nonumber \\
\mathcal{A}_{2} & = & \frac{1}{4J_{42}\Omega}\;,\qquad\mathcal{C}_{0}=\frac{1}{2J_{42}}\;,\nonumber \\
\mathcal{B}_{0} & = & \mathcal{B}_{1}\frac{J_{41}}{J_{31}}\;,\qquad\mathcal{B}_{1}=\frac{1}{\Lambda J_{21}}\;.
\end{eqnarray}

\subsection{Second order theory}

For the second order theory (i.e. the theory including the derivative
of the quantities in the first order theory), one needs to investigate
the following $3$-rd moment with the full distribution function $f(x,p)$

\begin{equation}
\partial_{\alpha}F^{\alpha\mu\nu}[f]=\int_{p}p^{\mu}p^{\nu}p^{\alpha}\partial_{\alpha}f=\int_{p}p^{\mu}p^{\nu}C[f]\equiv P^{\mu\nu}\;.
\end{equation}
The contractions of the indices in $F^{\alpha\mu\nu}$ and $P^{\mu\nu}$
give 
\begin{eqnarray}
F_{\nu}^{\mu\nu} & = & \int_{p}p^{2}p^{\mu}f=m^{2}j^{\mu}\;,\nonumber \\
P_{\mu}^{\mu} & = & m^{2}\partial_{\mu}j^{\mu}=0\;.
\end{eqnarray}
As a rank-$2$ tensor, $P^{\mu\nu}$ can be decomposed as 
\[
P^{\mu\nu}=Au^{\mu}u^{\nu}+B\Delta^{\mu\nu}+2Cu^{(\mu}q^{\nu)}+D\pi^{\mu\nu}\;,
\]
where $A,B,C,D$ are the integrals of the collision term. 

The rank-$2$ tensor $P^{\mu\nu}$ can be written in the form
\begin{equation}
P^{\mu\nu}=-X^{\mu\nu\alpha\beta}\epsilon_{\alpha\beta}\;,
\end{equation}
where $X^{\mu\nu\alpha\beta}$ is a rank-$4$ tensor. Recalling the
definition (\ref{eq:entropy_kin_1}), the entropy principle requires
\[
0\leq\partial_{\mu}s^{\mu}=-\int_{p}\psi^{\prime}p^{\mu}\partial_{\mu}f=-\int_{p}\epsilon_{\mu\nu}p^{\mu}p^{\nu}C[f]=\epsilon_{\mu\nu}X^{\mu\nu\alpha\beta}\epsilon_{\alpha\beta}\;,
\]
which implies that 
\begin{equation}
X^{\mu\nu\alpha\beta}=X^{\alpha\beta\mu\nu}\;.\label{eq:X-pro-1}
\end{equation}
By the symmetry (\ref{eq:X-pro-1}) and $g_{\mu\nu}X^{\mu\nu\alpha\beta}=g_{\alpha\beta}X^{\mu\nu\alpha\beta}=0$,
the general form of $X^{\mu\nu\alpha\beta}$ is 
\begin{eqnarray}
X^{\mu\nu\alpha\beta} & = & \frac{1}{3}A\left(-\frac{1}{3}\Delta^{\mu\nu}\Delta^{\alpha\beta}-3u^{\mu}u^{\nu}u^{\alpha}u^{\beta}+\Delta^{\mu\nu}u^{\alpha}u^{\beta}+\Delta^{\alpha\beta}u^{\mu}u^{\nu}\right)\nonumber \\
 &  & +\frac{1}{5}B\Delta^{\mu<\alpha}\Delta^{\beta>\nu}-4Cu^{(\mu}\Delta^{\nu)(\alpha}u^{\beta)}\;,\label{eq:X-full-01}
\end{eqnarray}
where $A,B,C$ are the integrals of the collision term $C[f]$ in
the Boltzmann equation (\ref{eq:Boltzmann-eq-01}). 

On the other hand, the tensor $\partial_{\alpha}F^{\alpha\mu\nu}[f]$
can be evaluated as 
\begin{eqnarray}
 &  & -X^{\mu\nu\alpha\beta}\epsilon_{\alpha\beta}\nonumber \\
 & = & \int_{p}p^{\mu}p^{\nu}p^{\alpha}f_{0}\Delta_{0}\partial_{\alpha}y\nonumber \\
 & = & J^{\mu\nu\alpha}\partial_{\alpha}(\alpha+\epsilon)-J^{\mu\nu\alpha\beta}\partial_{\alpha}(\beta_{\beta}+\epsilon_{\beta})+J^{\mu\nu\alpha\rho\sigma}\partial_{\alpha}\epsilon_{\rho\sigma}\;.\label{eq:14-m-1}
\end{eqnarray}

Inserting Eq.(\ref{eq:exp_epsilon_1}) and (\ref{eq:X-full-01}) into
Eq.(\ref{eq:14-m-1}) and calculating the derivatives of all quantities
step by step, the differential equations for the quantities in the
first order theory give the IS equations which are identical with
Eqs. (\ref{eq:2nd-equation-phe}). The details of the calculation
can be found in Ref. \cite{Israel:1979wp,Muronga:2006zw}.

\subsection{Complete IS equations\label{sub:Complete-IS-equations}}

In the work \cite{Israel:1979wp,Muronga:2006zw}, the authors neglected
the terms $\sigma^{\mu\nu}\equiv P^{\mu\nu\alpha\beta}\partial_{\alpha}u_{\beta}$,
$\theta$ and $\partial_{\mu}\alpha$. However, it was pointed out
in Ref. \cite{Betz:2008me,Betz:2009zz} that these terms are in the
same order as others. 

Generally, there are three length scales in an effective theory for
long distance limit of a given theory. The first one is the microscopic
length scale $\ell_{micro}$. In a weakly coupled theory with well-defined
quasi-particles, this quantity is equal to the inter-particle distance.
The second one is the mesoscopic length scale $\ell_{meso}$ which
is identical with the mean free path $l_{mfp}$ in the dilute gas
limit. The third one is the macroscopic length scale $\ell_{macro}\sim\epsilon/\left|\partial\epsilon\right|$
describing the variety of the macroscopic quantities (e.g. the energy
density $\epsilon$). Thus, $\ell_{macro}^{-1}$ is proportional to
the gradients of the conserved quantities. If the so-called Knudsen
number $K\equiv\ell_{meso}/\ell_{macro}$ is sufficiently small, the
expansion in terms of $K$ is equivalent to a gradient expansion,
i.e., the expansion in terms of powers of $l_{mfp}\partial_{\mu}$. 

The ratios of the quantities in the first order theory to the energy
density are proportional to $K$. For example, the ratio of bulk viscous
pressure to the energy density reads 
\begin{equation}
\frac{\Pi_{NS}}{\epsilon}\sim\frac{\zeta\partial\cdot u}{\epsilon}\sim\frac{\zeta}{s}\frac{\partial}{T}\sim l_{mfp}\partial_{\mu}\equiv K\;,\label{eq:Pi_e-01}
\end{equation}
where the fundamental relation of thermodynamics, $\epsilon+P=Ts+\mu n$
has been used and 
\begin{equation}
\frac{\zeta}{s}\sim\frac{\eta}{s}\sim\frac{1}{\lambda_{th}\sigma}\frac{1}{T^{3}}\sim\frac{1}{\sigma n}\frac{1}{\lambda_{th}}\sim\frac{l_{mfp}}{\lambda_{th}}\;,
\end{equation}
with the help of $\eta\sim(\lambda_{th}\sigma)^{-1}$, $l_{mfp}\sim(\sigma n)^{-1}$.
Here $\sigma$ is the average cross section and $\lambda_{th}\sim T^{-1}$
is the thermal wavelength. Note here the result (\ref{eq:Pi_e-01})
is independent on the ratio $\zeta/s$, i.e., the expansion in terms
of $K$ (or the gradient expansion ) is available in both weakly and
strongly coupled theories. The ratios of the correlations from the
second order theory to the energy density are also proportional to
the $K$. For example, the ratio of the $\tau_{\Pi}\dot{\Pi}$ with
the relaxation time $\tau_{\Pi}$ for the bulk viscous pressure to
the energy density reads 
\begin{equation}
\frac{\tau_{\Pi}\dot{\Pi}}{\epsilon}\sim\frac{\Pi}{\epsilon}\tau_{\Pi}u\cdot\partial\sim K\frac{\zeta}{T^{4}}\partial\sim K\frac{\zeta}{s}\lambda_{th}\partial\sim K^{2}\;,
\end{equation}
where the estimation of $\tau_{\Pi}$ is employed in Ref. \cite{Israel:1979wp,Betz:2010cx}
\begin{equation}
\frac{\tau_{\Pi}}{\zeta}=\beta_{0}\sim\frac{1}{T^{4}}\;.\label{eq:beta-0-01}
\end{equation}

It can be proved that the terms $\sigma^{\mu\nu}$ and 
\begin{equation}
\omega^{\mu\nu}\equiv\Delta^{\mu\alpha}\Delta^{\nu\beta}(\partial_{\alpha}u_{\beta}-\partial_{\beta}u_{\alpha})\;,\label{eq:omega-mu-nu}
\end{equation}
give the same contribution as $\partial\cdot u\sim\partial_{\mu}$
in the power series of $K$. Therefore, the complete IS equations
should contain these terms. Taking account of these terms, the complete
IS equations is obtained in Ref. \cite{Betz:2008me,Betz:2009zz} 
\begin{eqnarray}
\pi^{\mu\nu} & = & \pi_{NS}^{\mu\nu}-\tau_{\pi}\dot{\pi}^{<\mu\nu>}+2l_{\pi q}\nabla^{<\mu}q^{\nu>}\nonumber \\
 &  & +2\tau_{\pi q}q^{<\mu}\dot{u}^{\nu>}+2\tau_{\pi}\pi_{\lambda}^{<\mu}\omega^{\nu>\lambda}-2\eta\hat{\delta_{2}}\pi^{\mu\nu}\theta\nonumber \\
 &  & -2\tau_{\pi}\pi_{\lambda}^{<\mu}\sigma^{\nu>\lambda}-2\lambda_{\pi q}q^{<\mu}\nabla^{\nu>}\alpha+2\lambda_{\pi\Pi}\Pi\sigma^{\mu\nu}\;,\nonumber \\
q^{\mu} & = & q_{NS}^{\mu}-\tau_{q}\Delta^{\mu\nu}\dot{q}_{\nu}+l_{q\Pi}\nabla^{\mu}\Pi-l_{q\pi}\Delta^{\mu\nu}\partial^{\lambda}\pi_{\nu\lambda}\nonumber \\
 &  & \tau_{q\Pi}\Pi\dot{u}^{\mu}-\tau_{q\pi}\pi^{\mu\nu}\dot{u}_{\nu}+\tau_{q}\omega^{\mu\nu}q_{\nu}-\frac{\kappa}{\beta}\hat{\delta}_{1}q^{\mu}\theta\nonumber \\
 &  & -\lambda_{qq}\sigma^{\mu\nu}q_{\nu}+\lambda_{q\Pi}\Pi\nabla^{\mu}\alpha+\lambda_{q\pi}\pi^{\mu\nu}\nabla_{\nu}\alpha\;,\nonumber \\
\Pi & = & \Pi_{NS}-\tau_{\Pi}\dot{\Pi}-l_{\Pi q}\partial\cdot q\nonumber \\
 &  & +\tau_{\Pi q}q\cdot\dot{u}-\zeta\hat{\delta}_{0}\Pi\theta\nonumber \\
 &  & +\lambda_{\Pi q}q\cdot\nabla\alpha+\lambda_{\Pi\pi}\pi^{\mu\nu}\sigma_{\mu\nu}\;,\label{eq:complete}
\end{eqnarray}
where $\pi_{NS}^{\mu\nu},\; q_{NS}^{\mu},\;\Pi_{NS}$ are given by
Eq.(\ref{eq:NS-equation-01}) and in the dilute gases limit the values
of the transport coefficients in the second order theory are given
in Ref. \cite{Betz:2010cx}. These values can also be evaluated by
other theories, such as Boltzmann equation and AdS/CFT duality. The
details for computing the transport coefficients in kinetic approach
will be shown in Chap. \ref{chap:Transport} and by the AdS/CFT duality
in Chap. \ref{chap:Applications-of-AdS/CFT}. 

Note that in dilute gases limit, the ratios of the transport coefficients
in the second order theory to the quantities in the first order theory
are only the function of $\alpha$ and $\beta$. For example, as shown
in Eq.(\ref{eq:beta-0-01}), $\beta_{0}$ will be determined by the
macroscopic state , and therefore will be changed with the evolution
of the fluid. It is a quite different treatment in the simulation
for the viscous fluid dynamics, for their simplicity, the value of
$\eta$ and the relaxation time $\tau_{\pi}$ are fixed, see e.g.
Ref. \cite{Song2008,Song2008a}.





\chapter{Causality and stability\label{cha:Causality-ans-stability}}

The first order theory does not obey the causality \cite{Hiscock1983,Hiscock1985,Hiscock1987,Olson:1989dp,Olson:1989eu}.
The causality cannot be satisfied automatically. Therefore, there
has to be constraint condition for the transport coefficients, which
is called asymptotic causality condition \cite{Denicol:2008ha,Pu:2009fj,Pu:2010zz}.
On the other hand, stability is intimately related to causality \cite{Denicol:2008ha,Pu:2009fj,Pu:2010zz}.
A causal theory will be stable.

In this chapter, we investigate the causality and stability of relativistic
dissipative fluid dynamics in the absence of conserved charges \cite{Pu:2009fj}.
In a linear stability analysis of the rest frame, we obtain the asymptotic
causality condition and find that the equations of relativistic dissipative
fluid dynamics are always stable. In a Lorentz-boosted frame, we find
the equations of fluid dynamics are stable if the asymptotic causality
condition is fulfilled. The group velocity may exceed the velocity
of light in a certain finite range of wave numbers. However, we demonstrate
that this does not violate causality, if the asymptotic causality
condition is fulfilled. Finally, we compute the characteristic velocities
and show that they remain below the velocity of light if we choose
the parameters fulfilled the asymptotic causality condition. The similar
discussion for the dissipative currents can be found in Ref. \cite{Pu:2010zz}.

\section{General discussion}

To demonstrate the causal problem in the first order theory, for simplicity
only the heat conductivity is taken into account in Eq. (\ref{eq:eom-general})
for the ideal fluid. In this case, Eq.(\ref{eq:EOM_2}) becomes 
\begin{eqnarray}
0 & = & \partial_{\mu}j^{\mu}\nonumber \\
 & = & -J_{20}\dot{\beta}+J_{21}\beta\theta+J_{10}\dot{\alpha}+\partial_{\mu}\nu^{\mu}\;.\label{eq:EOM_2-1}
\end{eqnarray}
Substituting Eq.(\ref{eq:EOM_2-1}) into Eq.(\ref{eq:EOM_1}) yields
\begin{eqnarray}
\dot{\beta} & = & \frac{1}{D_{20}}[(J_{31}J_{10}-J_{21}J_{20})\beta\theta-J_{20}\partial_{\mu}\nu^{\mu}]\;,\nonumber \\
\dot{\alpha} & = & \frac{1}{D_{20}J_{20}}[(J_{31}J_{20}^{2}-J_{30}J_{21}J_{20})\beta\theta-J_{20}\partial_{\mu}\nu^{\mu}]\;,
\end{eqnarray}
where the second line in the massless limit, i.e. $J_{30}/J_{31}=J_{20}/J_{21}=3$,
becomes 
\[
\dot{\alpha}+\frac{\kappa}{D_{20}}[\partial^{2}-(u\cdot\partial)^{2}]\alpha=0\;.
\]
In the local rest frame, the above equation can be written in the
form 
\begin{equation}
\partial_{t}\alpha=-\frac{\kappa}{D_{20}}\nabla^{2}\alpha\;.\label{eq:alpha-causal-01}
\end{equation}
If the chemical potential $\mu$ is fixed, Eq.(\ref{eq:alpha-causal-01})
is actually the heat conduction equation
\begin{equation}
\frac{\partial T}{\partial t}=D\frac{\partial^{2}T}{\partial x^{2}}\;,\label{eq:heat-conductivity-01}
\end{equation}
which gives the dispersion relation 
\begin{equation}
\omega=iDk^{2}\;.
\end{equation}
This implies the system is acausal since that the group speed of the
signal is proportional to the wave-number $k$ \cite{Denicol:2008ha}. 

To avoid this problem, the simplest way is to introduce a second order
time derivative $\partial_{t}^{2}T$ in Eq.(\ref{eq:heat-conductivity-01}).
Then Eq.(\ref{eq:heat-conductivity-01}) becomes a standard wave equation
\begin{equation}
\tau_{q}\frac{\partial^{2}T}{\partial t^{2}}+\frac{\partial T}{\partial t}=D\frac{\partial^{2}T}{\partial x^{2}}\;,
\end{equation}
where $\tau_{q}$ has the dimension of the time. However, there must
be some constraint condition for $\tau_{q}$ and $D$ since if $\tau_{q}\rightarrow0$
the system will be acausal again.

\section{Causality in local rest frame\label{sec:Causality-LRF}}

The approaches to formulate a second order theory of relativistic
fluid dynamics is not unique, different approaches differ only by
non-linear second order terms (see e.g. Ref. \cite{Israel:1979wp,Betz:2008me,Baier2008,Denicol2009,Koide2007,Jou1999,Jou1988,Carter1991,Denicol:2008ha}).
These differences will vanish since a linear analysis is applied here.
In that case, the evolution equations of the dissipative quantities
are given by 
\begin{eqnarray}
\tau_{\Pi}\,\frac{d}{d\tau}\Pi+\Pi & = & -\zeta\,\partial_{\mu}u^{\mu}\;,\nonumber \\
\tau_{\pi}\, P^{\mu\nu\alpha\beta}\,\frac{d}{d\tau}\pi_{\alpha\beta}+\pi^{\mu\nu} & = & 2\eta\, P^{\mu\nu\alpha\beta}\,\partial_{\alpha}u_{\beta}\;,\nonumber \\
\tau_{q}\Delta^{\mu\nu}\frac{d}{d\tau}\nu_{\nu}+\nu^{\mu} & = & \frac{n^{2}\kappa T^{2}}{(\epsilon+P)^{2}}\nabla^{\mu}\alpha\;.\label{eq:simplify-IS}
\end{eqnarray}

The investigation of the causality and stability for a hydrostatic
background in exclusively the low- and high-wave-number limit is given
by Hiscock, Lindblom and Olson \cite{Hiscock1985,Hiscock1987}. However,
they did not find a generic anomalous behavior of the group velocity.
The analysis of the causality and stability for the dissipative fluid
with the bulk viscous pressure only has been done in Ref. \cite{Denicol:2008ha},
where they point out the relation between causality and stability.
Since the analysis in the fluid with bulk viscosity only is similar
to that with shear viscosity only \cite{Denicol:2008ha}. In this
chapter, we will make a discussion on the properties of the fluid
with shear stress tensor only in the absence of the conserved charges. 

For convenience, the following dimensionless parameters will be used
\begin{equation}
a=\frac{\eta}{s}\;,\qquad b=\frac{(\epsilon+P)\tau_{\pi}}{\eta}=\frac{\tau_{\pi}T}{a}\;,
\end{equation}
where $\epsilon+P=Ts$ has been used in the absence of conserved charges.
A $D$ dimensional ($D\geq3$) system will be considered. The symmetric
rank-four projector is in the form 
\begin{equation}
P^{\mu\nu\alpha\beta}=\frac{1}{2}\left(\Delta^{\mu\alpha}\Delta^{\nu\beta}+\Delta^{\mu\beta}\Delta^{\nu\alpha}\right)-\frac{1}{D-1}\Delta^{\mu\nu}\Delta^{\alpha\beta}\;.
\end{equation}

A perturbation around the hydrostatic equilibrium state is introduced
\begin{equation}
X=X_{0}+\delta Xe^{i\omega t-ikx}\;,\label{eq:X-defin}
\end{equation}
where $X=\epsilon,\pi^{\mu\nu},u^{\mu}$ with $\epsilon_{0}=\textrm{const}$,
$\pi_{0}^{\mu\nu}=0$ and $u^{\mu}=(1,0,0,...)$. In the linear approximation,
the perturbation quantities $\delta X$ are chosen to be 
\begin{eqnarray}
\delta X & = & (\delta\varepsilon,\delta u^{1},\delta\pi^{11},\delta u^{2},\delta\pi^{12},\ldots,\delta u^{D-1},\delta\pi^{1(D-1)},\delta\pi^{22},\delta\pi^{33},\ldots,\nonumber \\
 &  & \;\;\delta\pi^{(D-2)(D-2)},\delta\pi^{23},\delta\pi^{24},\ldots,\delta\pi^{2(D-1)},\delta\pi^{34},\ldots,\delta\pi^{(D-2)(D-1)})^{T}\;,
\end{eqnarray}
constrained by the normalization condition $u^{\mu}u_{\mu}=1$, the
traceless condition $\pi_{\mu}^{\mu}=0$ and the orthogonality condition
$u_{\mu}\pi^{\mu\nu}=0$. The linearized fluid-dynamical equations
including the evolution equations (\ref{eq:simplify-IS}) can be written
as 
\begin{equation}
A\delta X=0\;,\label{eq:AX}
\end{equation}
where the matrix $A$ is in the form 
\begin{equation}
A=\left(\begin{array}{cccc}
T & 0 & 0 & 0\\
0 & B & 0 & 0\\
G & 0 & C & 0\\
0 & 0 & 0 & E
\end{array}\right)\;,
\end{equation}
with 
\begin{alignat}{2}
 & T=\left(\begin{array}{ccc}
i\omega & f_{1} & 0\\
-ikc_{s}^{2} & f_{2} & -ik\\
0 & \Gamma & f
\end{array}\right)\;, & \quad & G=\left(\begin{array}{ccc}
0 & \Gamma_{2} & 0\\
 & \ldots\\
0 & \Gamma_{2} & 0
\end{array}\right)_{(D-3)\times3}\;,\nonumber \\
 & B={\rm diag}(B_{0},\ldots,B_{0})_{(D-2)\times(D-2)}\;, &  & B_{0}=\left(\begin{array}{cc}
f_{2} & -ik\\
\Gamma_{1} & f
\end{array}\right)\;,\nonumber \\
 & C={\rm diag}(f,\ldots,f)_{(D-3)\times(D-3)}\;,\nonumber \\
 & E={\rm diag}(f,\ldots,f)_{\frac{1}{2}(D-2)(D-3)\times\frac{1}{2}(D-2)(D-3)}\;,
\end{alignat}
 where $c_{s}=\sqrt{\partial P/\partial\varepsilon}$ is the speed
of sound and for convenience, the following abbreviations have been
used 
\begin{alignat}{2}
 & f=i\omega\,\tau_{\pi}+1\;, & \quad & f_{1}=-ik\,(\epsilon+P)\;,\nonumber \\
 & f_{2}=i\omega\,(\epsilon+P)\;, &  & \Gamma=-ik\,\frac{2(D-2)}{D-1}\,\eta\;,\nonumber \\
 & \Gamma_{1}=-ik\,\eta\;, &  & \Gamma_{2}=ik\,\frac{2}{D-1}\,\eta\;.
\end{alignat}
The determinant of the matrix $A$ must vanish to avoid the trivial
solutions of Eq.(\ref{eq:AX}). Solving $\det A=0$ gives \begin{subequations}
\begin{eqnarray}
f & = & 0\;,\label{eqn:rest_eq1}\\
\det B=\left(\det B_{0}\right)^{D-2} & = & 0\;,\label{eqn:rest_eq2}\\
\det T=\det\left(\begin{array}{ccc}
i\omega & f_{1} & 0\\
-ik\, c_{s}^{2} & f_{2} & -ik\\
0 & \Gamma & f
\end{array}\right) & = & 0\;.\label{eqn:rest_eq3}
\end{eqnarray}
\end{subequations}These gives the dispersion relation for $\omega(k)$.

\begin{figure}
\begin{centering}
\includegraphics{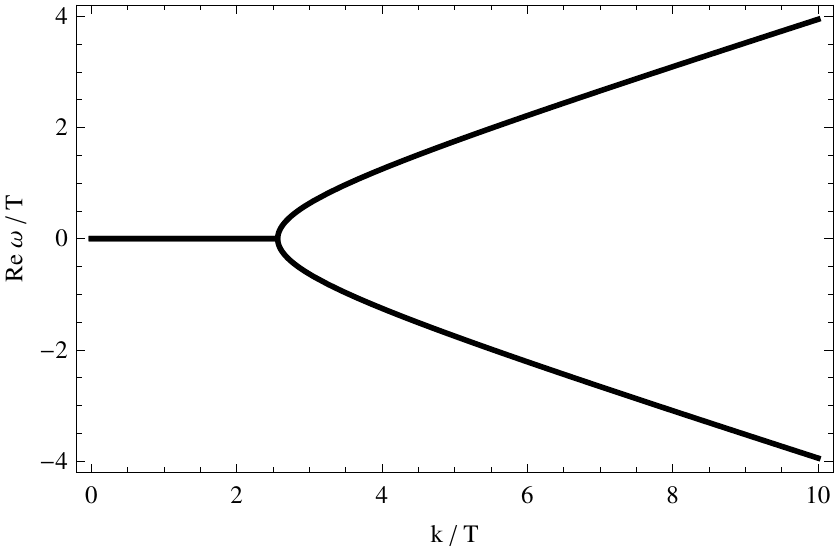}
\par\end{centering}

\begin{centering}
\includegraphics{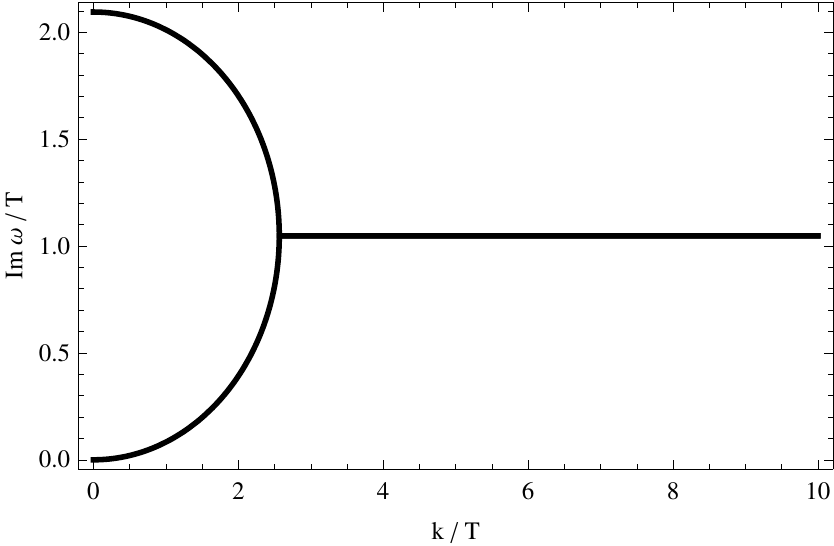}
\par\end{centering}

\caption{\label{fig:shear-LRF}The real parts (left panel) and the imaginary
parts (right panel) of solutions for the shear modes obtained from
Eq.(\ref{eqn:rest_eq2}). The parameters are $a=\frac{1}{4\pi}\,,\; b=6\,,\; c_{s}^{2}=\frac{1}{3}$
for the 3+1-dimensional case, $D=4$.}

\end{figure}

A nonpropagating mode with the degeneracy $(D-3)[1+(D-2)/2]$ is given
by the solution of Eq.(\ref{eqn:rest_eq1}), 
\begin{equation}
\omega=\frac{i}{\tau_{\pi}}\;.\label{eq:non-pro-1}
\end{equation}
The so-called shear modes with the degeneracy $2(D-2)$ are given
by the solutions of Eq.(\ref{eqn:rest_eq2}),
\begin{equation}
\omega=\frac{1}{2\tau_{\pi}}\left(i\pm\sqrt{\frac{4\,\eta\,\tau_{\pi}}{\varepsilon+P}\, k^{2}-1}\right)\;,\label{eq:dispersion-shear-LRF}
\end{equation}
when $k$ is larger than the critical wave number 
\begin{equation}
k_{c}=\sqrt{\frac{\varepsilon+P}{4\,\eta\,\tau_{\pi}}}\equiv\frac{\sqrt{b}}{2\,\tau_{\pi}}\;.\label{eq:rest-kc}
\end{equation}
The numerical result for Eq.(\ref{eq:dispersion-shear-LRF}) is shown
in Fig. \ref{fig:shear-LRF}.

\begin{figure}[tbph]
\begin{centering}
\includegraphics{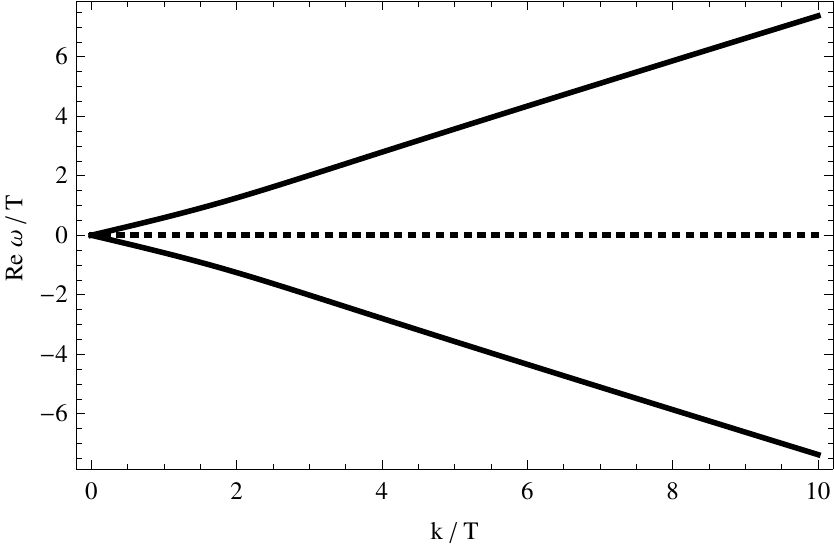}
\par\end{centering}

\begin{centering}
\includegraphics{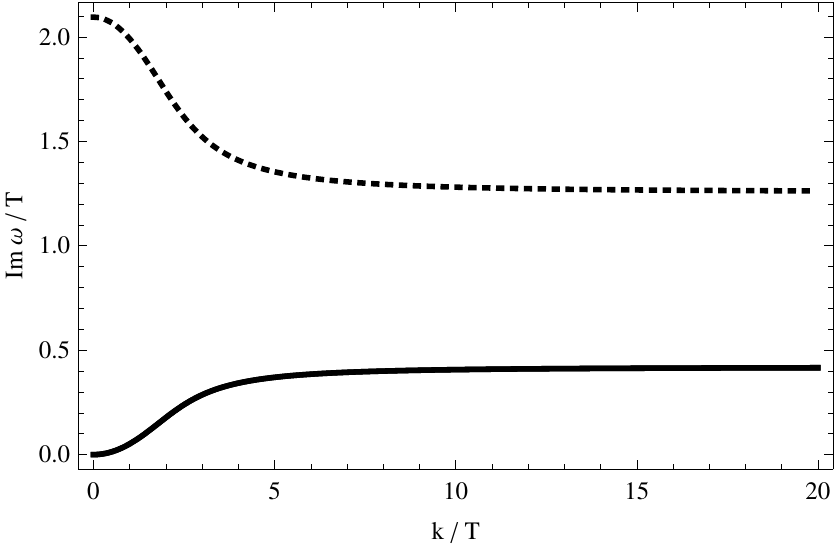}
\par\end{centering}

\caption{The real parts (left panel) and the imaginary parts (right panel)
of the solutions for the sound modes (full lines) and the nonpropagating
mode (dashed line) obtained from Eq.(\ref{eqn:rest_eq3}). The parameters
are $a=\frac{1}{4\pi}\,,\; b=6\,,\; c_{s}^{2}=\frac{1}{3}$ for the
3+1-dimensional case, $D=4$.\label{fig:sound-LRF}}
\end{figure}

The solutions of Eq.(\ref{eqn:rest_eq3}) lead to another nonpropagating
mode and two propagating modes or the sound modes. The analytic solutions
in the limit of small wavenumber $k$ are 
\begin{equation}
\omega=\left\{ \begin{array}{l}
{\displaystyle \frac{i}{\tau_{\pi}}\;,}\\
{\displaystyle \pm\, k\, c_{s}+i\,\frac{\Gamma_{s}}{2}\, k^{2}\;,}
\end{array}\right.\label{eq:sound-modes-small-k}
\end{equation}
while for large wavenumber, we have 
\begin{equation}
\omega=\left\{ \begin{array}{l}
{\displaystyle \frac{i}{\tau_{\pi}}\,\left[1+\frac{\Gamma_{s}}{\tau_{\pi}c_{s}^{2}}\right]^{-1}\;,}\\[0.3cm]
{\displaystyle \pm\, k\, c_{s}\sqrt{1+\frac{\Gamma_{s}}{\tau_{\pi}c_{s}^{2}}}+\frac{i}{2\tau_{\pi}}\,\left[1+\frac{\tau_{\pi}c_{s}^{2}}{\Gamma_{s}}\right]^{-1}\;,}
\end{array}\right.\label{eq:sound-modes-large-k}
\end{equation}
where 
\begin{equation}
\Gamma_{s}\equiv\frac{2(D-2)}{D-1}\,\frac{\eta}{\varepsilon+P}\equiv\frac{2(D-2)}{D-1}\,\frac{\tau_{\pi}}{b}\;,
\end{equation}
is the sound attenuation length. The numerical solutions of Eq.(\ref{eqn:rest_eq3})
are shown in Fig. \ref{fig:sound-LRF}. The nonpropagating as well
as other propagating modes are stable around the hydrostatic equilibrium
state since all imaginary parts are positive. It is identical to the
conclusion of Hiscock and Lindblom \cite{Hiscock1983,Hiscock1985,Hiscock1987}
and others \cite{Denicol:2008ha,Pu:2009fj,Pu:2010zz}. 

\begin{figure}
\begin{centering}
\includegraphics{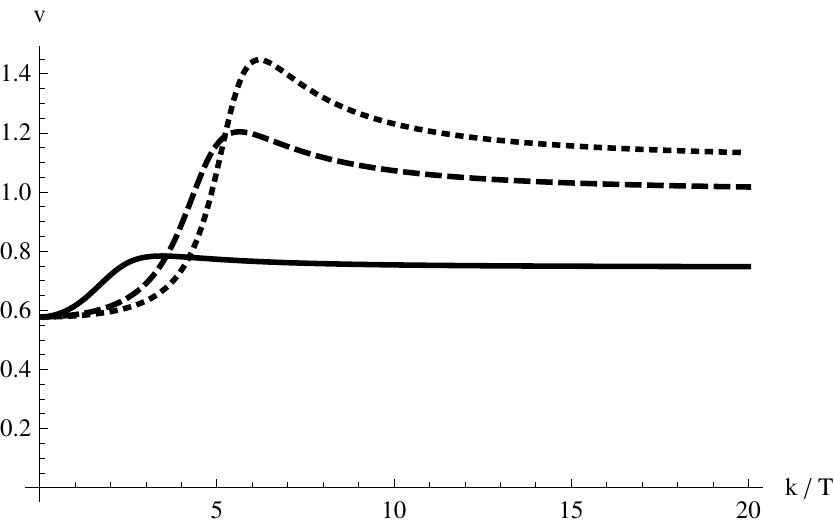}
\par\end{centering}

\caption{\label{fig:sound-v-LRF}The group velocity for sound modes $v_{g}$
for $a=1/(4\pi)\,,\; D=4\,,\; c_{s}^{2}=\frac{1}{3}$, and $b=6$
(full line), $b=2$ (dashed line), as well as $b=1.5$ (dotted line).
If the asymptotic causality condition (\ref{eq:asym-casual-cond})
is fulfilled, the group velocity will be always smaller than the speed
of light. Otherwise, the causality is violated. }
\end{figure}

\begin{figure}
\begin{centering}
\includegraphics{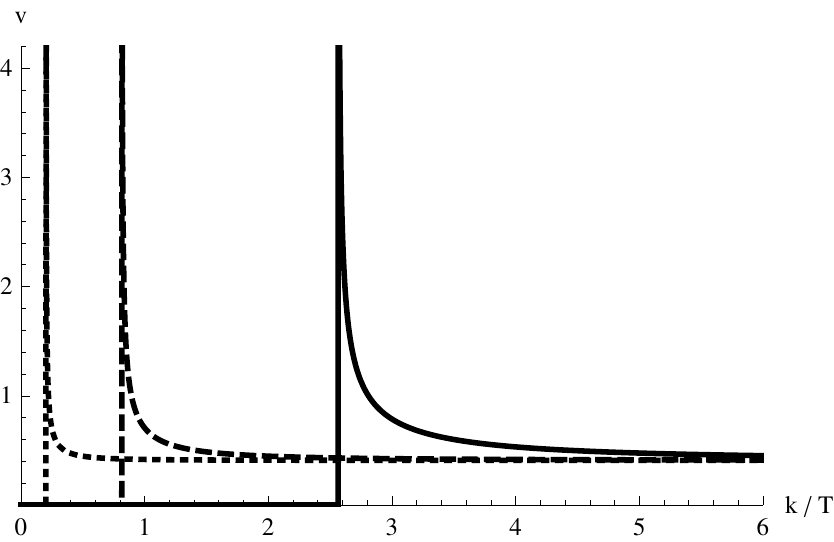}
\par\end{centering}

\caption{\label{fig:shear-v-LRF}The group velocity of shear modes for $D=4\,,\; b=6\,,\; c_{s}^{2}=\frac{1}{3}$,
and $a=1/(4\pi)$ (full line), $a=1/4$ (dashed line), as well as
$a=1$ (dotted line). The group velocity is smaller than the speed
of light in the large $k$ limit since the asymptotic causality condition
(\ref{eq:asym-casual-cond}) is fullfiled. }
\end{figure}

The causality requies that the group velocity 
\[
v_{g}=\frac{\partial\textrm{Re}\omega}{\partial k}\;,
\]
must be less than the speed of light. For the nonpropagating modes
${\rm Re}\,\omega=0$, the causality is associated with the behavior
of the imaginary part \cite{Denicol:2008ha}. Generally speaking,
a $k^{2}$ dependence of any nonpropagating mode can be considered
to violate the causality. In that case, the results in Eq. (\ref{eq:non-pro-1},\ref{eq:sound-modes-small-k},\ref{eq:sound-modes-large-k})
show that two nonpropagating modes are causal. For the sound modes,
the group velocity is shown in Fig. \ref{fig:sound-v-LRF}. The group
velocity for the shear modes (\ref{eq:dispersion-shear-LRF}) is shown
in Fig. \ref{fig:shear-v-LRF}. Unfortunately, there are divergences
near the critical wave number $k_{c}$. The details for these divergences
will be discussed in Sec. \ref{sec:Divergences}.

Taking solutions for all modes into account, it is found that the
causal condition is determined by the behavior of group velocity in
large $k$ limit,
\begin{equation}
v_{g}^{as}=v_{g,{\rm sound}}^{{\rm as}}=\lim_{k\rightarrow\infty}\frac{\partial Re\,\omega}{\partial k}=c_{s}\,\sqrt{1+\frac{\Gamma_{s}}{\tau_{\pi}c_{s}^{2}}}\;.
\end{equation}
Consequently, the asymptotic causality condition reads
\begin{equation}
\frac{\Gamma_{s}}{\tau_{\pi}}\le1-c_{s}^{2}\;\;\Longleftrightarrow\;\;\frac{1}{b}\le\frac{D-1}{2(D-2)}(1-c_{s}^{2})\;.\label{eq:asym-casual-cond}
\end{equation}
For conformal fluids, the above causality condition is always satisfied
since 
\begin{equation}
b=2(2-\ln2)\simeq2.614>2\;,
\end{equation}
where $c_{s}^{2}=1/(D-1)$ and the values of $a=\eta/s\simeq1/(4\pi)$
and $\tau_{\pi}=(2-\ln2)/(2\pi T)$ are derived from the AdS/CFT correspondence
\cite{Baier2008,Heller2007,Pu:2010zza}. 

The stability of a propagating mode is associated with the the behavior
of imaginary part. If the imaginary part of the frequency in this
mode is always positive, the damping amplitude of perturbations will
decrease with the evolution and the system is therefore stable. 

Note that in local rest frame the stability of relativistic dissipative
fluid dynamics is not affected by the causality. The stable propagating
modes with acausal parameters can be observed in Figs. \ref{fig:shear-LRF},
\ref{fig:sound-LRF}. For instance, an acausal fluid with $D=4$ and
$b=1$ is demonstrated to be stable for the shear modes in Fig. \ref{fig:shear-LRF}
and sound modes in Fig. \ref{fig:sound-LRF}, respectively.

\section{Stability in Lorentz-boost frame}

In order to demonstrate the intimate relation between causality and
stability, it is necessary to consider the system in a moving frame.
For simplicity, the space-time dimension is restricted to be $D=4$.

\subsection{Boost along $x$ direction}

The perturbation of the fluid velocity in a frame boosted with the
constant speed $V$ along the $x$ direction is given by 
\begin{equation}
u^{\prime\;\mu}=u_{0}^{\prime\;\mu}+\delta u^{\prime\;\mu}\; e^{i\omega t-ikx}\;,
\end{equation}
 where
\begin{alignat}{1}
u_{0}^{\prime\;\mu}=\gamma_{V}(1,V,0,0)\;,\quad & \delta u^{\prime\;\mu}=(V\gamma_{V}\delta u^{x},\gamma_{V}\delta u^{x},\delta u^{y},\delta u^{z})\;.
\end{alignat}
The linearized fluid-dynamical equations are given by $AX=0$ with
\begin{equation}
X=(\delta\varepsilon,\delta u^{x},\delta\pi^{xx},\delta u^{y},\delta\pi^{xy},\delta u^{z},\delta\pi^{xz},\delta\pi^{yy},\delta\pi^{yz})^{T}\;,
\end{equation}
and 
\begin{equation}
A=\left(\begin{array}{cccc}
T_{1} & 0 & 0 & 0\\
0 & B_{1} & 0 & 0\\
G_{1} & 0 & C_{1} & 0\\
0 & 0 & 0 & E_{1}
\end{array}\right)\;,
\end{equation}
where 
\begin{eqnarray}
T_{1} & = & \gamma_{V}^{2}\left(\begin{array}{ccc}
T_{11} & T_{12} & \;\; i\gamma_{V}^{-2}V(\omega V-k)\\
T_{13} & T_{14} & \;\; i\gamma_{V}^{-2}(\omega V-k)\\
0 & \frac{4}{3}i\eta\gamma_{V}(\omega V-k) & \;\;\gamma_{V}^{-2}F
\end{array}\right)\;,\nonumber \\
B_{1} & = & {\rm diag}(B_{01},B_{01})\;,\qquad B_{01}=\left(\begin{array}{cc}
i\gamma_{V}(\omega-kV)(\epsilon+P) & \;\; i(\omega V-k)\\
i\eta\gamma_{V}^{2}(\omega V-k) & \;\; F
\end{array}\right)\;,\nonumber \\
G_{1} & = & \left(\frac{}{}0\qquad-\frac{2}{3}i\eta\gamma_{V}(\omega V-k)\qquad0\right)\;,\nonumber \\
C_{1} & = & E_{1}=F\;,
\end{eqnarray}
with
\begin{eqnarray}
T_{11} & = & i\omega(1+V^{2}c_{s}^{2})-ikV(1+c_{s}^{2})\;,\nonumber \\
T_{12} & = & i[2\omega V-k(1+V^{2})](\epsilon+P)\;,\nonumber \\
T_{13} & = & i\omega V(1+c_{s}^{2})-ik(V^{2}+c_{s}^{2})\;,\nonumber \\
T_{14} & = & i\left[\omega(1+V^{2})-2kV\right](\epsilon+P)\;,\nonumber \\
F & = & i\gamma_{V}(\omega-kV)\tau_{\pi}+1\;.
\end{eqnarray}
The determinant of the matrix $A$ is $\det A=\det T_{1}\times\det B_{1}\times F^{2}$.
From $F=0$, two trivial propagating modes are found 
\begin{equation}
\omega=\frac{i}{\gamma_{V}\tau_{\pi}}+kV\;,
\end{equation}
which correspond to nonpropagating modes in the local rest frame.
From $\det B_{1}=0$, four modes corresponding to the shear modes
are observed
\begin{equation}
\omega_{\pm}=\frac{1}{2a(b-V^{2})\gamma_{V}}\left[iT-2a(1-b)kV\gamma_{V}\pm\sqrt{-T^{2}+4iakTV\gamma_{V}^{-1}+4a^{2}bk^{2}\gamma_{V}^{-2}}\right]\;.\label{eq:shear-3+1}
\end{equation}
On the other hand, the sound modes are given by the following equation
\begin{align}
 & c_{s}^{2}(\epsilon+P)\left[1-i\gamma_{V}\tau_{\pi}(kV-\omega)\right]\left\{ k^{2}\left[V^{2}+(V-1)^{2}V\gamma_{V}^{2}+1\right]\right.\nonumber \\
 & +\left.2kV\omega\left[(V-1)V\gamma_{V}^{2}-1\right]+V^{2}\omega^{2}-c_{s}^{-2}(\omega-kV)^{2}\right\} \nonumber \\
 & +\frac{4}{3}i\gamma_{V}\eta(k-V\omega)^{2}\left\{ kV\left[c_{s}^{2}\gamma_{V}^{2}V(1-V)-1\right]+\omega\right\} =0\;.
\end{align}

\begin{figure}[tbph]
\begin{centering}
\includegraphics{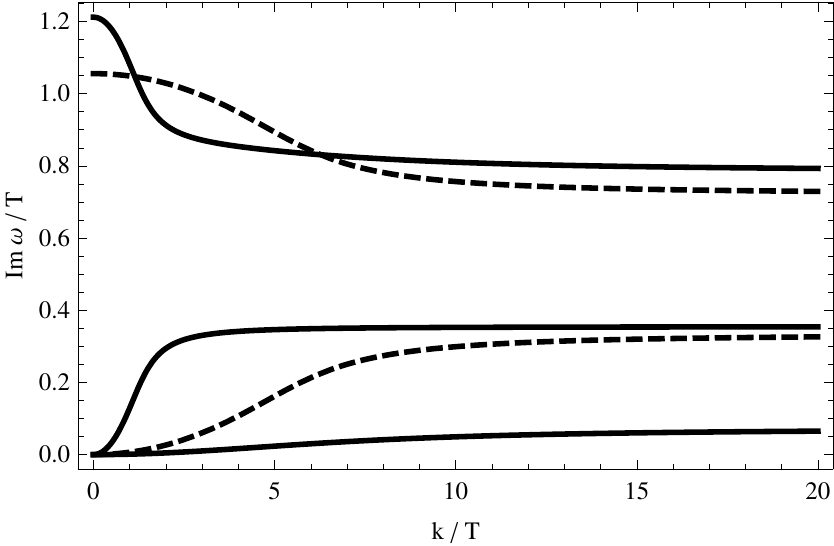}
\par\end{centering}

\begin{centering}
\includegraphics{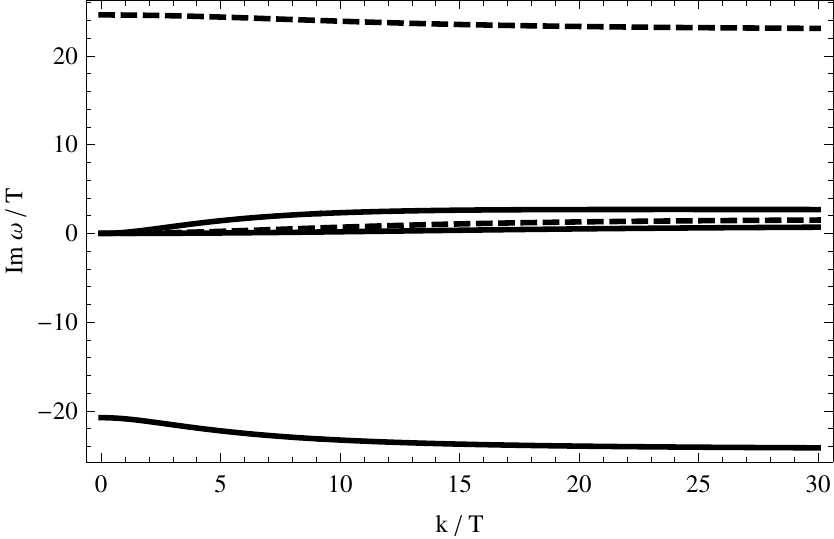}
\par\end{centering}

\caption{\label{fig:3+1-boost-im}The imaginary parts of $\omega$ for a boost
in $x$ direction with velocity $V=0.9$ . In the upper panel, the
parameter set $a=\frac{1}{4\pi},\, b=6,\, c_{s}^{2}=\frac{1}{3}$
fulfills the asymptotic causality condition (\ref{eq:asym-casual-cond}),
while in the lower panel $a=\frac{1}{4\pi},\, b=1,\, c_{s}^{2}=\frac{1}{3}$
violates this condition. The dashed lines are for the shear modes,
while the solid lines are for the sound modes.}
\end{figure}

In contrast to the results in the local rest frame, the appearance
of negative imaginary parts (i.e. the theory becomes unstable) could
be observed in the right panel of Fig. \ref{fig:3+1-boost-im} where
the parameter set violates the asymptotic causality condition (\ref{eq:asym-casual-cond}).

\subsection{Boost along $y$ direction}

The velocity in a frame boosted with the speed $V$ along $y$ direction
is given by 
\begin{equation}
u_{0}^{\prime\;\mu}=\gamma_{V}(1,0,V,0)\;,\quad\delta u^{\prime\;\mu}=(V\gamma_{V}\delta u^{y},\delta u^{x},\gamma_{V}\delta u^{y},\delta u^{z})\;.
\end{equation}
The matrix $A$ is 
\begin{equation}
A=\left(\begin{array}{cccc}
T_{2} & H_{1} & H_{2} & 0\\
H_{3} & B_{2} & H_{4} & H_{5}\\
G_{2} & H_{6} & C_{2} & 0\\
0 & H_{7} & 0 & E_{2}
\end{array}\right)\;,
\end{equation}
with
\begin{align}
 & T_{2}=\left(\begin{array}{ccc}
i\omega\gamma_{V}^{2}(1+c_{s}^{2}V^{2}) & -ik\gamma_{V}(\varepsilon+P) & 0\\
-ikc_{s}^{2} & i\omega\gamma_{V}(\varepsilon+P) & -ik\\
0 & -\frac{4}{3}ik\eta & F_{1}
\end{array}\right)\;,\nonumber \\
 & H_{1}=\left(\begin{array}{cccc}
2i\omega V(\varepsilon+P)\gamma_{V}^{2} & -ikV & 0 & 0\\
0 & i\omega V & 0 & 0\\
-\frac{2}{3}i\omega V\eta\gamma_{V} & 0 & 0 & 0
\end{array}\right)\;,\nonumber \\
 & H_{3}=\left(\begin{array}{ccc}
i\omega V\gamma_{V}^{2}(1+c_{s}^{2}) & -ikV\gamma_{V}(\varepsilon+P) & 0\\
0 & i\omega V\gamma_{V}^{2}\eta & 0\\
0 & 0 & 0\\
0 & 0 & 0
\end{array}\right)\;,\nonumber \\
 & B_{2}=\left(\begin{array}{cccc}
i\omega\gamma_{V}^{2}(1+V^{2})(\varepsilon+P) & -ik & 0 & 0\\
-ik\gamma_{V}\eta & F_{1} & 0 & 0\\
0 & 0 & i\omega\gamma_{V}(\varepsilon+P) & -ik\\
0 & 0 & -ik\eta & F_{1}
\end{array}\right)\;,\nonumber \\
 & H_{6}=\left(\begin{array}{cccc}
\frac{4}{3}i\omega V\gamma_{V}^{3}\eta & 0 & 0 & 0\end{array}\right)\;,\nonumber \\
 & H_{7}=\left(\begin{array}{cccc}
0 & 0 & i\omega V\gamma_{V}^{2}\eta & 0\end{array}\right)\;,
\end{align}
and 
\begin{align}
 & H_{2}=\left(\begin{array}{ccc}
i\omega V^{2} & 0 & 0\end{array}\right)^{T}\;,\nonumber \\
 & H_{4}=\left(\begin{array}{cccc}
i\omega V & 0 & 0 & 0\end{array}\right)^{T}\;,\nonumber \\
 & H_{5}=\left(\begin{array}{cccc}
0 & 0 & i\omega V & 0\end{array}\right)^{T}\;,\nonumber \\
 & G_{2}=\left(\begin{array}{ccc}
0 & \frac{2}{3}ik\gamma_{V}^{2}\eta & 0\end{array}\right)\;,\nonumber \\
 & C_{2}=E_{2}=F_{1}\;,
\end{align}
where $F_{1}=i\omega\gamma_{V}\tau_{\pi}+1$. 

Solving the equation $\det A=0$ gives all modes. The nonpropagating
mode takes the same form as in the local rest frame, 
\begin{equation}
\omega=\frac{i}{\gamma_{V}\tau_{\pi}}\;.
\end{equation}
The shear modes are given by the solutions 
\begin{equation}
\omega_{\pm}=\frac{1}{2a(b-V^{2})\gamma_{V}}\left[i\, T\pm\sqrt{-T^{2}+4a^{2}bk^{2}-4a^{2}k^{2}V^{2}}\right]\;,
\end{equation}
and the equations for the bulk modes are
\begin{align}
 & 3c_{s}^{2}(\epsilon+P)(-i+\gamma_{V}\tau_{\pi}\omega)(k^{2}+V^{2}\gamma_{V}^{2}\omega^{2})\nonumber \\
 & +\gamma_{V}\omega\left\{ 4k^{2}\eta+\gamma_{V}\omega\left[3i(\varepsilon+P)+4V^{2}\gamma_{V}\eta\omega-3(\varepsilon+P)\gamma_{V}\tau_{\pi}\omega\right]\right\} =0\;,
\end{align}
respectively. 

\begin{figure}[tbph]
\begin{centering}
\includegraphics{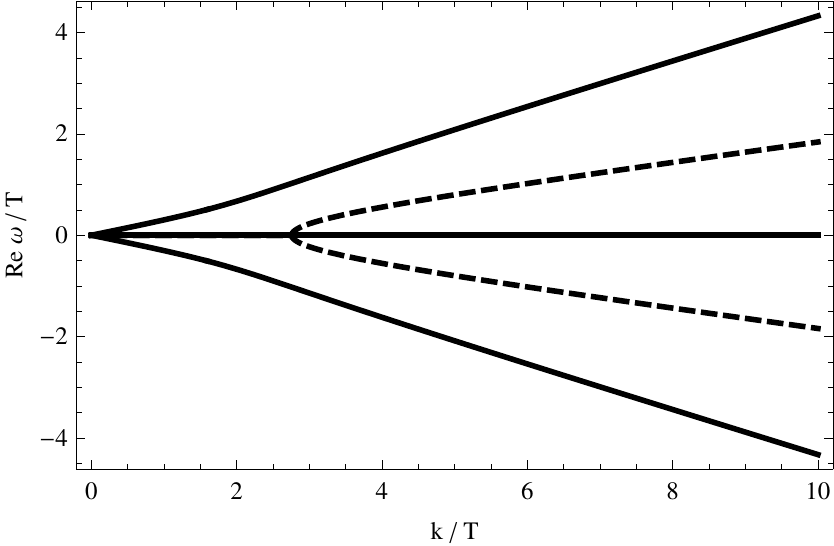}
\par\end{centering}

\begin{centering}
\includegraphics{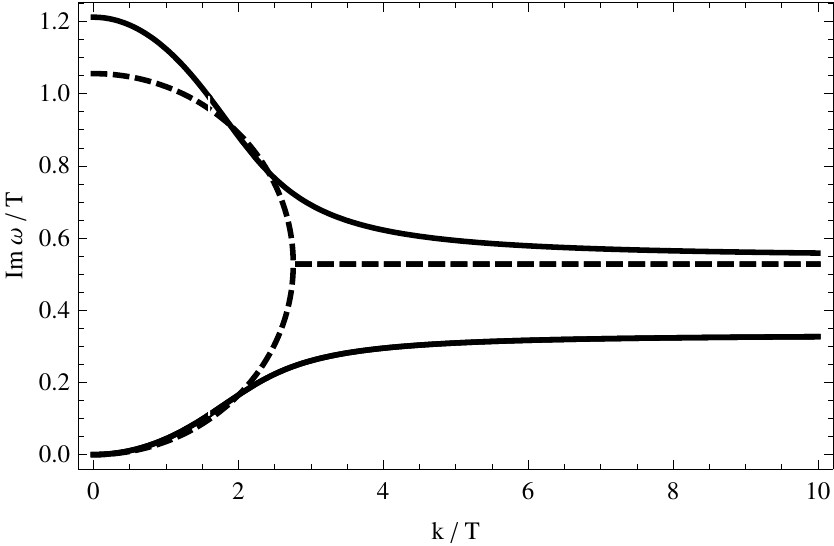}
\par\end{centering}

\caption{\label{fig:3+1-boost-y}The real and imaginary parts for the shear
modes (dashed lines) and sound modes (solid lines), for a Lorentz
boost in $y$ direction with $a=\frac{1}{4\pi},\, b=6,\, c_{s}^{2}=\frac{1}{3},\, V=0.9\;,D=4$.}
\end{figure}

The numerical results are shown in Fig. \ref{fig:3+1-boost-y} with
the conclusion same as the case of boost along $x$ direction. It
is confirmed that if the asymptotic causality condition (\ref{eq:asym-casual-cond})
is fulfilled, the system will be causal and stable. However, the reverse
is not true. A stable theory may also violate the asymptotic causality
condition (\ref{eq:asym-casual-cond}). 

\section{Divergences in shear modes\label{sec:Divergences}}

As mentioned in Sec. \ref{sec:Causality-LRF} (see also Fig. \ref{fig:shear-v-LRF}),
there are divergences near the critical wave number $k_{c}$. However,
the analysis for the stability of the fluid in a moving frame implies
that these divergences do not affect the stability. Moreover, a lot
of work \cite{Denicol:2008ha,Pu:2009fj,Pu:2010zz} show that the causality
of theory is guaranteed if the group velocity in large $k$ limit
is smaller than the speed of light. This problem has also been studied
in the classical electrodynamics (e.g. see \cite{Jackson}). The divergent
group velocity may become superluminal when one analyzes the propagating
modes of electromagnetic waves in some special material. However,
such kind of divergences is considered to be unphysical. 

The similar analysis for the divergent group velocity indicates that
the divergences in shear modes do not affect the causality of the
theory. The perturbation $\delta X$ in Eq.(\ref{eq:X-defin}) is
given by 
\begin{equation}
\delta X(x,t)=\sum_{j}\int d\omega\,\widetilde{\delta X}_{j}(\omega)\, e^{i\omega t-ik_{j}(\omega)x}\;,\label{eq:dX-01}
\end{equation}
where $j$ denotes the different modes and $k_{j}(\omega)$ is the
inverted $\omega(k)$ of the respective mode. The inverse Fourier
transform gives the components $\widetilde{\delta X}_{j}(\omega)$
\begin{equation}
\sum_{j}\widetilde{\delta X}_{j}(\omega)=\frac{1}{2\pi}\int_{-\infty}^{\infty}dt\,\delta X(0,t)\, e^{-i\omega t}\;,
\end{equation}
where $\delta X(0,t)=0$ for $t<0$ as a result of the assumption
that there is no change in a fluid-dynamical variable before $t=0$.
It is found that the $\sum_{j}\widetilde{\delta X}_{j}(\omega)$ is
analytic in the lower half of the complex $\omega$ plane. If the
asymptotic causality condition is satisfied, the imaginary part of
$\omega$ is positive; therefore the singularities only arise in the
upper half-plane and the system will be stable. On the other hand,
if the asymptotic causality condition is violated, the singularities
may appear in the lower half-plane and the system is unstable. 

In order to demonstrate that the divergences of group velocity in
shear modes do not violate the causality, it is necessary to compute
the contour integrals (\ref{eq:dX-01}) in the complex $\omega$ plane.
To close the contour, the asymptotic behavior of the dispersion relations,
i.e., the behavior in large $k$ limit, must be known. In large $k$
limit, the exponential in Eq.(\ref{eq:dX-01}) becomes 
\begin{equation}
\exp[i\omega t-ik_{j}(\omega)x]\rightarrow\exp\left[-i\,\frac{\omega}{v_{gj}^{{\rm as}}}\,(x-v_{gj}^{{\rm as}}\, t)\right]\;,
\end{equation}
with 
\begin{equation}
\lim_{k\rightarrow\infty}\textrm{Re}\omega_{j}(k)=v_{gj}^{{\rm as}}\, k\;,
\end{equation}
where $v_{gj}^{{\rm as}}$ is given by Eq. (\ref{eq:sound-modes-large-k}). 

If $x>v_{gj}^{{\rm as}}\, t$, the integral contour must be closed
in the lower half-plane. If the asymptotic causality condition is
fulfilled, Eq.(\ref{eq:dX-01}) vanishes since there are no singularities
in the lower half-plane. If $x\leq v_{gj}^{{\rm as}}\, t$, the integral
contour must be closed in the upper half-plane. Therefore the value
of $\delta X(x,t)$ in Eq.(\ref{eq:dX-01}) will be nonzero because
of the singularities. On the condition that the asymptotic causality
condition is fulfilled , i.e., the asymptotic group velocity $v_{gj}^{as}$
is smaller than the speed of light, the locations $x$ lie within
the light cone. In that case, the system is causal. 

The conclusion is that the causality of theory as a whole is guaranteed
by the asymptotic causality condition (\ref{eq:asym-casual-cond}).

\section{Characteristic velocity\label{sec:Characteristic-velocity}}

\begin{figure}[tbph]
\begin{centering}
\includegraphics{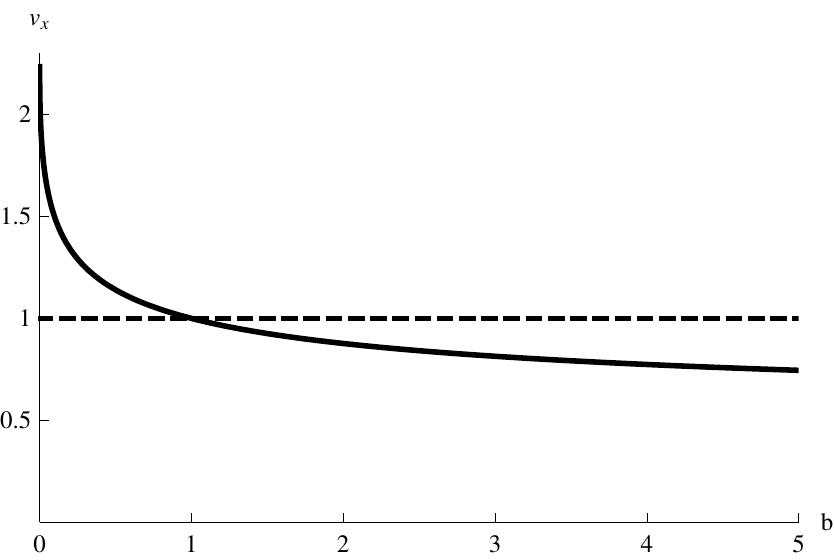}
\par\end{centering}

\begin{centering}
\includegraphics{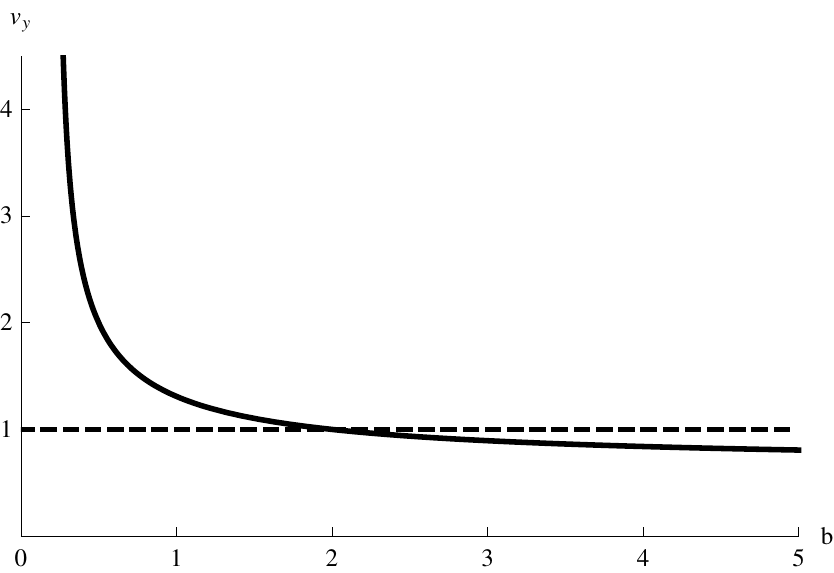}
\par\end{centering}

\caption{\label{fig:char-velocity-01}One of the five characteristic velocities
given by Eq.(\ref{eq:char-velocity-02}) with $u^{\mu}=(\sqrt{5}/2,\;1/2,\;0),\;\pi^{xx}=\pi^{xy}=0$,
and $c_{s}^{2}=1/2$.}
\end{figure}

The fluid-dynamical equations with nonlinear effect can be written
as 
\begin{equation}
\left(A_{ab}^{t}\partial_{t}+A_{ab}^{x}\partial_{x}+A_{ab}^{y}\partial_{y}\right)Y_{b}=B_{a}\;,\label{eq:char-velocity-01}
\end{equation}
with $Y_{b}^{T}=(\varepsilon,u^{x},u^{y},\pi^{xx},\pi^{xy})$ and
$B_{a}^{T}=(0,\;0,\;0,\;\pi^{xx},\;\pi^{xy})$. The expressions of
the matrix $A$ are given in the Appendix \ref{chap:Matrix-elements}.
The characteristic velocities are given by the solution of the following
equations \cite{Hiscock1983,Hiscock1985,Hiscock1987,Olson:1989dp,Olson:1989eu}
\begin{eqnarray}
\det(v_{x}A^{t}-A^{x}) & = & 0\;,\nonumber \\
\det(v_{y}A^{t}-A^{y}) & = & 0\;.\label{eq:char-velocity-02}
\end{eqnarray}
In the local rest frame, the characteristic velocities are given by
\begin{equation}
v_{x}=v_{y}=\left\{ \begin{array}{l}
0\;,\\
{\displaystyle \pm\sqrt{\frac{1}{b}}\;,}\\
{\displaystyle \pm\sqrt{\frac{1}{b}+c_{s}^{2}}\;.}
\end{array}\right.\;,
\end{equation}
The numerical results for the $b$ dependence of one of the five characteristic
velocities are shown in Fig. \ref{fig:char-velocity-01}. 

The characteristic velocity which includes all the non-linear effect
of the fluid dynamics also show that the system will be causal if
the asymptotic causality condition is fulfilled. 

\section{Discussion}

In this chapter, the analysis of causality and stability of the fluid
dynamical equations is performed. Considering a linear analysis, the
so-called asymptotic causality condition (\ref{eq:asym-casual-cond})
is obtained. The divergences in shear modes are also observed. However,
the analysis from the contour integrals show that the causality as
a whole is determined by the asymptotic causality condition. The stability
is found to be intimately related with the causality of the system
in a Lorentz boosted frame. The system will be always stable if the
asymptotic causality condition (\ref{eq:asym-casual-cond}) is fulfilled.
More work on this topic could be found in Ref. \cite{Pu:2009fj} for
the competition of bulk and shear viscosity, Ref. \cite{Pu:2010zz}
for the analysis of the fluid dynamics with the heat conductivity
only. 

From the asymptotic causality condition, it is found that the NS equation
is acausal if the relaxation time for shear viscosity goes to zero.
On the other hand, it also implies that the equations of second order
theory are not automatically causal by construction. It is easy to
check that the results from the Grad's 14 moments \cite{Israel:1979wp,Betz:2010cx}
as well as the results from the AdS/CFT fulfill the asymptotic causality
condition (\ref{eq:asym-casual-cond}).



\chapter{Transport coefficients by Boltzmann equation\label{chap:Transport}}

In chapter 2, we have introduced IS theory of hydrodynamics \cite{Israel:1979wp,Betz:2010cx,Muronga:2006zw}.
There are many transport coefficients in the theory. These coefficients
cannot be determined in hydrodynamics but can only be determined in
the underlying microscopic theory. The collision term in the Boltzmann
equation has the microscopic nature since it is given by the invariant
amplitudes of microscopic processes. In this chapter we will discuss
about the procedure of computing these coefficients in the Boltzmann
approach based on Ref. \cite{Chen:2009sm,Chen:2010xk,Chen2010a,Chen2010}. 

In a leading order expansion of the coupling constant, there are an
infi{}nite number of diagrams in non-equilibrium field theory \cite{Jeon1995,Jeon:1995zm}.
However, it is proven that the summation of the leading order diagrams
in a weakly coupled $\phi^{4}$ theory \cite{Jeon1995,Jeon:1995zm,Carrington:1999bw,Wang:1999gv,Hidaka:2010gh}
or in hot QED \cite{Gagnon:2007qt} is equivalent to solving the linearized
Boltzmann equation with temperature-dependent particle masses and
scattering amplitudes. The conclusion is expected to hold in weakly
coupled systems and can as well be used to compute the leading order
transport coeffi{}cients in QCD-like theories \cite{Arnold2003a,Chen:2009sm,Chen:2010xk,Arnold:2007pg,Arnold2001},
hadronic gases \cite{Prakash:1993bt,Dobado:2001jf,Dobado:2003wr,Chen:2006iga,Chen:2007xe,Itakura:2007mx}.

\section{Order expansion\label{sec:Order-expansion}}

The basic feature of relativistic Boltzmann equation (\ref{eq:Boltzmann-eq-01})
is shown in Sec. \ref{sec:BE}. In some case, we can add the external
field $F^{\mu\nu}=\partial^{\mu}A^{\nu}-\partial^{\nu}A^{\mu}$ to
the Boltzmann equation 
\begin{eqnarray}
\frac{df(x,p)}{dt} & = & \frac{p^{\mu}}{E_{p}}\left(\partial_{\mu}-qF^{\mu\nu}\frac{\partial}{\partial p^{\nu}}\right)f=\mathcal{C}[f]\;,\label{eq:boltzmann-F}
\end{eqnarray}
which can be written in a classical form. Here $q$ is the conserved
charge, $qF^{\mu\nu}\frac{p^{\mu}}{E_{p}}=q(\mathbf{E}+\mathbf{v}^{\prime}\times\mathbf{B})$
is the classical Lorentz force with $\mathbf{v}^{\prime}\equiv\mathbf{p}/E_{p}$
the velocity of a particle. 

The distribution function $f_{0}(x,p)$ in an equilibrium state is
given in Eq.(\ref{eq:f0-01}). It is straightforward to extend it
to an off-equilibrium state is 
\begin{equation}
f(x,p)=\frac{1}{e^{\beta(u\cdot p-\mu q)+\chi(x,p)}-a}\;,\label{eq:complete-distribution}
\end{equation}
where $\chi\sim\partial_{\mu}u_{\nu}\sim O(K)$ is a small quantity.
The complete distribution function $f(x,p)$ can be expanded in terms
of the power series of $K$ around an equilibrium state

\begin{eqnarray}
f(x,p) & = & f_{0}(x,p)+f_{1}(x,p)+O(K^{2})\;,\label{eq:f-expansion}
\end{eqnarray}
where 
\begin{equation}
f_{1}=-\chi(x,p)f_{0}(x,p)[1+af_{0}(x,p)]\;,
\end{equation}
Substituting Eq.(\ref{eq:f-expansion}) into the Boltzmann equation
(\ref{eq:boltzmann-F}) yields the Boltzmann equation in the $i$-th
order 
\begin{equation}
\frac{df_{i}}{dt}=C_{i}[f]\;,\label{eq:Boltzmann_2}
\end{equation}
where 
\begin{equation}
C_{i}[f]=\sum_{n=1}^{i}C^{(n)}[f_{i-n}]\;,\label{eq:C-normal-exp}
\end{equation}
with $C^{(n)}$ the $n$-th order derivative of the collision term
$C[f]$. Note that the energy conservation in the process of two particles
scattering $ab\rightarrow cd$, i.e. $E_{a}+E_{b}=E_{c}+E_{d}$ leads
to the vanishing of $C[f_{0}]$, 
\begin{eqnarray*}
C[f_{0}] & = & \frac{1}{2}\int_{abc}d\Gamma_{ab\rightarrow cp}[f_{0}^{a}f_{0}^{b}(1+af_{0}^{c})(1+af_{0}^{p})-(1+af_{0}^{a})(1+af_{0}^{b})f_{0}^{c}f_{0}^{p}]\\
 & = & \frac{1}{2}\int_{abc}d\Gamma_{ab\rightarrow cp}(1+af_{0}^{c})(1+af_{0}^{p})(1+af_{0}^{a})(1+af_{0}^{b})\\
 &  & \quad\times\left[\frac{f_{0}^{a}f_{0}^{b}}{(1+af_{0}^{a})(1+af_{0}^{b})}-\frac{f^{c}f^{p}}{(1+af_{0}^{c})(1+af_{0}^{p})}\right]\\
 & = & 0\;.
\end{eqnarray*}
Here we have used 
\begin{equation}
\frac{f_{0}^{a}f_{0}^{b}}{(1+af_{0}^{a})(1+af_{0}^{b})}=\frac{f_{0}^{c}f_{0}^{p}}{(1+af_{0}^{c})(1+af_{0}^{p})}\;,
\end{equation}

The derivative of the $f_{0}(x,p)$ reads 
\begin{eqnarray}
\frac{df_{0}}{dt} & = & -f^{eq}[1+af^{eq}(x,p)]\left\{ \frac{p^{\mu}}{E_{p}}\partial_{\mu}\left[\beta(u\cdot p-\mu q)\right]\right.\nonumber \\
 &  & \left.\vphantom{\frac{p^{j}}{E_{P}}}+[q(\mathbf{E}+\mathbf{v}^{\prime}\times\mathbf{B})]\cdot\nabla_{p}\left[\beta(u\cdot p-\mu q)\right]\right\} \;.
\end{eqnarray}
It is convenient to evaluate the above equation in the local rest
frame (i.e. $u^{0}=1$, $\mathbf{u}=0$), 
\begin{eqnarray}
\left.\frac{df_{0}}{dt}\right|_{\textrm{LRF}} & = & -f^{eq}[1+af^{eq}(x,p)]\left[E_{p}\partial_{t}\beta+p^{i}\left(\partial_{i}\beta+\beta\partial_{t}u_{i}\right)+\frac{p^{i}p^{j}}{E_{p}}\beta\partial_{\iota}u_{j}\right.\nonumber \\
 &  & \left.+\frac{p^{\mu}}{E_{P}}q\partial_{\mu}\alpha+\beta[q(\mathbf{E}+\mathbf{v}^{\prime}\times\mathbf{B})]\cdot\frac{\mathbf{p}}{E_{p}}\right]\;.
\end{eqnarray}
Using the Eq.(\ref{eq:eom-ex-massless}) in the local rest frame,
and the following identity 

\begin{equation}
p^{i}p^{j}\partial_{j}u_{i}=\left(p^{i}p^{j}-\frac{1}{3}\delta^{ij}\mathbf{p}^{2}\right)\sigma_{ij}+\frac{1}{3}\mathbf{p}^{2}\nabla\cdot\mathbf{u}\;,
\end{equation}
with $\sigma_{ij}=\frac{1}{2}\left(\partial_{j}u_{i}+\partial_{i}u_{i}\right)-\frac{1}{3}\delta_{ij}\nabla\cdot\mathbf{u}$,
we obtain 
\begin{eqnarray}
\left.\frac{df_{0}}{dt}\right|_{\textrm{LRF}} & = & -f_{0}[1+af_{0}(x,p)]\left[\left(\frac{n}{\epsilon+P}+\frac{q}{E_{p}}\right)\left(\mathbf{p}\cdot\nabla\alpha+\mathbf{p}\cdot\frac{\mathbf{E}}{T}\right)\right.\nonumber \\
 &  & +\quad\left.\frac{\beta}{E_{p}}\left(p^{i}p^{j}-\frac{1}{3}\delta^{ij}\mathbf{p}^{2}\right)\sigma_{ij}\right]\;,
\end{eqnarray}
where the particles are assumed to be massless and therefore the bulk
viscous pressure is vanished. 

Because of the Boltzmann equation (\ref{eq:Boltzmann_2}) the dissipative
term $\chi$ in Eq.(\ref{eq:complete-distribution}) can be written
as 
\begin{eqnarray}
\chi & = & \frac{1}{T}\left[A_{1}(p)\left(p^{i}p^{j}-\frac{1}{3}\delta^{ij}\mathbf{p}^{2}\right)\sigma_{ij}\right.\nonumber \\
 &  & \left.+A_{2}(p)\left(\frac{n}{\epsilon+P}+\frac{q}{E_{p}}\right)\left(\mathbf{p}\cdot\nabla\alpha+\mathbf{p}\cdot\frac{\mathbf{E}}{T}\right)\right]\;,\label{eq:chi-BE-01}
\end{eqnarray}
which indicates the fact that the heat conductivity is equal to the
electric conductivity. By considering the contribution from the external
fields, the $\nu^{\mu}$ in Eq.(\ref{eq:NS-equation-01}) now becomes
\begin{equation}
\nu^{\mu}=\kappa T\Delta^{\mu\nu}\left(\partial_{\nu}\alpha+\frac{E_{\nu}}{T}\right)\;.
\end{equation}

\section{Shear viscosity}

To show the details of computing transport coefficients via the Boltzmann
equation, we choose the shear viscosity as an example. For a review
of the shear viscosity, see e.g. Ref. \cite{Schafer:2009dj}.

\subsection{Non-relativistic system}

Suppose the fluid is following along the $y$ direction with fluid
velocity $v_{y}(x)$ which is a function of the transverse position
$x$. The friction force per unit area felt in the $yz$ plane is
proportional to the gradient of $v_{y}$ along $x$, 
\begin{equation}
\frac{F}{A}=\eta\partial_{x}v_{y}\;,\label{eq:def-shear-01}
\end{equation}
where $\eta$ is the shear viscosity and its inverse is called the
fluidity. In the molecular theory of dilute gases, one can estimate
the value of shear viscous coefficient $\eta$ as Maxwell did. The
number of particles which are moving through the unit area in $yz$
plane in unit time is 
\begin{equation}
\Delta N=v_{x}fdv_{x}dv_{y}dv_{z}\;,
\end{equation}
with $f$ the distribution function. The total momentum transferred
in unit time gives the force
\begin{equation}
\frac{F}{A}=-\int_{-\infty}^{\infty}mv_{y}\Delta N\;,
\end{equation}
which vanishes if using the equilibrium Maxwell distribution function.
It implies that the shear viscous effect is a dissipative phenomenon
in an off-equilibrium state. In order to evaluate this effect, an
off-equilibrium distribution function $f=f_{0}+f_{1}$, where $f_{0}$
is the equilibrium one and $f_{1}$ is the fluctuation near the equilibrium
state, is considered. The linearized Boltzmann equation $\partial_{t}f+\mathbf{v}\cdot\boldsymbol{\partial}f=0$
gives 
\begin{equation}
v_{x}\partial_{x}f_{0}=-\frac{1}{\tau_{0}}f_{1}\;,\label{eq:temp-eq-BE-01}
\end{equation}
where the formula in the right hand side is given by the assumption
$\partial_{t}(f-f^{0})=-(f-f^{0})/\tau_{0}$ with $\tau_{0}$ the
time in which the system comes back to the equilibrium state. The
solution of Eq.(\ref{eq:temp-eq-BE-01}) is 
\begin{equation}
f_{1}=-\tau_{0}v_{x}\partial_{x}f_{0}=-\tau_{0}v_{x}\frac{\partial f_{0}}{\partial v_{y}}\frac{dv_{y}}{dx}\;.
\end{equation}
Substituting the above solution into the expression (\ref{eq:def-shear-01})
yields the shear viscosity
\begin{equation}
\eta=-\int mv_{x}^{2}v_{y}\tau_{0}\frac{\partial f_{0}}{\partial v_{y}}dv_{x}dv_{y}dv_{z}=m\tau_{0}\int v_{x}^{2}f_{0}dv\;.
\end{equation}
More simplification will be taken. In non-relativistic statistic physics,
it is known that 
\[
m\int v_{x}^{2}f_{0}dv=nm\overline{v_{x}^{2}}=nk_{B}T\;,
\]
where the $k_{B}$ is the Boltzmann constant and $n$ is the particle
number density. Thus, the shear viscosity will be 
\begin{equation}
\eta=\tau_{0}nk_{B}T\sim\frac{l_{mfp}}{\bar{v}}nk_{B}T\propto\sqrt{T}\;,
\end{equation}
where $l_{mfp}$ is the mean free path which is proportion to $n^{-1}$.
This result indicates that in non-relativistic case the shear viscosity
is only a function of temperature $T$ and is independent on the number
density $n$. By using the approximation that $\bar{v}_{x}\simeq\frac{1}{3}\bar{v}$
and $\bar{v}_{x}^{2}=\overline{v_{x}^{2}}$, the famous formula given
by Maxwell is obtained 
\begin{equation}
\eta=\frac{1}{3}npl_{mfp}\;.\label{eq:non-shear-gas}
\end{equation}

Instead of the picture of quasi-particles, Frenkel and etc.\cite{Frenkel1955}
gave a simple picture for the motion of liquid molecules. The shear
viscosity in a liquid is given by
\begin{equation}
\eta\simeq hne^{E/(k_{B}T)}\;,\label{eq:non-shear-liquid}
\end{equation}
where $h$ is the Planck constant and the collision time of the molecules
was assumed to be $h/(k_{B}T)$ which is the shortest timescale in
the liquid. In contrast to the results (\ref{eq:non-shear-gas}),
the shear viscosity in the liquid decreases with temperature. Therefore
the value of the shear viscosity must be minimum in the critical point
of liquid-gas phase transition. 

From Eq.(\ref{eq:non-shear-gas}) and Eq.(\ref{eq:non-shear-liquid}),
the ratio $\eta/n$ (or $\eta/\rho$, the kinetic viscosity with $\rho=mn$)
is found to be a good quantity to describe the minimum value in the
critical point. In QGP created by heavy ion collisions only net number
of quarks is well-defined. The exact number of quarks and gluons is
unknown. In that case, the ratio $\eta/n$ is not good enough to describe
the properties of the fireball. The entropy density $s$ is well-defined
and proportional to the $n$. Therefore, one can choose $\eta/s$
around the phase transition to replace $\eta/n$. The uncertainty
relation gives $pl_{mfp}\simeq\hbar$, thus, $\eta/s\gtrsim\hbar/k_{B}\sim1$
\cite{Danielewicz:1984ww} which is an estimate of $1/(4\pi)$ from
the AdS/CFT correspondence \cite{Policastro2001,Son2007b}.

\subsection{Relativistic system and variational approach}

In this section we will introduce the variational method in the Boltzmann
approach to shear viscosity. We now turn to a relativistic system.
A good example is the shear viscosity for a quark gluon system \cite{Chen:2009sm,Chen:2010xk,Chen2010,Chen2010a}. 

The shear viscous term in Eq.(\ref{eq:chi-BE-01}) can be rewritten
as
\begin{equation}
\chi=\beta B_{ij}\sigma_{ij}\;,\label{eq:chi-BE-02}
\end{equation}
where the heat and electric conductivities are ignored and
\begin{equation}
B_{ij}=A_{1}(p)\left|\mathbf{p}\right|^{2}I_{ij}\;,\label{eq:Bij-BE-01}
\end{equation}
with 
\begin{equation}
I_{ij}=\frac{1}{\left|\mathbf{p}\right|^{2}}\left(p^{i}p^{j}-\frac{1}{3}\delta^{ij}\mathbf{p}^{2}\right)\;.
\end{equation}
Substituting Eq.(\ref{eq:chi-BE-02}) into the expression of $T^{\mu\nu}$
up to the first order of the power series of $K$ yields 
\begin{equation}
\eta=\frac{1}{10T}\int_{p}\frac{1}{E_{p}}f_{p}\Delta_{p}\left|\mathbf{p}\right|^{2}I_{ij}B_{ij}(p)\equiv(S,B)\;,\label{eq:shear_B_01}
\end{equation}
where $S$ and $B$ are matrices with the components $S_{ij}=\left|\mathbf{p}\right|^{2}I_{ij}$. 

On the other hand, the linearized Boltzmann equation (\ref{eq:Boltzmann_2})
with $\chi$ in Eq.(\ref{eq:chi-BE-02}) reads
\begin{equation}
\left|\mathbf{p}\right|^{2}I_{ij}=\frac{E_{p}}{2}\int_{123}d\Gamma_{12,3p}\Delta_{1}\Delta_{2}f_{3}(\Delta_{P})^{-1}\left[B_{ij}(p)+B_{ij}(k_{3})-B_{ij}(k_{2})-B_{ij}(k_{1})\right]\;,\label{eq:I_ij_1}
\end{equation}
which involves collision terms. The above equation can be written
in a compact form 
\begin{equation}
S=CB\;,\label{eq:scb-BE-01}
\end{equation}
where the matrix $C$ is determined by Eq.(\ref{eq:I_ij_1}). 

By using Eq.(\ref{eq:scb-BE-01}), Eq.(\ref{eq:shear_B_01}) becomes

\begin{eqnarray}
\eta & = & \frac{1}{20T}\int_{123p}d\Gamma_{12,3p}\Delta_{1}\Delta_{2}f_{3}f_{p}\left[B_{ij}(p)+B_{ij}(k_{3})-B_{ij}(k_{2})-B_{ij}(k_{1})\right]B_{ij}(p)\nonumber \\
 & = & \frac{1}{80T}\int_{123p}d\Gamma_{12,3p}\Delta_{1}\Delta_{2}f_{3}f_{p}\left[B_{ij}(p)+B_{ij}(k_{3})-B_{ij}(k_{2})-B_{ij}(k_{1})\right]^{2}\nonumber \\
 & = & (B,CB)\;,\label{eq:shear_B_02}
\end{eqnarray}
which implies that the matrix $C$ is positive. Then we obtain
\begin{equation}
(S,B)=(B,CB)\;.\label{eq:equal_B_01}
\end{equation}

A straightforward way of computing the shear viscosity is to employ
the solution to Eq. (\ref{eq:scb-BE-01}) in Eq.(\ref{eq:shear_B_01}).
However, there are two problems. The first one is from the numerical
technique. The numerical errors in this kinds of integration equations
will lead to some kinds of the divergent behavior \cite{Arnold2003a,Arnold2000}.
The second problem is the solutions of Eq.(\ref{eq:scb-BE-01}) might
not fulfill the Eq.(\ref{eq:equal_B_01}) since the solutions of integration
equations are not unique. 

Instead of solving Eq.(\ref{eq:equal_B_01}) directly, the variational
method is normally used. Eq.(\ref{eq:shear_B_02}) can be rewritten
as 
\begin{eqnarray}
\eta & = & 2(S,B)-(B,CB)\nonumber \\
 & = & (S,C^{-1}S)-(A,CA)\;,\label{eq:variation-BE-01}
\end{eqnarray}
where $A=B-C^{-1}S$. If Eq.(\ref{eq:I_ij_1}) is not fulfilled, $\eta\leq(S,C^{-1}S)$
because of positive $C$. In numerical calculations, one needs to
obtain the maximum value of Eq.(\ref{eq:variation-BE-01}) \cite{Arnold2003a,Arnold2000}. 

In variational approach, we solve Eq. (\ref{eq:equal_B_01}) instead
of Eq. (\ref{eq:scb-BE-01}). The critical step is to find a good
form of $A_{1}(p)$ to make $\eta$ as large as possible (see e.g.
\cite{Chen:2009sm,Chen:2010xk,Chen2010,Chen2010a,Dobado:2001jf}).
As an assumption, one could expand $A_{1}(p)$ by a set of orthogonal
polynomials
\begin{equation}
A_{1}(p)=\left|\mathbf{p}\right|^{y}\sum_{r=0}^{r_{max}}b_{r}B^{(r)}(\beta p)\;.\label{eq:Br-BE-01}
\end{equation}
where $B^{(r)}(\beta p)$ is a polynomial up to $(\beta p)^{r}$ and
$b_{r}$ is its coefficient. Here $y$ is a constant to make the numerical
error get the fastest convergence \cite{Chen2010,Dobado:2001jf}.
In the numerical calculations, $y$ is chosen to be $1$ or $2$.
The orthogonal condition is set to be 
\begin{equation}
\frac{1}{15T}\int_{p}\frac{\left|\mathbf{p}\right|^{2+y}}{E_{p}}f_{p}\Delta_{p}B^{(r)}(\beta p)B^{(s)}(\beta p)=\mathcal{B}^{(r)}\delta_{r,s}\;,\label{eq:orthogonal_conidtion_01}
\end{equation}
where $\mathcal{B}^{(r)}$ is constant depending on the integrals
in Eq. (\ref{eq:orthogonal_conidtion_01}). Without loss of generality,
we can assume $B^{(0)}=1$. Substituting Eq.(\ref{eq:Br-BE-01}) into
Eq.(\ref{eq:shear_B_02}) yields 
\begin{eqnarray}
\eta_{test} & = & \sum_{r,s}b_{r}b_{s}\widetilde{C}_{rs}\equiv<b|\widetilde{C}|b>\;,\label{eq:shear_test_b_01}
\end{eqnarray}
where $|b>=(b_{0},b_{1},...,b_{r_{max}})^{T}$, the inner product
is defined as $<A|B>=AB$ and 
\begin{eqnarray*}
\widetilde{C}_{rs} & = & \frac{1}{80T}\int_{123p}d\Gamma_{12,3p}\Delta_{1}\Delta_{2}f_{3}f_{p}\\
 & \times & \sum_{m,n=1}^{4}(-1)^{m+n}I_{ij}(\mathbf{p}_{m}\cdot\mathbf{p}_{n})\left(\left|\mathbf{p}_{n}\right|\left|\mathbf{p}_{m}\right|\right)^{y}B^{(r)}(\beta p_{m})B^{(s)}(\beta p_{m})\;,
\end{eqnarray*}
where $p_{i}=(p,k_{1},k_{2},k_{3})$ and $\widetilde{C}$ is found
to be a positive constant matrix into Eq.(\ref{eq:shear_B_01}). Inserting
Eq.(\ref{eq:Br-BE-01}), we have 
\begin{eqnarray}
\eta_{test} & = & \sum_{r}b_{r}S_{r}=<S|b>\;,\label{eq:shear_test_b_02}
\end{eqnarray}
where $|S>=(S_{0},S_{1},...,S_{r_{max}})^{T}$ with the $r$-th component
\begin{equation}
S_{r}=\frac{1}{15T}\int_{p}\frac{1}{E_{p}}f_{p}\Delta_{p}\left|\mathbf{p}\right|^{2+y}B^{(r)}(\beta p)\;.\label{eq:sr-BE-01}
\end{equation}
From Eq.(\ref{eq:shear_test_b_01}) and (\ref{eq:shear_test_b_02}),
$|b>=\widetilde{C}^{-1}|S>$, we obtain 
\begin{eqnarray*}
\eta_{test} & = & <S|\widetilde{C}^{-1}|S>\\
 & = & \sum_{r,s}\int_{p}\left(\frac{1}{15T}\frac{1}{E_{p}}f_{p}\Delta_{p}\right)^{2}\left|\mathbf{p}\right|^{4+2y}B^{(r)}(\beta p)B^{(s)}(\beta p)\widetilde{C}_{rs}^{-1}(p)\;.
\end{eqnarray*}

According to the orthogonality condition (\ref{eq:orthogonal_conidtion_01}),
only $B^{(0)}$ will survive in Eq.(\ref{eq:shear_test_b_02}), then
we finally obtain 
\begin{equation}
\eta_{test}=\int\frac{1}{15T}\frac{1}{E_{p}}f_{p}\Delta_{p}\left|\mathbf{p}\right|^{2+y}b_{0}=b_{0}\mathcal{B}^{(0)}=(S_{0})^{2}(\widetilde{C}^{-1})_{00}\;.\label{eq:shear_b_03}
\end{equation}
Moreover, it is proved by the authors of Ref. \cite{Chen2010,Chen2010a}
that the value of $\eta_{test}$ will increase with $r_{max}$ increasing.
Therefore, the value of $r_{max}$ depends on the numerical precision.

\section{An example: shear viscosity of a gluon plasma}

Recently perturbative QCD calculation of $\eta/s$ of a gluon plasma
has raised wide attention. Xu and Greiner (XG) used a parton cascade
model to calculate $\eta/s$ \cite{Xu:2007jv,Xu2008}. They claimed
that the dominant contribution comes from the inelastic $gg\leftrightarrow ggg$
(23) process instead of the elastic $gg\rightarrow gg$ (22) process:
the 23 process is 7 times more important than 22. This result is in
sharp contrast to AMY's result \cite{Arnold2003a,Arnold2000} where
the 23 process only gives $\sim10\%$ correction to the 22 process.

Both XG and AMY use kinetic theory for their calculations. The main
differences are \cite{Chen:2009sm,Chen:2010xk} (i) XG uses a parton
cascade model \cite{Xu2005} to solve the Boltzmann equation and,
for technical reasons, gluons are treated as a classical gas instead
of a bosonic gas. On the other hand, AMY solves the Boltzmann equation
for a bosonic gas. (ii) AMY approximates the $Ng\leftrightarrow(N+1)g$
processes, $N=2,3,4\ldots$, by the $g\leftrightarrow gg$ splitting
in the collinear limit where the two gluon splitting angle is higher
order. XG uses the soft gluon bremsstrahlung limit where one of the
gluon momenta in the final state of $gg\rightarrow ggg$ is soft but
it can have a large splitting angle with its mother gluon.

In an earlier attempt to resolve the discrepancy between XG's and
AMY's results \cite{Chen:2009sm}, a Boltzmann equation computation
of $\eta$ is carried out without taking the classical gluon approximation
(like AMY's approach) but the soft gluon bremsstrahlung limit is applied
to the 23 matrix element. It was found that the classical gas approximation
does not cause a significant error in $\eta/s$. However, the result
is sensitive to whether the soft gluon bremsstrahlung limit is imposed
on the phase space or not. If this limit is imposed, the result is
closer to AMY's; if not, the result is closer to XG's. This raises
the concern whether this approximation is good for computing $\eta$.

This issue has been settled in Ref. \cite{Chen:2010xk} by using the
exact amplitude for the 23 process, which removes both the soft gluon
bremsstrahlung approximation and the collinear approximation to the
23process. The result of Ref. \cite{Chen:2010xk} shows that the contribution
from the 23 process lies between AMY's and XG's result but more close
to AMY's. So the 23 process is less important than the 22 one in most
range of coupling constant. This is consistent to the perturbative
approach where higher order processes are only perturbation to lower
order processes.



\chapter{Applications of AdS/CFT duality\label{chap:Applications-of-AdS/CFT}}

Inspired by the great success of computing the ratio $\eta/s$ of
a strongly coupled super Yang-Mills plasma by AdS/CFT duality, many
work \cite{Heller2007,Janik:2006ft,Sin:2006pv,Janik2006,Janik:2006gp,Yee:2009wn}
appear on the market about applying AdS/CFT correspondence to relativistic
hydrodynamics with Bjorken boost invariance \cite{Bjorken:1982qr}.
People try to establish a well-defined gravity dual to relativistic
fluid dynamical by AdS/CFT duality. 

In Sec. \ref{sec:Shear-viscosity-late}, we investigate the shear
viscosity $\eta$ of strongly coupled super Yang-Mills (SYM) plasma
in late time of hydrodynamic evolution with Bjorken scaling via AdS/CFT
duality. We obtain the metric $g_{\mu\nu}$ in a proper time dependent
$AdS_{5}$ space via holographic renormalization, whose boundary condition
is given by energy-momentum tensor of the QGP with transverse expansion
or radial flow. With this metric we compute $\eta$ of fluids in 0+1
and 1+1 dimension without and with radial flow. We find the ratio
$\eta/s=1/(4\pi)$ in 0+1 dimension consistent with the KSS bound
if next-to-leading terms in proper time are included in the equation
of motion for metric perturbations. For 1+1 dimension the result is
unchanged in the leading order of transverse rapidity \cite{Pu:2010zza}.

In Sec. \ref{sec:High-spin-baryon}, we consider a string-junction
holographic model of a probe baryon in the finite-temperature AdS
background. We investigate the screening length for a high spin baryon.
By defining the screening length as the critical separation of quarks,
we compute the $\omega$ (spin) dependence of the baryon screening
length numerically and find that baryons with high spin dissociate
more easily. Finally, we discuss the Regge-like relation between the
angular momentum $J$ and the total energy $E^{2}$ for baryons \cite{Li:2008py}.

\section{Shear viscosity in late time\label{sec:Shear-viscosity-late}}

\subsection{Kubo relation}

The Kubo relation can be obtained by the statistical analysis \cite{kubo1984,Lifshitz}.
Here a simple way given to derive the Kubo formula given based on
Ref. \cite{Baier2008} which is associated with the AdS/CFT duality. 

The action with a source $J_{a}(x)$ and its operator $O_{a}(x)$
is written as

\[
S=S_{0}+\int_{x}J_{a}(x)O_{a}(x)\;.
\]
The perturbation of $J_{a}(x)$ gives 

\begin{equation}
\left\langle O_{a}(x)\right\rangle =-\int_{y}G_{ab}^{R}(x-y)J_{b}(x)\;,\label{eq:source-Green-function}
\end{equation}
with the help of the linear response theory. Here $G_{ab}^{R}$ is
the retarded Green's function defined by
\begin{equation}
iG_{ab}^{R}(x-y)=\theta(x^{0}-y^{0})\left\langle [O_{a}(x),O_{b}(y)]\right\rangle \;.
\end{equation}

It is known that the metric $g^{\mu\nu}$ as a source is coupled to
$T^{\mu\nu}$ as an operator in the general theory of relativity.
For the sake of simplicity, the metric $g^{\mu\nu}$ is considered
as a homogeneous one $\delta_{ij}$ with a perturbation $h_{ij}(t)\ll1$.
Moreover, $h_{ij}$ is assumed to be traceless $h_{ii}=0$. In the
local rest frame, i.e. $u^{\mu}=(1,0,0,0)$, the shear viscous tensor
defined in Eq.(\ref{eq:NS-equation-01}) with covariant derivatives
reads

\begin{equation}
\sigma_{xy}=2\eta\Gamma_{xy}^{0}=\eta\partial_{0}h_{xy}\;.\label{eq:shear-kubo-01}
\end{equation}
Considering a perturbation in the form of the plane wave $h_{ij}=(h_{0})_{ij}e^{i\omega t}$
with constant $h_{0}$ yields
\begin{equation}
\left\langle \sigma_{xy}\right\rangle =i\omega\eta+O(\omega^{2})\;.
\end{equation}
Using Eq.(\ref{eq:source-Green-function}), one obtains 

\begin{equation}
\left\langle \sigma_{xy}\right\rangle =-\int G_{xy,ij}^{R}h_{ij}=-\int G_{xy,xy}^{R}h_{xy}\;.\label{eq:shear-kubo-02}
\end{equation}
Substituting Eq.(\ref{eq:shear-kubo-02}) into Eq.(\ref{eq:shear-kubo-01})
and taking the low frequency limit, i.e. the fluid dynamical limit, 

\begin{equation}
\eta=\underset{\omega\rightarrow0}{\lim}\frac{1}{\omega}\int dtdxe^{i\omega t}\theta(t)<[T_{xy}(x),T_{xy}(0)]>\;.
\end{equation}
Then we derive the Kubo formula
\begin{equation}
\eta=-\underset{\omega\rightarrow0}{\lim}\frac{1}{\omega}\mathsf{\mathrm{Im}}G_{xy,xy}^{R}(\omega,0)\;,\label{eq:kubo01}
\end{equation}
where in the low wave-number limit the Fourier transform of the retarded
Green function gives
\begin{equation}
G_{xy,xy}^{R}(\omega,0)=-i\underset{k\rightarrow0}{\lim}\int dtdxe^{-ik_{\mu}x^{\mu}}\theta(t)<[T_{xy}(t,x),T_{xy}(0,0)]>\;.
\end{equation}

\subsection{Hydrodynamics with Bjorken boost invariance}

In heavy ion collisions, it is assumed that all particles are created
in the same proper time after the collisions. In the laboratory frame,
one will observe in the center region of the collisions that the particles
moving fast are produced much earlier than those moving slowly due
to the Lorentz transformation. So one can assume that the space rapidity
$\eta=\frac{1}{2}\ln\left(\frac{t+z}{t-z}\right)$ is equal to the
momentum rapidity $y=\frac{1}{2}\ln\left(\frac{p_{0}+p_{z}}{p_{0}-p_{z}}\right)$,
for collisions along the $z$-direction. After some simplification,
the following formula is obtained 
\begin{equation}
\tanh y\equiv\frac{p_{z}}{p_{0}}=v_{z}=\frac{z}{t}\;,
\end{equation}
where $\tau=\sqrt{t^{2}-z^{2}}$ is the transverse proper time. 

Bjorken \cite{Bjorken:1982qr} suggested that all the thermal quantities
should be independent of the rapidity $y$. It is convenient to use
the coordinates $(\tau,y)$ instead of $(t,z)$. Using the coordinates
$(\tau,y)$ and the equation of state $\epsilon=3P$, the thermal
quantities of the ideal fluid can be solved as 
\begin{equation}
x=x_{0}\left(\frac{\tau_{0}}{\tau}\right)^{m}\;,\label{eq:Bjorken-quantities-01}
\end{equation}
where $x_{0}$ is the initial value for $x$ in the proper time $\tau_{0}$,
and for $x=\epsilon,P$, $m=4/3$, for $x=T$, $m=1/3$, and for $x=s,n$,
$m=1$. 

\begin{figure}
\begin{centering}
\includegraphics[scale=0.5]{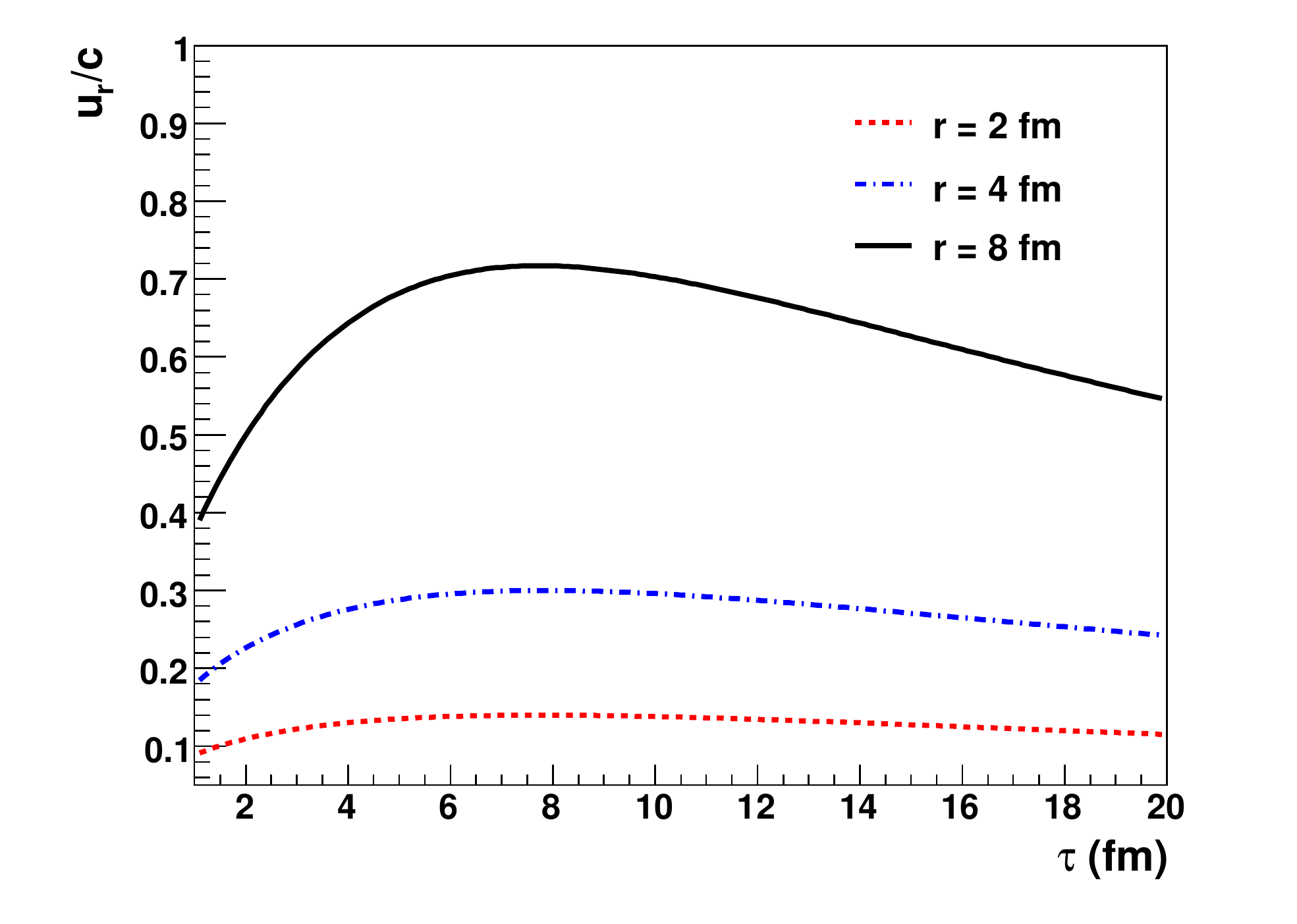}
\par\end{centering}

\begin{centering}
\includegraphics[scale=0.5]{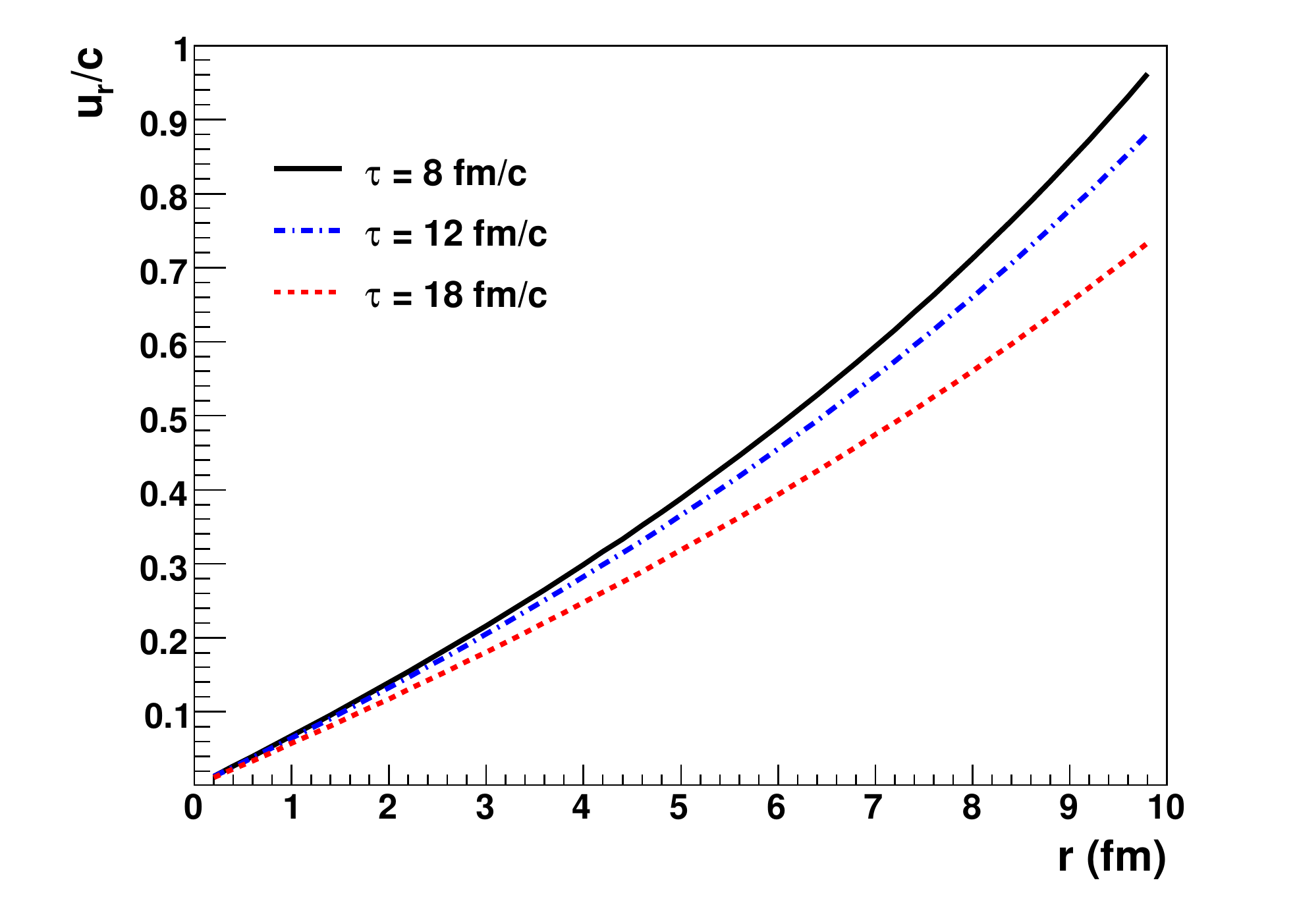}
\par\end{centering}

\caption{The radial velocity $u_{r}=\sinh\alpha$ as functions of $r$ and
$\tau$.\label{fig:radial-velocity}}

\end{figure}

\begin{figure}
\begin{centering}
\includegraphics[scale=0.5]{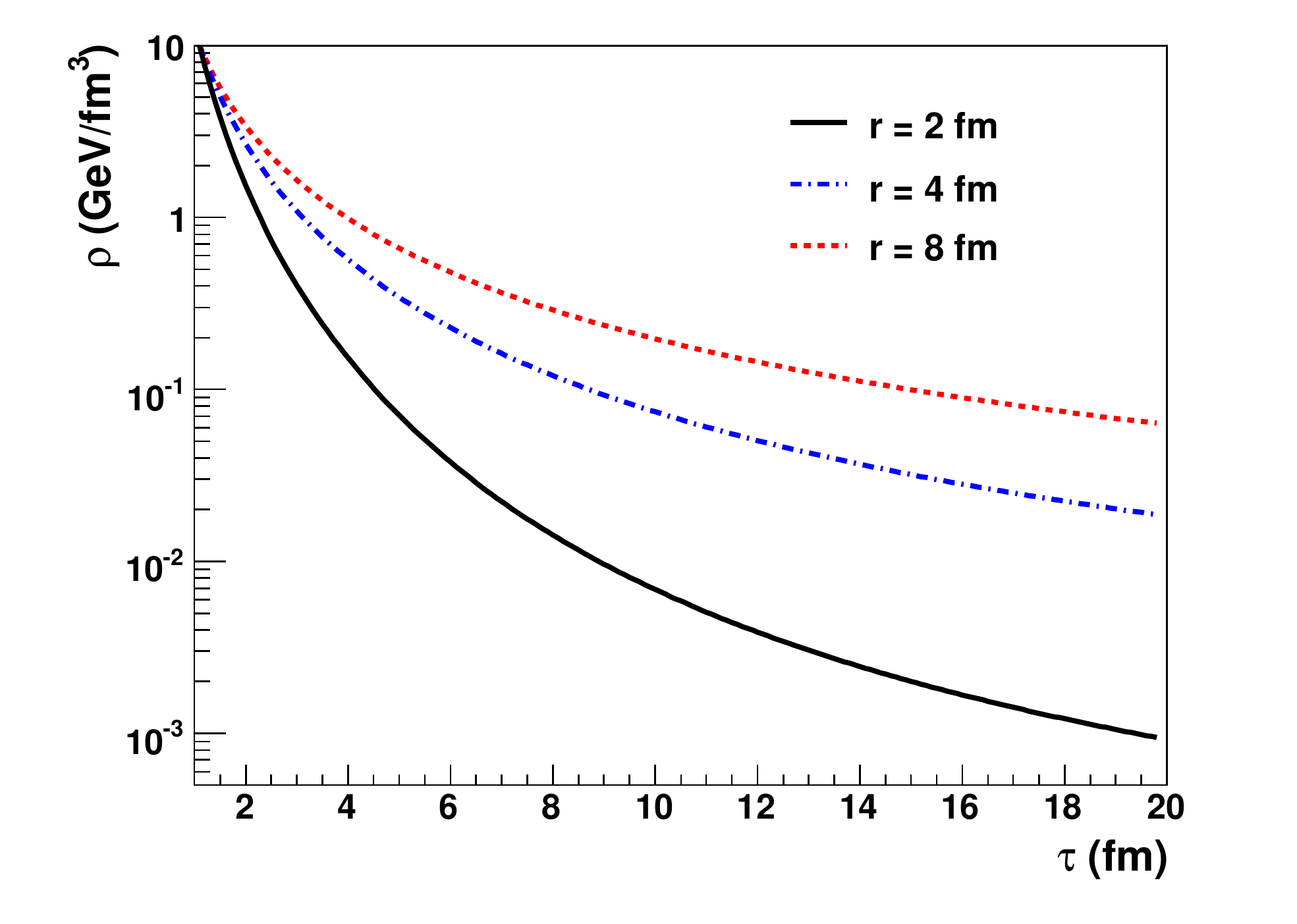}
\par\end{centering}

\caption{The energy density as a function of $r$ and $\tau$. \label{fig:energy-density-conformal}}

\end{figure}

In order to describe the evolution of the fluid with transverse expansion,
besides $(\tau,y)$ the cylindrical coordinates will also be used
\begin{eqnarray}
x^{\mu} & = & (\tau,y,r,\theta)\;,
\end{eqnarray}
with $r$ and $\theta$ are radius and azimuthal angle in transverse
plane. The velocity of the fluid cells can be parametrized as 
\begin{eqnarray}
u^{\mu} & = & \frac{dX^{\mu}}{d\tau_{f}}=\frac{1}{\sqrt{1-R^{2}(\tau,r)/\tau^{2}}}\left[1,0,\frac{R(\tau,r)}{\tau},0\right]\nonumber \\
 & \equiv & \left[\cosh\alpha(\tau,r),0,\sinh\alpha(\tau,r),0\right]\;,
\end{eqnarray}
where the total proper time $\tau_{f}$ is given by $\tau_{f}=\sqrt{\tau^{2}-r^{2}}$
and $R(\tau,r)$ is an unknown function which has to be determined
by the evolution equations of the fluid. For simplicity, the total
baryon number density is assumed to vanish. The energy-momentum tensor
reads 
\begin{eqnarray}
T_{\mu\nu} & = & \left(\begin{array}{cccc}
-\frac{\epsilon}{3}+\frac{4}{3}\epsilon\cosh^{2}\alpha & 0 & -\frac{4}{3}\epsilon\sinh\alpha\cosh\alpha & 0\\
0 & \frac{\rho}{3}\tau^{2} & 0 & 0\\
-\frac{4}{3}\epsilon\sinh\alpha\cosh\alpha & 0 & \frac{\epsilon}{3}+\frac{4}{3}\epsilon\sinh^{2}\alpha & 0\\
0 & 0 & 0 & \frac{\epsilon}{3}r^{2}
\end{array}\right)\;.\label{eq:EM-tensor-conformal-01}
\end{eqnarray}
The conservation equations are 
\begin{equation}
\nabla_{\mu}T^{\mu\nu}\equiv\partial_{\mu}T^{\mu\nu}+\Gamma_{\mu\sigma}^{\mu}T^{\sigma\nu}+\Gamma_{\mu\sigma}^{\nu}T^{\mu\sigma}=0\;,\label{eq:EM-consevation-conformal-01}
\end{equation}
where $\Gamma_{\mu\sigma}^{\mu}$ are Christoffel symbols for the
metric $\ensuremath{\widetilde{g}_{\mu\nu}^{(0)}=\mathrm{diag}(-1,\tau^{2},1,r^{2})}$.
Solving Eqs. (\ref{eq:EM-consevation-conformal-01}) leads to 
\begin{eqnarray}
\partial_{\tau}\ln\epsilon & = & -\frac{2}{\tau}\frac{2\cosh^{2}\alpha}{2\cosh^{2}\alpha+1}\left(1+\frac{R}{\tau}+\partial_{\tau}R\right)\;,\nonumber \\
\partial_{\tau}\ln R & = & \frac{1}{\tau}\frac{R/\tau+2+2(1-\partial_{r}R)\cosh^{2}\alpha}{2\cosh^{2}\alpha+1}\;.\label{eq:evolution-equation-01}
\end{eqnarray}
With a given initial condition $R(\tau_{0},r)=\xi\tau_{0}r$, where
$\xi$ is set to 0.05 $\mathrm{fm}^{-1}$ given by Ref. \cite{Kolb:2003dz,Kolb:2001qz},
the numerical results of the Eqs. (\ref{eq:evolution-equation-01})
are shown in Fig. \ref{fig:radial-velocity} and are identical to
the results in Ref. \cite{Kolb:2003dz,Kolb:2001qz}. The results indicate
that in this case the radial velocity is proportional to the distance,
i.e. $u_{r}\sim\xi r$, and is observed to rise sharply at the early
time and fall with increasing transverse proper time. The evolution
of the energy density shown in Fig. \ref{fig:energy-density-conformal}
is found to damp in the power series of $\tau^{4/3}$ if the transverse
expansion is negligible.

\subsection{Holographic renormalization}

Generally, the AdS metric can be written in the form of Fefferman
coordinates \cite{Fefferman1985} 
\begin{eqnarray}
ds^{2} & = & g_{MN}dX^{M}dX^{N}=\frac{1}{z^{2}}\left[\widetilde{g}_{\mu\nu}(x,z)dx^{\mu}dx^{\nu}+dz^{2}\right]\;,\label{eq:ds-01}
\end{eqnarray}
where $X^{M}=(x^{\mu},z)$ and $\widetilde{g}_{\mu\nu}(x,z)$ is the
metric tensor in the 4-dimensional space. In Eq.(\ref{eq:ds-01}),
there is divergences at the boundary $z\rightarrow0$. To cancel this
divergence, i.e. to renormalize the metric $\widetilde{g}_{\mu\nu}$
at the boundary $z\rightarrow0$, it is straightforward to assume
that $\widetilde{g}_{\mu\nu}\sim z^{2}f(x,z)$ where $f(x,z)$ is
a function free of singularities near the boundary. The authors of
\cite{Skenderis2002a,Karch:2005ms} obtain the form of $f(x,z)$ 
\begin{equation}
\widetilde{g}_{\mu\nu}(x,z)=\sum_{n=0}z^{2n}\widetilde{g}_{\mu\nu}^{(2n)}(x)\;,\label{eq:g-mu-nu-conformal}
\end{equation}
and prove that $\widetilde{g}_{\mu\nu}$ is the metric for a conformal
theory. In the expansion (\ref{eq:g-mu-nu-conformal}), $\widetilde{g}_{\mu\nu}^{(0)}(x)$
is just the metric in the 4-dimensional flat space. The term $\widetilde{g}_{\mu\nu}^{(2)}(x)$
can be proved to vanish in a conformal theory. The term $\widetilde{g}_{\mu\nu}^{(4)}(x)$
is associated with the energy-momentum tensor of the conformal theory,
\begin{equation}
\widetilde{g}_{\mu\nu}^{(4)}(x)\propto\left\langle T_{\mu\nu}\right\rangle \;,\label{eq:g-mu-nu-02}
\end{equation}
where the factor $\frac{N_{c}^{2}}{2\pi^{2}}=1$ with $N_{c}$ the
number of colors. Higher order terms are given by the Einstein equation
\begin{equation}
R_{MN}-\frac{1}{2}g_{MN}R+6g_{MN}=0\;,\label{eq:Einstein-eq-01}
\end{equation}
with the given initial condition $\widetilde{g}_{\mu\nu}^{(0)}(x)$
and $\widetilde{g}_{\mu\nu}^{(4)}(x)$, the cosmology constant $\Lambda=-\frac{d(d-2)}{2}=-6$.
Here $R_{MN}$ and $R$ are the curvature tensor and scalar of AdS
space, respectively. In this case, one finds that a conformal system
is located at the boundary of an AdS space.

\subsection{AdS metric with radial flow}

The authors of Ref. \cite{Janik:2006ft,Sin:2006pv,Janik2006,Janik:2006gp}
suggest that one can use the holographic renormalization to regard
the energy-momentum tensor of a boost invariant fireball as a boundary
condition for $\widetilde{g}_{\mu\nu}$ in Eq.(\ref{eq:g-mu-nu-conformal}).
By solving the Einstein equation (\ref{eq:Einstein-eq-01}), an AdS
metric with the properties of the boost invariant fireball can be
obtained. 

In cylindrical coordinates, $\widetilde{g}_{\mu\nu}(x,z)$ can be
cast in the form 
\[
\widetilde{g}_{\mu\nu}(x,z)dx^{\mu}dx^{\nu}=-Ad\tau^{2}+B\tau^{2}dy^{2}+C\left(dr^{2}+r^{2}d\theta^{2}\right)+2Dd\tau dr\;,
\]
where $d\tau dr$ is off-diagonal element. It is too complicated to
solve the AdS metric directly in the above form. A simplification
is necessary. As argued in the last subsection, the radial velocity
is negligible compared to the transverse one. Therefore, one can assume
$\alpha$ in Eq.(\ref{eq:EM-tensor-conformal-01}) is small, so $T^{\mu\nu}$
can be expanded in $\alpha$, 
\begin{eqnarray}
T_{\mu\nu} & = & \left(\begin{array}{cccc}
\rho & 0 & 0 & 0\\
0 & \tau^{2}\frac{1}{3}\rho & 0 & 0\\
0 & 0 & \frac{1}{3}\rho & 0\\
0 & 0 & 0 & r^{2}\frac{1}{3}\rho
\end{array}\right)+\alpha\left(\begin{array}{cccc}
0 & 0 & -\frac{4}{3}\rho & 0\\
0 & 0 & 0 & 0\\
-\frac{4}{3}\rho & 0 & 0 & 0\\
0 & 0 & 0 & 0
\end{array}\right)\nonumber \\
 &  & +\alpha^{2}\left(\begin{array}{cccc}
\frac{4}{3}\rho & 0 & 0 & 0\\
0 & 0 & 0 & 0\\
0 & 0 & \frac{4}{3}\rho & 0\\
0 & 0 & 0 & 0
\end{array}\right)+O\left(\alpha^{3}\right)\;.\label{eq:EM-tensor-conformal-expan-01}
\end{eqnarray}
The off-diagonal terms proportional to $\alpha$ only arise in the
next-to-leading order of the expansion. Therefore, the modification
from the radial flow should be also in the order of $O(\alpha)$.
Solving the Einstein equation (\ref{eq:Einstein-eq-01}) with the
boundary condition given by Eq.(\ref{eq:g-mu-nu-02}) and (\ref{eq:EM-tensor-conformal-expan-01})
yields 
\begin{equation}
g_{MN}=\frac{1}{z^{2}}\left(\begin{array}{ccccc}
-\frac{(1-a)^{2}}{1+a} & 0 & \frac{4}{3}\alpha\frac{(1-a)^{2}}{1+a} & 0 & 0\\
0 & \tau^{2}(1+a) & 0 & 0 & 0\\
\frac{4}{3}\alpha\frac{(1-a)^{2}}{1+a} & 0 & 1+a & 0 & 0\\
0 & 0 & 0 & r^{2}(1+a) & 0\\
0 & 0 & 0 & 0 & 1
\end{array}\right)\;.
\end{equation}
with $a=\frac{\epsilon_{0}z^{4}}{3\tau^{4/3}}$ and $\epsilon_{0}$
is the same as the one in Eq.(\ref{eq:Bjorken-quantities-01}). Note
here $a$ is set to a scaling and higher order contribution $\tau^{-1}$
is neglected in the late time $\tau\rightarrow\infty$ limit. 

For $\alpha=0$, the metric is 
\begin{equation}
ds^{2}=-\frac{(1-z^{4}/z_{H}^{4})^{2}}{1+z^{4}/z_{H}^{4}}\frac{d\tau^{2}}{z^{2}}+\frac{1+z^{4}/z_{H}^{4}}{z^{2}}[\tau^{2}dy^{2}+dr^{2}+r^{2}d\theta^{2}]+\frac{dz^{2}}{z^{2}}\;,\label{eq:metric-a0-01}
\end{equation}
with the horizon of the black hole 
\begin{equation}
z_{H}=\left(\frac{\epsilon_{0}}{3}\right)^{-1/4}\tau^{1/3}\;.
\end{equation}
The standard D$3$ black AdS metric can be obtained from Eq.(\ref{eq:metric-a0-01})
\begin{equation}
ds^{2}=-\frac{1-\widetilde{z}^{4}/\widetilde{z}_{H}^{4}}{\widetilde{z}^{2}}d\tau^{2}+\frac{d\mathbf{x}^{2}}{\widetilde{z}^{2}}+\frac{1}{1-\widetilde{z}^{4}/\widetilde{z}_{H}^{4}}\frac{d\widetilde{z}^{2}}{\widetilde{z}^{2}}\;,
\end{equation}
with the replacement 
\[
z\rightarrow\widetilde{z}=\frac{z}{\sqrt{1+\frac{z^{4}}{z_{H}^{4}}}},\qquad z_{H}\rightarrow\widetilde{z}_{H}=\frac{z_{H}}{\sqrt{2}}\;.
\]
The Hawking temperature can also be obtained 
\begin{equation}
T_{H}=\frac{1}{\pi\widetilde{z}_{H}}=\frac{\sqrt{2}}{\pi z_{H}}\;.\label{eq:Hawking-temp-01}
\end{equation}

\subsection{Evolution of $\eta/s$}

To compute the retarded Green function, the action of gravity is necessary.
Generally, the action in the AdS space reads 
\begin{equation}
I_{5D}=\frac{N^{2}}{8\pi^{2}R^{3}}\int d^{5}x(\mathcal{R}_{5D}-2\Lambda)\;,
\end{equation}
where the contributions from the matter term $g_{\mu\nu}T^{\mu\nu}$
in AdS space is ignored. Considering a perturbation $\ensuremath{g_{\mu\nu}=g_{\mu\nu}^{(0)}+h_{\mu\nu}}$
with $\ensuremath{h_{z\mu}=h_{\mu z}=0}$ yields 
\begin{eqnarray}
I_{5D} & = & \frac{N^{2}}{8\pi^{2}}\int d^{5}x\sqrt{-g}(\mathcal{R}_{\mu\nu}-\frac{1}{2}g_{\mu\nu}\mathcal{R}+\Lambda g_{\mu\nu})h^{\mu\nu}\nonumber \\
 & = & \frac{N^{2}}{8\pi^{2}}\int d^{5}x\sqrt{-g}(\mathcal{R}_{\mu\nu}^{(0)}-\frac{1}{2}g_{\mu\nu}^{(0)}\mathcal{R}_{(0)}+\Lambda g_{\mu\nu}^{(0)})h^{\mu\nu}\nonumber \\
 &  & +\frac{N^{2}}{8\pi^{2}}\int d^{5}x\sqrt{-g}(\delta\mathcal{R}_{\mu\nu}-\frac{1}{2}h_{\mu\nu}\mathcal{R}_{(0)}-\frac{1}{2}g_{\mu\nu}^{(0)}\delta\mathcal{R}+\Lambda h_{\mu\nu})h^{\mu\nu}\nonumber \\
 & = & \frac{N^{2}}{8\pi^{2}}\int d^{5}x\sqrt{-g}(\delta\mathcal{R}_{\mu\nu}-\frac{1}{2}g_{\mu\nu}^{(0)}\delta\mathcal{R}+4h_{\mu\nu})h^{\mu\nu}\;,
\end{eqnarray}
where $\ensuremath{\mathcal{R}_{0}=-20}$ and $\ensuremath{\Lambda=-6}$
have been used. For most AdS metric, the $h_{2}^{1}\equiv\phi$ is
decoupled with others 
\begin{eqnarray}
I_{5D} & \approx & \frac{N^{2}}{8\pi^{2}}\int d^{5}x\sqrt{-g}(\delta R_{12}+4h_{12})h^{12}\nonumber \\
 & = & \frac{N^{2}}{8\pi^{2}}\int d^{5}x\sqrt{-g}\left(-\frac{1}{2}g^{\mu\nu}\partial_{\mu}\phi\partial_{\nu}\phi+\cdots\right)\;.\label{eq:action-conformal-02}
\end{eqnarray}

In $0+1$-dimensional case, i.e. the radial flow vanishes, the metric
gives 
\begin{equation}
ds^{2}=\frac{1}{z^{2}}\left\{ -\frac{(1-a)^{2}}{1+a}d\tau^{2}+(1+a)[\tau^{2}dy^{2}+dx_{1}^{2}+dx_{2}^{2}]\right\} +\frac{dz^{2}}{z^{2}}\;,
\end{equation}
which is identical to Ref. \cite{Janik:2006ft,Sin:2006pv,Janik2006,Janik:2006gp}.
The equation of motion for $\phi$ is given by
\begin{equation}
\partial_{\mu}(\sqrt{-g}g^{\mu\nu}\partial_{\nu}\phi)=0\;.\label{eq:EOM-conformal-01}
\end{equation}
Assuming 
\begin{eqnarray*}
\phi(\tau,y,z) & = & \phi_{0}(y,\tau)f(z)\\
 & = & \int\frac{d\omega dp_{3}}{(2\pi)^{2}}\exp(-i\omega\tau\cosh y+ip_{3}\tau\sinh y)\phi_{0}(\omega,p_{3})f_{p}(z)\;,
\end{eqnarray*}
and taking the limit $\tau\rightarrow\infty$ while keeping $a$ as
a scaling yields 
\begin{equation}
(1-a^{2})\frac{d^{2}f_{p}}{dz^{2}}+f_{p}\frac{(1+a)^{2}}{1-a}\omega^{2}\cosh^{2}y-f_{p}(1-a)(-\omega\sinh y+p_{3}\cosh y)^{2}=0\;.
\end{equation}
In central rapidity $y\simeq0$ and static limit $p_{3}=0$, the solution
of the above equation is 
\begin{equation}
f_{p}(z)=(1-z')^{\pm i\omega'}\;,
\end{equation}
with 
\begin{equation}
z\rightarrow z'=\left(\frac{\rho_{0}}{3}\right)^{1/4}\tau^{-1/3}z\;,\quad\omega'\rightarrow\frac{1}{2\sqrt{2}}\left(\frac{\rho_{0}}{3}\right)^{-1/4}\tau^{1/3}\omega\;.
\end{equation}
In this case, the action (\ref{eq:action-conformal-03}) becomes 
\begin{eqnarray}
I_{5D} & = & -\frac{N^{2}}{16\pi^{2}}\int dx^{1}dx^{2}\int\frac{d\omega dp_{3}}{(2\pi)^{2}}\phi_{0}(-\omega,-p_{3})\phi_{0}(\omega,p_{3})\nonumber \\
 &  & \times\left.\frac{1-a^{2}}{z^{3}}f_{-p}(z)\frac{df_{p}(z)}{dz}\right|_{z=z_{H}}\nonumber \\
 & \equiv & \int\frac{d^{4}p}{(2\pi)^{4}}\phi_{0}(-\omega,-p)\phi_{0}(\omega,p)F(\omega)\;,\label{eq:action-conformal-03}
\end{eqnarray}
where 
\begin{eqnarray}
F(\omega) & = & -\frac{N^{2}}{16\pi^{2}}\left.\frac{1-a^{2}}{z^{3}}f_{-p}(z)\frac{df_{p}(z)}{dz}\right|_{z=z_{H}}\nonumber \\
 & = & -\frac{N^{2}}{16\pi^{2}}\frac{8}{3}\rho_{0}\tau^{-4/3}\lim_{z\rightarrow z_{H}}(1-z')f_{-p}(z)\frac{df_{p}(z)}{dz'}\nonumber \\
 & = & i\omega\frac{\sqrt{2}N^{2}}{8\pi^{2}}\left(\frac{\rho_{0}}{3}\right)^{3/4}\tau^{-1}\;.
\end{eqnarray}
The retarded Green function $G^{R}$ can be written in the form 
\[
G^{R}\propto F(p)+F(-p)\;.
\]
The negative modes of the function $F(p)$ is $F(-p)=F^{*}(p)$. Thus,
$G^{R}\propto\textrm{Re }F(p)$ and the imaginary part of retarded
Green functions in CFT via the AdS/CFT duality is found to be zero.
The authors of Ref. \cite{Son2002d} assume that the imaginary part
of $G^{R}$ is given by $2\textrm{Im }F(p)$. In this case, the retarded
Green function reads 
\begin{equation}
G_{12,12}^{R}(\omega,0)=-2F(\omega)=-i\frac{\sqrt{2}}{3^{3/4}\times4\pi^{2}}N^{2}\rho_{0}^{3/4}\tau^{-1}\omega\;.
\end{equation}
The shear viscosity can be obtained via the Kubo relation (\ref{eq:kubo01})
\begin{equation}
\eta=\frac{\rho_{0}^{3/4}\sqrt{N}}{6^{3/4}\sqrt{\pi}}\frac{1}{\tau}\;.\label{eq:shear-conformal-02}
\end{equation}
On the other hand, from Eq.(\ref{eq:Hawking-temp-01}) the entropy
per transverse area and rapidity reads 
\begin{equation}
S=\left(\frac{N^{2}}{2\pi}\right)^{1/4}\left(\frac{\pi}{3}\right)^{3/4}2\sqrt{2}\rho_{0}^{3/4}=\frac{N^{2}}{2}\pi^{2}T^{3}\tau\;.
\end{equation}
Finally, the famous KSS bound is obtained 
\begin{equation}
\frac{\eta}{s}=\frac{1}{4\pi}\;.
\end{equation}

The metric in the $1+1$-dimensional AdS space is written as 
\begin{eqnarray}
ds^{2} & = & \frac{1}{z^{2}}\left\{ -\frac{(1-a)^{2}}{1+a}dt^{2}+(1+a)(dx_{3}dx_{3}+dx_{i}dx_{i})\right.\nonumber \\
 &  & \quad\left.+\frac{8}{3}\alpha\frac{(1-a)^{2}}{1+a}\frac{1}{r}x_{i}dx_{i}dt+\frac{dz^{2}}{z^{2}}\right\} \;,
\end{eqnarray}
where for convenience the rectangular transverse coordinates $(x_{1},x_{2})$
is used instead of cylindrical ones $(r,\theta)$. The equation of
motion for $\phi$ is found to be the same as Eq.(\ref{eq:EOM-conformal-01}).
Using the same assumption as in $0+1$-dimensional case $\phi(\tau,y,z)=\phi_{0}(y,\tau)f(z)$
yields 
\begin{equation}
\frac{d^{2}f_{p}}{dz^{2}}-\frac{3+5a}{z(1-a^{2})}\frac{df_{p}}{dz}+\frac{1+a}{(1-a)^{2}}\omega^{2}f_{p}-\frac{i}{1+a}\frac{2\alpha}{3r}\omega f_{p}=0\;,
\end{equation}
whose solutions are 
\begin{eqnarray}
f_{p} & = & (1-z')^{i\omega'}z_{H}^{i\omega'}\left(\frac{\alpha\omega}{3r}\right)^{i\omega'}\left[\frac{(-1)^{i3\omega'/4}2^{-i\omega'}}{\Gamma(1+i\omega')}\right.\nonumber \\
 &  & \left.+\frac{i(-1)^{i3\omega'/4}2^{-2-i\omega'}}{(1+i\omega')\Gamma(1+i\omega')}(1-z')^{2}z_{H}^{2}\frac{\alpha\omega}{3r}\right]\;,\nonumber \\
f_{-p} & = & f_{p}(i\omega'\rightarrow-i\omega')\;.
\end{eqnarray}
In this case, $F(\omega)$ in Eq.(\ref{eq:action-conformal-03}) becomes
\begin{eqnarray}
F(\omega) & = & -\frac{N^{2}}{16\pi^{2}}\left.\frac{1-a^{2}}{z^{3}}f_{-p}(z)\frac{df_{p}(z)}{dz}\right|_{z=z_{H}}\nonumber \\
 & = & -\frac{N^{2}}{16\pi^{2}}\frac{8}{3}\rho_{0}\tau^{-4/3}\lim_{z\rightarrow z_{H}}(1-z')f_{-p}(z)\frac{df_{p}(z)}{dz'}\nonumber \\
 & = & i\omega\frac{\sqrt{2}N^{2}}{8\pi^{2}}\left(\frac{\rho_{0}}{3}\right)^{3/4}\tau^{-1}\;,
\end{eqnarray}
where the derivative of $f_{p}(z)$ is given by 
\begin{eqnarray}
\frac{df_{p}(z)}{dz'} & = & -i\omega'(1-z')^{i\omega'-1}z_{H}^{i\omega'}\left(\frac{\alpha\omega}{3r}\right)^{i\omega'}\left[\frac{(-1)^{i3\omega'/4}2^{-i\omega'}}{\Gamma(1+i\omega')}\right.\nonumber \\
 &  & \left.+\frac{i(-1)^{i3\omega'/4}2^{-2-i\omega'}}{(1+i\omega')\Gamma(1+i\omega')}(1-z')^{2}z_{H}^{2}\frac{\alpha\omega}{3r}\right]\nonumber \\
 &  & -2(1-z')^{i\omega'+1}z_{H}^{i\omega'+2}\left(\frac{\alpha\omega}{3r}\right)^{i\omega'+1}\frac{i(-1)^{i3\omega'/4}2^{-2-i\omega'}}{(1+i\omega')\Gamma(1+i\omega')}\;,
\end{eqnarray}
Finally, the ratio $\eta/s$ turns out to be 
\begin{equation}
\frac{\eta}{s}=\frac{1}{4\pi}+O(\alpha^{2})\;.\label{eq:shear_s_late_1+1_01}
\end{equation}

\subsection{Discussion}

We have derived a time dependent metric dual to fl{}uid in 1+1 dimension
with Bjorken scaling and radial fl{}ow in late time via holographic
renormalization. We assume the transverse expansion is small and can
be treated as a perturbation. In that case, we solve the Einstein
equation and obtain the new AdS metric. With this metric we calculate
the ratio $\eta/s$ in late time limit. We have shown that the ratio
for fl{}uids in 1+1 dimension is the same as in 0+1 dimension in the
leading order of transverse rapidity. In 1+1 dimension one can introduce
the shear viscosity of the next-to-leading order in the stress tensor
in late time solution {[}9,11,12{]}. We found that the correction
to the shear viscosity is of the next-to-leading order as shown in
Eq.(\ref{eq:shear_s_late_1+1_01}). Compared Eq.(\ref{eq:shear-conformal-02})
with $\eta\sim\epsilon l_{\mathrm{mfp}}\sim g^{-2}T^{3}$,we see that
the coupling is as strong at the beginning as at the late time, i.e.
the evolution does not influence the strength of the interaction. 

This technique can also be used to investigate the evolution of QGP
at the very early time with the the scale $a\sim z\tau^{+\left|s\right|}$,
see e.g. Ref. \cite{Kovchegov2007}. The local thermal equilibrium
for the fireball has not been established because of the anisotropic
evolution, i.e. the temperature which is a varying slowly function
of the coordinates is not well-defined. It will be of interest to
investigate the phenomena near the phase transition, e.g. evaluation
of the time for the local thermal equilibrium \cite{Kovchegov2007}.

\section{High spin baryon in hot plasma\label{sec:High-spin-baryon}}

\begin{figure}
\begin{centering}
\includegraphics[scale=0.45]{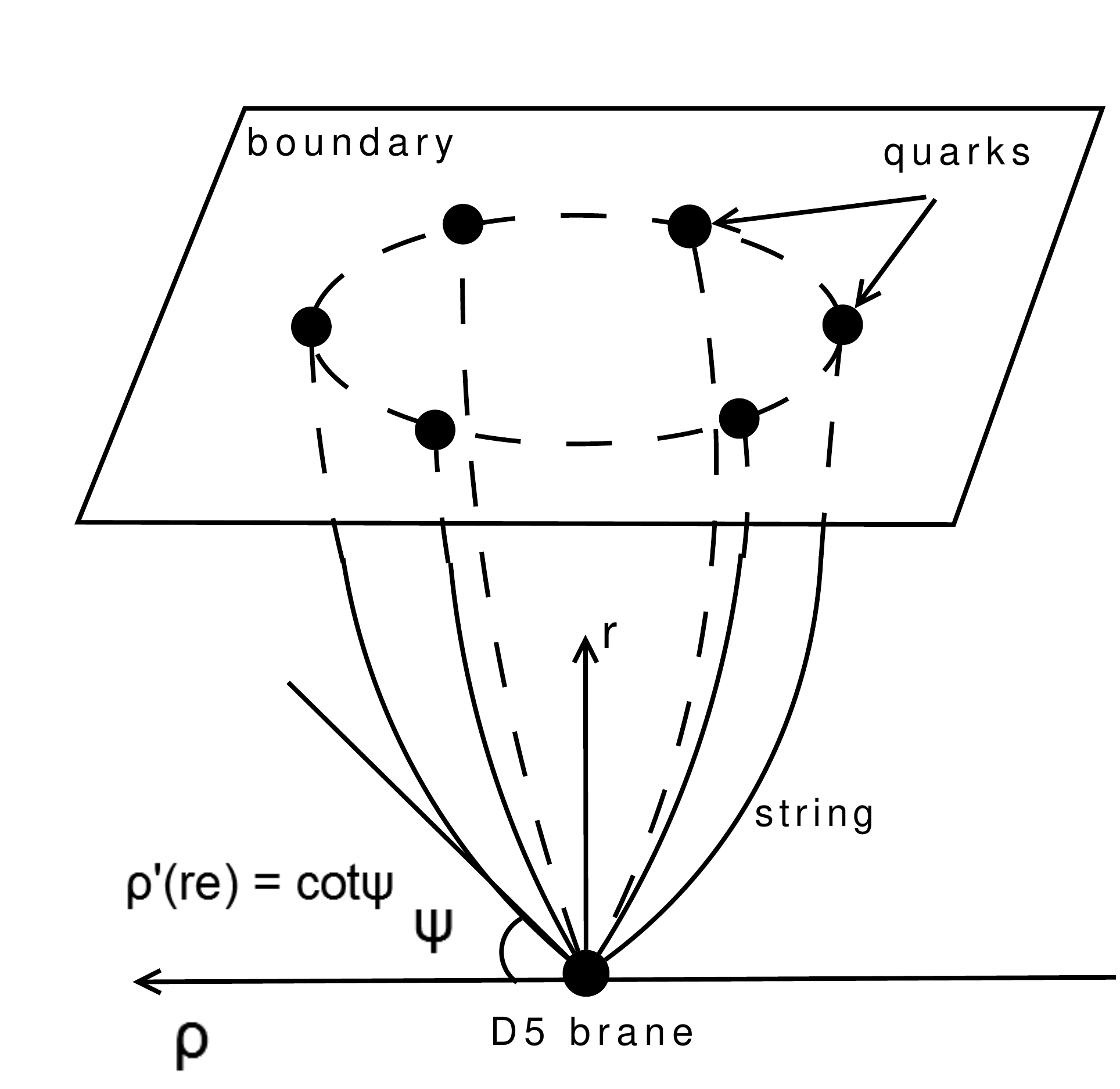}
\par\end{centering}

\begin{centering}
\includegraphics[scale=0.45]{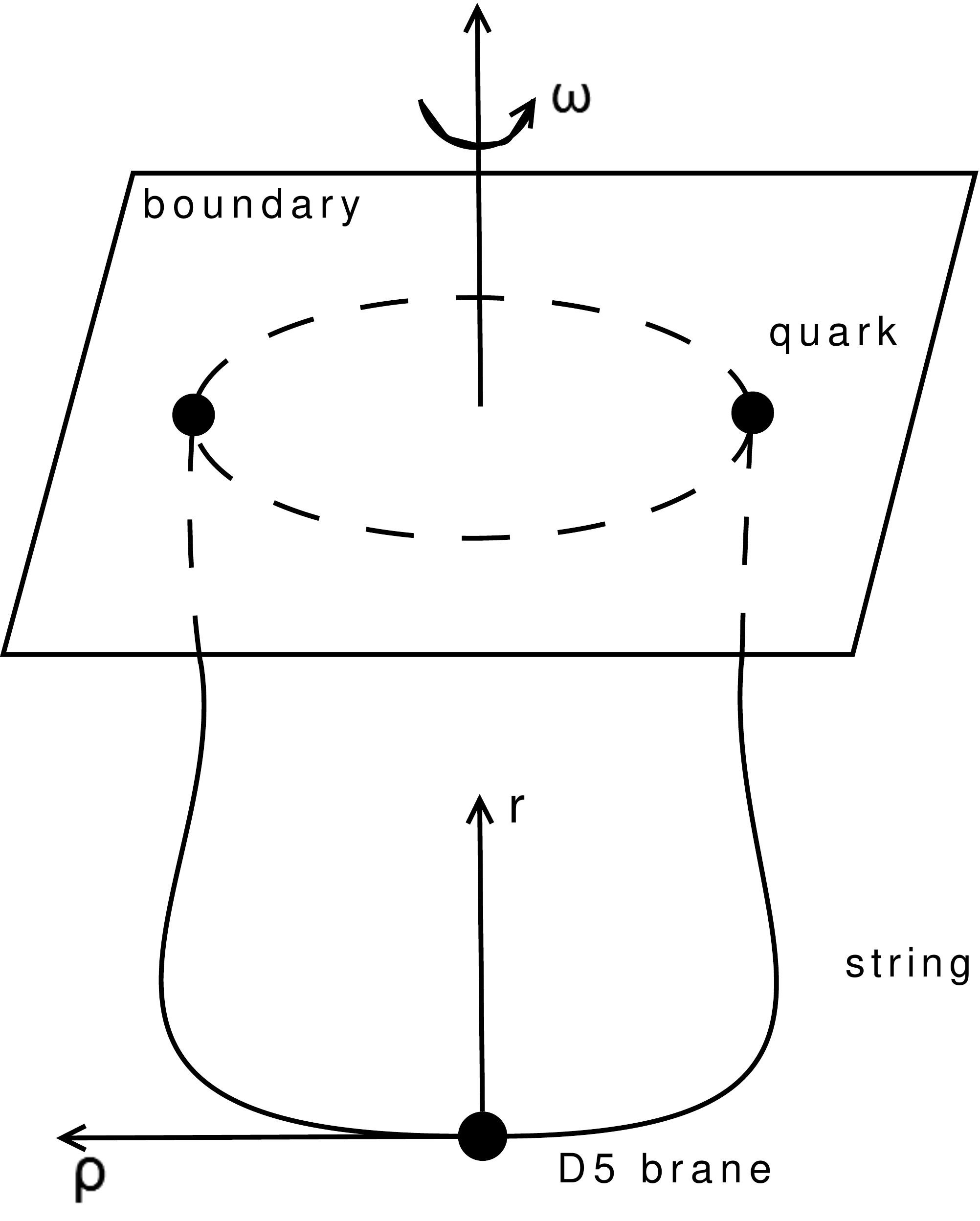}
\par\end{centering}

\caption{String and a brane as the boundary represent a baryon.\label{fig:baryons}}

\end{figure}

As mentioned in Sec. \ref{sub:Wilson-loops}, the Wilson loops via
AdS/CFT can be used to study the potential between quarks and anti-quarks.
Besides these, the velocity dependence of the screening length in
the QGP can also be learned about in a boost AdS metric \cite{Athanasiou:2008pz}.
In Ref. \cite{Li:2008py}, the rotation dependence of the screening
length for the baryons in the QGP is considered. The physical picture
for Wilson loops is similar as in Ref. \cite{Li:2008py}. They differs
on the end points of the open strings. In Wilson loops, the two end
points of open strings stay in the same brane while in Ref. \cite{Li:2008py}
they stay in a static brane as a boundary and a probe one (also see
Fig. \ref{fig:baryons}). The baryons live in the boundary of the
AdS space. And another D5 brane as a probe is in the bulk of the AdS
space as shown in Fig. \ref{fig:baryons}.

\subsection{Setup}

The AdS metric is given by 
\begin{equation}
ds^{2}=-f(r)dt^{2}+\frac{r^{2}}{R^{2}}dx_{3}^{2}+\frac{r^{2}}{R^{2}}\left(d\rho^{2}+\rho^{2}d\theta^{2}\right)+\frac{1}{f(r)}dr^{2}+R^{2}d\Omega_{5}^{2}\;,
\end{equation}
where $\rho$ and $\theta$ are in the $x_{1}-x_{2}$ plane. Mapping
the coordinates of the strings to the that of the space-time 
\begin{equation}
\tau=t\;,\qquad\sigma=r\;,\qquad\theta=\omega t\;,\qquad\rho=\rho(r)\;,
\end{equation}
and using the Nambu-Goto action 
\begin{equation}
S_{string}=\frac{1}{2\pi\alpha'}\int d\sigma d\tau\sqrt{-\det[h_{ab}]}\;,
\end{equation}
with $h_{ab}=g_{\mu\nu}\frac{\partial x^{\mu}\partial x^{\nu}}{\partial\sigma^{a}\partial\sigma^{b}}$
and $1/(2\pi\alpha')$ the string tension becomes 
\begin{equation}
S_{string}=\frac{\mathcal{T}}{2\pi\alpha'}\int_{r_{e}}^{r_{\Lambda}}dr\sqrt{-\left(\frac{r^{2}}{R^{2}}\rho^{2}\omega^{2}-f(r)\right)\left(\frac{1}{f(r)}+\frac{r^{2}}{R^{2}}\rho'^{2}(r)\right)}\;,\label{eq:action-string-01}
\end{equation}
where $\mathcal{T}$ is the total time and $r_{\Lambda},r_{e}$ are
the cut-off and end points of the strings, respectively. The total
action is given by 
\[
S_{total}=\sum_{i=1}^{N_{c}}S_{string}^{(i)}+S_{D5}\;,
\]
where the action for the probe D5 brane is 
\begin{equation}
S_{D5}=\frac{\mathcal{V}(r_{e})\mathcal{T}V_{5}}{(2\pi)^{5}\alpha'^{3}}\;,
\end{equation}
with $V_{5}$ the volume of the compact brane and $\mathcal{V}(r_{e})=\sqrt{-g_{00}}$
the potential for the brane at $r=r_{e}$. 

On the other hand, a constraint condition, the so-called force balance
condition (FBC), in the $r$-direction is also considered
\begin{equation}
\sum_{i=1}^{N_{c}}H^{(i)}\biggr|_{r_{e}}=\Sigma\;,\label{eq:FBC-01}
\end{equation}
with $H$ the Hamilton of the system and 
\begin{equation}
\Sigma\equiv\frac{2\pi\alpha^{\prime}}{\mathcal{T}}\frac{\partial S_{D5}}{\partial r_{e}}\;.
\end{equation}

\subsection{$\omega$ dependence}

From Eq.(\ref{eq:action-string-01}), the equation of motion for $\rho(r)$
is obtained
\begin{equation}
\left(\frac{\partial}{\partial\rho}-\frac{\partial}{\partial r}\frac{\partial}{\partial\rho^{\prime}}\right)\mathcal{L}=0\;,\label{eq:eom-string-01}
\end{equation}
where $\mathcal{L}$ is the Lagrangian of the action (\ref{eq:action-string-01}). 

Substituting Eq.(\ref{eq:action-string-01}) into Eq.(\ref{eq:FBC-01})
we can obtain the FBC as a boundary condition for the numerical calculations
\begin{equation}
\rho'(r_{e})=\frac{R}{r_{e}f(r_{e})^{1/2}}\sqrt{\frac{R^{2}(r_{e}^{4}-r_{0}^{4})r_{e}^{4}}{(r_{e}^{4}+r_{0}^{4})^{2}A^{2}}-1}\;,\label{eq:FBC-02}
\end{equation}
with 
\[
A=\frac{1}{N_{c}}\frac{V_{5}}{(2\pi)^{4}(\alpha^{\prime})^{2}}\;.
\]

\begin{figure}
\begin{centering}
\includegraphics[scale=0.5]{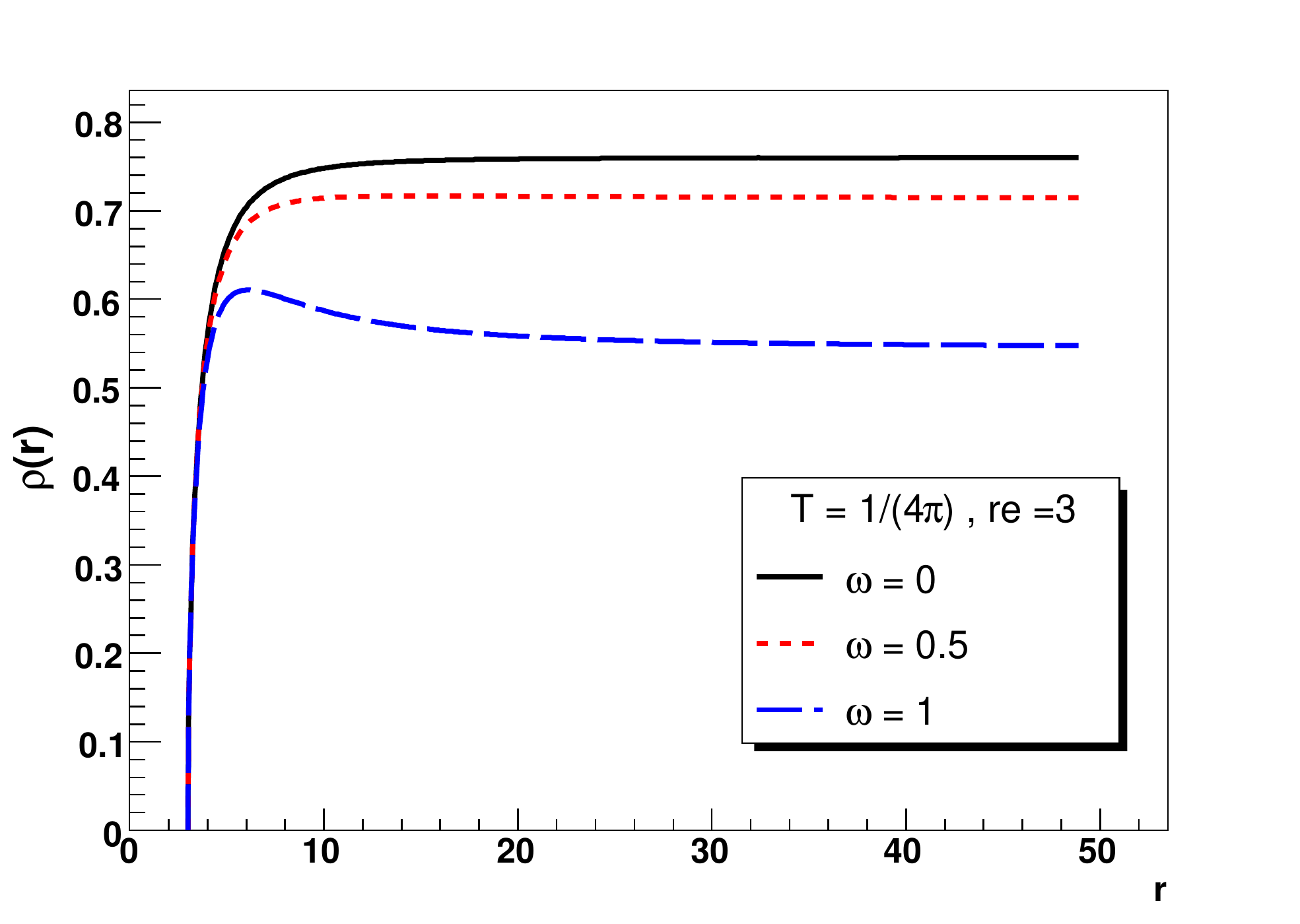}
\par\end{centering}

\begin{centering}
\includegraphics[scale=0.5]{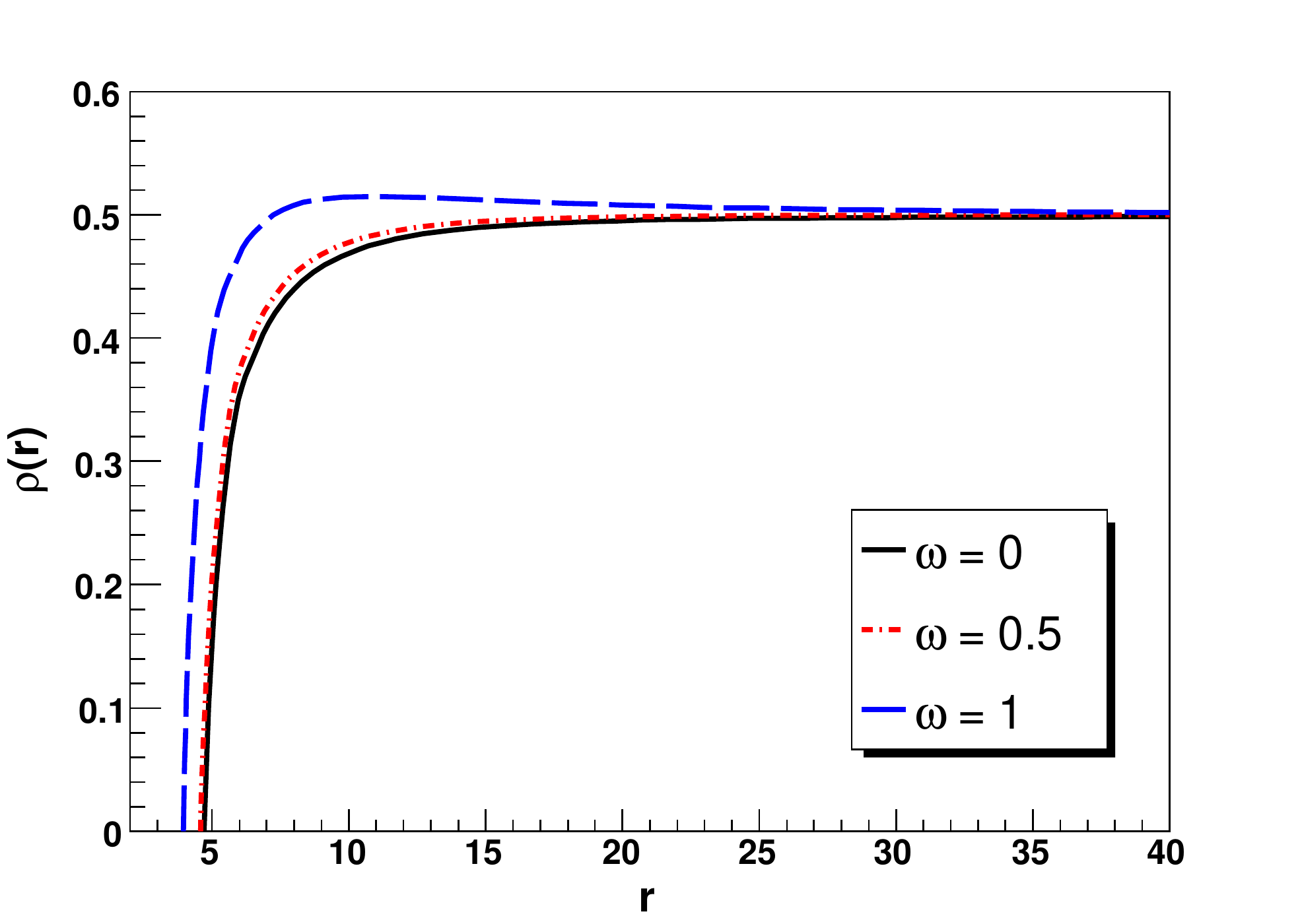}
\par\end{centering}

\caption{\label{fig:rho-01}Embedding function $\rho(r)$ as a function of
$\omega$ with fixed $r_{e}$ (upper plane) and $l_{q}$ with $r_{\Lambda}=100$
(lower plane). }
\end{figure}

\begin{figure}
\begin{centering}
\includegraphics[scale=0.5]{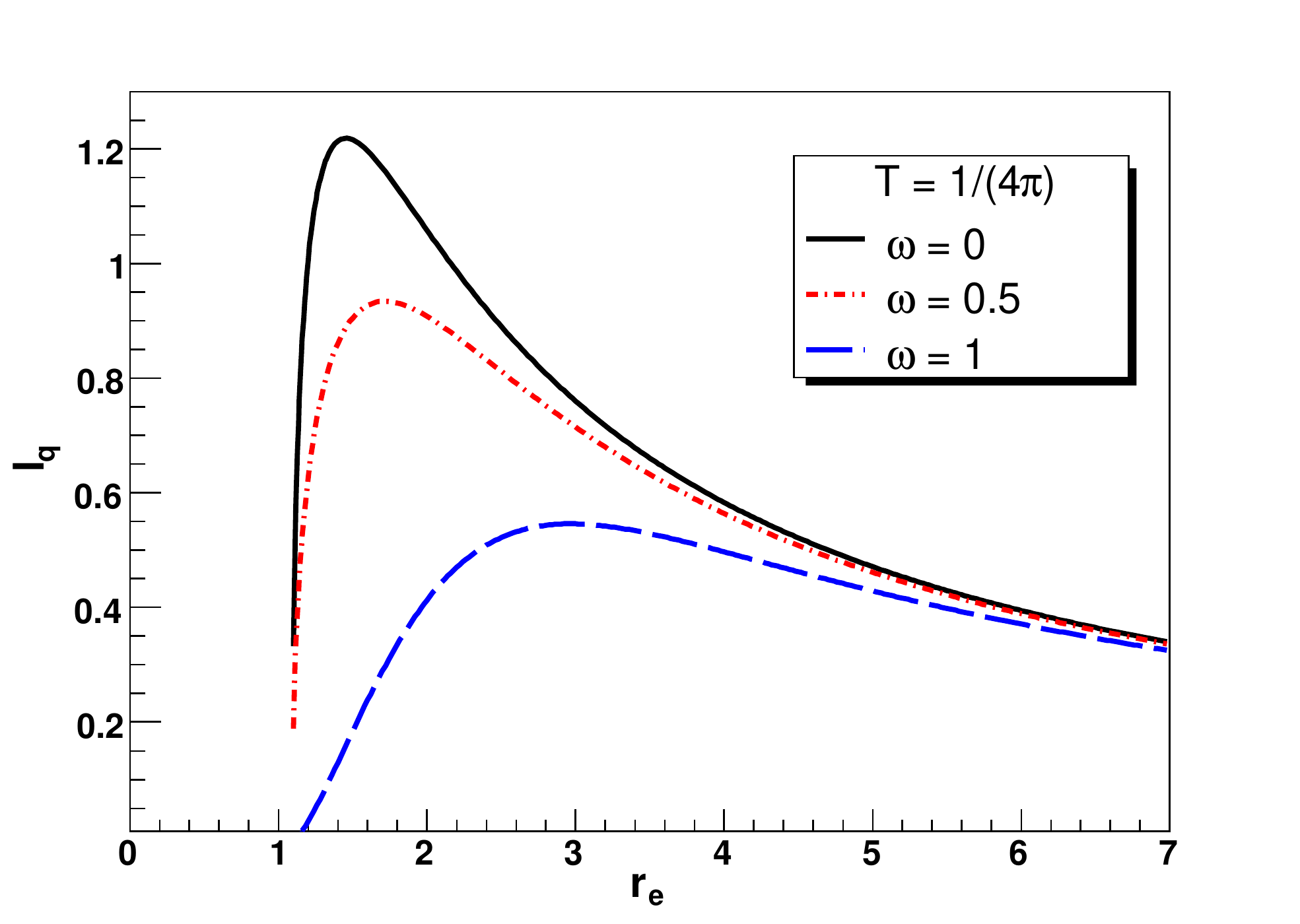}
\par\end{centering}

\caption{\label{fig:lq-re}Separation distance of quarks as a function of $r_{e}$
with $r_{\Lambda}=100$.}
\end{figure}

\begin{figure}
\begin{centering}
\includegraphics[scale=0.5]{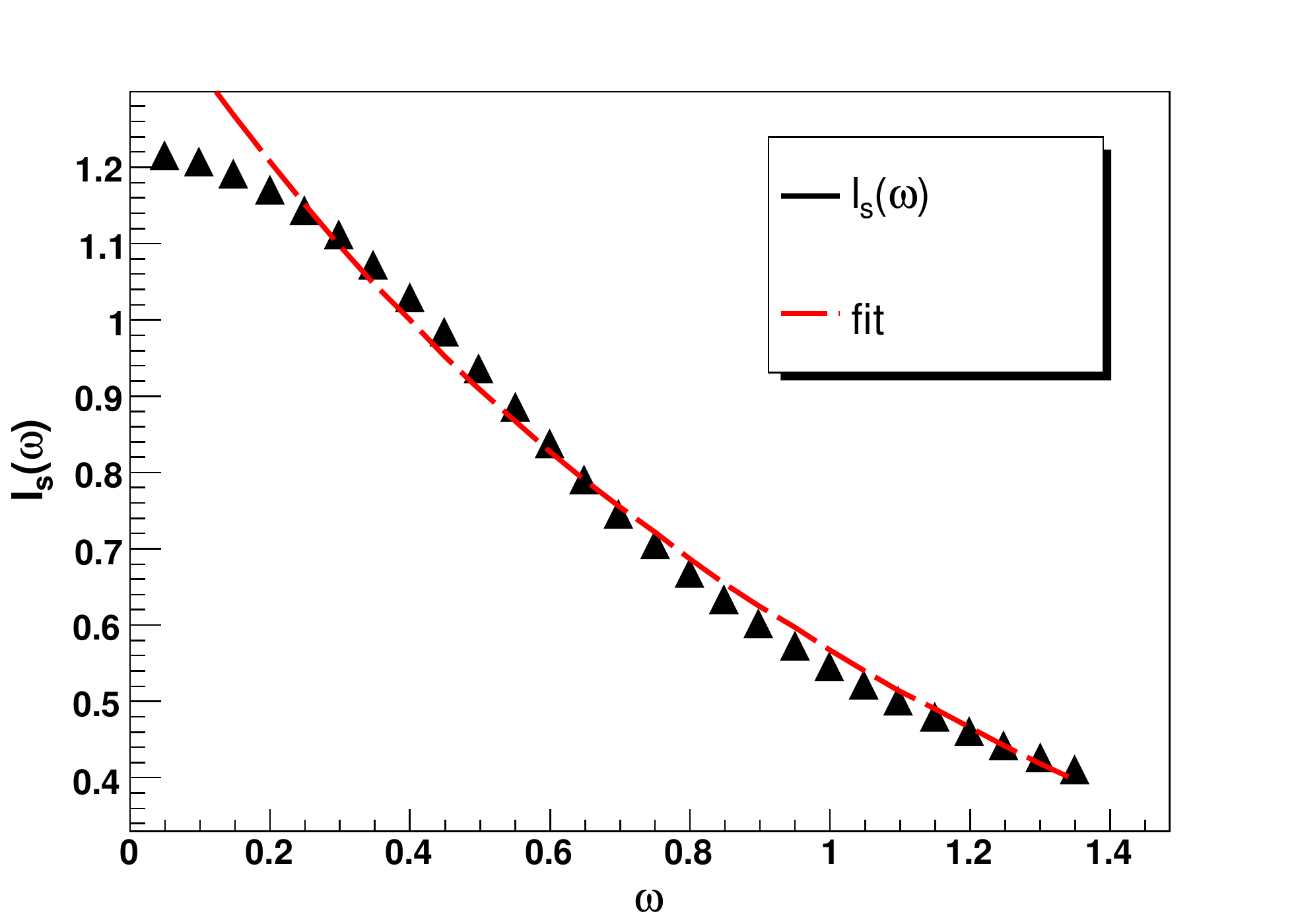}
\par\end{centering}

\begin{centering}
\includegraphics[scale=0.5]{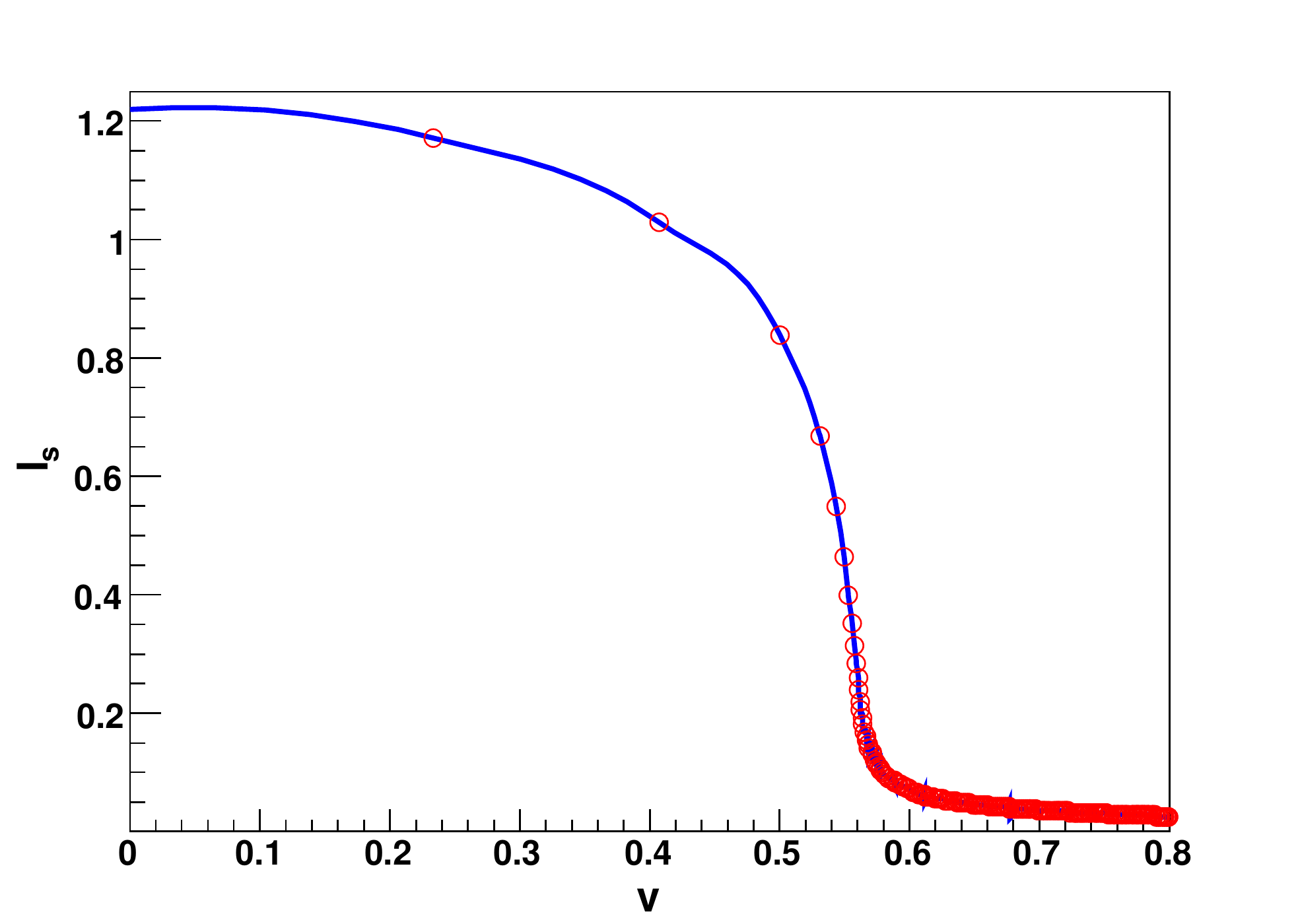}
\par\end{centering}

\caption{\label{fig:ls-01}Screening length as a function of spin of baryons
$\omega$ (upper plane) and linear velocity $v$ (lower plane).}
\end{figure}

With a given value of $r_{e}$, Eq.(\ref{eq:eom-string-01}) can be
solved with the boundary condition (\ref{eq:FBC-02}). The numerical
result is shown in Fig. \ref{fig:rho-01}. The separation distance
of quarks with a given $r_{e}$ is 
\begin{equation}
l_{q}=2\int_{r_{e}}^{r_{\Lambda}}\rho'(r)dr=2\rho(r_{\Lambda})\;.
\end{equation}
The result of $l_{q}$ as a function of $r_{e}$ is shown in Fig.
\ref{fig:lq-re}. The maximum value of $l_{q}$ is defined as the
screening length $l_{s}$ of the baryon $l_{s}$ as shown in Fig.
\ref{fig:ls-01}. We obtain the form of $l_{s}$ as 
\begin{equation}
l_{s}=\frac{a}{b\omega+c}-d\;,
\end{equation}
where it is drawn in Fig. \ref{fig:ls-01} with following parameters,
\begin{equation}
l_{s}=\frac{0.51}{0.14\omega+0.22}-0.85\;.
\end{equation}
In hot QGP, the screening length depends on the temperature and $\omega$
or spin. The smaller screening length of a baryon with higher spin
is easier to dissociate. The linear velocity of quarks $v\equiv l_{q}\omega/2$
is similar to Ref. \cite{Peeters:2006iu} if a drag force in the $x_{3}$
direction orthogonal to the rotating plane is added. 

\subsection{Regge behavior}

\begin{figure}
\begin{centering}
\includegraphics[scale=0.5]{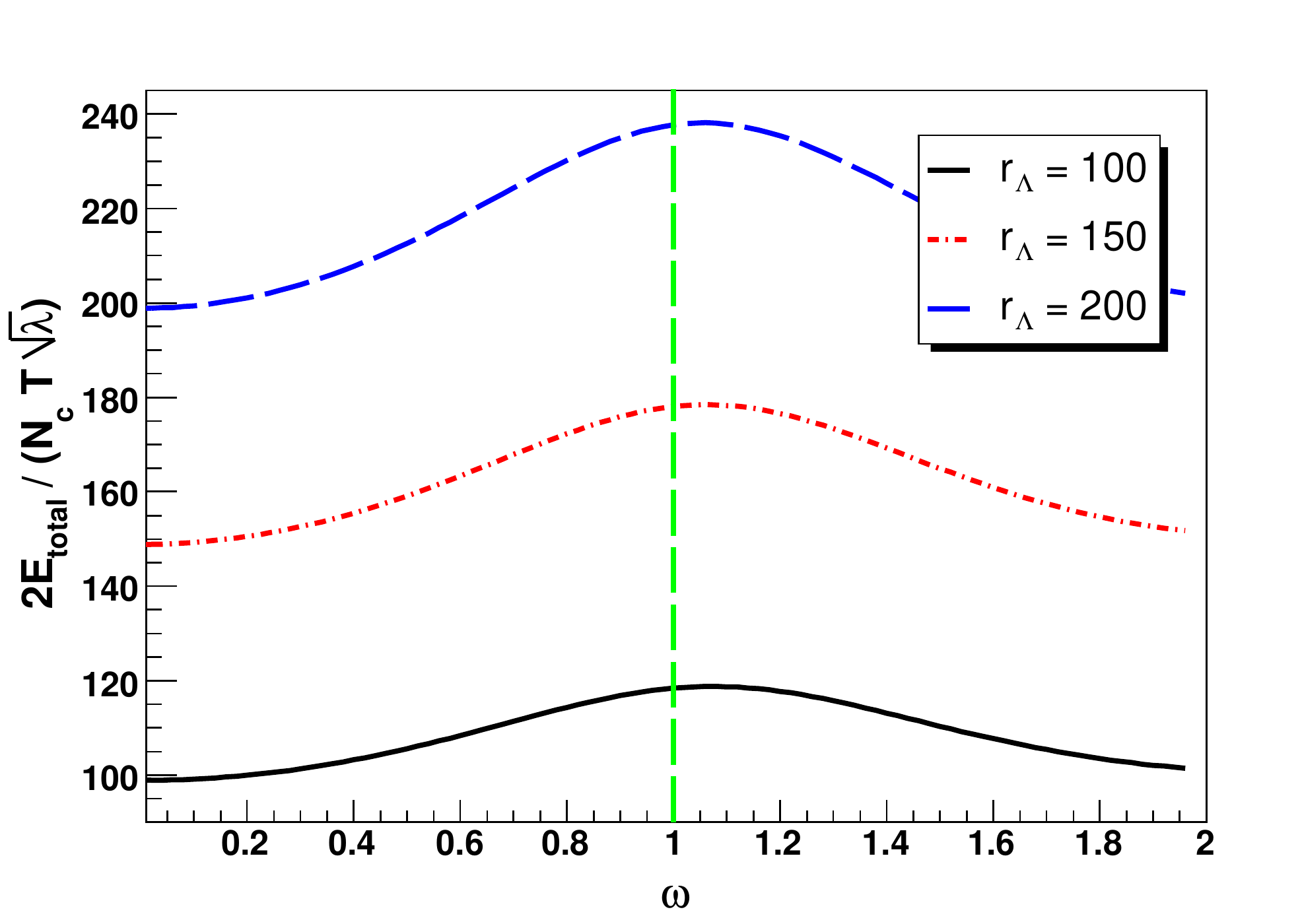}
\par\end{centering}

\begin{centering}
\includegraphics[scale=0.5]{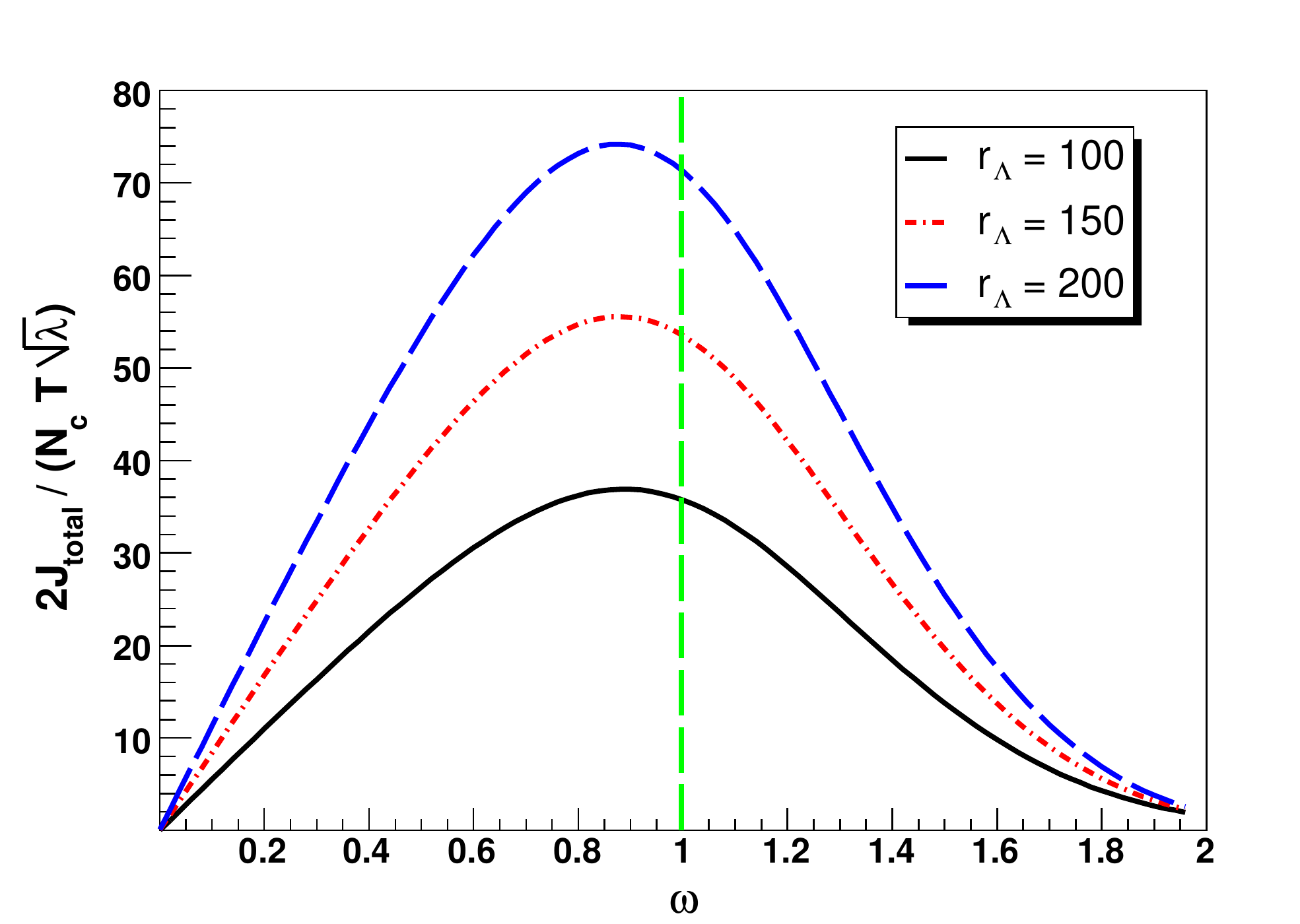}
\par\end{centering}

\caption{\label{fig:E-J-omega}$J_{total}$ and $E_{total}$ as functions of
$\omega$ with fixed $r_{e}$.}

\end{figure}

\begin{figure}
\begin{centering}
\includegraphics[scale=0.5]{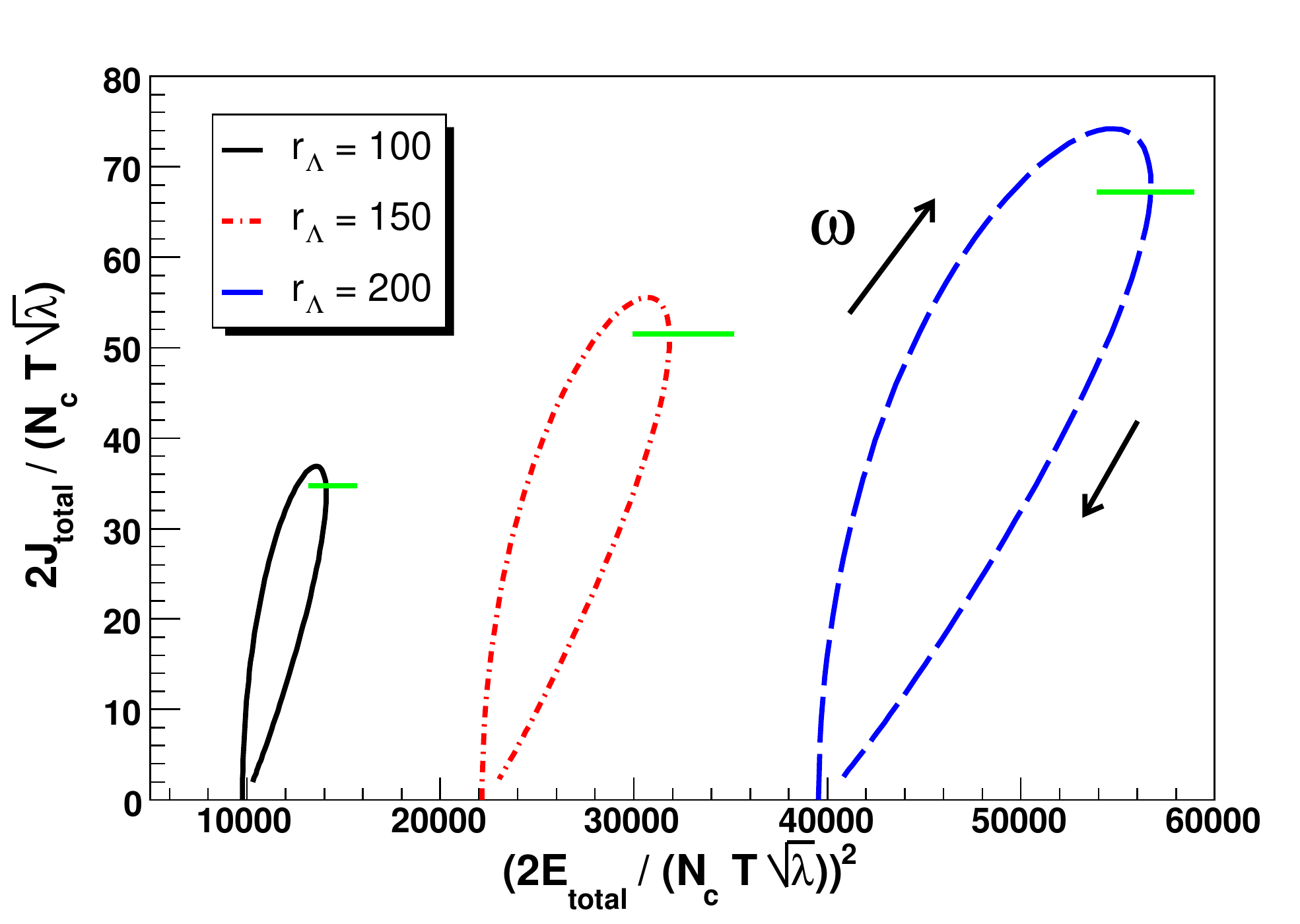}
\par\end{centering}

\caption{\label{fig:E-J-re}The relation between $J_{total}$ and $E_{total}$
with fixed $r_{e}$.}
\end{figure}

\begin{figure}
\begin{centering}
\includegraphics[scale=0.5]{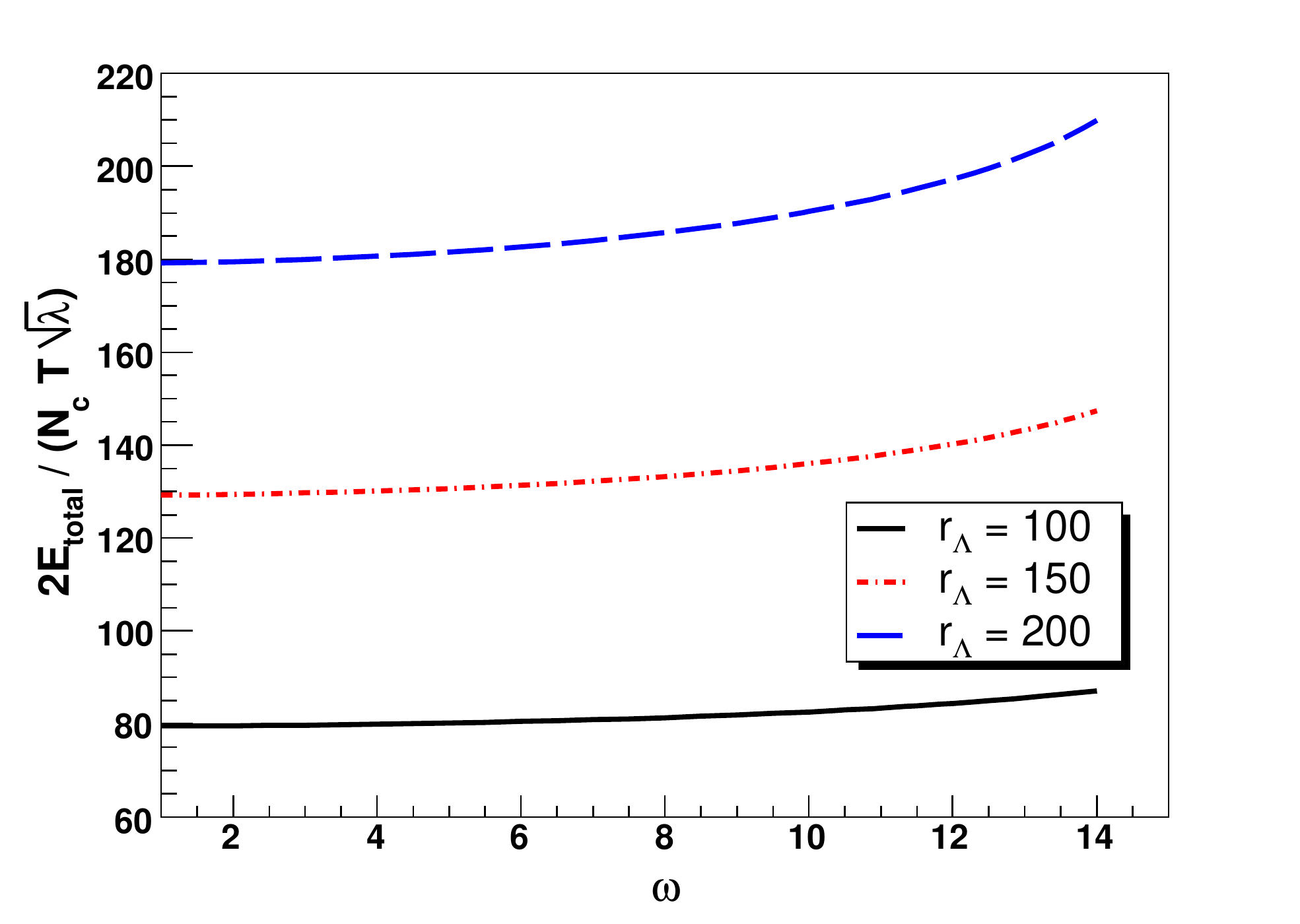}
\par\end{centering}

\begin{centering}
\includegraphics[scale=0.5]{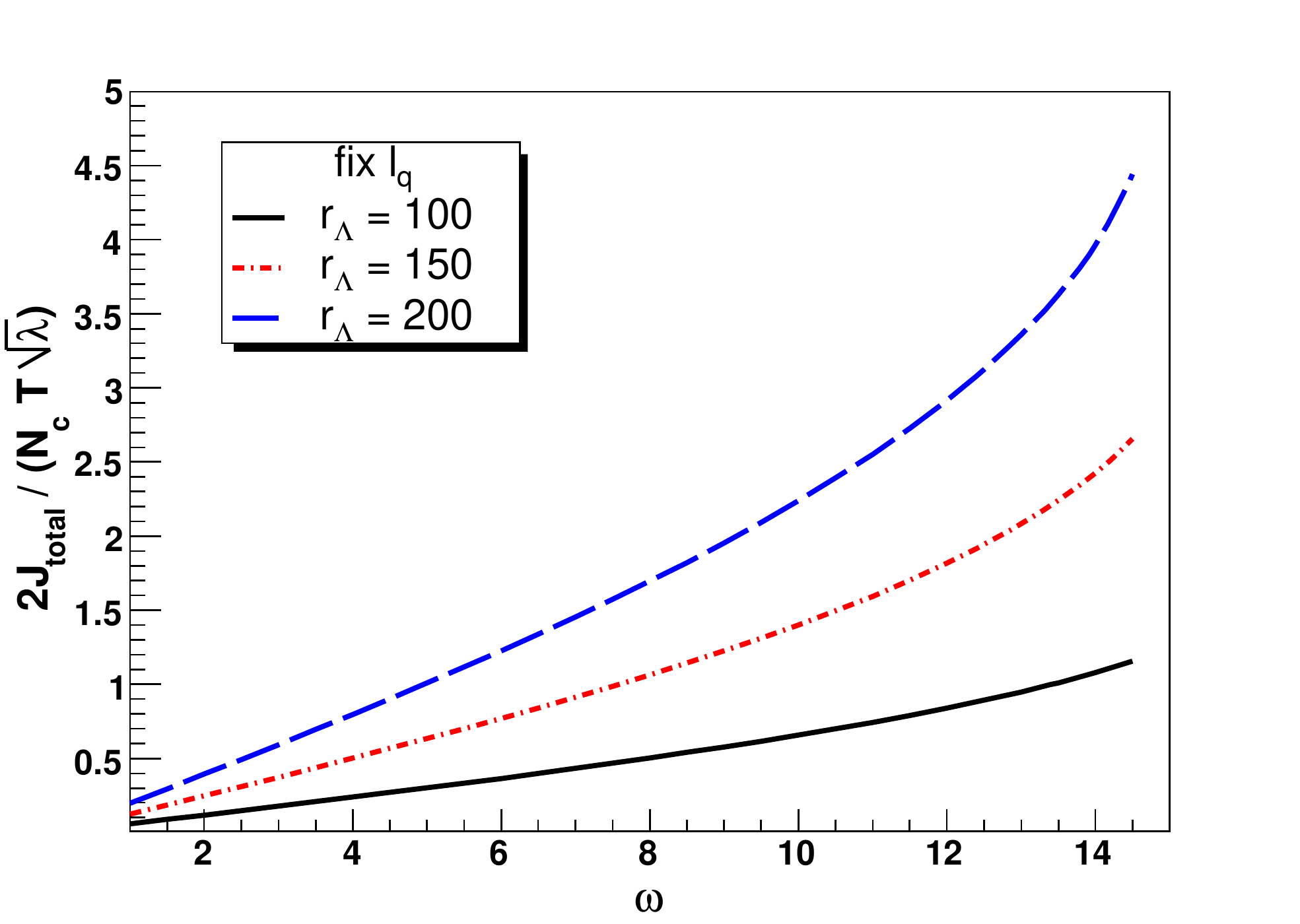}
\par\end{centering}

\caption{\label{fig:E-J-lq}$J_{total}$ and $E_{total}$ as functions of $\omega$
with fixed $l_{q}$.}

\end{figure}

\begin{figure}
\begin{centering}
\includegraphics[scale=0.5]{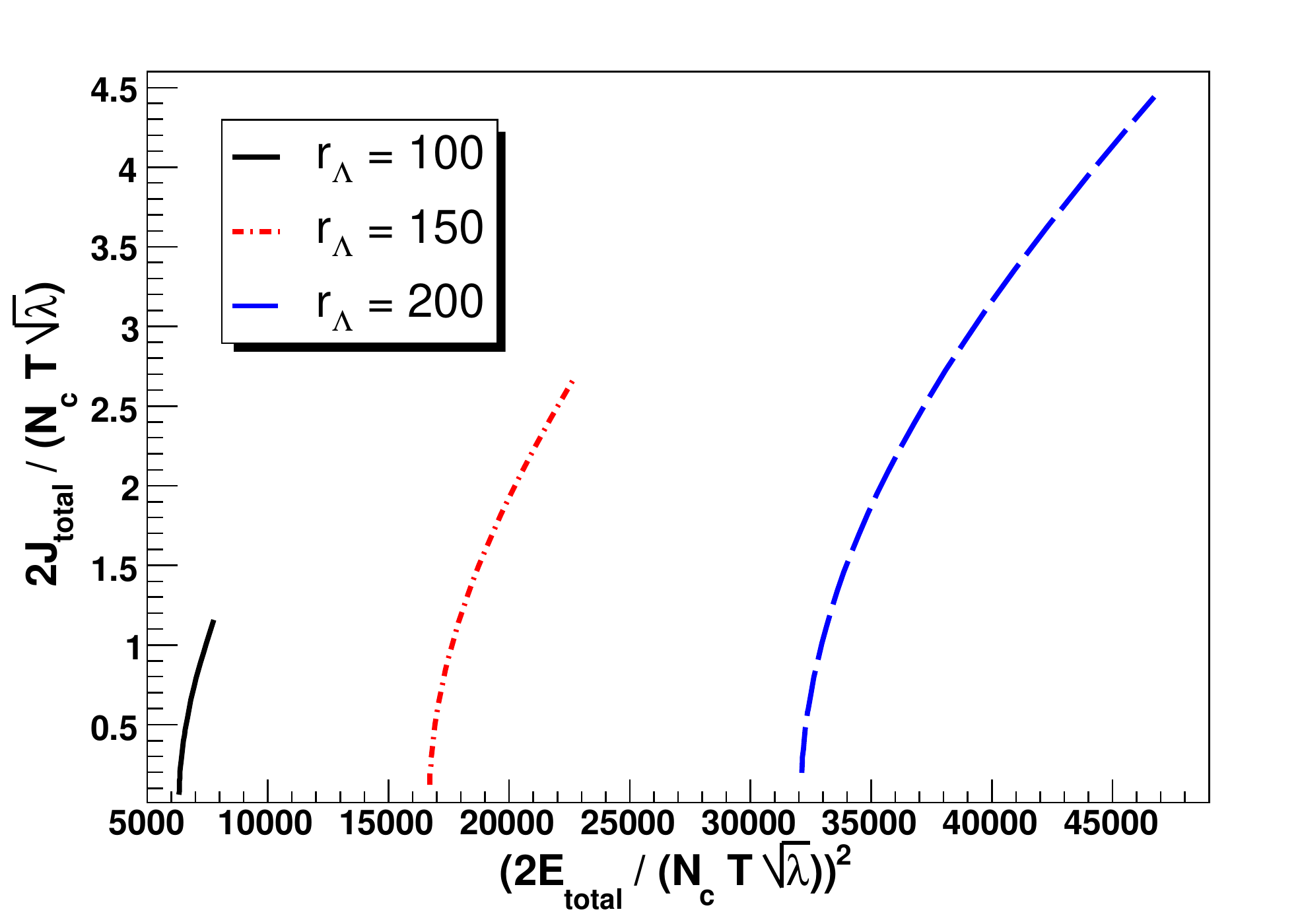}
\par\end{centering}

\caption{\label{fig:E-J-lq-1}The relation between $J_{total}$ and $E_{total}$
with fixed $l_{q}$.}
\end{figure}

At the time when the classic string theory was considered as a candidate
for the theory of the strong interaction, it was found that the angular
momentum of baryons $J$ is proportional to the square of energy $E^{2}$
as follows
\[
J\propto E^{2}\;,
\]
which is called the Regge behavior. In this model, the angular momentum
and energy of the strings read
\begin{eqnarray}
J_{string} & = & \frac{\partial\mathcal{L}}{\partial\omega}\nonumber \\
 & = & \frac{1}{2\pi\alpha'}\int_{r_{e}}^{r_{\Lambda}}dr\frac{\left(\frac{1}{f(r)}+\frac{r^{2}}{R^{2}}\rho'^{2}(r)\right)(\frac{r^{2}}{R^{2}}\rho^{2}\omega)}{\sqrt{-\left(\frac{r^{2}}{R^{2}}\rho^{2}\omega^{2}-f(r)\right)\left(\frac{1}{f(r)}+\frac{r^{2}}{R^{2}}\rho'^{2}(r)\right)}}\;,
\end{eqnarray}
and 
\begin{eqnarray}
E_{string} & = & \omega\frac{\partial\mathcal{L}}{\partial\omega}-\mathcal{L}\nonumber \\
 & = & \frac{1}{2\pi\alpha'}\int_{r_{e}}^{r_{\Lambda}}dr\frac{\left(\frac{1}{f(r)}+\frac{r^{2}}{R^{2}}\rho'^{2}(r)\right)f(r)}{\sqrt{-\left(\frac{r^{2}}{R^{2}}\rho^{2}\omega^{2}-f(r)\right)\left(\frac{1}{f(r)}+\frac{r^{2}}{R^{2}}\rho'^{2}(r)\right)}}\;.
\end{eqnarray}
The total angular momentum and energy of the system are given by 
\begin{eqnarray}
E_{total} & = & N_{c}E_{string}+E_{brane}\;,\nonumber \\
J_{total} & = & N_{c}J_{string}+J_{brane}\;,
\end{eqnarray}
where 
\begin{equation}
E_{brane}=\frac{\mathcal{V}(r_{e})V_{5}}{(2\pi)^{5}\alpha'^{3}}\;.
\end{equation}
The numerical results for $J_{total}$ and $E_{total}$ as functions
of $\omega$ with different cutoff $r_{\Lambda}$ are shown in the
upper and lower plane of Fig. \ref{fig:E-J-omega}, respectively.
The relation between $J_{total}$ and $E_{totoal}^{2}$ with the same
$\omega$ (i.e. fixed $r_{e}$) is shown in Fig. \ref{fig:E-J-re}.
In the case of the same separation distance of quarks, i.e. the fixed
value of $l_{q}$, the relation between $J_{total}$ and $E_{totoal}^{2}$
is shown in Fig. \ref{fig:E-J-lq} and \ref{fig:E-J-lq-1}.

\begin{figure}
\begin{centering}
\includegraphics[scale=0.5]{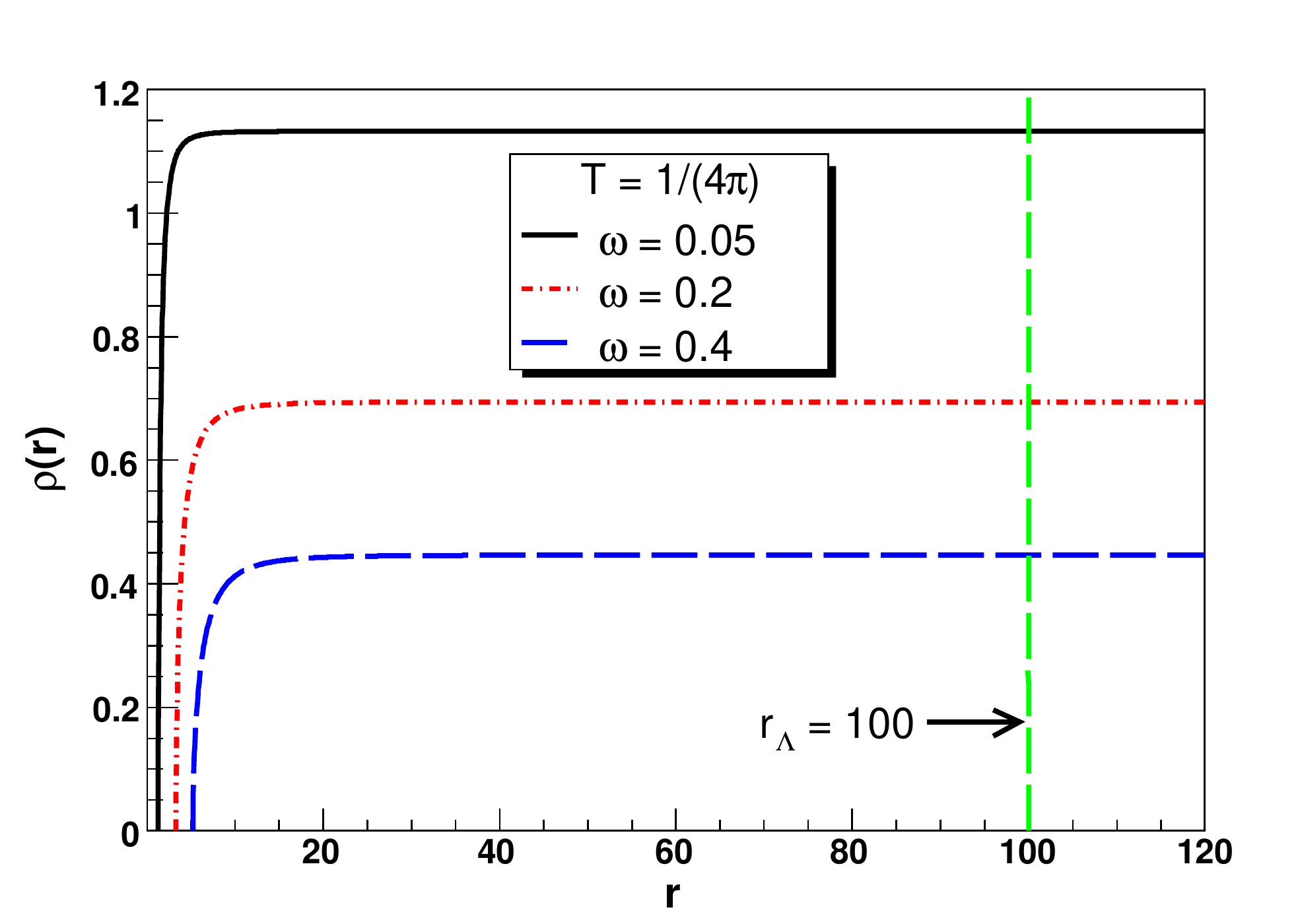}
\par\end{centering}

\caption{\label{fig:rho-orth}Embedding function $\rho(r)$ with orthogonal
boundary condition (\ref{eq:boundary-cond-02}).}
\end{figure}

Finally, another physical boundary condition is considered. In order
to pick up some configurations for baryons with the same constituents
but different $\omega$ or spin, the configurations satisfying the
following boundary condition are chosen, 
\begin{equation}
\frac{\partial\mathcal{L}}{\partial\rho^{\prime}}=0\;,
\end{equation}
which gives 
\begin{equation}
\frac{1}{4}\: l_{q}^{2}\:\omega^{2}=\left(1-\frac{r_{0}^{4}}{r_{\Lambda}^{4}}\right)\;,\label{eq:boundary-cond-01}
\end{equation}
or 
\begin{equation}
\rho^{\prime}(r_{\Lambda})=0\;.\label{eq:boundary-cond-02}
\end{equation}
It is observed that the end points of strings move with the speed
of light on the cutoff brane with the first condition (\ref{eq:boundary-cond-01}),
while the string is orthogonal to this brane with the condition (\ref{eq:boundary-cond-02}).
Unfortunately, it is not guaranteed to find out the points to fulfill
the boundary condition (\ref{eq:boundary-cond-01}) for small $\omega$.
The configurations for the strings fulfilling with condition (\ref{eq:boundary-cond-02})
are shown in Fig. \ref{fig:rho-orth}.

\begin{figure}
\begin{centering}
\includegraphics[scale=0.5]{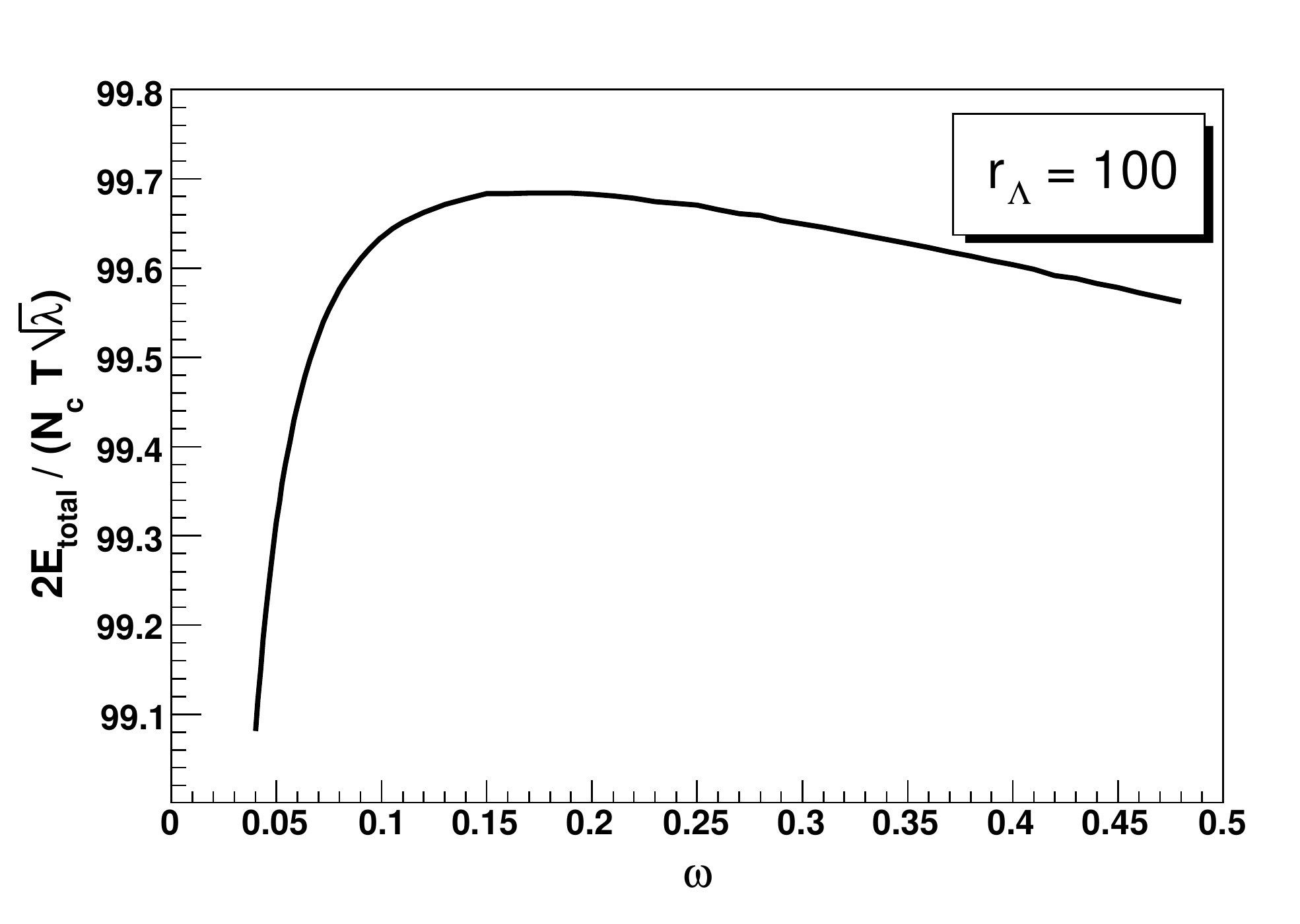}
\par\end{centering}

\begin{centering}
\includegraphics[scale=0.5]{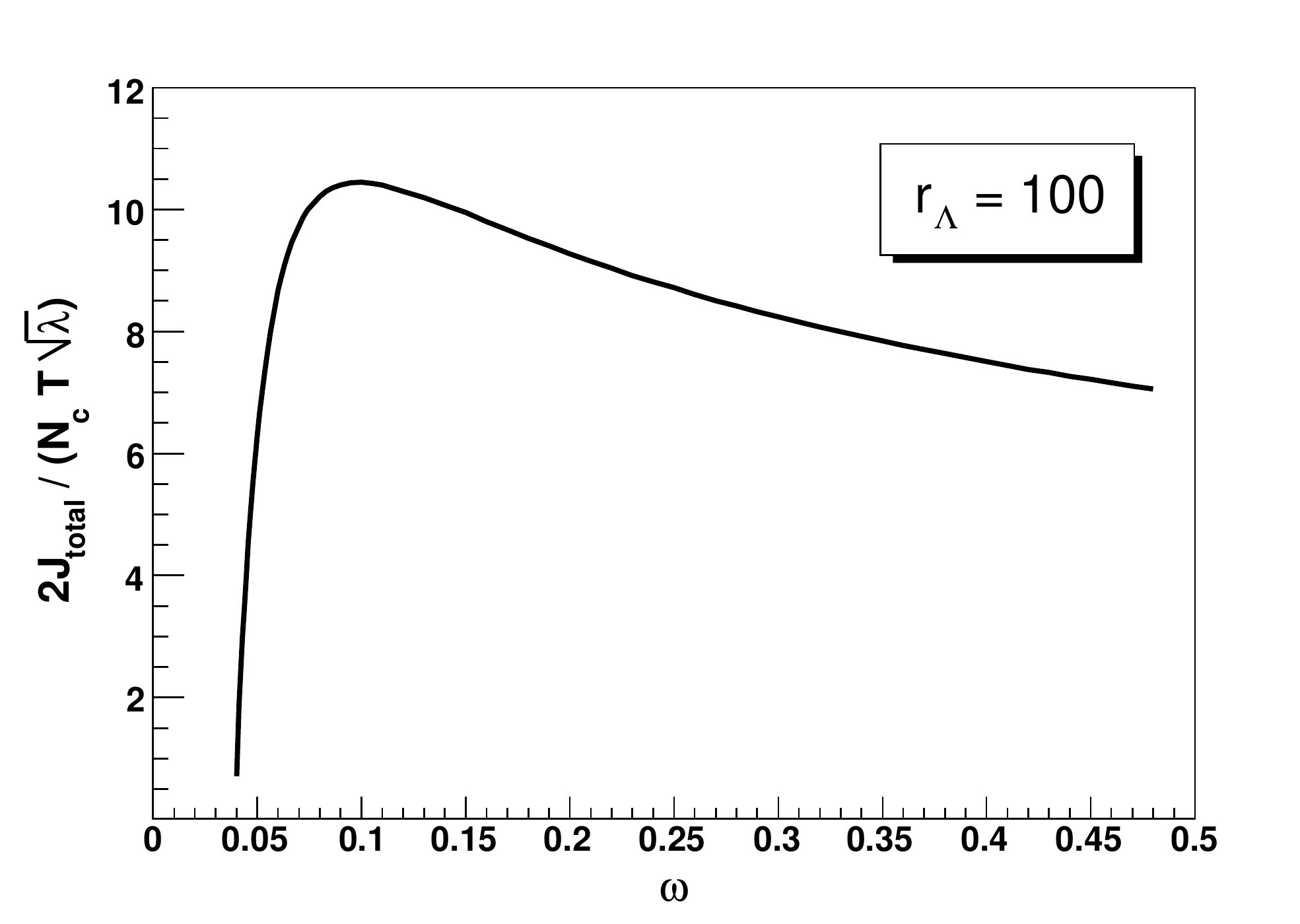}
\par\end{centering}

\caption{\label{fig:E-J-ph-01}Angular momentum and energy with the FBC (\ref{eq:FBC-03})
as functions of $\omega$.}

\end{figure}

\begin{figure}
\begin{centering}
\includegraphics[scale=0.5]{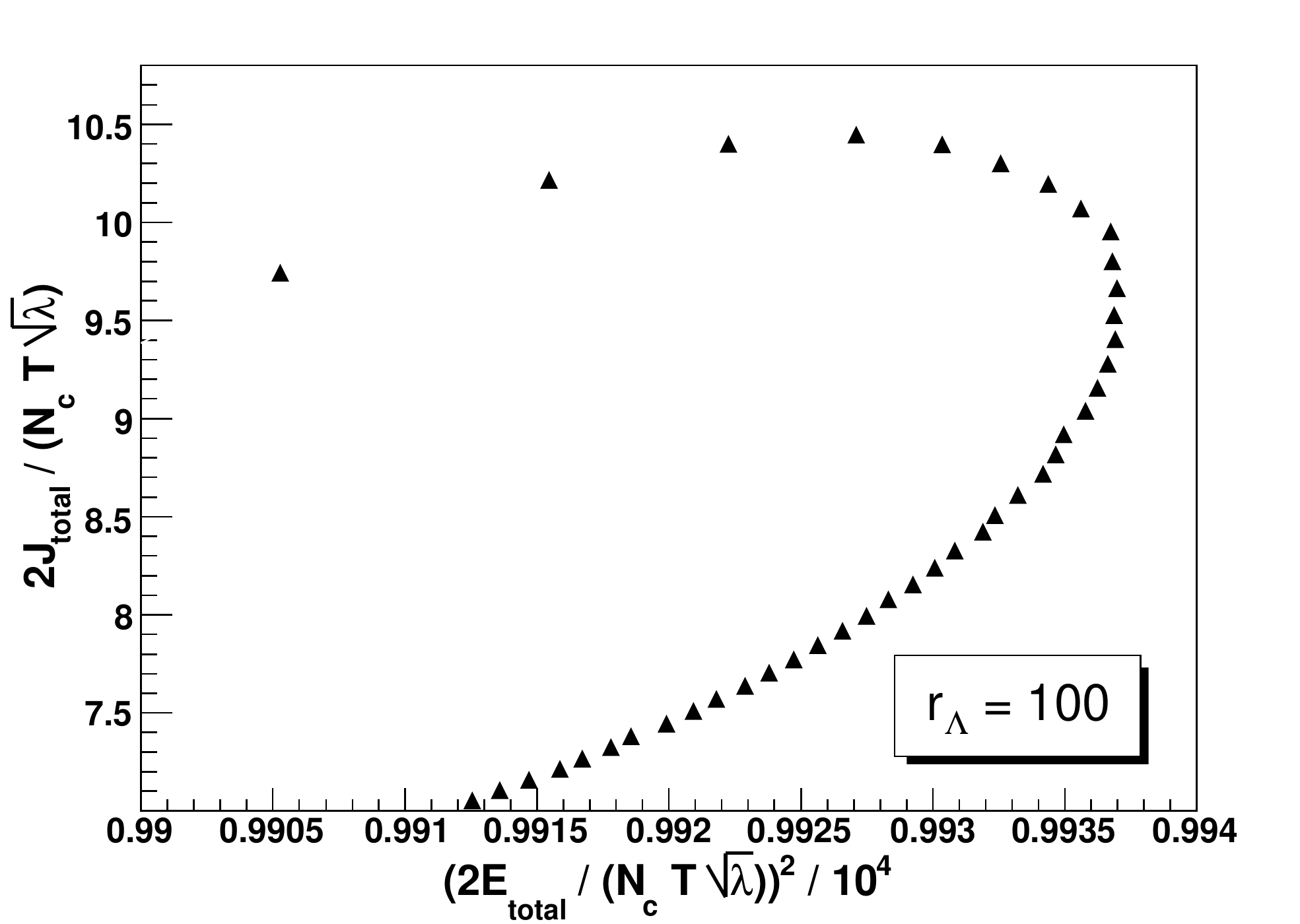}
\par\end{centering}

\caption{\label{fig:E-J-ph-02}Angular momentum and energy with the FBC (\ref{eq:FBC-03})
as functions of $\omega$.}
\end{figure}

In this case, the FBC in the $\rho$-direction becomes 
\begin{equation}
\frac{\partial E_{I}}{\partial\rho}=m_{q}\omega^{2}\rho\;,\label{eq:FBC-03}
\end{equation}
with $m_{q}=\frac{1}{2\pi\alpha^{\prime}}(r_{\Lambda}-r_{0})$ the
mass of quarks and $E_{I}$ the interaction potential defined as 
\begin{equation}
E_{I}=N_{c}E_{string}-E_{q}+E_{brane}\;,
\end{equation}
where 
\begin{equation}
E_{q}=\frac{N_{c}}{2\pi\alpha^{\prime}}\int_{r_{0}}^{r_{\Lambda}}dr\;,
\end{equation}
is the energy of free quarks. The $\omega$ dependence of angular
momentum and energy with the FBC (\ref{eq:FBC-03}) is shown in Fig.
\ref{fig:E-J-ph-01}. The relation between $J_{total}$ and $E_{totoal}^{2}$
with the FBC (\ref{eq:FBC-03}) is shown in Fig. \ref{fig:E-J-ph-02}.

\subsection{Conclusion}

We calculate the $\omega$ dependence of the baryon screening length
in strongly coupled hot plasma via AdS/CFT duality. In our model,
we consider a baryon in the bulk space as a probe and a baryon at
the boundary with high spin using the rotating strings. We obtain
the $\omega$ dependence of the embedding function of these strings
and their screening length. We also investigate the relation between
total angular momentum and energy of baryons and show the numerical
results in three different conditions. We find that these solutions
with orthogonal boundary condition (\ref{eq:FBC-03}) are the best
candidates for baryons with same constituents but different spins.



\chapter{Fluid dynamics with triangle anomaly\label{chap:vorticity}}

The RHIC data for collective flows have been well described by the
ideal and dissipative hydrodynamics \cite{Betz:2008me,Baier2008,Song2008a,Huovinen2001,Muronga2002,Kolb2003,Muronga2004,Romatschke2007}.
The correspondence of relativistic hydrodynamics to charged black-branes
was investigated by AdS/CFT duality \cite{Erdmenger2009,Banerjee:2008th}.
A new term associated with the axial anomalies was found in the first
order dissipative hydrodynamics (see, e.g., Ref. \cite{Torabian2009a}
about holographic hydrodynamics with multiple/non-Abelian symmetries,
or Ref. \cite{Rebhan2010} in Sakai-Sugimoto model). Recently the
new term has been derived in hydrodynamics with a triangle anomaly
\cite{Son2009}. A similar result was also obtained in microscopic
theory of the superfluid \cite{Lublinsky2010}. This problem is closely
related to the so-called Chiral Magnetic Effect (CME) in heavy ion
collisions \cite{Kharzeev2008,Fukushima2008,Fukushima2010a,Kharzeev2010}.
When two energetic nuclei pass each other a strong magnetic field
up to $10^{18}$ G is formed, which breaks local parity via axial
anomaly. This effect may be observed through charge separation. Hydrodynamics
in an external background field can be used to pin down the CME in
real time simulation. However the anomalous term in the charge current
breaks the second law of thermodynamics unless new terms of vorticity
and magnetic field are introduced in the charge and entropy currents
\cite{Son2009}. 

In this chapter, we try to provide a consistent description of the
kinetic equation with a triangle anomaly. We will derive the kinetic
equation to the next to leading order as well as the leading order
correction to the particle distribution function arising from anomaly.
These results are compatible with the entropy principle of the second
law of thermodynamics and the charge/energy-momentum conservation
equations. Most of the contents in this chapter are taken from Ref.
\cite{Pu:2010as}

\section{Constraining distribution function with anomaly compatible to second
law of thermodynamics \label{sec:2}}

In this and the next sections we will consider the most simple case
with one charge and one particle species (without anti-particles).
The relativistic Boltzmann equation for the on-shell phase-space distribution
$f(x,p)$ in a background electromagnetic field $F_{\mu\nu}$ is given
by Eq.(\ref{eq:boltzmann-F}). Note that $\mathcal{C}[f]$ contains
a normal collision term $\mathcal{C}_{0}[f]$ and a source term from
anomaly $\mathcal{C}_{A}[f]$, $\mathcal{C}[f]=\mathcal{C}_{0}[f]+\mathcal{C}_{A}[f]$.
We assume that $\mathcal{C}_{A}[f]$ is at most of the first order,
a small quantity. The necessity for the source term is to make the
charge conservation equation hold, 
\begin{equation}
\partial_{\mu}j^{\mu}=-CE^{\mu}B_{\mu}\equiv-CE\cdot B.\label{eq:charge-current}
\end{equation}
Here $j^{\mu}$ is the charge current and $E^{\mu}=u_{\nu}F^{\mu\nu}$
and $B_{\mu}=\frac{1}{2}\epsilon_{\mu\nu\alpha\beta}u^{\nu}F^{\alpha\beta}$
are electric and magnetic field vectors respectively, where $u_{\mu}$
is the fluid velocity and $\epsilon_{\mu\nu\alpha\beta}=-\epsilon^{\mu\nu\alpha\beta}=-1,1$
for the order of Lorentz indices $(\mu\nu\alpha\beta)$ is an even/odd
permutation of $(0123)$. However, the presence of the source term
should not influence the energy momentum conservation, 
\begin{equation}
\partial_{\mu}T^{\mu\nu}=F^{\nu\mu}j_{\mu}.\label{eq:em-cons1}
\end{equation}

One can verify that the equilibrium solution of the distribution function,
\begin{equation}
f_{0}=\frac{1}{\exp[\beta u_{\mu}(p^{\mu}-QF^{\mu\nu}x_{\nu})-\beta Q\mu_{0}]-e},
\end{equation}
satisfies the collisionless Boltzmann equation (\ref{eq:boltzmann-F})
in an external field for constant $\beta=1/T$ ($T$ is the local
temperature), $u_{\mu}$ and $\mu_{0}$ (local chemical potential
without electromagnetic field). Here $e=0,\pm1$ for Boltzmann, Bose
and Fermi distributions respectively. When $\beta$, $u_{\mu}$ and
$\mu_{0}$ are not constants but functions of space-time, the Boltzmann
equation (\ref{eq:boltzmann-F}) is not satisfied automatically. Note
that we can absorb $-Qx_{\nu}u_{\mu}F^{\mu\nu}=Qx\cdot E$ into $\mu_{0}$
so that $f_{0}$ has the form of an equilibrium distribution function, 

\begin{eqnarray}
f_{0}(x,p) & = & \frac{1}{e^{(u\cdot p-Q\mu)/T}-e},\label{eq:free-f}
\end{eqnarray}
where $\mu\equiv\mu_{0}-x\cdot E$. 

We assume that the distribution function $f$ in presence of an anomaly
is a solution of the Boltzmann equation with collision terms in Eq.
(\ref{eq:boltzmann-F}), where $\beta$, $u_{\mu}$ and $\mu$ are
functions of space-time. Generally $f(x,p)$ can be written in the
following form, 

\begin{eqnarray}
f(x,p) & = & \frac{1}{e^{(u\cdot p-Q\mu)/T+\chi(x,p)}-e}=f_{0}(x,p)+f_{1}(x,p),\label{eq:fxp-expansion}
\end{eqnarray}
where $f_{0}(x,p)$ is given in Eq. (\ref{eq:free-f}) and $f_{1}(x,p)$
is the first order deviation from it, 

\begin{eqnarray}
f_{1}(x,p) & = & -f_{0}(x,p)\left[1+ef_{0}(x,p)\right]\chi(x,p).\label{f1xp}
\end{eqnarray}
It is known that a magnetic field is closely related to a charge rotation
characterized by vorticity. So we introduce into the distribution
function terms associated with the vorticity-induced current $\omega_{\mu}=\frac{1}{2}\epsilon_{\mu\nu\alpha\beta}u^{\nu}\partial^{\alpha}u^{\beta}$
and the magnetic field 4-vector $B_{\mu}$ which are assumed to be
of the first order, which provide a leading order correction to the
particle distribution function. For simplicity we will neglect viscous
and diffusive effects throughout the paper, then the ordinary form
in the current scheme for $\chi(x,p)$ reads, 

\begin{eqnarray}
\chi(x,p) & = & \lambda(p)p\cdot\omega+\lambda_{B}(p)p\cdot B,\label{eq:deviation}
\end{eqnarray}
where $\lambda(p)$ and $\lambda_{B}(p)$ are functions of $\mu$,
$T$, and $u\cdot p$ and have mass dimension $-2$ and $-3$ respectively.
We will show that $\lambda(p)$ and $\lambda_{B}(p)$ must depend
on momentum otherwise they will contradict the entropy principle from
the second law of thermodynamics. 

Using Eq. (\ref{eq:fxp-expansion}) we can decompose the charge and
entropy currents and the stress tensor into equilibrium values and
the leading order (first order) corrections as $j^{\mu}=j_{0}^{\mu}+j_{1}^{\mu}$,
$S^{\mu}=S_{0}^{\mu}+S_{1}^{\mu}$ and $T^{\mu\nu}=T_{0}^{\mu\nu}+T_{1}^{\mu\nu}$
with 
\begin{eqnarray}
j_{0,1}^{\mu}(x) & = & q\int[dp]p^{\mu}f_{0,1}(x,p),\nonumber \\
S_{0}^{\mu}(x) & = & -\int[dp]p^{\mu}\psi(f_{0}),\nonumber \\
S_{1}^{\mu}(x) & = & -\int[dp]p^{\mu}\psi'(f_{0})f_{1},\nonumber \\
T_{0,1}^{\mu\nu}(x) & = & \int[dp]p^{\mu}p^{\nu}f_{0,1}(x,p),\label{eq:j1s1t1}
\end{eqnarray}
where we have defined $[dp]\equiv d_{g}\frac{d^{3}p}{(2\pi)^{3}(u\cdot p)}$
($d_{g}$ is the degeneracy factor), $\psi(f_{0})=f_{0}\ln(f_{0})-e(1+ef_{0})\ln(1+ef_{0})$
and $\psi'(f_{0})=\ln[f_{0}/(1+ef_{0})]=-(u\cdot p-Q\mu)/T$. Inserting
$f_{0}$ into the above formula, we obtain the charge and entropy
currents and the stress tensor in equilibrium, $j_{0}^{\mu}=nu^{\mu}$,
$S_{0}^{\mu}=su^{\mu}$ and $T_{0}^{\mu\nu}=(\epsilon+P)u^{\mu}u^{\nu}-Pg^{\mu\nu}$,
with the energy density $\varepsilon$, the pressure $P$, the particle
number density $n$ and the entropy density $s=(\epsilon+P-n\mu)/T$.
Using Eqs. (\ref{f1xp},\ref{eq:deviation},\ref{eq:j1s1t1}), we
obtain 

\begin{eqnarray}
j_{1}^{\mu} & = & \xi\omega^{\mu}+\xi_{B}B^{\mu},\nonumber \\
T_{1}^{\mu\nu} & = & DT\left(u^{\mu}\omega^{\nu}+u^{\nu}\omega^{\mu}\right)+D_{B}T\left(u^{\mu}B^{\nu}+u^{\nu}B^{\mu}\right),\nonumber \\
S_{1}^{\mu} & = & -\frac{\mu}{T}(\xi\omega^{\mu}+\xi_{B}B^{\mu})+(D\omega^{\mu}+D_{B}B^{\mu}),\label{eq:j-t-s1}
\end{eqnarray}
where 
\begin{eqnarray}
\xi & = & -QJ_{21}^{\lambda}\equiv\frac{1}{3}Q\int[dp][(p\cdot u)^{2}-m^{2}]f_{0}(1+ef_{0})\lambda(p),\nonumber \\
\xi_{B} & = & -QJ_{21}^{\lambda_{B}}\equiv\frac{1}{3}Q\int[dp][(p\cdot u)^{2}-m^{2}]f_{0}(1+ef_{0})\lambda_{B}(p),\nonumber \\
D & = & -\frac{J_{31}^{\lambda}}{T}\equiv\frac{1}{3T}\int[dp][(p\cdot u)^{2}-m^{2}](p\cdot u)f_{0}(1+ef_{0})\lambda(p),\nonumber \\
D_{B} & = & -\frac{J_{31}^{\lambda_{B}}}{T}\equiv\frac{1}{3T}\int[dp][(p\cdot u)^{2}-m^{2}](p\cdot u)f_{0}(1+ef_{0})\lambda_{B}(p).\label{eq:db}
\end{eqnarray}

On the other hand, $\xi$, $\xi_{B}$, $D$ and $D_{B}$ as functions
of $\mu$ and $T$ can be determined by the second law of thermodynamics
or the entropy principle together with Eq. (\ref{eq:charge-current}-\ref{eq:em-cons1}).
The entropy production rate is given by 
\begin{eqnarray}
\partial_{\mu}\left(su^{\mu}-\frac{\mu}{T}\nu^{\mu}\right) & = & \frac{\pi^{\mu\nu}}{T}\partial_{\mu}u_{\nu}-\nu^{\mu}\left(\partial_{\mu}\frac{\mu}{T}+\frac{E^{\mu}}{T}\right)-C\frac{\mu}{T}E\cdot B\;,\label{eq:entorpy_normal}
\end{eqnarray}
where $C$ is a constant whose sign is arbitrary.\textcolor{black}{{}
}We find that $\partial_{\mu}(su^{\mu}+S_{1}^{\mu})$ cannot be positive
definite unless we make a shift to introduce a new entropy current
$\tilde{S}^{\mu}$ as follows, 
\begin{eqnarray}
\tilde{S}^{\mu} & = & su^{\mu}+S_{1}^{\mu}-(D\omega^{\mu}+D_{B}B^{\mu})=su^{\mu}-\frac{\mu}{T}\left(\xi\omega^{\mu}+\xi_{B}B^{\mu}\right)\nonumber \\
 & = & \frac{1}{T}(Pu^{\mu}-\mu j^{\mu}+u_{\lambda}T^{\lambda\mu})-(D\omega^{\mu}+D_{B}B^{\mu})
\end{eqnarray}
We will use the thermodynamic relation 
\begin{equation}
\partial_{\mu}(Pu^{\mu})=j_{0}^{\mu}\partial_{\mu}\overline{\mu}-T_{0}^{\lambda\mu}\partial_{\mu}u_{\lambda}\label{eq:d-pu}
\end{equation}
and the identities 
\begin{eqnarray}
u^{\mu}u^{\lambda}\partial_{\mu}\omega_{\lambda} & = & \frac{1}{2}\partial_{\mu}\omega^{\mu},\nonumber \\
u^{\mu}u^{\lambda}\partial_{\mu}B_{\lambda} & = & \partial_{\mu}B^{\mu}-2\omega^{\rho}E_{\rho},\nonumber \\
\partial_{\mu}\omega^{\mu} & = & -\frac{2}{\epsilon+P}(n\omega^{\mu}E_{\mu}+\omega^{\mu}\partial_{\mu}P),\nonumber \\
\partial_{\mu}B^{\mu} & = & 2\omega^{\rho}E_{\rho}-\frac{1}{\epsilon+P}(nB_{\lambda}E^{\lambda}+B^{\mu}\partial_{\mu}P),\label{eq:id-oeb}
\end{eqnarray}
to evaluate $\partial_{\mu}\tilde{S}^{\mu}$. We have used the shorthand
notation $\overline{\mu}\equiv\mu/T$ in Eq. (\ref{eq:d-pu}). Following
the same procedure as in Ref. \cite{Son2009}, we obtain 
\begin{eqnarray}
\partial_{\mu}\widetilde{S}^{\mu} & = & \omega^{\mu}\left[\xi^{SS}\partial_{\mu}\overline{\mu}-\partial_{\mu}D+\frac{2D}{\epsilon+P}\partial_{\mu}P\right]\nonumber \\
 &  & +B^{\mu}\left[\xi_{B}^{SS}\partial_{\mu}\overline{\mu}-\partial_{\mu}D_{B}+\frac{D_{B}}{\epsilon+P}\partial_{\mu}P\right]\nonumber \\
 &  & +E\cdot\omega\left[\frac{1}{T}\xi^{SS}+\frac{2nD}{\epsilon+P}-2D_{B}\right]\nonumber \\
 &  & +E\cdot B\left[\frac{1}{T}\xi_{B}^{SS}+C\frac{\mu_{A}}{T}+\frac{nD_{B}}{\epsilon+P}\right].\label{eq:ds-tilde}
\end{eqnarray}
where we have defined 
\begin{eqnarray}
\xi^{SS} & = & \frac{DTn}{\epsilon+P}-\xi,\;\xi_{B}^{SS}=\frac{D_{B}Tn}{\epsilon+P}-\xi_{B}.\label{eq:ss-xi}
\end{eqnarray}
For the constraint $\partial_{\mu}\widetilde{S}^{\mu}\geq0$ to hold,
we impose that all quantities inside the square brackets should vanish.
We finally obtain 
\begin{equation}
D=\frac{1}{3}C\frac{\mu^{3}}{T},\; D_{B}=\frac{1}{2}C\frac{\mu^{2}}{T},\;\xi=-C\frac{sT\mu^{2}}{\epsilon+P},\;\xi_{B}=-C\frac{sT\mu}{\epsilon+P}.\label{eq:entropy-d}
\end{equation}
Using Eqs. (\ref{eq:ss-xi},\ref{eq:entropy-d}), one can verify that
the values of $\xi^{SS}$ and $\xi_{B}^{SS}$ are identical to Ref.
\cite{Son2009}. The difference between our values in Eq. (\ref{eq:entropy-d})
and those in Ref. \cite{Son2009} arises from the fact that we do
not use the Landau frame while the authors of Ref. \cite{Son2009}
do. By equating Eq. (\ref{eq:db}) and (\ref{eq:entropy-d}), we obtain
equations for $\lambda$ and $\lambda_{B}$, 
\begin{eqnarray}
 &  & QJ_{21}^{\lambda}=-\xi,\; J_{31}^{\lambda}=-DT,\; QJ_{21}^{\lambda_{B}}=-\xi_{B},\; J_{31}^{\lambda_{B}}=-D_{B}T.\label{eq:constraint-lambda}
\end{eqnarray}
Equation (\ref{eq:constraint-lambda}) forms a complete set of constraints
for $\lambda$ and $\lambda_{B}$. We note that $\lambda$ and $\lambda_{B}$
must depend on momentum in general. If $\lambda$ and $\lambda_{B}$
are constants, we would obtain 
\begin{equation}
\frac{\xi}{DT}=\frac{\xi_{B}}{D_{B}T}=Q\frac{J_{21}}{J_{31}},
\end{equation}
which contradict Eq. (\ref{eq:entropy-d}) from the entropy principle. 

We can expand $\lambda(p)$ and $\lambda_{B}(p)$ in powers of $u\cdot p$,
\begin{equation}
\lambda(p)=\sum_{i=0}\lambda_{i}(u\cdot p)^{i},\;\lambda_{B}(p)=\sum_{i=0}\lambda_{i}^{B}(u\cdot p)^{i}.\label{eq:lambda-expand}
\end{equation}
So we obtain the following expressions 
\begin{equation}
J_{n1}^{\lambda}=\sum_{i=0}\lambda_{i}J_{i+n,1},\; J_{n1}^{\lambda_{B}}=\sum_{i=0}\lambda_{i}^{B}J_{i+n,1},\label{eq:j-expand}
\end{equation}
for $n=2,3$. Here the functions $J_{nq}$ are integrals defined in
Ref. \cite{Israel:1979wp,Muronga2002}, 
\begin{equation}
J_{nq}=(-1)^{q}\frac{1}{(2q+1)!!}\int\frac{d^{3}p}{(2\pi)^{3}(u\cdot p)}[(u\cdot p)^{2}-m^{2}]^{q}(u\cdot p)^{n-2q}f_{0}(1+ef_{0}),\label{eq:jnk}
\end{equation}
Using Eqs. (\ref{eq:lambda-expand},\ref{eq:j-expand}) in Eq. (\ref{eq:constraint-lambda}),
we can constrain the coefficients $\lambda_{i}$ and $\lambda_{i}^{B}$.
If we expand both $\lambda(p)$ and $\lambda_{B}(p)$ to the first
power of $u\cdot p$, we can completely fix the coefficients $\lambda_{0,1}$
and $\lambda_{0,1}^{B}$ from Eq. (\ref{eq:constraint-lambda}) since
we have two equations for $\lambda_{0,1}$ and two for $\lambda_{0,1}^{B}$,
\begin{eqnarray}
\left(\begin{array}{cc}
QJ_{21} & QJ_{31}\\
J_{31} & J_{41}
\end{array}\right)\left(\begin{array}{c}
\lambda_{0}\\
\lambda_{1}
\end{array}\right) & = & \left(\begin{array}{c}
-\xi\\
-DT
\end{array}\right),\label{eq:lambda01-1p}
\end{eqnarray}
whose solutions to $\lambda_{0,1}$ are 
\begin{eqnarray}
\left(\begin{array}{c}
\lambda_{0}\\
\lambda_{1}
\end{array}\right) & = & \frac{1}{Q(J_{21}J_{41}-J_{31}^{2})}\left(\begin{array}{c}
-\xi J_{41}+DTQJ_{31}\\
\xi J_{31}-DTQJ_{21}
\end{array}\right).\label{eq:sol-lambda01-1p}
\end{eqnarray}
The equations and solutions for $\lambda_{0,1}^{B}$ are in the same
form as Eqs. (\ref{eq:lambda01-1p},\ref{eq:sol-lambda01-1p}) with
replacements $\lambda_{0,1}\rightarrow\lambda_{0,1}^{B}$, $\xi\rightarrow\xi_{B}$
and $D\rightarrow D_{B}$. 

In massless limit, the integrals of $J_{nk}$ are well defined as
the Poly logarithm functions $\textrm{Li}_{n}(z)$, 
\begin{equation}
-\textrm{Li}_{s}(-z)=\frac{1}{\Gamma(s)}\int_{0}^{\infty}\frac{t^{s-1}}{e^{t}/z+1}dt.\quad\textrm{execpt for }z\leq-1
\end{equation}
When $\mu$ is small, the Poly logarithm functions can be expanded
in $\frac{\mu}{T}$, so the leading order contribution of the solution
(\ref{eq:sol-lambda01-1p}) is then 
\begin{alignat}{2}
 & \lambda_{0}\approx-CG_{1}\frac{\mu^{2}}{T^{4}}\;, & \qquad & \lambda_{1}\approx CG_{2}\frac{\mu^{2}}{T^{5}}\;,\nonumber \\
 & \lambda_{0}^{B}\approx\frac{\lambda}{\mu}\;, &  & \lambda_{1}^{B}\approx\frac{\lambda_{1}}{\mu}\;,
\end{alignat}
where $G_{1}$ and $G_{2}$ are two constants, $G_{1}\equiv607500\pi^{2}\zeta(5)/(d_{g}G_{0})$
and $G_{2}\equiv1260\pi^{6}/(d_{g}G_{0})$ with $G_{0}\equiv455625\zeta(3)\zeta(5)-49\pi^{8}$.
We notice that the $D$ terms in Eq. (\ref{eq:sol-lambda01-1p}) are
negligible, so the solutions are proportional to $\xi$.

\section{One charge with particle/anti-particle \label{sec:4}}

We now consider one charge case but add anti-particles to the system.
We will calculate $\lambda(p)$ and $\lambda_{B}(p)$. Since there
are particles and anti-particles, we recover the index $Q=\pm1$ in
particle distribution function, $f\rightarrow f^{Q}$ and $f_{0,1}\rightarrow f_{0,1}^{Q}$.
In Eq. (\ref{eq:j1s1t1}), summations over $Q$ should be added. In
Eq. (\ref{eq:db}) we also have to add the index $Q$ to $J_{n1}^{\lambda}$
and $J_{n1}^{\lambda_{B}}$: $J_{n1}^{\lambda}\rightarrow J_{n1}^{\lambda,Q}$
and $J_{n1}^{\lambda_{B}}\rightarrow J_{n1}^{\lambda_{B},Q}$ ($n=2,3$),
and add summations over $Q$ into the formula of $\xi$, $\xi_{B}$,
$D$, $D_{B}$. Inserting $f_{0}^{Q}$ into Eq. (\ref{eq:j1s1t1}),
we obtain the charge and entropy currents and the stress tensor in
equilibrium, $j_{0}^{\mu}=nu^{\mu}$, $S_{0}^{\mu}=su^{\mu}$ and
$T_{0}^{\mu\nu}=(\epsilon+P)u^{\mu}u^{\nu}-Pg^{\mu\nu}$, with the
particle number density $n\equiv\sum_{Q}Qn_{Q}$, the energy density
$\epsilon=\sum_{Q}\epsilon_{Q}$, the pressure $P=\sum_{Q}P_{Q}$,
and the entropy density $s=\sum_{Q}s_{Q}=(\epsilon+P-n\mu)/T$. The
solutions to $\xi,\xi_{B},D,D_{B}$ are the same as in Eq. (\ref{eq:entropy-d}).
Then Eq. (\ref{eq:constraint-lambda}) is modified to 
\begin{eqnarray}
 &  & J_{21}^{\lambda,+}-J_{21}^{\lambda,-}=-\xi,\; J_{31}^{\lambda,+}+J_{31}^{\lambda,+}=-DT,\nonumber \\
 &  & J_{21}^{\lambda_{B},+}-J_{21}^{\lambda_{B},-}=-\xi_{B},\; J_{31}^{\lambda_{B},+}+J_{31}^{\lambda_{B},-}=-D_{B}T.
\end{eqnarray}
Eq. (\ref{eq:j-expand}) now becomes 
\begin{equation}
J_{n1}^{\lambda,Q}=\sum_{i=0}\lambda_{i}J_{i+n,1}^{Q},\; J_{n1}^{\lambda_{B},Q}=\sum_{i=0}\lambda_{i}^{B}J_{i+n,1}^{Q},
\end{equation}
In the case of minimal number of coefficients we can completely fix
the coefficients $\lambda_{0,1}$ by solving following system of equations,
\begin{eqnarray}
\left(\begin{array}{cc}
\delta J_{21} & \delta J_{31}\\
\sigma J_{31} & \sigma J_{41}
\end{array}\right)\left(\begin{array}{c}
\lambda_{0}\\
\lambda_{1}
\end{array}\right) & = & \left(\begin{array}{c}
-\xi\\
-DT
\end{array}\right),\label{eq:lambda01}
\end{eqnarray}
where we have used the shorthand notation, $\delta J_{n1}\equiv J_{n1}^{+}-J_{n1}^{-}$
and $\sigma J_{n1}\equiv J_{n1}^{+}+J_{n1}^{-}$. The solutions to
$\lambda_{0,1}$ are 
\begin{eqnarray}
\left(\begin{array}{c}
\lambda_{0}\\
\lambda_{1}
\end{array}\right) & = & \frac{1}{\Delta}\left(\begin{array}{c}
-\xi(\sigma J_{41})+DT(\delta J_{31})\\
\xi(\sigma J_{31})-DT(\delta J_{21})
\end{array}\right),\label{eq:sol-lambda01}
\end{eqnarray}
where $\Delta=(\delta J_{21})(\sigma J_{41})-(\sigma J_{31})(\delta J_{31})$.
The equations and solutions for $\lambda_{0,1}^{B}$ are in the same
form as Eqs. (\ref{eq:lambda01},\ref{eq:sol-lambda01}) with replacements
$\lambda_{0,1}\rightarrow\lambda_{0,1}^{B}$, $\xi\rightarrow\xi_{B}$
and $D\rightarrow D_{B}$. 

Using integration by parts in Appendix B, $\delta J_{nk}$ and $\sigma J_{nk}$
for $n=2,3,4$ and $k=1$ are given by 
\begin{alignat}{2}
 & \delta J_{21}=\frac{d_{g}T^{4}}{6}\left(\bar{\mu}+\frac{\bar{\mu}^{3}}{\pi^{2}}\right)\;, & \quad & \sigma J_{21}=\frac{d_{g}T^{4}}{\pi^{2}}\left[-\textrm{Li}_{3}(-e^{\bar{\mu}})-\textrm{Li}_{3}(-e^{-\bar{\mu}})\right]\;\nonumber \\
 & \delta J_{31}=\frac{4d_{g}T^{5}}{\pi^{2}}\left[-\textrm{Li}_{4}(-e^{\bar{\mu}})-\textrm{Li}_{4}(-e^{-\bar{\mu}})\right]\;, & \quad & \sigma J_{31}=d_{g}T^{5}\left[\frac{7\pi^{2}}{90}+\frac{\bar{\mu}^{2}}{3}+\frac{\bar{\mu}^{4}}{6\pi^{2}}\right]\;\nonumber \\
 & \delta J_{41}=d_{g}T^{6}\left[\frac{7\pi^{2}}{18}\bar{\mu}+\frac{5}{9}\bar{\mu}^{3}+\frac{\bar{\mu}^{5}}{6\pi^{2}}\right]\;, & \quad & \sigma J_{41}=\frac{20d_{g}T^{6}}{\pi^{2}}\left[-\textrm{Li}_{5}(-e^{\bar{\mu}})-\textrm{Li}_{5}(-e^{-\bar{\mu}})\right]\;.
\end{alignat}
For massless fermions and small $\overline{\mu}$, we have, 
\begin{eqnarray}
\sigma J_{21} & \approx & 9\zeta(3)G,\;\sigma J_{31}\approx\frac{7\pi^{4}}{15}GT,\;\sigma J_{41}\approx225\zeta(5)GT^{2},\nonumber \\
\delta J_{21} & \approx & \pi^{2}\overline{\mu}G,\;\delta J_{31}\approx36\zeta(3)\overline{\mu}GT,\;\delta J_{41}\approx\frac{7\pi^{4}}{3}\overline{\mu}GT^{2},\nonumber \\
\Delta & \approx & \frac{\pi^{2}}{5}G^{2}G_{0}T^{2}\overline{\mu}.
\end{eqnarray}
where $G\equiv\frac{d_{g}T^{4}}{6\pi^{2}}$ and $G_{0}\equiv-84\pi^{2}\zeta(3)+1125\zeta(5)$.
The solutions have very simple form,

\begin{alignat}{2}
 & \lambda_{0}\approx CG_{1}\frac{\mu}{T^{3}}\;, & \qquad & \lambda_{1}\approx-CG_{2}\frac{\mu}{T^{4}}\;,\nonumber \\
 & \lambda_{0}^{B}\approx\frac{\lambda}{\mu}\;, &  & \lambda_{1}^{B}\approx\frac{\lambda_{1}}{\mu}\;,\label{eq:coeff-one-c}
\end{alignat}
where we have used two constants, $G_{1}\equiv\frac{6750\pi^{2}\zeta(5)}{d_{g}G_{0}}$,
$G_{2}\equiv\frac{14}{d_{g}G_{0}}$. We notice that the $D$ terms
in Eq. (\ref{eq:sol-lambda01}) are negligible, so the solutions are
proportional to $\xi$. Note that the quantity $\frac{n\mu}{\epsilon+P}\sim\overline{\mu}^{2}$
is also small and we have dropped it, since $n\mu\approx\frac{d_{g}}{6\pi^{2}}T^{4}\overline{\mu}^{2}$
and $\epsilon+P=\frac{4}{3}\epsilon\approx d_{g}\frac{7\pi^{2}}{90}T^{4}$.

\section{With two charges and particle/antiparticle \label{sec:5}}

As an example for the case of two charges, we consider adding to the
system the chirality or an axial $U(1)$ charge to particles. Then
there are two currents, one for each chirality, or equivalently, for
the $U(1)/U_{A}(1)$ charge. For simplicity we assume that there is
an anomaly for the axial charge current but no anomaly for the charge
one. There are distribution functions for right-hand and left-hand
particles, $f_{a}^{Q}(x,p)$ ($a=R,L$), with chemical potentials
$\mu_{R.L}=\mu\pm\mu_{A}$. As an extension to Eq. (\ref{eq:deviation}),
the corrections $\chi_{a}(x,p)$ in $f_{a}^{Q}(x,p)$ are now $\chi_{a}(x,p)=\lambda_{a}p^{\mu}\omega_{\mu}+\lambda_{aB}p^{\mu}B_{\mu}$.
Instead of right-hand and left-hand quantities $X_{a}$, we can equivalently
use $X=X_{R}+X_{L}$ and $X_{A}=X_{R}-X_{L}$, where $X=\lambda,\lambda_{B},\xi,\xi_{B},n,s,\epsilon,P,j_{\mu}$.
The distribution functions $f_{a}^{Q}(x,p)$ satisfy two separate
Boltzmann equations of the following form, 
\begin{equation}
p^{\mu}\left(\frac{\partial}{\partial x^{\mu}}-QF_{\mu\nu}\frac{\partial}{\partial p_{\nu}}\right)f_{a}^{Q}(x,p)=\mathcal{C}_{aQ}[f_{b}^{Q'}].\label{eq:boltzmann-two-charge}
\end{equation}
The $U(1)/U_{A}(1)$ charge and entropy currents and the stress tensor
in equilibrium are give by: $j_{0}^{\mu}=nu^{\mu}$, $j_{A0}^{\mu}=n_{A}u^{\mu}$,
$S_{0}^{\mu}=su^{\mu}$ and $T_{0}^{\mu\nu}=(\epsilon+P)u^{\mu}u^{\nu}-Pg^{\mu\nu}$,
with the particle number density $n\equiv\sum_{aQ}Qn_{aQ}$ ($a=R,L$;
$Q=\pm1$), the energy density $\epsilon=\sum_{aQ}\epsilon_{aQ}$,
the pressure $P=\sum_{aQ}P_{aQ}$, and the entropy density $s=\sum_{aQ}s_{aQ}=(\epsilon+P-\sum_{a}n_{a}\mu_{a})/T$. 

Similar to Eqs. (\ref{eq:djx},\ref{eq:constraint1}), the $U(1)/U_{A}(1)$
charge conservation equations (\ref{eq:charge-current}) can be derived
as 
\begin{eqnarray}
\partial_{\mu}j^{\mu}(x) & = & \int[dp](\mathcal{C}_{R,+}-\mathcal{C}_{R,+}+\mathcal{C}_{L,+}-\mathcal{C}_{L,+})=0,\label{eq:djx-1}\\
\partial_{\mu}j_{A}^{\mu}(x) & = & \int[dp](\mathcal{C}_{R,+}-\mathcal{C}_{R,-}-\mathcal{C}_{L,+}+\mathcal{C}_{L,+})=-CE^{\mu}B_{\mu},\label{eq:djx-2}
\end{eqnarray}
where we have used Eq. (\ref{eq:property-1}). The energy and momentum
conservation equation reads, 
\begin{eqnarray}
\partial_{\mu}T^{\mu\nu}-F^{\nu\mu}j_{\mu} & = & \int[dp]p^{\nu}(\mathcal{C}_{R,+}+\mathcal{C}_{R,-}+\mathcal{C}_{L,+}+\mathcal{C}_{L,-})=0,\label{eq:em-two}
\end{eqnarray}
where we have used Eq. (\ref{eq:property-2}). 

Similar to Eq. (\ref{eq:db}), we can express $\xi_{a}$, $\xi_{aB}$,
$D$ and $D_{B}$ in terms of $\lambda_{a}$ and $\lambda_{aB}$ as, 

\begin{eqnarray}
 &  & \xi_{a}=-(J_{21}^{\lambda_{a},+}-J_{21}^{\lambda_{a},-}),\;\;-DT=J_{31}^{\lambda_{R},+}+J_{31}^{\lambda_{R},-}+J_{31}^{\lambda_{L},+}+J_{31}^{\lambda_{L},-},\nonumber \\
 &  & \xi_{aB}=-(J_{21}^{\lambda_{aB},+}-J_{21}^{\lambda_{aB},-}),\;\;-D_{B}T=J_{31}^{\lambda_{RB},+}+J_{31}^{\lambda_{RB},-}+J_{31}^{\lambda_{LB},+}+J_{31}^{\lambda_{LB},-},\label{eq:xi-d}
\end{eqnarray}
for $a=R,L$. We have the following first order corrections, 
\begin{eqnarray}
j_{1}^{\mu} & = & j_{R1}^{\mu}+j_{L1}^{\mu}=\sum_{a=R,L}(\xi_{a}\omega^{\mu}+\xi_{aB}B^{\mu})=\xi\omega^{\mu}+\xi_{B}B^{\mu},\nonumber \\
j_{A1}^{\mu} & = & j_{R1}^{\mu}-j_{L1}^{\mu}=(\xi_{R}-\xi_{L})\omega^{\mu}+(\xi_{RB}-\xi_{LB})B^{\mu}=\xi_{A}\omega^{\mu}+\xi_{AB}B^{\mu},\nonumber \\
T_{1}^{\mu\nu} & = & DT\left(u^{\mu}\omega^{\nu}+u^{\nu}\omega^{\mu}\right)+D_{B}T\left(u^{\mu}B^{\nu}+u^{\nu}B^{\mu}\right),\nonumber \\
S_{1}^{\mu} & = & -\sum_{a=R,L}\frac{\mu_{a}}{T}(\xi_{a}\omega^{\mu}+\xi_{aB}B^{\mu})+(D\omega^{\mu}+D_{B}B^{\mu})\nonumber \\
 & = & -\sum_{i=\mathrm{null},A}\frac{\mu_{i}}{T}(\xi_{i}\omega^{\mu}+\xi_{iB}B^{\mu})+(D\omega^{\mu}+D_{B}B^{\mu}).\label{eq:j-t-s2}
\end{eqnarray}
It can be verified that the entropy current in Eq. (\ref{eq:j-t-s2})
cannot satisfy $\partial_{\mu}S^{\mu}\geq0$ unless $C=0$. In order
to ensure the positivity of $\partial_{\mu}S^{\mu}$ in presence of
an anomaly, one would have to substract the vector $Q^{\mu}=D\omega^{\mu}+D_{B}B^{\mu}$
from $S^{\mu}$. With $U(1)$ and $U_{A}(1)$ charges, we have 
\begin{eqnarray}
\widetilde{S}^{\mu} & = & S^{\mu}-Q^{\mu}=su^{\mu}-\sum_{a=R,L}\overline{\mu}_{a}(\xi_{a}\omega^{\mu}+\xi_{aB}B^{\mu})\nonumber \\
 & = & \frac{1}{T}(Pu^{\mu}-\sum_{a=R,L}\mu_{a}j_{a}^{\mu}+u_{\lambda}T^{\lambda\mu})-Q^{\mu}.\label{eq:entropy-current}
\end{eqnarray}
In the same way as in Sec. \ref{sec:2}, the divergence of the entropy
current can be evaluated as, 
\begin{eqnarray}
\partial_{\mu}\widetilde{S}^{\mu} & = & \omega^{\mu}\left[\sum_{a=R,L}\partial_{\mu}\overline{\mu}_{a}\left(\frac{n_{i}TD}{\epsilon+P}-\xi_{a}\right)-\partial_{\mu}D+\frac{2D}{\epsilon+P}\partial_{\mu}P\right]\nonumber \\
 &  & +B^{\mu}\left[\sum_{a=R,L}\partial_{\mu}\overline{\mu}_{a}\left(\frac{n_{i}TD_{B}}{\epsilon+P}-\xi_{aB}\right)-\partial_{\mu}D_{B}+\frac{D_{B}}{\epsilon+P}\partial_{\mu}P\right]\nonumber \\
 &  & +E\cdot\omega\left[\sum_{a=R,L}\left(\frac{n_{a}D}{\epsilon+P}-\frac{\xi_{a}}{T}\right)+\frac{2nD}{\epsilon+P}-2D_{B}\right]\nonumber \\
 &  & +E\cdot B\left[\sum_{a=R,L}\left(\frac{n_{a}D_{B}}{\epsilon+P}-\frac{\xi_{aB}}{T}\right)+C\frac{\mu_{A}}{T}+\frac{nD_{B}}{\epsilon+P}\right],\label{eq:div-s}
\end{eqnarray}
following Eqs. (\ref{eq:djx-1}-\eqref{eq:em-two}). By imposing all
quantities inside the square brackets to vanish we can solve $\xi_{a}$,
$\xi_{aB}$, $D$, $D_{B}$ as follows,
\begin{alignat}{2}
D & =-C\mu_{A}T\;, & \quad & D_{B}=-C\mu_{A}\overline{\mu}\;,\nonumber \\
\xi & =-2C\mu\mu_{A}\left(1-\frac{n\mu}{\epsilon+P}\right)\;, & \quad & \xi_{A}=-C\mu^{2}\left(1-\frac{2n_{A}\mu_{A}}{\epsilon+P}\right)\;,\nonumber \\
\xi_{B} & =-C\mu_{A}\left(1-\frac{n\mu}{\epsilon+P}\right)\;, & \quad & \xi_{AB}=-C\mu\left(1-\frac{n_{A}\mu_{A}}{\epsilon+P}\right)\;,\label{eq:coefficient-anomaly-2}
\end{alignat}
In small $\mu_{A}$ limit, Eq.(\ref{eq:coefficient-anomaly-2}) will
be 
\begin{alignat}{2}
D=-C\mu_{A}T\;, & \quad\xi=-2C\mu\mu_{A}\;, & \quad\xi_{B}=-C\mu_{A}\;,\nonumber \\
D_{B}=-C\mu_{A}\overline{\mu}, & \quad\xi_{A}=-C\mu^{2}\;, & \quad\xi_{AB}=-C\mu\;,\label{eq:solution1}
\end{alignat}
which is identical to the result of Ref. \cite{Fukushima2008,Kharzeev2010}.
Here we have assumed that all integral constants are vanishing. Note
that Eq. (\ref{eq:solution1}) is the result of the entropy principle
in the hydrodynamic approach which was also obtained in Ref. \cite{Sadofyev:2010pr,Amado:2011zx,Kalaydzhyan:2011vx}. 

Equivalently we can use $\xi_{R}=(\xi+\xi_{A})/2$ and $\xi_{L}=(\xi-\xi_{A})/2$
to determine $\lambda_{a}$, $\lambda_{aB}$ $(a=R,L)$ via solving
a system of equations (\ref{eq:xi-d}), where the first/second line
(each has three equations) is for $\lambda_{R,L}/\lambda_{RB,LB}$.
For minimal number of coefficients we can determine the values of
these coefficients completely, $\lambda_{R,L}$ and $\lambda_{RB,LB}$
can be expanded to the zeroth or first power of $u\cdot p$. For example,
if we expand $\lambda_{R}$ to the zeroth power, then we have to expand
$\lambda_{L}$ to the first power, and vice versa. Suppose we take
the former case, $\lambda_{R}=\lambda_{R0}$ and $\lambda_{L}=\lambda_{L0}+\lambda_{L1}(u\cdot p)$,
we can solve $\lambda_{R,L}$ as 
\begin{eqnarray}
\lambda_{R0} & = & -\xi_{R}\frac{1}{\delta J_{21}^{R}},\nonumber \\
\lambda_{L0} & = & -\frac{1}{\Delta}\xi_{L}(\sigma J_{41}^{L})+\frac{1}{\Delta}\left(DT-\xi_{R}\frac{\sigma J_{31}^{R}}{\delta J_{21}^{R}}\right)(\delta J_{31}^{L}),\nonumber \\
\lambda_{L1} & = & \frac{1}{\Delta}\xi_{L}(\sigma J_{31}^{L})-\frac{1}{\Delta}\left(DT-\xi_{R}\frac{\sigma J_{31}^{R}}{\delta J_{21}^{R}}\right)(\delta J_{21}^{L}),
\end{eqnarray}
where $\Delta=(\delta J_{21}^{L})(\sigma J_{41}^{L})-(\delta J_{31}^{L})(\sigma J_{31}^{L})$.
The solutions to $\lambda_{RB,LB}$ take the same form as above with
replacements $\lambda_{a}\rightarrow\lambda_{aB}$, $\xi_{a}\rightarrow\xi_{aB}$
and $D\rightarrow D_{B}$ if we assume the same expansion as $\lambda_{R,L}$:
$\lambda_{RB}=\lambda_{RB,0}$ and $\lambda_{LB}=\lambda_{LB,0}+\lambda_{LB,1}(u\cdot p)$. 

For massless fermions and small $\overline{\mu}_{R,L}$ (or equivalently
small $\overline{\mu}$ and $\overline{\mu}_{A}$), we obtain
\begin{alignat}{3}
 & \sigma J_{21}^{L,R}\simeq9\zeta(3)G\;, & \quad & \sigma J_{31}^{L,R}\simeq\frac{7\pi^{4}}{15}GT\;, & \quad & \sigma J_{41}^{L,R}\simeq225\zeta(5)GT^{2}\;,\nonumber \\
 & \delta J_{21}^{L,R}\simeq\pi^{2}\bar{\mu}_{L,R}G\;, &  & \delta J_{31}^{L,R}\simeq36\zeta(3)\bar{\mu}_{L,R}GT\;, &  & \delta J_{41}^{L,R}\simeq\frac{7\pi^{4}}{3}\bar{\mu}_{L,R}GT^{2}\;,\nonumber \\
 & \Delta=\frac{\pi^{2}}{5}G^{2}G_{0}T^{2}\bar{\mu}_{L}\;,
\end{alignat}
 where $G$ and $G_{0}$ are the same as in the former section. Then
the solutions to $\lambda_{R,L}$ are 
\begin{alignat}{2}
 & \lambda_{R0}\simeq-\frac{3C}{d_{g}}\frac{\mu}{T^{3}}\frac{\mu+2\mu_{A}}{\mu+\mu_{A}}\;, & \quad & \lambda_{L0}\simeq\frac{3C}{d_{g}}\frac{\mu}{T^{3}}\left(G_{4}\frac{\mu+2\mu_{A}}{\mu+\mu_{A}}+G_{5}\frac{\mu-2\mu_{A}}{\mu-\mu_{A}}\right)\;,\nonumber \\
 & \lambda_{L1}\simeq\frac{14C}{d_{g}G_{0}}\frac{\mu}{T^{4}}\frac{\mu\mu_{A}}{\mu^{2}-\mu_{A}^{2}}\;, &  & \lambda_{RB,0}\simeq-\frac{3C}{d_{g}}\frac{1}{T^{3}}\;,\nonumber \\
 & \lambda_{LB,0}\simeq\frac{3C}{d_{g}}\frac{1}{T^{3}}\;, &  & \lambda_{RB,1}\simeq\frac{30\pi^{2}C}{d_{g}G_{0}}\frac{\mu\mu_{A}}{T^{6}}\;,
\end{alignat}
where $G_{4}\equiv-\frac{1}{G_{0}}84\pi^{2}\zeta(3)$ and $G_{5}\equiv\frac{1}{G_{0}}1125\zeta(5)$.
The coefficient ratios of $\lambda_{ai}/\lambda_{aB,i}$ ($i=0,1$)
are proportional to $\mu$ times dimensionless factors, 
\begin{eqnarray}
\frac{\lambda_{R0}}{\lambda_{RB,0}} & \approx & \mu\frac{\mu+2\mu_{A}}{\mu+\mu_{A}},\;\frac{\lambda_{L0}}{\lambda_{LB,0}}\approx\mu\left(G_{4}\frac{\mu+2\mu_{A}}{\mu+\mu_{A}}+G_{5}\frac{\mu-2\mu_{A}}{\mu-\mu_{A}}\right),\nonumber \\
\frac{\lambda_{L1}}{\lambda_{LB,1}} & \approx & \mu\frac{7}{15}\pi^{2}\frac{T^{2}}{\mu^{2}-\mu_{A}^{2}}.
\end{eqnarray}

Note that $\lambda_{LB,1}\ll T\lambda_{LB,0}\sim T\lambda_{RB,0}$,
so both $\lambda_{BR}(p)$ and $\lambda_{BL}(p)$ can be constants
at small $\overline{\mu}$ and $\overline{\mu}_{A}$ limit. This property
is quite different from the one-charge case in which $\lambda_{B}(p)$
must have momentum dependence in order to comply with the entropy
principle.

\section{Collision and anomalous source terms \label{sec:3}}

In this section we will show that a general form of $\lambda(p)$
and $\lambda_{B}(p)$ are compatible to the charge and energy-momentum
conservation equations (\ref{eq:charge-current}) and (\ref{eq:em-cons1}).
We will also derive equations for the collision and anomalous source
terms. For simplicity we consider the single charge case without anti-particles. 

The charge conservation equation (\ref{eq:charge-current}) can be
derived from the Boltzmann equation (\ref{eq:boltzmann-F}) as 
\begin{eqnarray}
\partial_{\mu}j^{\mu}(x) & = & \int\frac{d^{3}p}{(2\pi)^{3}E_{p}}p^{\mu}\partial_{\mu}f(x,p)=\int\frac{d^{3}p}{(2\pi)^{3}E_{p}}p^{\mu}F_{\mu\nu}\frac{\partial f}{\partial p_{\nu}}+\int\frac{d^{3}p}{(2\pi)^{3}E_{p}}\mathcal{C}[f]\nonumber \\
 & = & \int\frac{d^{3}p}{(2\pi)^{3}E_{p}}\mathcal{C}[f],\label{eq:djx}
\end{eqnarray}
where have used the identity
\begin{align}
 & \int\frac{d^{3}p}{2E_{p}}p^{\mu}F_{\mu\nu}\frac{\partial}{\partial p_{\nu}}f\nonumber \\
= & \int d^{4}p\theta(p^{0})\delta(p^{2}-m^{2})p^{\mu}F_{\mu\nu}\frac{\partial}{\partial p_{\nu}}f\nonumber \\
= & -\int d^{4}p\theta(p^{0})\delta(p^{2}-m^{2})p^{\mu}F_{\mu\nu}\frac{\partial}{\partial p_{\nu}}f\nonumber \\
= & -\int d^{4}pf[\theta(p^{0})\delta(p^{2}-m^{2})F^{\mu\nu}\delta_{\mu\nu}\nonumber \\
 & +\theta(p^{0})F^{\mu\nu}p_{\mu}2g_{\alpha\beta}p^{\alpha}\delta_{\beta}^{\nu}\delta^{\prime}(p)\nonumber \\
 & +\delta(p^{0})F^{0\mu}p_{\mu}\delta(p^{2}-m^{2})]\nonumber \\
= & 0\;.\label{eq:property-1}
\end{align}
 Then the momentum integral of the collision term must obey 
\begin{equation}
\int\frac{d^{3}p}{(2\pi)^{3}E_{p}}\mathcal{C}[f]=-CE\cdot B,\label{eq:constraint1}
\end{equation}
so that Eq. (\ref{eq:charge-current}) can hold. The energy and momentum
conservation equation (\ref{eq:em-cons1}) can be derived from the
Boltzmann equation (\ref{eq:boltzmann-F}) as 
\begin{eqnarray}
\partial_{\mu}T^{\mu\nu} & = & \int\frac{d^{3}p}{(2\pi)^{3}E_{p}}p^{\mu}p^{\nu}\partial_{\mu}f=\int\frac{d^{3}p}{(2\pi)^{3}E_{p}}p^{\nu}p^{\alpha}F_{\alpha\beta}\frac{\partial f}{\partial p_{\beta}}+\int\frac{d^{3}p}{(2\pi)^{3}E_{p}}p^{\nu}\mathcal{C}[f]\nonumber \\
 & = & F^{\nu\mu}j_{\mu}+\int\frac{d^{3}p}{(2\pi)^{3}E_{p}}p^{\nu}\mathcal{C}[f],
\end{eqnarray}
where we have used

\begin{align}
 & \int\frac{d^{3}p}{(2\pi)^{3}E_{p}}F^{\mu\alpha}p_{\mu}p_{\nu}\frac{\partial f}{\partial p^{\alpha}}\nonumber \\
= & \int d^{4}p\theta(p^{0})\delta(p^{2}-m^{2})F^{\mu\alpha}p_{\mu}p_{\nu}\frac{\partial f}{\partial p^{\alpha}}\nonumber \\
= & -\int d^{4}pf\frac{\partial}{\partial p^{\alpha}}[\theta(p^{0})\delta(p^{2}-m^{2})F^{\mu\alpha}p_{\mu}p_{\nu}]\nonumber \\
= & -\int d^{4}pf[\theta(p^{0})\delta(p^{2}-m^{2})F^{\mu\alpha}(\delta_{\mu\alpha}p_{\nu}+p_{\mu}\delta_{\nu\alpha})\nonumber \\
 & +\delta(p^{0})F^{0\mu}p_{\mu}p_{\nu}\delta(p^{2}-m^{2})\nonumber \\
 & +F^{\mu\alpha}p_{\mu}p_{\nu}\theta(p^{0})\delta^{\prime}(p^{2}-m^{2})2g^{\nu\rho}p_{\rho}\delta_{\nu\alpha}]\nonumber \\
= & F^{\nu\alpha}j_{\alpha}\;.\label{eq:property-2}
\end{align}
Then we require that the collision term must satisfy 
\begin{equation}
\int\frac{d^{3}p}{(2\pi)^{3}E_{p}}p^{\nu}\mathcal{C}[f]=0,\label{eq:constraint2}
\end{equation}
so that Eq. (\ref{eq:em-cons1}) can hold. 

We can expand $\mathcal{C}[f]$ to the second order as 
\begin{eqnarray}
\mathcal{C}[f] & = & \mathcal{C}_{0}[f_{0}+f_{1}+f_{2}]+\mathcal{C}_{A}[f]\approx\mathcal{C}_{1}+\mathcal{C}_{2}\label{eq:collision-expansion}
\end{eqnarray}
where we have used the property $\mathcal{C}_{0}[f_{0}]=0$, and defined
\begin{eqnarray}
\mathcal{C}_{1} & = & \left.\frac{d\mathcal{C}_{0}}{df}\right|_{f=f_{0}}f_{1}+\mathcal{C}_{A1},\nonumber \\
\mathcal{C}_{2} & = & \left.\frac{d\mathcal{C}_{0}}{df}\right|_{f=f_{0}}f_{2}+\frac{1}{2}\left.\frac{d^{2}\mathcal{C}_{0}}{df^{2}}\right|_{f=f_{0}}f_{1}^{2}+\mathcal{C}_{A2}.\label{eq:c1c2}
\end{eqnarray}
Note that the general form for the normal part of $\mathcal{C}_{1}$
is $\left.\frac{d\mathcal{C}_{0}}{df}\right|_{f=f_{0}}f_{1}=H_{\lambda}(u\cdot p)p\cdot\omega+H_{\lambda_{B}}(u\cdot p)p\cdot B$.
When inserting the distribution function (\ref{eq:fxp-expansion})
into the Boltzmann equation (\ref{eq:boltzmann-F}) and using Eq.
(\ref{eq:collision-expansion}), we obtain the Boltzmann equations
to the first and second order, 
\begin{eqnarray}
p^{\mu}\left(\frac{\partial}{\partial x^{\mu}}-F_{\mu\nu}\frac{\partial}{\partial p_{\nu}}\right)f_{0} & = & \mathcal{C}_{1},\label{eq:first-order}\\
p^{\mu}\left(\frac{\partial}{\partial x^{\mu}}-F_{\mu\nu}\frac{\partial}{\partial p_{\nu}}\right)f_{1} & = & \mathcal{C}_{2}.\label{eq:second-order}
\end{eqnarray}
From Eqs. (\ref{eq:c1c2},\ref{eq:first-order}) we can determine
the anomalous source term of the first order, 
\begin{equation}
\mathcal{C}_{A1}=-H_{\lambda}(u\cdot p)p\cdot\omega-H_{\lambda_{B}}(u\cdot p)p\cdot B-f_{0}(1+ef_{0})p^{\mu}\partial_{\mu}[(u\cdot p-\mu_{0})/T].\label{eq:ca1}
\end{equation}
By evaluating the left hand sides of Eq. (\ref{eq:second-order}),
we can fix $\mathcal{C}_{2}$ as follows 
\begin{eqnarray}
\mathcal{C}_{2} & = & f_{0}(1+ef_{0})\left\{ -[(1+2ef_{0})\lambda\partial_{\mu}\psi'(f_{0})+(\partial_{\mu}\lambda)]p^{\mu}p^{\nu}\omega_{\nu}\right.\nonumber \\
 &  & -[(1+2ef_{0})\lambda_{B}\partial_{\mu}\psi'(f_{0})+(\partial_{\mu}\lambda_{B})]p^{\mu}p^{\nu}B_{\nu}\nonumber \\
 &  & +\left[-\beta(1+2ef_{0})\lambda+\frac{d\lambda}{d(u\cdot p)}\right]p^{\mu}p^{\nu}E_{\mu}\omega_{\nu}+\lambda p^{\mu}F_{\mu\nu}\omega^{\nu}\nonumber \\
 &  & +\left[-\beta(1+2ef_{0})\lambda_{B}+\frac{d\lambda_{B}}{d(u\cdot p)}\right]p^{\mu}p^{\nu}E_{\mu}B_{\nu}+\lambda_{B}p^{\mu}F_{\mu\nu}B^{\nu}\nonumber \\
 &  & \left.-\lambda p^{\mu}p^{\nu}\partial_{\mu}\omega_{\nu}-\lambda_{B}p^{\mu}p^{\nu}\partial_{\mu}B_{\nu}\right\} \label{eq:second-kinetic-2}
\end{eqnarray}

Taking momentum integrals for Eqs. (\ref{eq:first-order},\ref{eq:second-order}),
we obtain the divergences of charge currents to the first and second
order, 
\begin{eqnarray}
\partial_{\mu}j_{0}^{\mu}(x) & = & \int[dp]\mathcal{C}_{1}=\int[dp]\mathcal{C}_{A1},\label{eq:id-1st}\\
\partial_{\mu}j_{1}^{\mu}(x) & = & \int[dp]\mathcal{C}_{2}.\label{eq:id-2nd}
\end{eqnarray}
where we have used the property, 
\begin{equation}
\int[dp]\left.\frac{d\mathcal{C}_{0}}{df}\right|_{f=f_{0}}f_{1}=\int[dp][H_{\lambda}(u\cdot p)p\cdot\omega+H_{\lambda_{B}}(u\cdot p)p\cdot B]=0.
\end{equation}
With Eq. (\ref{eq:j1s1t1}) and identities in Eq. (\ref{eq:id-oeb}),
the left hand side of Eq. (\ref{eq:id-2nd}) can be evaluated as,
\begin{eqnarray}
\partial_{\mu}j_{1}^{\mu}(x) & = & \omega\cdot E\left(2\xi_{B}-\frac{2n\xi}{\epsilon+P}\right)-B\cdot E\frac{n\xi_{B}}{\epsilon+P}\nonumber \\
 &  & +\omega^{\mu}\left(\partial_{\mu}\xi-\frac{2\xi}{\epsilon+P}\partial_{\mu}P\right)+B^{\mu}\left(\partial_{\mu}\xi_{B}-\frac{\xi_{B}}{\epsilon+P}\partial_{\mu}P\right).\label{eq:lhs-j1}
\end{eqnarray}
In order for Eq. (\ref{eq:charge-current}) to be satisfied, it is
required that $\partial_{\mu}j_{0}^{\mu}(x)$ must have the form 
\begin{eqnarray}
\partial_{\mu}j_{0}^{\mu}(x) & = & -CB\cdot E-\partial_{\mu}j_{1}^{\mu}(x)\nonumber \\
 & = & -\omega\cdot E\left(2\xi_{B}-\frac{2n\xi}{\epsilon+P}\right)-B\cdot E\left(C-\frac{n\xi_{B}}{\epsilon+P}\right)\nonumber \\
 &  & -\omega^{\mu}\left(\partial_{\mu}\xi-\frac{2\xi}{\epsilon+P}\partial_{\mu}P\right)-B^{\mu}\left(\partial_{\mu}\xi_{B}-\frac{\xi_{B}}{\epsilon+P}\partial_{\mu}P\right).\label{eq:lhs-j0}
\end{eqnarray}
We see that $\partial_{\mu}j_{0}^{\mu}(x)$ is not vanishing but of
the second order though it is superficially a first order quantity.
This is very important otherwise it would lead to $\partial_{\mu}j_{1}^{\mu}(x)=-CE\cdot B$
and give rise to incompatible results for $\xi,\xi_{B},D,D_{B}$ with
Ref. \cite{Son2009}. Therefore we should keep in mind that it is
the full charge current that satisfies charge conservation equation
(\ref{eq:charge-current}). We note that the expression of $\partial_{\mu}j_{0}^{\mu}(x)$
in Eq. (\ref{eq:lhs-j0}) also gives the momentum integral of the
anomalous source term of the first order, 
\begin{eqnarray}
\int[dp]\mathcal{C}_{A1} & = & -\omega\cdot E\left(2\xi_{B}-\frac{2n\xi}{\epsilon+P}\right)-B\cdot E\left(C-\frac{n\xi_{B}}{\epsilon+P}\right)\nonumber \\
 &  & -\omega^{\mu}\left(\partial_{\mu}\xi-\frac{2\xi}{\epsilon+P}\partial_{\mu}P\right)-B^{\mu}\left(\partial_{\mu}\xi_{B}-\frac{\xi_{B}}{\epsilon+P}\partial_{\mu}P\right).\label{eq:ca1-int}
\end{eqnarray}

For the energy and momentum conservation we obtain, 
\begin{eqnarray}
\partial_{\mu}T_{0}^{\mu\nu}-F^{\nu\mu}j_{\mu}^{0} & = & \int[dp]p^{\nu}\mathcal{C}_{1},\label{eq:en-1st}\\
\partial_{\mu}T_{1}^{\mu\nu}-F^{\nu\mu}j_{\mu}^{1} & = & \int[dp]p^{\nu}\mathcal{C}_{2},\label{eq:en-2nd}
\end{eqnarray}
With Eqs. (\ref{eq:j1s1t1},\ref{eq:id-oeb}), we can evaluate the
left hand side of Eq. (\ref{eq:en-2nd}) as, 
\begin{eqnarray}
\partial_{\mu}T_{1}^{\mu\nu}-F^{\nu\mu}j_{\mu}^{1} & = & \omega^{\nu}[DT\partial\cdot u+u\cdot\partial(DT)]+B^{\nu}[D_{B}T\partial\cdot u+u\cdot\partial(D_{B}T)]\nonumber \\
 &  & +u^{\nu}\{\omega\cdot E(2D_{B}T-\frac{2DTn}{\epsilon+P})-E\cdot B\frac{D_{B}Tn}{\epsilon+P}+\omega\cdot[\partial(DT)-\frac{2DT}{\epsilon+P}\partial P]\nonumber \\
 &  & +B\cdot[\partial(D_{B}T)-\frac{D_{B}T}{\epsilon+P}\partial P]\}\nonumber \\
 &  & +DT(u\cdot\partial\omega^{\nu}+\omega\cdot\partial u^{\nu})+D_{B}T(u\cdot\partial B^{\nu}+B\cdot\partial u^{\nu})\nonumber \\
 &  & -F^{\nu\mu}(\xi\omega_{\mu}+\xi_{B}B_{\mu}),\label{eq:en-ob1}
\end{eqnarray}
where one can verify that each term in the right hand side is of second
order. The left hand side of Eq. (\ref{eq:en-1st}) is then given
by 
\begin{equation}
\partial_{\mu}T_{0}^{\mu\nu}-F^{\nu\mu}j_{\mu}^{0}=-(\partial_{\mu}T_{1}^{\mu\nu}-F^{\nu\mu}j_{\mu}^{1}),\label{eq:en-ob0}
\end{equation}
which is also a second order quantity though $p^{\nu}\mathcal{C}_{1}=p^{\nu}p^{\mu}\left(\frac{\partial}{\partial x^{\mu}}-F_{\mu\sigma}\frac{\partial}{\partial p_{\sigma}}\right)f_{0}$
is superficially of first order. One might question the validity of
Eq. (\ref{eq:id-oeb}) which follows $\partial_{\mu}T_{0}^{\mu\nu}-F^{\nu\mu}j_{\mu}^{0}=0$,
but it is not a problem here since this equation really holds at the
first order or the leading order but not true at the second order.
It is essential that the energy momentum equation (\ref{eq:em-cons1})
hold for the full quantities $T^{\mu\nu}$ and $j_{\mu}$, and not
for $T_{0,1}^{\mu\nu}$ and $j_{\mu}^{0,1}$ separately, otherwise
the results would be contradictory to those of Ref. \cite{Son2009}
following the entropy principle of the second law of thermodynamics. 

We now obtain the momentum integral of $p^{\nu}\mathcal{C}_{1,2}$
from Eqs. (\ref{eq:en-1st},\ref{eq:en-2nd}) with $\partial_{\mu}T_{0,1}^{\mu\nu}-F^{\nu\mu}j_{\mu}^{0,1}$
given by Eqs.(\ref{eq:en-ob1},\ref{eq:en-ob0}).

\section{Discussions and conclusions \label{sec:6}}

We have shown that induced terms related to the vorticity and magnetic
field in the charge and entropy currents from a triangle anomaly can
be derived in kinetic theory by introducing correction terms to the
phase space distribution function at the first order. We demonstrated
that the anomalous source terms are necessary to ensure that the equations
for the charge and energy-momentum conservation are satisfied and
that the correction terms of distribution functions are compatible
to these equations. 

As examples for the correction terms of distribution functions, we
focus on the massless fermionic system in three cases for small $\mu/T$,
with one charge {[}$U(1)${]} and one particle species (without anti-particles),
with one charge {[}$U(1)${]} and particles/anti-particles, and with
two charges {[}$U(1)\times U_{A}(1)${]}. In the latter two cases,
the coefficients for $\omega$ and $B$ terms in distribution functions
are found to be proportional to $C\mu/T^{3}$ and $C/T^{3}$ respectively.
In the two-charges case the coefficients can be constants or independent
of momentum, such a property is impossible for the one-charge case
since it is not allowed by the entropy principle. 

In the two-charges case, we assumed that there is an anomaly for the
axial charge current but no anomaly for the charge one. The coefficients
of correction terms for the charge/axial-charge currents and energy-momentum
tensor have a very simple and symmetric form at small $\mu/T$ and
$\mu_{A}/T$ limit: $\xi\approx2C\mu\mu_{A}$, $\xi_{A}\approx C\mu^{2}$,
$\xi_{B}\approx C\mu_{A}$, $\xi_{AB}\approx C\mu$, $DT=-C\mu_{A}\mu^{2}$,
and $D_{B}T=-C\mu_{A}\mu$. This means that similar to the CME an
axial anomaly can induce a residual charge current which is proportional
to the magnetic field and the axial chemical potential \cite{Fukushima2008}. 

We have a few comments about our results. In our evaluation of the
correction terms of distribution functions, we have assumed that $\lambda(p)$
and $\lambda_{B}(p)$ are identical to particles and anti-particles.
Alternatively we can assume that they have an opposite sign for particles
to anti-particles. We can not tell which case is correct due to lack
of deeper knowledge about these anomalous term at a microscopic level.
Similarly we have assumed that $\lambda(p)$ and $\lambda_{B}(p)$
have different values for right-handed particles from left-handed
ones. One can also assume that they have the same or opposite values
for right-handed and left-handed particles. In the current framework
one can not tell which is correct. Such a situation is like what happens
in an effective theory when many effective candidates point to a unique
microscopic theory. We also note that the solutions to $\lambda(p)$
and $\lambda_{B}(p)$ given in this paper are for the cases where
the number of unknown coefficients in $\lambda(p)$ and $\lambda_{B}(p)$
is equal to that of constraining equations. It is possible that $\lambda(p)$
and $\lambda_{B}(p)$ can be expanded to higher powers of $(p\cdot u)$
and then have larger number of unknown coefficients than that of constraining
equations. In this case the constraining equations just provide constraints
for $\lambda(p)$ and $\lambda_{B}(p)$ from the second law of thermodynamics.


\appendix

\chapter{Matrix elements in Eq.(\ref{eq:char-velocity-01})\label{chap:Matrix-elements}}

For the sake of simplicity, the velocity of the fluid is parametrized
as 
\[
u^{\mu}=(\cosh\theta,\sinh\theta\cos\phi,\sinh\theta\sin\phi)\;.
\]
The matrix elements of $A_{ab}^{x}$ are 
\begin{eqnarray*}
A_{11}^{x} & = & \left(c_{s}^{2}+1\right)\sinh\theta\cosh\theta\cos\phi\;,\\
A_{12}^{x} & = & \frac{1}{2}\text{sech}^{3}\theta\left\{ 2\sinh^{2}\theta\left[(2w+\pi^{xx})\sin^{2}\phi+3w\cos^{2}\phi-\pi^{xy}\sin\phi\cos\phi\right]\right.\\
 & + & \left.w\sinh^{4}\theta(\cos(2\phi)+3)+w+\pi^{xx}\frac{}{}\right\} \;,\\
A_{13}^{x} & = & \text{sech}^{3}\theta\{\sinh\theta\cos\phi\left[(w-\pi^{xx})\sin\phi+\pi^{xy}\cos\phi\right]\\
 &  & +w\sinh^{4}\theta\sin\phi\cos\phi+\pi^{xy}\}\;,\\
A_{14}^{x} & = & \tanh\theta\cos\phi\;,\\
A_{15}^{x} & = & \tanh\theta\sin\phi\;,\\
A_{21}^{x} & = & \left(c_{s}^{2}+1\right)\sinh^{2}\theta\cos^{2}\phi+c_{s}^{2}\;,\\
A_{22}^{x} & = & 2w\sinh\theta\cos\phi\;,\\
A_{24}^{x} & = & A_{35}^{x}=1\;,\\
A_{31}^{x} & = & \left(c_{s}^{2}+1\right)\sinh^{2}\theta\sin\phi\cos\phi\;,\\
A_{32}^{x} & = & w\sinh\theta\sin\phi\;,\\
A_{33}^{x} & = & w\sinh\theta\cos\phi\;,\\
A_{42}^{x} & = & \text{sech}^{2}\theta\left\{ \frac{}{}\sinh^{4}\theta\cos^{2}\phi\left[\eta+\tau_{\pi}\pi^{xx}\cos(2\phi)-\tau_{\pi}\pi^{xx}+\tau_{\pi}\pi^{xy}\sin(2\phi)\right]\right.\\
 & + & \left.\sinh^{2}\theta\left[2(\eta-\tau_{\pi}\pi^{xx})\cos^{2}\phi+\eta\sin^{2}\phi\right]+\eta\frac{}{}\right\} \;,\\
A_{43}^{x} & = & -2\tau_{\pi}\tanh^{2}\theta\cos^{2}\phi\left[\sinh^{2}\theta\cos\phi(\pi^{xy}\cos\phi-\pi^{xx}\sin\phi)+\pi^{xy}\right]\;,\\
A_{44}^{x} & = & A_{55}^{x}=\tau_{\pi}\sinh\theta\cos\phi\;,
\end{eqnarray*}
\begin{eqnarray*}
A_{52}^{x} & = & \frac{\tanh^{2}\theta\cos\phi}{2(\sinh^{2}\theta\cos^{2}\phi+1)}\left\{ -2\sinh^{2}\theta(\pi^{xx}\sin^{3}\phi+2\pi^{xx}\sin\phi\cos^{2}\phi\right.\\
 &  & +\pi^{xy}\cos^{3}\phi)+\sinh^{4}\theta\sin^{2}(2\phi)(\pi^{xy}\cos\phi-2\pi^{xx}\sin\phi)\\
 &  & \left.-2\pi^{xx}\sin\phi-2\pi^{xy}\cos\phi\frac{}{}\right\} \;,\\
A_{53}^{x} & = & \frac{1}{2}\text{sech}^{2}\theta\left\{ 2\sinh^{4}\theta\cos^{2}\phi[\eta-\tau_{\pi}\pi^{xx}\cos(2\phi)+\tau_{\pi}\pi^{xx}\right.\\
 &  & -\tau_{\pi}\pi^{xy}\sin(2\phi)]+\sinh^{2}\theta[(\eta+\tau_{\pi}\pi^{xx})\cos(2\phi)\\
 &  & \left.+3\eta+\tau_{\pi}\pi^{xx}-\tau_{\pi}\pi^{xy}\sin(2\phi)]+2\eta\right\} \;.
\end{eqnarray*}
The matrix elements of $A_{ab}^{t}$ are given by
\begin{eqnarray*}
A_{11}^{t} & = & \frac{1}{2}\left[\left(c_{s}^{2}+1\right)\cosh(2\theta)-c_{s}^{2}+1\right]\;,\\
A_{12}^{t} & = & \frac{2\sinh\theta}{\left(\sinh^{2}\theta\cos^{2}\phi+1\right)^{2}}\left\{ \sinh^{2}\theta\cos\phi\left(2w\cos^{2}\phi+\pi^{xx}\sin^{2}\phi-\pi^{xy}\sin\phi\cos\phi\right)\right.\\
 &  & +\left.w\sinh^{4}\theta\cos^{5}\phi+(w+\pi^{xx})\cos\phi+\pi^{xy}\sin\phi\right\} \;,\\
A_{13}^{t} & = & 2\sinh\theta\left(w\sin\phi+\frac{\pi^{xy}\cos\phi-\pi^{xx}\sin\phi}{\sinh^{2}\theta\cos^{2}\phi+1}\right)\;,\\
A_{14}^{t} & = & \frac{\cos(2\phi)}{\text{csch}^{2}\theta+\cos^{2}\phi}\;,\\
A_{15}^{t} & = & \frac{\sin(2\phi)}{\text{csch}^{2}\theta+\cos^{2}\phi}\;,\\
A_{21}^{t} & = & \left(c_{s}^{2}+1\right)\sinh\theta\cosh\theta\cos\phi\;,\\
A_{31}^{t} & = & \left(c_{s}^{2}+1\right)\sinh\theta\cosh\theta\sin\phi\;,\\
A_{22}^{t} & = & \frac{\text{sech}^{3}\theta}{2}\left\{ \frac{}{}2\sinh^{2}\theta\left[(2w+\pi^{xx})\sin^{2}\phi+3w\cos^{2}\phi-\pi^{xy}\sin\phi\cos\phi\right]\right.\\
 &  & +\left.w\sinh^{4}\theta\left[\cos(2\phi)+3\right]+2w+2\pi^{xx}\frac{}{}\right\} \;,\\
A_{23}^{t} & = & \text{sech}^{3}\theta\{\sinh^{2}\theta\cos\phi\left[w\sinh^{2}\theta\sin\phi+(w-\pi^{xx})\sin\phi+\pi^{xy}\cos\phi\right]\\
 &  & +\pi^{xy}\}\;,\\
A_{24}^{t} & = & \tanh\theta\cos\phi\;,\\
A_{25}^{t} & = & \tanh\theta\sin\phi\;,\\
A_{32}^{t} & = & \frac{\text{sech}^{3}\theta}{\left(\sinh^{2}\theta\cos^{2}\phi+1\right)^{2}}\left\{ \sinh^{2}\theta[(w+3\pi^{xx})\sin\phi\cos\phi+3\pi^{xy}\sin^{2}\phi\right.\\
 &  & +2\pi^{xy}\cos^{2}\phi+]+\sinh^{4}\theta[3(w+\pi^{xx})\sin\phi\cos^{3}\phi\\
 &  & +(w+5\pi^{xx})\sin^{3}\phi\cos\phi+2\pi^{xy}\sin^{4}\phi+\pi^{xy}\cos^{4}\phi]\\
 &  & \left.+\frac{1}{16}\sinh^{6}\theta\left[10\sin(2\phi)+\sin(4\phi)\right][(w-\pi^{xx})\cos(2\phi)\right.\\
 &  & \left.+w+\pi^{xx}-\pi^{xy}\sin(2\phi)]+w\sinh^{8}\theta\sin\phi\cos^{5}\phi+\pi^{xy}\frac{}{}\right\} \;,
\end{eqnarray*}

\begin{eqnarray*}
A_{33}^{t} & = & \frac{\text{sech}^{3}\theta}{8\left(\sinh^{2}\theta\cos^{2}\phi+1\right)}\left\{ \sinh^{4}\theta\left[4(w+2\pi^{xx})\cos(2\phi)+(\pi^{xx}-w)\cos(4\phi)\right.\right.\\
 &  & \left.+21w-9\pi^{xx}+10\pi^{xy}\sin(2\phi)+\pi^{xy}\sin(4\phi)\right]\\
 &  & \left.+4\sinh^{2}\theta[6w+2\pi^{xx}\cos(2\phi)-4\pi^{xx}+3\pi^{xy}\sin(2\phi)]\right.\\
 &  & \left.-4w\sinh^{6}\theta\cos^{2}\phi\left[\cos(2\phi)-3\right]+8w-8\pi^{xx}\right\} \;,\\
A_{34}^{t} & = & -\frac{\tanh\theta\sin\phi\left(\sinh^{2}\theta\sin^{2}\phi+1\right)}{\sinh^{2}\theta\cos^{2}\phi+1}\;,\\
A_{35}^{t} & = & \frac{\tanh\theta\cos\phi}{2\sinh^{2}\theta\cos^{2}\phi+2}\left\{ \frac{}{}2-\sinh^{2}\theta[\cos(2\phi)-3]\right\} \;,\\
A_{42}^{t} & = & \tanh\theta\cos\phi\{\sinh\theta\{2\sin\phi\left[(\eta-\tau_{\pi}\pi^{xx})\sin\phi+\tau_{\pi}\pi^{xy}\cos\phi\right]\\
 &  & +\eta\cos^{2}\phi\}+\eta-2\tau_{\pi}\pi^{xx}\}\;,\\
A_{43}^{t} & = & -\tanh\theta\{\sinh^{2}\theta\cos^{2}\phi\left[(\eta-2\tau_{\pi}\pi^{xx})\sin\phi+2\tau_{\pi}\pi^{xy}\cos\phi\right]\\
 &  & +\eta\sin\phi+2\tau_{\pi}\pi^{xy}\cos\phi\}\;,\\
A_{44}^{t} & = & A_{55}^{t}=\tau_{\pi}\cosh\theta\;,\\
A_{52}^{t} & = & \frac{\tanh\theta}{4\sinh^{2}\theta\cos^{2}\phi+4}\left\{ -2\sinh^{2}\theta\{\sin\phi\left[-2\eta+\tau_{\pi}\pi^{xx}\cos(2\phi)+3\tau_{\pi}\pi^{xx}\right]\right.\\
 &  & +2\tau_{\pi}\pi^{xy}\cos^{3}\phi\}+\sinh^{4}\theta\sin^{2}(2\phi)\left[(\eta-2\tau_{\pi}\pi^{xx})\sin\phi+2\tau_{\pi}\pi^{xy}\cos\phi\right]\\
 &  & \left.+4(\eta-\tau_{\pi}\pi^{xx})\sin\phi-4\tau_{\pi}\pi^{xy}\cos\phi\right\} \;,\\
A_{53}^{t} & = & \tanh\theta\left\{ \frac{}{}\sinh^{2}\theta\left[\eta\cos^{3}\phi+\tau_{\pi}\pi^{xx}\sin\phi\sin(2\phi)-2\tau_{\pi}\pi^{xy}\sin\phi\cos^{2}\phi\right]\right.\\
 & + & \left.(\eta+\tau_{\pi}\pi^{xx})\cos\phi-\tau_{\pi}\pi^{xy}\sin\phi\frac{}{}\right\} \;.
\end{eqnarray*}
The matrix elements of $A_{ab}^{y}$ are
\begin{eqnarray*}
A_{11}^{y} & = & \left(c_{s}^{2}+1\right)\sinh\theta\cosh\theta\sin\phi\;,\\
A_{21}^{y} & = & \left(c_{s}^{2}+1\right)\sinh^{2}\theta\sin\phi\cos\phi\;,\\
A_{12}^{y} & = & \frac{\text{sech}^{3}\theta}{\left(\sinh^{2}\theta\cos^{2}\phi+1\right)^{2}}\left\{ \frac{}{}\sinh^{2}\theta[(w+3\pi^{xx})\sin\phi\cos\phi+3\pi^{xy}\sin^{2}\phi\right.\\
 &  & +2\pi^{xy}\cos^{2}\phi]+\sinh^{4}\theta[3(w+\pi^{xx})\sin\phi\cos^{3}\phi\\
 &  & +(w+5\pi^{xx})\sin^{3}\phi\cos\phi+2\pi^{xy}\sin^{4}\phi+\pi^{xy}\cos^{4}\phi]\\
 &  & +\frac{1}{16}\sinh^{6}\theta[10\sin(2\phi)+\sin(4\phi)][(w-\pi^{xx})\cos(2\phi)\\
 &  & \left.+w+\pi^{xx}-\pi^{xy}\sin(2\phi)]+w\sinh^{8}\theta\sin\phi\cos^{5}\phi+\pi^{xy}\frac{}{}\right\} \;,\\
A_{13}^{y} & = & \frac{\text{sech}^{3}\theta}{8\left(\sinh^{2}\theta\cos^{2}\phi+1\right)}\left\{ \frac{}{}\sinh^{4}\theta[4(w+2\pi^{xx})\cos(2\phi)+(\pi^{xx}-w)\cos(4\phi)\right.\\
 &  & +21w-9\pi^{xx}+10\pi^{xy}\sin(2\phi)+\pi^{xy}\sin(4\phi)]\\
 &  & +4\sinh^{2}\theta[6w+2\pi^{xx}\cos(2\phi)-4\pi^{xx}+3\pi^{xy}\sin(2\phi)]\\
 &  & \left.-4w\sinh^{6}\theta\cos^{2}\phi[\cos(2\phi)-3]+8w-8\pi^{xx}\frac{}{}\right\} \;,\\
A_{14}^{y} & = & -\frac{\tanh\theta\sin\phi\left(\sinh^{2}\theta\sin^{2}\phi+1\right)}{\sinh^{2}\theta\cos^{2}\phi+1}\;,
\end{eqnarray*}
\begin{eqnarray*}
A_{15}^{y} & = & \frac{\tanh\theta\cos\phi}{2\sinh^{2}\theta\cos^{2}\phi+2}\left\{ \frac{}{}2-\sinh^{2}\theta\left[\cos(2\phi)-3\right]\right\} \;,\\
A_{22}^{y} & = & w\sinh\theta\sin\phi\;,\\
A_{23}^{y} & = & w\sinh\theta\cos\phi\;,\\
A_{25}^{y} & = & 1\;,\\
A_{31}^{y} & = & \left(c_{s}^{2}+1\right)\sinh^{2}\theta\sin^{2}\phi+c_{s}^{2}\;,\\
A_{32}^{y} & = & \frac{2\sinh\theta\left[\sinh^{2}\theta\sin\phi\cos\phi(\pi^{xx}\sin\phi-\pi^{xy}\cos\phi)+\pi^{xx}\cos\phi+\pi^{xy}\sin\phi\right]}{\left(\sinh^{2}\theta\cos^{2}\phi+1\right)^{2}}\;,\\
A_{33}^{y} & = & 2\sinh\theta\left(w\sin\phi+\frac{\pi^{xy}\cos\phi-\pi^{xx}\sin\phi}{\sinh^{2}\theta\cos^{2}\phi+1}\right)\;,\\
A_{34}^{y} & = & -\frac{\sinh^{2}\theta\sin^{2}\phi+1}{\sinh^{2}\theta\cos^{2}\phi+1}\;,\\
A_{35}^{y} & = & \frac{\sin(2\phi)}{\text{csch}^{2}\theta+\cos^{2}\phi}\;,\\
A_{42}^{y} & = & \tanh^{2}\theta\sin\phi\cos\phi\{\sinh^{2}\theta[2\eta+\tau_{\pi}\pi^{xx}\cos(2\phi)-\tau_{\pi}\pi^{xx}+\tau_{\pi}\pi^{xy}\sin(2\phi)]\\
 &  & +2\eta-2\tau_{\pi}\pi^{xx}\}\;,\\
A_{43}^{y} & = & -\frac{\text{sech}^{2}\theta}{2}\left\{ \frac{}{}2\sinh^{4}\theta\cos^{2}\phi[\eta+\tau_{\pi}\pi^{xx}\cos(2\phi)-\tau_{\pi}\pi^{xx}+\tau_{\pi}\pi^{xy}\sin(2\phi)]\right.\\
 &  & \left.+\sinh^{2}\theta\left\{ \eta[\cos(2\phi)+3]+2\tau_{\pi}\pi^{xy}\sin(2\phi)\right\} +2\eta\frac{}{}\right\} \;,\\
A_{44}^{y} & = & A_{55}^{y}=\tau_{\pi}\sinh\theta\sin\phi\;,\\
A_{52}^{y} & = & \frac{\tanh^{2}\theta}{8(\sinh^{2}\theta\cos^{2}\phi+1)}\left\{ \frac{}{}\sinh^{2}\theta\left[(\tau_{\pi}\pi^{xx}-\eta)\cos(4\phi)+9\eta+4\tau_{\pi}\pi^{xx}\cos(2\phi)\right.\right.\\
 &  & \left.-5\tau_{\pi}\pi^{xx}-8\tau_{\pi}\pi^{xy}\sin\phi\cos^{3}\phi\right]+2\sinh^{4}\theta\sin^{2}(2\phi)[\eta+\tau_{\pi}\pi^{xx}\cos(2\phi)\\
 &  & \left.-\tau_{\pi}\pi^{xx}+\tau_{\pi}\pi^{xy}\sin(2\phi)]+4[4\eta+\tau_{\pi}\pi^{xx}\cos(2\phi)-\tau_{\pi}\pi^{xx}\right.\\
 &  & \left.-\tau_{\pi}\pi^{xy}\sin(2\phi)]+8\eta\text{csch}^{2}\theta\frac{}{}\right\} \;,\\
A_{53}^{y} & = & \tau_{\pi}\tanh^{2}\theta\sin\phi\left[\sinh^{2}\theta\sin(2\phi)(\pi^{xx}\sin\phi-\pi^{xy}\cos\phi)+\pi^{xx}\cos\phi-\pi^{xy}\sin\phi\right]\;,
\end{eqnarray*}

Here $w=\epsilon+P$. All other elements vanish.

\chapter{Moments of distribution function\label{sec:Jnk-Ink}}

\section{Basic properties}

In order to compute macroscopic quantities (energy density, pressure,
etc.) via the microscopic distributions, the following momentum moments
are of great help. More details can be found in Ref. \cite{Israel:1979wp}
with the metric $g^{\mu\nu}=\textrm{diag}\{-,+,+,+\}$ and Ref. \cite{Muronga:2006zw,Muronga:2006zx}
with metric same as this thesis. 

There are two kinds of moments, 
\begin{eqnarray}
I^{\alpha_{1}\alpha_{2}...\alpha_{n}} & = & \int_{p}f_{0}p^{\alpha_{1}}...p^{\alpha_{n}}\;,\nonumber \\
J^{\alpha_{1}\alpha_{2}...\alpha_{n}} & = & \int_{p}f_{0}(1+af_{0})p^{\alpha_{1}}...p^{\alpha_{n}}\;,\label{eq:IJ-full}
\end{eqnarray}
which are called the $n$-th and auxiliary moments, respectively.
Similar to the decomposition of the energy-momentum tensor $T^{\mu\nu}$
with respect to an arbitrary velocity $u^{\mu}$, the tensor decomposition
of above moments reads, 

\begin{eqnarray}
I^{\alpha_{1}\alpha_{2}...\alpha_{n}} & = & \sum_{q=0}^{[n/2]}a_{nq}I_{nq}U_{(q)}^{\alpha_{1}\alpha_{2}...\alpha_{n}}\;,\nonumber \\
J^{\alpha_{1}\alpha_{2}...\alpha_{n}} & = & \sum_{q=0}^{[n/2]}a_{nq}J_{nq}U_{(q)}^{\alpha_{1}\alpha_{2}...\alpha_{n}}\;,\label{eq:IJ-expan}
\end{eqnarray}
where $a_{nq}$ is given by $a_{nq}=\left(\begin{array}{c}
n\\
2q
\end{array}\right)(2q-1)!!$ and $U_{(q)}^{\alpha_{1}\alpha_{2}...\alpha_{n}}$ is the rank-$n$
projection operator 
\begin{equation}
U_{(q)}^{\alpha_{1}\alpha_{2}...\alpha_{n}}=\Delta^{(\alpha_{1}\alpha_{2}}...\Delta^{\alpha_{2q-1}\alpha_{2q}}u^{\alpha_{2q+1}}...u^{\alpha_{n})}\;.
\end{equation}
with 
\begin{eqnarray}
\left(\begin{array}{c}
n\\
k
\end{array}\right) & = & \frac{n!}{(n-k)!k!}=\frac{n(n-1)...(n-k+1)}{k!}\;.\nonumber \\
\Delta^{(\alpha_{1}...\alpha_{2q}}u^{\alpha_{2q+1}}...u^{\alpha_{n})} & = & \frac{2^{q}!q!(n-2q)!}{n!}\nonumber \\
 &  & \times\sum_{\textrm{permutations}}\Delta^{\alpha_{1}\alpha_{2}}...\Delta^{\alpha_{2q-1}\alpha_{2q}}u^{\alpha_{2q+1}}...u^{\alpha_{n}}\;.
\end{eqnarray}
These projectors possess the orthogonality property
\begin{equation}
U_{(q)}^{\alpha_{1}\alpha_{2}...\alpha_{n}}U_{(l)\alpha_{1}\alpha_{2}...\alpha_{n}}=\frac{(2q+1)!(n-2q)!}{n!}\delta_{ql}\;.\label{eq:orth-U}
\end{equation}
Using Eq.(\ref{eq:IJ-full}, \ref{eq:IJ-expan}, \ref{eq:orth-U}),
$I_{nq}$ and $J_{nq}$ are given by
\begin{eqnarray}
I_{nq} & = & (-1)^{q}\frac{1}{(2q+1)!!}\int_{p}[p^{2}-(u\cdot p)^{2}]^{q}(u\cdot p)^{n-2q}f_{0}\;,\nonumber \\
J_{nq} & = & (-1)^{q}\frac{1}{(2q+1)!!}\int_{p}[p^{2}-(u\cdot p)^{2}]^{q}(u\cdot p)^{n-2q}f_{0}\Delta_{0}\;.\label{eq:IJ-nk}
\end{eqnarray}
By using the identities 
\begin{eqnarray}
nU_{(q)}^{\alpha_{1}\alpha_{2}...\alpha_{n-1}\lambda} & = & (n-2q)U_{(q)}^{\alpha_{1}\alpha_{2}...\alpha_{n-1}}u^{\lambda}+2qU_{(q)}^{(\alpha_{1}\alpha_{2}...\alpha_{n-2}}\Delta^{\alpha_{n-1})\lambda}\;,\nonumber \\
U_{(q+1)}^{\alpha_{1}\alpha_{2}...\alpha_{n}} & = & U_{(q)}^{(\alpha_{1}\alpha_{2}...\alpha_{n-2}}\Delta^{\alpha_{n-1}\alpha_{n-2})}\;,
\end{eqnarray}
and Eq.(\ref{eq:IJ-expan}), it is easy to obtain 
\begin{eqnarray}
I_{n+2,q} & = & m^{2}I_{nq}+(2q+3)I_{n+2,q+1}\;,\nonumber \\
J_{n+2,q} & = & m^{2}J_{nq}+(2q+3)J_{n+2,q+1}\;.\label{eq:ide-I-J-1}
\end{eqnarray}

These integrals are Lorentz boost invariant, and therefore can be
evaluated in local rest frame for simplicity, In that case, these
integrals (\ref{eq:IJ-nk}) will be 
\begin{eqnarray}
I_{nq} & = & \frac{4\pi d_{g}T^{n+2}}{(2q+1)!!(2\pi)^{3}}\int_{0}^{\infty}dxx^{2(q+1)}(z^{2}+x^{2})^{n/2-q-1/2}(e^{\sqrt{x^{2}+z^{2}}-\alpha}-a)^{-1}\;,\nonumber \\
J_{nq} & = & \frac{4\pi d_{g}T^{n+2}}{(2q+1)!!(2\pi)^{3}}\int_{0}^{\infty}dxx^{2(q+1)}(z^{2}+x^{2})^{n/2-q-1/2}\frac{e^{\sqrt{x^{2}+z^{2}}-\alpha}}{(e^{\sqrt{x^{2}+z^{2}}-\alpha}-a)^{2}}\nonumber \\
 & = & \frac{1}{\beta}I_{n-1,q-1}+\frac{n-2q}{\beta}I_{n-1,q}\;,\label{eq:I_J_1}
\end{eqnarray}
where $p^{2}=m^{2}$ with $m$ the mass of the particles, $u\cdot p=E_{p}=\sqrt{m^{2}+\left|\mathbf{p}\right|^{2}}$
and two new variables are defined as 
\begin{equation}
x=\frac{\left|\overrightarrow{p}\right|}{T}\;,\qquad z=\frac{m}{T}\;.
\end{equation}
Taking variation of Eq. (\ref{eq:I_J_1}) gives
\begin{eqnarray}
dI_{nk} & = & J_{nk}d\alpha-J_{n+1,k}d\beta\;,\nonumber \\
dJ_{nk} & = & \quad\frac{1}{\beta}[J_{n-1,k-1}+(n-2k)J_{n-1,k}]d\alpha\nonumber \\
 &  & -\frac{1}{\beta}[J_{n,k-1}+(n+1-2k)J_{n,k}]d\beta\;,\label{eq:IJ-der}
\end{eqnarray}
which are used to compute the equations of motion for the fluid. 

The integrals of $I_{nk}$ and $J_{nk}$ can also be written in familiar
form
\begin{eqnarray}
I_{nk} & = & \frac{4\pi d_{g}m^{n+2}}{(2q+1)!!(2\pi)^{3}}\sum_{l=0}^{\frac{1}{2}n-k}a_{nkl}z^{-(k+l+1)}\times\begin{cases}
\mathcal{K}_{k+l+1}(\alpha,z), & n\textrm{ is even}\\
\mathcal{L}_{k+l+2}(\alpha,z), & n\textrm{ is odd}
\end{cases}\nonumber \\
J_{nk} & = & \frac{4\pi d_{g}m^{n+2}}{(2q+1)!!(2\pi)^{3}}\sum_{l=0}^{\frac{1}{2}n-k}a_{nkl}z^{-(k+l+1)}\nonumber \\
 &  & \times\begin{cases}
\mathcal{L}_{k+l+1}(\alpha,z), & n\textrm{ is even }\\
\mathcal{K}_{k+l}(\alpha,z)+\frac{2(k+l+1)}{z}\mathcal{K}_{k+l+1}(\alpha,z), & n\textrm{ is odd }
\end{cases}
\end{eqnarray}
where the coefficients $a_{nkl}$ are defined by 
\begin{equation}
a_{nkl}=\frac{(2k+2l+1)!!}{(2k+1)!!}\left(\begin{array}{c}
\frac{1}{2}n-k\\
l
\end{array}\right)\;,
\end{equation}
and
\begin{eqnarray}
\mathcal{K}_{n}(\alpha,z) & = & \frac{1}{(2n-1)!!}\frac{1}{z^{n}}\int_{0}^{\infty}\frac{dxx^{2n}(x^{2}+z^{2})^{-1/2}}{e^{\sqrt{x^{2}+z^{2}}-\alpha}-a}\;,\nonumber \\
\mathcal{L}_{n}(\alpha,z) & = & \frac{1}{(2n-1)!!}\frac{1}{z^{n+1}}\int_{0}^{\infty}\frac{dxx^{2n}}{e^{\sqrt{x^{2}+z^{2}}-\alpha}-a}\;.
\end{eqnarray}
Similar to Eq. (\ref{eq:I_J_1}), $\mathcal{K}_{n}$ is also associated
with $\mathcal{L}_{n}$, 
\begin{eqnarray}
\frac{\partial}{\partial\alpha}\mathcal{K}_{n} & = & \mathcal{L}_{n}\;,\nonumber \\
\frac{\partial}{\partial\alpha}\mathcal{L}_{n+1} & = & -z^{n}\frac{\partial}{\partial z}(z^{-n}\mathcal{K}_{n})\;.
\end{eqnarray}

\section{Comparison with fluid dynamics}

We can express energy-momentum tensor and conserved charge currents
of an ideal fluid in terms of momentum integrals
\begin{alignat}{1}
j^{\mu}=\int_{p}p^{\mu}f=I_{10}u^{\mu}\;, & \quad T^{\mu\nu}=\int_{p}p^{\mu}p^{\nu}f=I_{20}u^{\mu}u^{\nu}-I_{21}\Delta^{\mu\nu}\;.
\end{alignat}
Recalling the definitions of the macroscopic quantities of fluid dynamical
equations, one obtained 
\begin{equation}
I_{10}=n\;,\quad I_{20}=\epsilon\;,\quad I_{21}=P\;.
\end{equation}
After using Eq.(\ref{eq:I_J_1}) for $n=2,3,4$, more quantities can
be determined
\begin{alignat}{2}
 & J_{21}=\frac{I_{10}}{\beta}=\frac{n}{\beta}\;, & \quad & J_{31}=\frac{I_{20}+I_{21}}{\beta}=\frac{\epsilon+P}{\beta}\;,\nonumber \\
 & J_{41}=\frac{I_{30}+2I_{31}}{\beta}\;, &  & J_{42}=\frac{I_{31}}{\beta}\;.
\end{alignat}

From Eqs.(\ref{eq:IJ-der}) the differentials of $I_{nk}$ and $J_{nk}$
are 
\begin{eqnarray}
dn & = & J_{10}d\alpha-J_{20}d\beta\;,\nonumber \\
d\epsilon & = & J_{20}d\alpha-J_{30}d\beta\;,\label{eq:ne-ab}
\end{eqnarray}
and 
\[
dP=J_{21}d\alpha-J_{31}d\beta=\frac{n}{\beta}d\alpha-\frac{\epsilon+P}{\beta}d\beta\;,
\]
which is identical to the Gibbs relation $dP=sdT+\mu dn$. 

All thermal quantities can be chosen as functions of $\alpha$ and
$\beta$. Alternatively, one can also use $\epsilon$ and $n$ as
the thermal variables via the following identities
\begin{eqnarray}
d\alpha & = & \frac{1}{D_{20}}(-J_{20}d\epsilon+J_{30}dn)\;,\nonumber \\
d\beta & = & \frac{1}{D_{20}}\left(-J_{10}d\epsilon+J_{20}dn\right)\;,\nonumber \\
dP & = & \frac{1}{D_{20}}[(-J_{21}J_{20}+J_{31}J_{10})d\epsilon+(J_{21}J_{30}-J_{31}J_{20})dn]\;,\label{eq:thermal-1}
\end{eqnarray}
where 
\begin{equation}
D_{nk}\equiv J_{n+1,k}J_{n-1,k}-J_{nk}^{2}\;.\label{eq:Dnk-01}
\end{equation}
Moreover, one can also use $n$ and $s/n$ as thermal variables with
the replacement of $\alpha$ and $\beta$ by
\begin{eqnarray}
d\alpha & = & \frac{1}{D_{20}}\left[\left(J_{30}-\frac{J_{31}}{J_{21}}J_{20}\right)dn-J_{20}nTd\left(\frac{s}{n}\right)\right]\;,\nonumber \\
d\beta & = & \frac{1}{D_{20}}\left[\left(J_{20}-\frac{J_{31}}{J_{21}}J_{10}\right)dn-J_{10}nTd\left(\frac{s}{n}\right)\right]\;.\label{eq:a_b_n_s}
\end{eqnarray}

\section{Massless limit}

In massless limit ($p^{2}=m^{2}=0$), $u\cdot p=E_{p}=\left|\mathbf{p}\right|$,
$p^{\nu}p^{\mu}\Delta_{\mu\nu}=p^{2}-(u\cdot p)^{2}=-\left|\mathbf{p}\right|^{2}$
and Eq. (\ref{eq:ide-I-J-1}) become 
\begin{eqnarray}
I_{nq} & = & (2q+3)I_{n,q+1}\;,\nonumber \\
J_{nq} & = & (2q+3)J_{n,q+1}\;.\label{eq:massless_1}
\end{eqnarray}
For $n=2$, the equation of state for the ideal gas can be obtained
\begin{equation}
\epsilon=3P\;.
\end{equation}
Moreover, Eq. (\ref{eq:I_J_1}) becomes 
\begin{eqnarray}
I_{nk} & = & \frac{4\pi d_{g}T^{n+2}}{(2k+1)!!(2\pi)^{3}}\int_{0}^{\infty}dxx^{n+1}\frac{1}{e^{x-\phi}-a}\;,\nonumber \\
J_{nk} & = & \frac{4\pi d_{g}T^{n+2}}{(2k+1)!!(2\pi)^{3}}\int_{0}^{\infty}dxx^{n+1}\frac{e^{x-\phi}}{(e^{x-\phi}-a)^{2}}\;.\label{eq:I-J-massless-single}
\end{eqnarray}

For a Bosonic gas ($a=1$), if the chemical potential is also vanishing
$\alpha=0$, Eq. (\ref{eq:I-J-massless-single}) can be worked out
\begin{eqnarray}
I_{nk} & = & \frac{4\pi d_{g}T^{n+2}}{(2k+1)!!(2\pi)^{3}}\Gamma(n+2)\zeta(n+2)\;,\nonumber \\
J_{nk} & = & \frac{4\pi d_{g}T^{n+2}}{(2k+1)!!(2\pi)^{3}}\Gamma(n+2)\zeta(n+1)\;,\label{eq:bose_massless}
\end{eqnarray}
with the help of 
\begin{eqnarray}
\int_{0}^{\infty}dxx^{n+1}\frac{1}{e^{x}-a} & = & \Gamma(n+2)\zeta(n+2)\;,\nonumber \\
\int_{0}^{\infty}dxx^{n+1}\frac{e^{x}}{(e^{x}-a)^{2}} & = & \Gamma(n+2)\zeta(n+1)\;,
\end{eqnarray}
where $\Gamma(n)$ is the Gamma function and $\zeta(n)$ is the Rieman
Zeta function. 

For a Fermionic gas ($a=-1$), it is necessary to consider a system
with both particles and anti-particles. In this case, Eq. (\ref{eq:I-J-massless-single})
is modified to add the contributions from anti-particles 
\begin{eqnarray}
I_{nk}^{\pm} & = & \frac{4\pi d_{g}T^{n+2}}{(2k+1)!!(2\pi)^{3}}\int_{0}^{\infty}dxx^{n+1}\left[\frac{1}{e^{x-\alpha}-a}+(-1)^{n}\frac{1}{e^{x+\alpha}-a}\right]\;,\nonumber \\
J_{nk}^{\pm} & = & \frac{4\pi d_{g}T^{n+2}}{(2k+1)!!(2\pi)^{3}}\int_{0}^{\infty}dxx^{n+1}\left[\frac{e^{x-\alpha}}{(e^{x-\alpha}-a)^{2}}+(-1)^{n+1}\frac{e^{x+\alpha}}{(e^{x+\alpha}-a)^{2}}\right]\;,\label{eq:I-J-massless}
\end{eqnarray}
where the sign $(-1)^{n}$ makes the differences between the \emph{charges}
of particles and anti-particles. Through the integration by parts
\begin{align*}
 & \int_{0}^{\infty}dxx^{n+1}\frac{e^{x-\alpha}}{(e^{x-\alpha}+1)^{2}}\\
= & -\left.\frac{x^{n+1}}{e^{x-\alpha}+1}\right|_{-\alpha}^{\infty}+\int_{0}^{\infty}dx\frac{(n+1)x^{n}}{e^{x-\alpha}+1}\;,
\end{align*}
the $J_{nk}^{\pm}$ will be related to $I_{nk}^{\pm}$ via
\begin{eqnarray}
J_{nk}^{\pm} & = & \frac{(n+1)}{\beta}I_{n-1,k}^{\pm}\;,\label{eq:I_J_2}
\end{eqnarray}
After some calculations, the integrals $I_{nk}^{\pm}$ can be worked
out 
\begin{align}
I_{nk}^{\pm} & =\frac{4\pi d_{g}T^{n+2}}{(2k+1)!!(2\pi)^{3}}\left[\frac{\phi^{n+2}}{n+2}+2\sum_{j=0}^{[\frac{n+2}{2}]}\left(\begin{array}{c}
n+1\\
2j+1
\end{array}\right)\right.\nonumber \\
 & \quad\left.\times\left(1-\frac{1}{2^{2j+1}}\right)\Gamma(2j+2)\zeta(2j+2)\alpha^{n-2j}\right]\;,\label{eq:I-massless-fermi}
\end{align}
where the binomial expansions have been used
\begin{align*}
 & (x+\alpha)^{n}+(-1)^{n-1}(x-\alpha)^{n}\\
= & \sum_{k=0}^{n}\left(\begin{array}{c}
n\\
k
\end{array}\right)\left[1+(-1)^{k+1}\right]\alpha^{n-k}x^{k}\\
= & 2\sum_{j=0}^{[\frac{n+1}{2}]}\left(\begin{array}{c}
n\\
2j+1
\end{array}\right)\alpha^{n-2j-1}x^{2j+1}\;.
\end{align*}
The value of $J_{nk}^{\pm}$ is obtained through Eq. (\ref{eq:I_J_2}).

\section{Results for QGP of massless quarks and gluons}

If we neglected he mass of quarks and gluons for simplicity, in the
ultra-relativistic QGP the integrals of moments can be worked out,
\begin{alignat}{2}
\negthinspace & I_{10}=\frac{g_{Q}T^{3}}{6}\left(\alpha+\frac{\alpha^{3}}{\pi^{2}}\right)\;, & \quad & I_{20}=T^{4}\left[\left(g_{G}+\frac{7}{4}g_{Q}\right)\frac{\pi^{2}}{30}+g_{Q}\left(\frac{\alpha^{4}}{8\pi^{2}}\right)\right]\;,\nonumber \\
 & I_{30}=g_{Q}T^{5}\left(\frac{7\pi^{2}}{30}\alpha+\frac{\alpha^{3}}{3}+\frac{\alpha^{5}}{10\pi^{2}}\right)\;, &  & I_{40}=T^{6}\left[\left(g_{G}+\frac{31}{16}g_{Q}\right)\frac{4\pi^{2}}{63}\right.\nonumber \\
 &  &  & \qquad\left.+\frac{g_{Q}}{12}\left(7\pi^{2}\alpha^{2}+5\alpha^{4}+\frac{\alpha^{6}}{\pi^{2}}\right)\right]\;,\nonumber \\
 & J_{10}=\frac{1}{2}T^{3}\left[\frac{1}{3}\left(g_{G}+g_{Q}\right)+\frac{g_{Q}}{\pi^{2}}\alpha^{2}\right]\;, &  & J_{20}=\frac{g_{Q}}{2}T^{4}\left(\alpha+\frac{\alpha^{3}}{\pi^{2}}\right)\;,\nonumber \\
 & J_{30}=T^{5}\left[\left(g_{G}+\frac{7}{4}g_{Q}\right)\frac{2\pi^{2}}{15}\right. &  & J_{40}=g_{Q}T^{6}\left(\frac{7\pi^{2}}{6}\alpha+\frac{5}{3}\alpha^{3}+\frac{\alpha^{5}}{2\pi^{2}}\right)\;,\nonumber \\
 & \qquad\qquad\left.+g_{Q}\left(\alpha^{2}+\frac{\alpha^{4}}{2\pi^{2}}\right)\right]\;,
\end{alignat}
and 
\begin{align}
 & I_{n1}=\frac{1}{3}I_{n0}\;, & \; & J_{n1}=\frac{1}{3}J_{n0}\;,\nonumber \\
 & I_{n2}=\frac{1}{15}I_{n0}\;, &  & J_{n2}=\frac{1}{15}J_{n0}\;,
\end{align}
with the help of Eq. (\ref{eq:massless_1}) where $g_{G}=N_{s}(N_{c}^{2}-1)$
and $g_{Q}=N_{s}N_{c}N_{f}$ denotes the degrees of freedom for gluons
and quarks, respectively, with $N_{s}$ the number of spin states,
$N_{c}$ the number of color charges, and $N_{f}$ the number of quark
flavors. These results are given in Ref. \cite{Muronga:2006zw,Muronga:2006zx}.

\section{Equations of motion for an ideal fluid}

As shown in Eq. (\ref{eq:eom-01}), the equation of motion of fluid
dynamics are obtained from the conservation equations. However, the
time evolution of the temperature and chemical potential are not known
from these equations. The momentum moments of distribution are helpful
to express the time derivative of the temperature and chemical potential
in terms of thermodynamic quantities and fluid velocity. 

The equations of motion for an ideal fluid can be written as 
\begin{eqnarray}
0 & = & u_{\nu}\partial_{\mu}T^{\mu\nu}\nonumber \\
 & = & -J_{30}\dot{\beta}+J_{31}\beta\theta+J_{20}\dot{\alpha}\;,\label{eq:EOM_1}\\
0 & = & \Delta_{\mu\alpha}\partial_{\nu}T^{\mu\nu}\nonumber \\
 &  & (\Delta_{\alpha}^{\nu}\partial_{\nu}\beta+\beta\dot{u}_{\alpha})J_{31}-(\Delta_{\alpha}^{\nu}\partial_{\nu}\alpha)J_{21}\;,\label{eq:EOM_3}\\
0 & = & \partial_{\mu}j^{\mu}\nonumber \\
 & = & -J_{20}\dot{\beta}+J_{21}\beta\theta+J_{10}\dot{\alpha}\;.\label{eq:EOM_2}
\end{eqnarray}
Substituting Eq. (\ref{eq:EOM_2}) into Eqs. (\ref{eq:EOM_1}, \ref{eq:EOM_3})
yields 
\begin{eqnarray}
\dot{\beta} & = & \frac{1}{D_{20}}(J_{31}J_{10}-J_{21}J_{20})\beta\theta\;,\nonumber \\
\dot{\alpha} & = & \frac{1}{D_{20}J_{20}}(J_{31}J_{20}^{2}-J_{30}J_{21}J_{20})\beta\theta\;,\nonumber \\
\Delta_{\mu}^{\nu}\partial_{\nu}\alpha & = & \frac{J_{31}}{J_{21}}(\Delta_{\mu}^{\nu}\partial_{\nu}\beta+\beta\dot{u}_{\mu})\;.\label{eq:eom-general}
\end{eqnarray}
In massless limit, i.e., $J_{30}/J_{31}=J_{20}/J_{21}=3$ given by
Eq. (\ref{eq:massless_1}), Eq. (\ref{eq:eom-general}) becomes
\begin{eqnarray}
\dot{\beta} & = & \frac{1}{3}\beta\theta\;,\nonumber \\
\dot{\alpha} & = & 0\;,\nonumber \\
\partial_{\mu}\alpha & = & \frac{\epsilon+P}{n}\left(\partial_{\mu}\beta-u_{\mu}\dot{\beta}+\beta\dot{u}_{\mu}\right)\;,
\end{eqnarray}

In an external electromagnetic field $F^{\mu\nu}=\partial^{\mu}A^{\nu}-\partial^{\nu}A^{\mu}$,
there is a source term for the energy-momentum production, 
\begin{equation}
\partial_{\mu}T^{\mu\nu}=F^{\nu\lambda}j_{\lambda}\;,
\end{equation}
which can also be obtained from the quantum field theory in Chap.
\ref{chap:vorticity}. In this case, substituting the source term
into Eqs. (\ref{eq:EOM_1}, \ref{eq:EOM_3}) yields 
\begin{eqnarray}
-J_{30}\dot{\beta}+J_{31}\beta\theta+J_{20}\dot{\alpha} & = & 0\;,\nonumber \\
(\Delta_{\alpha}^{\nu}\partial_{\nu}\beta+\beta\dot{u}_{\alpha})J_{31}-(\Delta_{\alpha}^{\nu}\partial_{\nu}\alpha)J_{21} & = & nE_{\alpha}\;,
\end{eqnarray}
with $E^{\mu}\equiv F^{\mu\nu}u_{\nu}$ and $u\cdot E=0$. The equations
in present of the external field become
\begin{eqnarray}
\dot{\beta} & = & \frac{1}{D_{20}}(J_{31}J_{10}-J_{21}J_{20})\beta\theta\;,\nonumber \\
\dot{\alpha} & = & \frac{1}{D_{20}J_{20}}(J_{31}J_{20}^{2}-J_{30}J_{21}J_{20})\beta\theta\;,\nonumber \\
\Delta_{\mu}^{\nu}\partial_{\nu}\alpha & = & \frac{J_{31}}{J_{21}}(\Delta_{\mu}^{\nu}\partial_{\nu}\beta+\beta\dot{u}_{\mu})-\frac{nE_{\mu}}{J_{21}}\;.\label{eq:eom-ex-general}
\end{eqnarray}
which in the massless limit become 
\begin{eqnarray}
\dot{\beta} & = & \frac{1}{3}\beta\theta\;,\nonumber \\
\dot{\alpha} & = & 0\;,\nonumber \\
\partial_{\mu}\alpha & = & \frac{\epsilon+P}{n}\left(\partial_{\mu}\beta-u_{\mu}\dot{\beta}+\beta\dot{u}_{\mu}\right)-\frac{E_{\mu}}{T}\;.\label{eq:eom-ex-massless}
\end{eqnarray}

\bibliographystyle{bib/utcaps}
\phantomsection\addcontentsline{toc}{chapter}{\bibname}\bibliography{bib/main}

\end{document}